\documentclass[12pt]{article}
\usepackage{amsmath,amssymb,amsfonts,color,graphicx,cite,color,feynarts,soul}
\input paperdef

\graphicspath{{figs/}}

\evensidemargin \oddsidemargin
\allowdisplaybreaks

\begin{document}
\thispagestyle{empty}

\def\thefootnote{\fnsymbol{footnote}}

\begin{flushright}
KA--TP--15--2012 %\\
%arXiv:yymm.nnnn [hep-ph]
\end{flushright}

\vspace{0.5cm}

\begin{center}

{\large\sc {\bf Heavy Scalar Tau Decays in the Complex MSSM:}}

\vspace{0.4cm}

{\large\sc {\bf A Full One-Loop Analysis}}

\vspace{1cm}

{\sc
S.~Heinemeyer$^{1}$%
\footnote{email: Sven.Heinemeyer@cern.ch}%
~and C.~Schappacher$^{2}$%
\footnote{email: cs@particle.uni-karlsruhe.de}%
}

\vspace*{.7cm}

{\sl
$^1$Instituto de F\'isica de Cantabria (CSIC-UC), Santander,  Spain

\vspace*{0.1cm}

$^2$Institut f\"ur Theoretische Physik,
Karlsruhe Institute of Technology, \\
D--76128 Karlsruhe, Germany

}

\end{center}

\vspace*{0.1cm}

\begin{abstract}
\noindent
We evaluate all two-body decay modes of the heavy scalar tau in
the Minimal Supersymmetric Standard Model with complex parameters (cMSSM) 
and no generation mixing. The evaluation is based on a full one-loop 
calculation of all decay channels, also including hard and soft QED 
radiation. 
The renormalization of the relevant sectors is briefly reviewed.
The dependence of the heavy scalar tau decay on the relevant 
cMSSM parameters is analyzed numerically, including also the decay to 
Higgs bosons and another scalar lepton or to a tau and the 
lightest neutralino.
We find sizable contributions to many partial decay widths and  
branching ratios. They are mostly of \order{5-10\%} of the tree-level 
results, but can go up to $20\%$.
These contributions are potentially important for the correct
interpretation of scalar tau decays at the LHC and, if kinematically
allowed, at the ILC or CLIC.
The evaluation of the branching ratios of the heavy scalar tau
will be  implemented into the Fortran code {\tt FeynHiggs}.
\end{abstract}
%\pacs{}

\def\thefootnote{\arabic{footnote}}
\setcounter{page}{0}
\setcounter{footnote}{0}

\newpage

\newcommand{\decayhn}{\Stauzm \to \Stauem h_n}
\newcommand{\decayh}{\Stauzm \to \Stauem h_1}
\newcommand{\decayH}{\Stauzm \to \Stauem h_2}
\newcommand{\decayA}{\Stauzm \to \Stauem h_3}
\newcommand{\decayZ}{\Stauzm \to \Stauem Z}
\newcommand{\decayNe}{\Stauzm \to \tau^- \neu1}
\newcommand{\decayNi}[1]{\Stauzm \to \tau^- \neu{#1}}
\newcommand{\decayNz}{\Stauzm \to \tau^- \neu2}
\newcommand{\decayNd}{\Stauzm \to \tau^- \neu3}
\newcommand{\decayNv}{\Stauzm \to \tau^- \neu4}
\newcommand{\decayNk}{\Stauzm \to \tau^- \neu{k}}
\newcommand{\decayCj}{\Stauzm \to \nu_\tau \cham{j}}
\newcommand{\decayCme}{\Stauzm \to \nu_\tau \cham1}
\newcommand{\decayCmz}{\Stauzm \to \nu_\tau \cham2}
\newcommand{\decayCmj}{\Stauzm \to \nu_\tau \cham{j}}
\newcommand{\decayHm}{\Stauzm \to \tilde{\nu}_\tau H^-}
\newcommand{\decayW}{\Stauzm \to \tilde{\nu}_\tau W^-}
\newcommand{\decayxy}{\Stauzm \to {\rm xy}}

\newcommand{\SE}{${\cal S}$}

%%%%%%%%%%%%%%%%%%%%%%%%%%%%%%%%%%%%%%%%%%%%%%%%%%%%%%%%%%%%%%%%%%%%%%%%%%%%%%%
%%%%%%%%%%%%%%%%%%%%%%%%%%%%%%%%%%%%%%%%%%%%%%%%%%%%%%%%%%%%%%%%%%%%%%%%%%%%%%%

\section{Introduction}

Beside the Higgs boson search another important task at the LHC 
is to search for physics effects beyond the Standard Model (SM), 
where the Minimal Supersymmetric Standard Model (MSSM)~\cite{mssm} is 
one of the leading candidates. 
Two related important tasks are investigating the mechanism of electroweak
symmetry breaking, as well as the production and measurement of 
the properties of Cold Dark Matter (CDM).
The most frequently investigated models for electroweak symmetry
breaking are the Higgs mechanism within the SM and within the MSSM.
The latter also offers a natural candidate for CDM, the
Lightest Supersymmetric Particle (LSP), i.e.\ the lightest 
neutralino,~$\neu{1}$~\cite{EHNOS}.
Supersymmetry (SUSY) predicts two scalar partners for all SM fermions as well
as fermionic partners to all SM bosons.
Contrary to the case of the SM, in the MSSM 
two Higgs doublets are required.
This results in five physical Higgs bosons instead of the single Higgs
boson in the SM. These are the light and heavy $\cp$-even Higgs bosons, $h$
and $H$, the $\cp$-odd Higgs boson, $A$, and the charged Higgs bosons,
$H^\pm$.
In the MSSM with complex parameters (cMSSM) the three neutral Higgs
bosons mix~\cite{mhiggsCPXgen,mhiggsCPXRG1,mhiggsCPXFD1}, 
giving rise to the states $\He, \Hz, \Hd$.
The tree-level input parameters are the charged Higgs boson mass, $\MHp$
and $\tb$, the ratio of the two vacuum expectation values.

If SUSY is realized in nature and the scalar quarks and/or the gluino
are in the kinematic reach of the LHC, it is expected that these
strongly interacting particles are copiously produced. 
The primarily produced strongly interacting particles subsequently
decay via cascades to 
SM particles and (if $R$-parity conservation is assumed, as we do) the
LSP. One step in these decay chains%
\footnote{
  Scalar taus can also be produced directly at the LHC, see for instance
  \citere{LHCStau}, where, however, only cross sections for a lighter
  stau where evaluated numerically.
}%
~is often the decay of a scalar tau, $\Stau_{1,2}$,  
to a SM particle and the LSP, or as a `competing process' the
scalar taus decay to another SUSY particle accompanied by a
SM particle. Also neutral and charged Higgs bosons can be produced this way.
Via these decays some characteristics of the LSP and/or Higgs bosons can be
assessed, see, e.g., \citeres{atlas,cms} and references therein. 
At any future $e^+e^-$ collider (such as ILC or CLIC)
a precision determination of the properties of the observed particles is
expected~\cite{teslatdr,ilc,clic}. 
(For combined LHC/ILC analyses and further prospects see \citere{lhcilc}.) 
Thus, if kinematically accessible, the pair production of scalar taus
with a subsequent decay to the LSP and/or Higgs bosons 
can yield important information about the lightest neutralino and 
the Higgs sector of the model.

In order to yield a sufficient accuracy, one-loop corrections to
the various scalar tau decay modes have to be considered.
We take into account all two-body decay modes of the heavy scalar tau,
$\Stauzm$, in the MSSM with complex parameters (cMSSM),
but we neglect flavor violation effects. 
More specifically, we calculate the full one-loop corrections to the 
partial decay widths%
\footnote{
  It should be noted that the purely loop induced decay channels 
  $\Stauzm \to \Stauem \ga$ have been neglected because they yield 
  exactly zero, see \refse{sec:calc} for further details.
}
\begin{align}
\label{ststphi}
&\Ga(\Stauzm \to \Stauem h_n) \qquad (n = 1,2,3)~, \\
\label{ststZ}
&\Ga(\Stauzm \to \Stauem Z)~, \\
\label{sttneu}
&\Ga(\Stauzm \to \tau^- \neu{k}) \qquad (k = 1,2,3, 4)~, \\
\label{stsnH}
&\Ga(\Stauzm \to \Sneut H^-)~, \\
\label{stsnW}
&\Ga(\Stauzm \to \Sneut W^-)~, \\
\label{stnucha}
&\Ga(\Stauzm \to \nu_\tau \cham{j}) \qquad (j = 1,2)~,
\end{align}
where $\neu{k}$ denotes the neutralinos, $\cha{j}$
the charginos, $\tau$ and $\nu_\tau$ the tau and tau-neutrino 
and $Z$ and $W^{\pm}$ the SM gauge bosons.
The total decay width is defined as the sum of the partial decay
widths (\ref{ststphi}) to (\ref{stnucha}), where for a given parameter
point several channels may be kinematically forbidden.

As explained above, 
we are especially interested in the branching ratios (BR) of the
decays involving a Higgs boson, \refeqs{ststphi}, (\ref{stsnH})
as part of an evaluation of a Higgs production cross section, or
involving the LSP, \refeq{sttneu} as part of the measurement
of CDM properties at the LHC or a future $e^+e^-$~collider.
Consequently, it is not necessary to investigate three- or four-body decay
modes. These only play a significant role once the two-body modes
are kinematically forbidden, and thus the relevant BR's are zero. 

We also concentrate on the decays of $\Stauzm$ and do not
investigate $\aStauz$ decays. 
In the presence of complex phases this
would lead to somewhat different results. However, such an analysis of
$\cp$-violating effects is beyond the scope of this paper.

Scalar tau decays have been investigated in many analyses over the last
decade. Most of them were restricted to tree-level evaluations. 
Existing loop corrections are restricted to the MSSM with real 
parameters (rMSSM). First tree-level results for stau decays in the 
rMSSM were published in \citeres{stau-tree-r1,stau-tree-r2,stau-tree-r3}. 
Corresponding tree-level results are implemented in 
{\tt SDECAY}~\cite{sdecay}. 
Tree-level results in the cMSSM can be found in 
\citere{stau-tree-c1,stau-tree-c2,stau-tree-c3}. 
An analysis on three-body decays is given in \citere{stau-3body}.
Complete one-loop corrections to sfermion decays in the 
rMSSM involving SM fermions were presented in \citere{sferm_f_chi_full}. 
However, no explicit numerical results for the full one-loop 
corrections to stau decays are included in this paper. 
Full one-loop corrections to sfermion decays involving 
SM gauge bosons in the rMSSM are presented in \citere{sferm_f_V_full}. 
However, again no numerical results for scalar tau decays are included. 
One-loop corrections to stau decays in the rMSSM, derived
in a pure \DRbar\ scheme (see below) have been made available in the
program package {\tt SFOLD}~\cite{sfold}.

Several methods have been discussed in the literature to
extract the complex parameters of the model from experimental
measurements. A determination of the trilinear Stau-Higgs coupling,
$\Atau$,  in the rMSSM from heavy MSSM
Higgs decays was presented in \citere{Atau-det}.
$\cp$-even observables to extract its phase, $\phiatau$,
have been analyzed in
\citere{stau-CPe-phiAtau}. $\cp$-odd observables for this determination
are investigated in \citere{stau-CPo-phiAtau} (with more details on 
the specific LHC analysis in \citere{stau-CPasym}).
Depending on the realized cMSSM parameter space and on some further
assumptions on the LHC performance, it seems to be possible to obtain
limits on, e.g., $|\Atau|$ and $\phiatau$.

In this paper we present
for the first time a full one-loop calculation for all two-body decay
channels of the heavier scalar tau in the cMSSM (with no generation
mixing), taking into account soft and hard QED radiation.
In \refse{sec:cMSSM} we briefly review the renormalization of  all
relevant sectors of the cMSSM. Details about the calculation can be
found in \refse{sec:calc}, and the numerical results for all decay
channels are presented in \refse{sec:numeval}. The conclusions can be
found in \refse{sec:conclusions}.
The results will be  implemented into the Fortran code 
{\tt FeynHiggs}~\cite{feynhiggs,mhiggslong,mhiggsAEC,mhcMSSMlong}.

%%%%%%%%%%%%%%%%%%%%%%%%%%%%%%%%%%%%%%%%%%%%%%%%%%%%%%%%%%%%%%%%%%%%%%%%%%%%%%%
%%%%%%%%%%%%%%%%%%%%%%%%%%%%%%%%%%%%%%%%%%%%%%%%%%%%%%%%%%%%%%%%%%%%%%%%%%%%%%%

\section{The relevant sectors of the complex MSSM}
\label{sec:cMSSM}

All the channels (\ref{ststphi}) -- (\ref{stnucha}) are calculated at the
one-loop level, including real QED radiation. This requires the
simultaneous renormalization of several sectors of the cMSSM. In
the following 
subsections we briefly review these sectors to make this article
self-contained. 
Details about the renormalization of most of the sectors can be found in
\citeres{SbotRen,Stop2decay,LHCxC,Gluinodecay}.

%%%%%%%%%%%%%%%%%%%%%%%%%%%%%%%%%%%%%%%%%%%%%%%%%%%%%%%%%%%%%%%%%%%%%%%%%%%%%%%

\subsection{The tau lepton/slepton sector of the cMSSM}
\label{sec:slepton}

For the evaluation of the one-loop contributions to the decay channels 
in \refeqs{ststphi} -- (\ref{stnucha}) a renormalization of the scalar tau
($\Stau$) and $\tau$-neutrino ($\Sneut$) sector is needed (we assume no
generation mixing). The stau and tau sneutrino mass matrices
$\matr{M}_{\Stau}$ and $\matr{M}_{\Sneut}$ read
\begin{align}\label{Sfermionmassenmatrix}
\matr{M}_{\Stau} &= \begin{pmatrix} 
\MstauL^2 + \mtau^2 + M_Z^2 c_{2 \beta} (I_\tau^3 - Q_\tau \sw^2) & 
 \mtau \Xtau^* \\[.2em]
 \mtau \Xtau &
\MstauR^2 + \mtau^2 +M_Z^2 c_{2 \beta} Q_\tau \sw^2
\end{pmatrix}~, \\[.5em]
\matr{M}_{\Sneut} &= \MstauL^2 
+ I_{\nu_\tau}^3 c_{2\be} M_Z^2
\end{align}
with
\begin{align}
\Xtau &= \Atau - \mu^* \tb~.
\end{align}
$\MstauL$ and $\MstauR$ are the soft SUSY-breaking mass
parameters, where $\MstauL$ is equal for all members of an
$SU(2)_L$ doublet.
$\mtau$ and $Q_\tau$ are, respectively, the mass and the charge of the
corresponding lepton, $I_{\tau/\nu_\tau}^3$ denotes the isospin of
$\tau/\nu_\tau$, and $\Atau$ is the trilinear soft-breaking parameter.
$\MZ$ and $\MW$ are the masses of the $Z$~and $W$~boson, 
$\cw = \MW/\MZ$, and $\sw = \sqrt{1 - \cw^2}$. Finally we use the
short-hand notations $c_{x} = \cos(x)$, $s_x = \sin(x)$.
The mass matrix $\matr{M}_{\Stau}$ can be diagonalized with the help of 
a unitary transformation ${\matr{U}}_{\Stau}$,
\begin{align}\label{transformationkompl}
\matr{D}_{\Stau} &= 
\matr{U}_{\Stau}\, \matr{M}_{\Stau} \, {\matr{U}}_{\Stau}^\dagger = 
\begin{pmatrix} \mstaue^2 & 0 \\ 0 & \mslz^2 \end{pmatrix}~, \qquad
{\matr{U}}_{\Stau}= 
\begin{pmatrix} U_{\Stau_{11}} & U_{\Stau_{12}} \\  
                U_{\Stau_{21}} & U_{\Stau_{22}} \end{pmatrix}~.
\end{align}
The mass eigenvalues depend only on $|\Xtau|$. 
The scalar tau masses will always be mass ordered, i.e.\
$\mstaue \le \mstauz$:
\begin{align}
\label{MSlep}
m_{\Stau_{1,2}}^2 &= \edz \KL M_{\Stau_L}^2 + M_{\Stau_R}^2 \KR
       + \mtau^2 + \edz I_\tau^3 c_{2\be} \MZ^2 \\
&\quad \mp \edz \sqrt{\KKL \MstauL^2 - \MstauR^2
       + \MZ^2 c_{2\be} (I_\tau^3 - 2 Q_\tau \sw^2) \KKR^2 + 4 \mtau^2 |\Xtau|^2}~, 
\non\\[.5em]
m_{\Sneut}^2 &= \MstauL^2 + I_{\nu_\tau}^3 c_{2\be} M_Z^2~.
\end{align}

%%%%%%%%%%%%%%%%%%%%%%%%%%%%%%%%%%%%%%%%%%%%%%%%%%%%%%%%%%%%%%%%%%%%%%%%%%%%%%%

\subsubsection{Renormalization}

The parameter renormalization can be performed as follows, 
\begin{align}
\matr{M}_{\Stau} \to \matr{M}_{\Stau} + \de\matr{M}_{\Stau}~, \qquad
\matr{M}_{\Sneut} \to \matr{M}_{\Sneut} + \de\matr{M}_{\Sneut} 
\end{align}
which means that the parameters in the mass matrix $\matr{M}_{\Stau}$ 
are replaced by the renormalized parameters and a counterterm. After the
expansion $\de\matr{M}_{\Stau}$ contains the counterterm part,
\begin{align}
\label{proc1a}
\de\matr{M}_{\Stau_{11}} &= \de M_{\Stau_L}^2 + 2 \mtau \de \mtau 
- M_Z^2 c_{2 \beta}\, Q_\tau \, \de \sw^2 + (I_\tau^3 - Q_\tau \sw^2) 
  ( c_{2 \beta}\, \de M_Z^2 + M_Z^2\, \de c_{2\beta})~, \\
\label{proc1b}
\de\matr{M}_{\Stau_{12}} &= (\Atau^*  - \mu \tb)\, \de \mtau 
+ \mtau (\de \Atau^* - \mu\, \de \tb - \tb \, \de \mu)~, \\
\label{proc1c}
\de\matr{M}_{\Stau_{21}} &=\de\matr{M}_{\Stau_{12}}^*~, \\
\label{proc1d}
\de\matr{M}_{\Stau_{22}} &= \de M_{\Stau_R}^2 
+ 2 \mtau \de \mtau +  M_Z^2 c_{2 \beta}\, Q_\tau \, \de \sw^2
+ Q_\tau \sw^2 ( c_{2 \beta}\, \de M_Z^2+ M_Z^2\, \de c_{2 \beta})~, \\
\label{proc1e}
\de\matr{M}_{\Sneut} &= \de M_{\Stau_L}^2 + I_{\nu_\tau}^3
(c_{2 \beta}\, \de M_Z^2 + M_Z^2\, \de c_{2\beta})~. 
\end{align}

Another possibility for the parameter renormalization of the staus is
to start out with the physical parameters which corresponds to
the replacement:
\begin{align} \label{proc2}
\matr{U}_{\Stau}\, \matr{M}_{\Stau} \, 
{\matr{U}}_{\Stau}^\dagger &\to\matr{U}_{\Stau}\, \matr{M}_{\Stau} \, 
{\matr{U}}_{\Stau}^\dagger + \matr{U}_{\Stau}\, \de \matr{M}_{\Stau} \, 
{\matr{U}}_{\Stau}^\dagger =
\begin{pmatrix} \mstaue^2 & Y_\tau \\ Y_\tau^* & \mstauz^2 \end{pmatrix} +
\begin{pmatrix}
\de \mstaue^2 & \de Y_\tau \\ \de Y_\tau^* & \de \mstauz^2
\end{pmatrix}
\end{align}
where $\de\mstaue^2$ and $\de\mstauz^2$ are the counterterms
of the stau mass squares. $\de Y_\tau$ is the counterterm%
\footnote{
  The unitary matrix $\matr{U}_{\Stau}$ can be expressed by a
  mixing angle and a corresponding phase. Then the counterterm  $\de
  Y_\tau$ can be related to the counterterms of the mixing angle and 
  the phase (see \citere{mhcMSSM2L}).
}%
~to the stau mixing parameter $Y_\tau$ (which vanishes at tree-level, 
$Y_\tau = 0$, and corresponds to the off-diagonal entries in 
$\matr{D}_{\Stau} = \matr{U}_{\Stau}\, \matr{M}_{\Stau}\, 
                   {\matr{U}}_{\Stau}^\dagger$, 
\refeq{transformationkompl}). 
Using \refeq{proc2} one can express $\de\matr{M}_{\Stau}$ 
by the counterterms $\de \mstaue^2$, $\de \mstauz^2$ and $\de Y_\tau$. 
Especially for $\de\matr{M}_{\Stau_{12}}$ one finds
\begin{align}\label{dMsq12physpar}
\de\matr{M}_{{\Stau}_{12}} &=
U^*_{\Stau_{11}} U_{\Stau_{12}}
(\de \mstaue^2 - \de \mstauz^2) +
U^*_{\Stau_{11}} U_{\Stau_{22}} \de Y_\tau + U_{\Stau_{12}}
U^*_{\Stau_{21}} \de Y_\tau^*~.
\end{align}
\refeqs{proc1b} and \eqref{dMsq12physpar} yield a relation between 
$\de Y_\tau$, $\de\Atau$ and $\de\mtau$, see below.

For the field renormalization the following procedure is applied,
\begin{align}
\begin{pmatrix} \Staue \\ \Stauz \end{pmatrix} &\to 
  \KL \id + \edz \de\matr{Z}_{\Stau} \KR 
  \begin{pmatrix} \Staue \\ \Stauz \end{pmatrix} 
~~{\rm with}~~
\de\matr{Z}_{\Stau} = \begin{pmatrix} 
                   \dZ{\Stau_{11}} & \dZ{\Stau_{12}} \\
                   \dZ{\Stau_{21}} & \dZ{\Stau_{22}} 
                   \end{pmatrix}~, \\
\Sneut &\to \KL 1 + \tedz \dZ{\Sneut} \KR \Sneut~.
\end{align}

This yields for the renormalized self-energies
\begin{align}
\hSi_{\Stau_{11}}(p^2) &= \Si_{\Stau_{11}}(p^2) 
  + \tedz (p^2 - \mstaue^2) (\dZ{\Stau_{11}} + \dZ{\Stau_{11}}^*)
  - \de\mstaue^2~, \\
\hSi_{\Stau_{12}}(p^2) &= \Si_{\Stau_{12}}(p^2)
  + \tedz (p^2 - \mstaue^2) \dZ{\Stau_{12}}
  + \tedz (p^2 - \mstauz^2) \dZ{\Stau_{21}}^* 
  - \de Y_\tau~, \\
\hSi_{\Stau_{21}}(p^2) &= \Si_{\Stau_{21}}(p^2)
  + \tedz (p^2 - \mstaue^2) \dZ{\Stau_{12}}^*
  + \tedz (p^2 - \mstauz^2) \dZ{\Stau_{21}} 
  - \de Y_\tau^*~, \\
\hSi_{\Stau_{22}}(p^2) &= \Si_{\Stau_{22}}(p^2) 
  + \tedz (p^2 - \mstauz^2) (\dZ{\Stau_{22}} + \dZ{\Stau_{22}}^*)
  - \de\mstauz^2~, \\
\hSi_{\Sneut}(p^2) &= \Si_{\Sneut}(p^2) 
  + \tedz (p^2 - \msneut^2) (\dZ{\Sneut} + \dZ{\Sneut}^*)
  - \de\msneut^2~.
\end{align}
In order to complete the tau lepton/slepton sector renormalization 
also for the corresponding lepton (i.e.\ the $\tau$ mass, $\mtau$, 
and the lepton fields $\tau_L$, $\tau_R$, $\nu_{\tau_L}$) renormalization 
constants have to be introduced: 
\begin{align}
\mtau &\to \mtau + \de \mtau~,\\
\tau_{L/R} &\to (1 + \tedz \dZ{\tau}^{L/R})\, \tau_{L/R}~, \\
\nu_{\tau_L} &\to (1 + \tedz \dZ{\nu_\tau})\, \nu_{\tau_L}~,
\end{align}
with $\de \mtau$ being the tau mass counterterm and $\dZ{\tau}^{L/R}$ 
being the $Z$~factors of the left/right-handed charged lepton fields; 
$\dZ{\nu_\tau}$ is the neutrino field renormalization.
Then the renormalized self energy $\hSi_{\tau}$ 
can be decomposed 
into left/right-handed and scalar left/right-handed parts, 
${\Si}_{\tau}^{L/R}$ and ${\Si}_{\tau}^{SL/SR}$, 
respectively,
while only the left-handed part exists for the self energy
$\hSi_{\nu_\tau}$ of the massless neutrino
\begin{align}\label{decomposition}
\hSi_{\tau} (p) &= \pslash\, {\omega}_{-} \hSi_\tau^L (p^2)
                   + \pslash\, {\omega}_{+} \hSi_\tau^R (p^2)
                   + {\omega}_{-} \hSi_\tau^{SL} (p^2) 
                   + {\omega}_{+} \hSi_\tau^{SR} (p^2)~, \\[.3em]
\hSi_{\nu_\tau} (p) &= \pslash\, {\omega}_{-} \hSi_{\nu_\tau}^L (p^2)
~,
\end{align}
where the components are given by
\begin{align}
\hSi_\tau^{L/R} (p^2) &= {\Si}_\tau^{L/R} (p^2) 
   + \frac{1}{2} (\dZ{\tau}^{L/R} + {\dZ{\tau}^{L/R}}^*)~, \\
\hSi_\tau^{SL} (p^2) &=  {\Si}_\tau^{SL} (p^2) 
   - \frac{\mtau}{2} (\dZ{\tau}^L + {\dZ{\tau}^R}^*) - \de \mtau~,  \\
\hSi_\tau^{SR} (p^2) &=  {\Si}_\tau^{SR} (p^2) 
   - \frac{\mtau}{2} (\dZ{\tau}^R + {\dZ{\tau}^L}^*) - \de \mtau~, \\[.3em]
\hSi_{\nu_\tau}^{L} (p^2) &= {\Si}_{\nu_\tau}^{L} (p^2) 
   + \frac{1}{2} (\dZ{\nu_\tau}^{L} + {\dZ{\nu_\tau}^{L}}^{\!\!*})~, 
\end{align}
and ${\omega}_{\pm} = \frac{1}{2}(\id \pm \gamma_5)$ 
are the right- and left-handed projectors, respectively.
It should be noted that 
$\wtre\hSi_{\tau}^{SR} (p^2) = (\wtre\hSi_{\tau}^{SL} (p^2))^*$ 
holds due to ${\cal CPT}$ invariance.
$\wtre$ denotes the real part with respect to contributions from the 
loop integral, but leaves the complex couplings unaffected.

%%%%%%%%%%%%%%%%%%%%%%%%%%%%%%%%%%%%%%%%%%%%%%%%%%%%%%%%%%%%%%%%%%%%%%%%%%%%%%%

\subsubsection{The tau neutrino/sneutrino sector}
\label{sec:sneutrino}

We follow closely the renormalization presented in
\citere{SbotRen,Stop2decay}, slightly modified to be applicable to the
tau/stau sector. 

\begin{itemize}

\item[(i)] The tau neutrino is defined on-shell (OS), 
yielding the one-loop field renormalization
\begin{align}
\label{dZnu}
\re \dZ{\nu_\tau} = - \wtre \Si_{\nu_\tau}(0)~, \qquad
\im \dZ{\nu_\tau} = 0~.
\end{align}

\item[(ii)]
The $\Sneut$ mass is defined OS,
\begin{align}
\wtre\hSi_{\Sneut}(\msneut^2) = 0~.
\end{align}
This yields for the tau sneutrino mass counter terms
\begin{align}
\de\msneut^2 = \wtre\Si_{\Sneut}(\msneut^2)~.
\end{align}

\item[(iii)]
Due to $m_{\nu_\tau} \equiv 0$ no off-diagonal parameters in the sneutrino mass
matrix have to be renormalized.

\item[(iv)]
The diagonal tau sneutrino $Z$~factor is determined OS such that 
the real part of the residuum of the propagator is set to unity, 
\begin{align}
\label{residuumSneutOS}
\wtre \hSi'_{\Sneut}(p^2)\big|_{p^2 = \msneut^2} = 0~.
\end{align}
with $\Si'(p^2) \equiv \frac{\partial \Si(p^2)}{\partial p^2}$.
This condition fixes the real part of the diagonal $Z$~factor to
\begin{align}
\re\,\dZ{\Sneut} = - \wtre \Si'_{\Sneut}(p^2)\big|_{p^2 = \msneut^2}~,
\end{align}
which is correct, since the imaginary part of the diagonal 
$Z$~factor does not contain any divergences and can be 
(implicitly) set to zero, 
\begin{align}
\im \dZ{\Sneut} &= 0~.
\end{align}
Including absorptive parts of self-energy type corrections into this
$Z$~factor leads to new combined factors $\cZ$ which are (in general) 
different for incoming particles/outgoing antiparticles (unbarred) and 
outgoing particles/incoming antiparticles (barred), 
see \citeres{Stop2decay,LHCxC} for more details.
The combined diagonal tau sneutrino $Z$~factors read
\begin{align}
\de\cZ_{\Sneut} = - \Si'_{\Sneut}(p^2)\big|_{p^2 = \msneut^2}~, \qquad
\de \bar\cZ_{\Sneut} = \de\cZ_{\Sneut}~.
\end{align}

\item[(v)]
Due to $m_{\nu_\tau} \equiv 0$ no off-diagonal field renormalization for the
tau sneutrino has to be performed.

\end{itemize}

%%%%%%%%%%%%%%%%%%%%%%%%%%%%%%%%%%%%%%%%%%%%%%%%%%%%%%%%%%%%%%%%%%%%%%%%%%%%%%%

\subsubsection{The tau/stau sector}

We choose the stau masses $\mstaue$, $\mstauz$ and the 
tau mass $\mtau$ as independent parameters.
Since we also require an independent renormalization of the scalar
neutrino, this requires an explicit restoration of the $SU(2)_L$
relation, achieved via a shift in the $M_{\Stau_L}$ parameter entering 
the $\Stau$~mass matrix (see also \citeres{stopsbot_phi_als,dr2lA}).
Requiring the $SU(2)_L$ relation
to be valid at the loop level induces the following shift in 
$M^2_{\Stau_L}(\Stau)$ 
\begin{align}
M_{\Stau_L}^2(\Stau) = M_{\Stau_L}^2(\Sneut) 
   + \de M_{\Stau_L}^2(\Sneut) - \de M_{\Stau_L}^2(\Stau)
\label{MStaushift}
\end{align}
with
\begin{align}
\de M_{\Stau_L}^2(\Stau) &= |U_{\Stau_{11}}|^2 \de\mstaue^2
   + |U_{\Stau_{12}}|^2 \de\mstauz^2
   - U_{\Stau_{22}} U_{\Stau_{12}}^* \de Y_\tau
   - U_{\Stau_{12}} U_{\Stau_{22}}^* \de Y_\tau^* - 2 \mtau \de\mtau \non \\
&\quad  + \MZ^2\, c_{2\be}\, Q_\tau\, \de \sw^2 
        - (I_\tau^3 - Q_\tau \sw^2) (c_{2\be}\, \de \MZ^2 + \MZ^2\, \de c_{2\be})~, 
\\[.5em]
\de M_{\Stau_L}^2(\Sneut) &= \de\msneut^2 
   - I_{\nu_\tau}^3(c_{2\be}\, \de \MZ^2 + \MZ^2\, \de c_{2\be})~.
\label{MStaushift-detail}
\end{align}
This choice avoids problems concerning UV- and IR-finiteness as
discussed in detail in \citere{SbotRen},
but also leads to shifts in both stau masses, 
which are therefore 
slightly shifted away from their on-shell values.
An additional shift in $M_{\Stau_R}$ recovers at least
one on-shell stau mass, which is now compatible with our choice
of independent parameters
\begin{align}
M_{\Stau_R}^2(\Stau_i) = \frac{\mtau^2\, |\Atau^* - \mu \tb|^2}
  {M_{\Stau_L}^2(\tilde{l}) + \mtau^2 
   + \MZ^2\, c_{2\be} (I_\tau^3 - Q_\tau \sw^2) - \mstaui^2} 
  - \mtau^2 - \MZ^2\, c_{2\be}\, Q_\tau\, \sw^2+ \mstaui^2~.
\label{backshift}
\end{align}
The choice of stau for this additional shift, which relates its mass to
the stau parameter $M_{\Stau_R}$, also represents a choice of scenario, 
with the chosen stau having a dominantly right-handed character.
A ``natural'' choice is to preserve the character of the staus in the
renormalization process.
With our choice of mass ordering, $\mstaue \le \mstauz$ (see
above), this suggests to recover 
$\mstaue$ for $M_{\Stau_L}^2 > M_{\Stau_R}^2$, and to recover
$\mstauz$ for the other mass hierarchy. Consequently, for our numerical
choice given below in \refta{tab:para}, we insert $\mstauz$ into
\refeq{backshift} and recover its original value from the
re-diagonalization after applying this shift.

\bigskip

For the tau/stau sector we can now employ a ``full'' on-shell
scheme, where the following renormalization conditions are imposed:
\begin{itemize}

\item[(i)] The tau mass is defined on-shell, yielding the one-loop
  counterterm $\de \mtau$:
\begin{align}\label{dmt}
\de \mtau &= \tedz \wtre \KKKL 
    \mtau \KKL\Si_\tau^L (\mtau^2) + \Si_\tau^R (\mtau^2) \KKR  
  + \KKL \Si_\tau^{SL} (\mtau^2) + \Si_\tau^{SR} (\mtau^2) \KKR \KKKR~,
\end{align}
referring to the Lorentz decomposition of the self energy 
${\hSi}_{\tau}(p)$, see \refeq{decomposition}.\\
The field renormalization constants are given by
\begin{align}
\dZ{\tau}^{L/R} &= - \wtre \Big\{ {\Si}_\tau^{L/R} (\mtau^2)
+ \mtau^2  \KL {{\Si}_\tau^{L}}'(\mtau^2) + {{\Si}_\tau^{R}}'(\mtau^2) \KR
+ \mtau^{} \KL {{\Si}_\tau^{SL}}'(\mtau^2) + {{\Si}_\tau^{SR}}'(\mtau^2) \KR
                            \non \\
&\quad \pm \frac{1}{2\, \mtau} 
       \KL {\Si}_\tau^{SL}(\mtau^2) - {\Si}_\tau^{SR}(\mtau^2) \KR \Big\}
\end{align}
with 
$\Si'(m^2) \equiv \frac{\partial \Si(p^2)}{\partial p^2}\big|_{p^2 = m^2}$.

\item[(ii)]
The stau masses are also determined via on-shell
conditions~\cite{mhiggslong,hr}, yielding  
\begin{align}
\label{dmsl}
\de\mstaui^2 &= \wtre\Si_{\Stau_{ii}}(\mstaui^2) \qquad (i = 1,2)~.
\end{align}

\item[(iii)]
The non-diagonal entry of \refeq{proc2} is fixed
as~\cite{mhiggsFDalbals,hr,SbotRen} 
\begin{align}
\de Y_\tau =  \tedz \wtre 
    \big\{ \Si_{\Stau_{12}}(\mstaue^2) + \Si_{\Stau_{12}}(\mstauz^2) \big\}~, 
\end{align}
which corresponds to two separate conditions in the case of a complex
$\de Y_\tau$.
The counterterm of the trilinear coupling $\de\Atau$ can be obtained from the
relation of \refeqs{proc1b} and~\eqref{dMsq12physpar},
\begin{align}
\de \Atau &= \frac{1}{\mtau}\bigl[U_{\Stau_{11}} U_{\Stau_{12}}^*
           (\de \mstaue^2 - \de \mstauz^2)
        +  U_{\Stau_{11}} U_{\Stau_{22}}^{*} \de Y_\tau^*
        + U_{\Stau_{12}}^{*} U_{\Stau_{21}} \de Y_\tau  
        - (\Atau - \mu^* \tb)\, \de\mtau  \bigr]  \non \\
&\quad  + (\de\mu^* \tb + \mu^* \dtanb)~.
\end{align}
So far undetermined are $\dtanb$ and $\de\mu$, which are defined 
via the Higgs sector and the chargino/neutralino sector, see
\citere{Stop2decay} for details.

\item[(iv)]
The diagonal scalar tau $Z$~factors are determined OS such that 
the real parts of the residua of the propagators are set to unity, 
\begin{align}
\label{residuumSlepOS}
\wtre \hSi'_{\Stau_{ii}}(p^2) \big|_{p^2 = \mstaui^2} = 0 
\qquad (i = 1,2)~.
\end{align}
This condition fixes 
the real parts of 
the diagonal $Z$~factors to
\begin{align}
\re\,\dZ{\Stau_{ii}} = 
- \wtre \Si'_{\Stau_{ii}}(p^2)\big|_{p^2 = \mstaui^2} 
\qquad (i = 1,2)~,
\end{align}
which is correct, since the imaginary parts of the diagonal $Z$~factors 
does not contain any divergences and can be (implicitly) set to zero,
\begin{align}
\im \dZ{\Stau_{ii}} &= 0 \qquad (i = 1,2)~.
\end{align}
Including absorptive parts of self-energy type corrections into these 
$Z$~factors leads to new combined factors $\cZ$
\begin{align}
\de\cZ_{\Stau_{ii}} = - \Si'_{\Stau_{ii}}(p^2)\big|_{p^2 = \mstaui}^2~, \qquad
\de\bar\cZ_{\Stau_{ii}} = \de\cZ_{\Stau_{ii}}~.
\end{align}

\item[(v)]
For the non-diagonal $Z$~factors we impose the condition that for
on-shell staus no transition from one stau to the other occurs, 
\begin{align}
\wtre\hSi_{\Stau_{12}}(\mstaui^2)  = 0~, \qquad 
\wtre\hSi_{\Stau_{21}}(\mstaui^2) = 0 \qquad (i = 1,2)~.
\end{align}
This yields
\begin{align}
\dZ{\Stau_{12}} = + 2 \frac{\wtre\Si_{\Stau_{12}}(\mstauz^2) - \de Y_\tau}
                       {(\mstaue^2 - \mstauz^2)}~, \qquad
\dZ{\Stau_{21}} = - 2 \frac{\wtre\Si_{\Stau_{21}}(\mstaue^2) - \de Y_\tau^*}
                       {(\mstaue^2 - \mstauz^2)}~.
\label{dZslepoffdiagOS}
\end{align}

Taking the absorptive parts of the self-energy type corrections into
account, the conditions change to
\begin{align}
\hSi_{\Stau_{12}}(\mstaui^2) = 0~, \qquad
\hSi_{\Stau_{21}}(\mstaui^2) = 0 \qquad (i = 1,2)~.
\end{align}
This yields the following combined field renormalization constants
for incoming particles/outgoing antiparticles (unbarred) and outgoing 
particles/incoming antiparticles (barred),
\begin{align}
\de\cZ_{\Stau_{12}}&= 
+ 2 \frac{\Si_{\Stau_{12}}(\mstauz^2) - \de Y_\tau}{(\mstaue^2 - \mstauz^2)}~,
& \de\bar\cZ_{\Stau_{12}} &= 
+ 2 \frac{\Si_{\Stau_{21}}(\mstauz^2) - \de Y_\tau^*}{(\mstaue^2 - \mstauz^2)}~, 
\\
\de\cZ_{\Stau_{21}} &= 
- 2 \frac{\Si_{\Stau_{21}}(\mstaue^2) - \de Y_\tau^*}{(\mstaue^2 - \mstauz^2)}~,
& \de\bar\cZ_{\Stau_{21}} &= 
- 2 \frac{\Si_{\Stau_{12}}(\mstaue^2) - \de Y_\tau}{(\mstaue^2 - \mstauz^2)}~.
\end{align}
\end{itemize}

%%%%%%%%%%%%%%%%%%%%%%%%%%%%%%%%%%%%%%%%%%%%%%%%%%%%%%%%%%%%%%%%%%%%%%%%%%%%%%%
%%%%%%%%%%%%%%%%%%%%%%%%%%%%%%%%%%%%%%%%%%%%%%%%%%%%%%%%%%%%%%%%%%%%%%%%%%%%%%%

\subsection{The Higgs and gauge boson sector of the cMSSM}
\label{sec:higgs}

The two Higgs doublets of the cMSSM are decomposed in the following way,
\begin{align}
\label{eq:higgsdoublets}
\cHe = \begin{pmatrix} H_{11} \\ H_{12} \end{pmatrix} =
\begin{pmatrix} v_1 + \tfrac{1}{\sqrt{2}} (\phi_1-i \chi_1) \\
  -\phi^-_1 \end{pmatrix}, \qquad %\notag \\ 
\cHz = \begin{pmatrix} H_{21} \\ H_{22} \end{pmatrix} = e^{i \xi}
\begin{pmatrix} \phi^+_2 \\ v_2 + \tfrac{1}{\sqrt{2}} (\phi_2+i
  \chi_2) \end{pmatrix}. 
\end{align}
Besides the vacuum expectation values $v_1$ and $v_2$, in 
\refeqs{eq:higgsdoublets} a possible new phase $\xi$ between the two
Higgs doublets is introduced. 
The Higgs potential $\VHiggs$ can be written in powers of the Higgs fields,
\begin{align}
\VHiggs &=  \ldots + T_{\phi_1}\, \phi_1  +T_{\phi_2}\, \phi_2 +
        T_{\chi_1}\, \chi_1 + T_{\chi_2}\, \chi_2 \non \\ 
&\quad - \edz \begin{pmatrix} \phi_1,\phi_2,\chi_1,\chi_2
        \end{pmatrix} 
\matr{M}_{\phi\phi\chi\chi}
\begin{pmatrix} \phi_1 \\ \phi_2 \\ \chi_1 \\ \chi_2 \end{pmatrix} -
\begin{pmatrix} \phi^{+}_1,\phi^{+}_2  \end{pmatrix}
\matr{M}^{\top}_{\phi^\pm\phi^\pm}
\begin{pmatrix} \phi^{-}_1 \\ \phi^{-}_2  \end{pmatrix} + \ldots~,
\end{align}
where the coefficients of the linear terms are called tadpoles and
those of the bilinear terms are the mass matrices
$\matr{M}_{\phi\phi\chi\chi}$ and $\matr{M}_{\phi^\pm\phi^\pm}$. 
After a rotation to the physical fields 
one obtains
\begin{align}
\label{VHiggs}
\VHiggs &=  \ldots + T_{h}\, h + T_{H}\, H + T_{A}\, A \non \\ 
&\quad  - \edz \begin{pmatrix} h, H, A, G 
        \end{pmatrix} 
\matr{M}_{hHAG}^{\rm diag}
\begin{pmatrix} h \\ H \\ A \\ G  \end{pmatrix} -
\begin{pmatrix} H^{+}, G^{+}  \end{pmatrix}
\matr{M}_{H^\pm G^\pm}^{\rm diag}
\begin{pmatrix} H^{-} \\ G^{-} \end{pmatrix} + \ldots~,
\end{align}
where the tree-level masses are denoted as
$\mh$, $\mH$, $\mA$, $\mG$, $\MHp$, $\mGp$.
With the help of a Peccei-Quinn
transformation~\cite{Peccei} $\mu$ and the complex soft SUSY-breaking
parameters in the Higgs sector can be 
redefined~\cite{MSSMcomplphasen} such that the complex phases
vanish at tree-level.

Concerning the renormalization we follow the usual approach where the
gauge-fixing term does not receive a net contribution from the
renormalization transformations. 
As input parameter we choose the mass of the charged Higgs boson, $\MHp$.
All details can be found in \citeres{Stop2decay,mhcMSSMlong}%
\footnote{
  Corresponding to the convention used in \fa/\fc, we exchanged in
  the charged part the positive Higgs fields with the negative ones, 
  which is in contrast to \citere{mhcMSSMlong}. 
  As we keep the definition of the matrix $\matr{M}_{\phi^\pm\phi^\pm}$ 
  used in \cite{mhcMSSMlong} the transposed matrix will appear in the 
  expression for $\matr{M}_{H^\pm G^\pm}^{\rm diag}$.
}%
~(see also \citere{Demir} for the alternative effective potential
approach and \citere{mhcMSSMother} for the renormalization group improved
effective potential approach including Higgs pole mass effects).

Including higher-order corrections the three neutral Higgs bosons can
mix~\cite{mhiggsCPXgen,mhiggsCPXRG1,mhiggsCPXFD1,mhcMSSMlong}, 
\begin{align}
\KL h, H, A \KR \quad \longrightarrow \quad \KL \He, \Hz, \Hd \KR~,
\end{align} 
where we define the loop corrected masses according to
\begin{align}
\MHe \le \MHz \le \MHd~.
\end{align}
A vertex with an external on-shell Higgs boson $h_{n}$ ($n = 1,2,3$)
is obtained from the decay widths to the tree-level Higgs bosons via the
complex matrix $\matr{Z}$~\cite{mhcMSSMlong},
\begin{align}
\Ga_{h_n} &=
[\matr{Z}]_{n1} \Ga_h +
[\matr{Z}]_{n2} \Ga_H +
[\matr{Z}]_{n3} \Ga_A + \ldots ~,
\label{eq:zfactors123}
\end{align}
where the ellipsis represents contributions from the mixing with the
Goldstone boson and the $Z$~boson, see \refse{sec:calc}.
It should be noted that the `rotation' with $\matr{Z}$ is not a
unitary transformation, see \citere{mhcMSSMlong} for details.

Also the charged Higgs boson appearing as an external particle in a
stau decay has to obey the proper on-shell conditions. This leads to
an extra $Z$~factor,
\begin{align}
\hat Z_{H^-H^+} = 
   \KKL 1 + \re \hSip_{H^-H^+}(p^2)\big|_{p^2 = \MHp^2} \KKR^{-1}~.
\end{align}
Expanding to one-loop order yields the $Z$~factor that has to be applied
to the process with an external charged Higgs boson,
\begin{align}
\sqrt{\hat Z_{H^-H^+}} = 1 + \frac{1}{2} \de\hat Z_{H^-H^+} 
\end{align}
with 
\begin{align}
\label{dhZHpHm}
\de\hat Z_{H^-H^+} = - \re\hSip_{H^-H^+}(p^2)\big|_{p^2 = \MHp^2} = 
- \re\Sip_{H^-H^+}(\MHp^2) - \dZ{H^-H^+}~.
\end{align}
As for the neutral Higgs bosons, there are contributions from the 
mixing with the Goldstone boson and the $W$~boson.
This $Z$~factor is by definition UV-finite. However, it contains
IR-divergences that cancel with the soft photon contributions from 
the loop diagrams, see \refse{sec:calc}.

For the renormalization of $\tb$ and the Higgs field
renormalization the \DRbar\ scheme is
chosen~\cite{mhcMSSMlong,Stop2decay}. This leads to the introduction
of the scale $\mu_R$, which will be fixed later to the
mass of the decaying particle.

%%%%%%%%%%%%%%%%%%%%%%%%%%%%%%%%%%%%%%%%%%%%%%%%%%%%%%%%%%%%%%%%%%%%%%%%%%%%%%%

\subsection{The chargino/neutralino sector of the cMSSM}
\label{sec:chaneu}

The mass eigenstates of the charginos can be determined from the matrix
\begin{align}
  \matr{X} =
  \begin{pmatrix}
    \MTwo & \sqrt{2} \sinb \MW \\
    \sqrt{2} \cosb \MW & \mu
  \end{pmatrix}.
\end{align}
In addition to the higgsino mass parameter $\mu$ it contains the 
soft breaking term $\MTwo$, which can also be complex in the cMSSM.
The rotation to the chargino mass eigenstates is done by transforming
the original wino and higgsino fields with the help of two unitary 
2$\times$2 matrices $\matr{U}$ and $\matr{V}$,
\begin{align}
\label{eq:charginotransform}
\tilde{\chi}^-_i = 
\begin{pmatrix} \psi^L_i
   \\ \overline{\psi}^R_i \end{pmatrix}
\quad \text{with} \quad \psi^L_{i} = U_{ij} \begin{pmatrix} \tilde{W}^-
  \\ \tilde{H}^-_1 \end{pmatrix}_{j} \quad \text{and} \quad
 \psi^R_{i} = V_{ij} \begin{pmatrix} \tilde{W}^+
  \\ \tilde{H}^+_2 \end{pmatrix}_{j}~,
\end{align}
where the $i$th mass eigenstate can be expressed in terms of either the Weyl
spinors $\psi^L_i$ and $\psi^R_i$ or the Dirac spinor $\tilde{\chi}^-_i$.
These rotations lead to the diagonal mass matrix
\begin{align}
\matr{M}_{\cham{}} = 
  \matr{V}^* \, \matr{X}^{\top} \, \matr{U}^{\dagger} =
  \matr{diag}(m_{\tilde{\chi}^\pm_1}, m_{\tilde{\chi}^\pm_2})~.
\end{align}
{}From this relation, it becomes clear that the mass ordered
chargino masses $\mcha{1} < \mcha{2}$ can be determined as the 
(real and positive) singular values of $\matr{X}$.
The singular value decomposition of $\matr{X}$ also yields results for 
$\matr{U}$ and~$\matr{V}$.

A similar procedure is used for the determination of the neutralino masses and
mixing matrix, which can both be calculated from the mass matrix
\begin{align}
  \matr{Y} =
  \begin{pmatrix}
    \MOne                  & 0                & -\MZ \, \sw \cosb
    & \MZ \, \sw \sinb \\
    0                      & \MTwo            & \quad \MZ \, \cw \cosb
    & -\MZ \, \cw \sinb \\
    -\MZ \, \sw \cosb      & \MZ \, \cw \cosb & 0
    & -\mu             \\
    \quad \MZ \, \sw \sinb & -\MZ \, \cw \sinb & -\mu                   & 0
  \end{pmatrix}.
\end{align}
This symmetric matrix contains the additional complex soft-breaking
parameter $\MOne$. 
The diagonalization of the matrix
is achieved by a transformation starting from the original
bino/wino/higgsino basis,
\begin{align}
\tilde{\chi}^0_{k} = \begin{pmatrix} \psi^0_{k} \\ \overline{\psi}^0_{k} 
                   \end{pmatrix} \qquad \text{with} \qquad 
\psi^0_{k} = N_{kl}\, 
         (\tilde{B}^0, \tilde{W}^0, \tilde{H}^0_1,\tilde{H}^0_2)_{l}^{\top}~, 
\\
\matr{M}_{\neu{}} = \matr{N}^* \, \matr{Y} \, \matr{N}^{\dagger} =
\matr{diag}(\mneu{1}, \mneu{2}, \mneu{3}, \mneu{4})~,
\end{align}
where $\psi^0_{k}$ denotes the two component Weyl spinor and
$\tilde{\chi}^0_{k}$ 
the four component Majorana spinor of the $k$th neutralino field.
The unitary 4$\times$4 matrix $\matr{N}$ and the physical neutralino
masses result from a numerical Takagi factorization of $\matr{Y}$.
The symmetry of $\matr{Y}$ permits the non-trivial condition of using 
only one matrix $\matr{N}$ for its diagonalization, in contrast to the 
chargino case shown above. 

\noindent
Concerning the renormalization we use the results of
\citere{Stop2decay,dissAF,imim,diplTF,dissTF}.  
Since the chargino masses $\mcha{1}, \mcha{2}$ and the lightest neutralino 
mass $\mneu{1}$ have been chosen as independent parameters the one-loop 
masses of the heavier neutralinos are obtained from the tree-level ones 
with the shifts
\begin{align}
\De \mneu{k} = -\re \KKL \mneu{k} \hat\Si_{\neu{}}^{L}(\mneu{k}^2) 
               + \hat\Si_{\neu{}}^{SL}(\mneu{k}^2) \KKR_{kk} 
               \qquad (k = 2,3,4)~,
\label{Deltamneu}
\end{align}
where the renormalized self energies of the neutralino have been decomposed 
into their left/right-handed and scalar left/right-handed parts as in 
Eq.~(\ref{decomposition}).
$\De \mneu{1} = 0$ is just the real part of one of 
our renormalization conditions.
Special care has to be taken in the regions of the cMSSM parameter space 
where the gaugino-higgsino mixing in the chargino sector is maximal,
i.e.\ where $\mu \approx M_2$. 
Here $\delta M_2$ (see Eq.~(180) in \citere{Stop2decay}) and 
$\delta \mu$ (see Eq.~(181) in \citere{Stop2decay}) diverge as 
$(U^*_{11}U^*_{22}V^*_{11}V^*_{22} - U^*_{12}U^*_{21}V^*_{12}V^*_{21})^{-1}$
and the loop calculation does not yield a reliable result.
An analysis of various renormalization schemes was recently
published in \citere{onshellCNmasses}, where this kind of divergences
were discussed.%
\footnote{
  Similar divergences appearing in the on-shell renormalization 
  in the sbottom sector, occurring for ``maximal sbottom mixing'', 
  have been observed and discussed in \citeres{SbotRen,Stop2decay}.
}%
~In \citere{onshellCNmasses} it was furthermore emphasized that 
in the case of the renormalization of two chargino and one neutralino
mass always the most bino-like neutralino has to be renormalized in order
to find a numerically stable result (see also \citere{Baro}).
In our numerical set-up, see \refse{sec:numeval}, 
the lightest neutralino is nearly always rather bino-like.
If required, however, it would be trivial to change our prescription 
from the lightest neutralino to any other neutralino.%
\footnote{
  In \citere{onshellCNmasses} it was also suggested that the
  numerically most stable result is obtained via the renormalization 
  of one chargino and two neutralinos.
  However, in our approach, this choice leads to IR divergences, 
  since the chargino mass changes (from the tree-level mass to the 
  one-loop pole mass) by a finite shift due to the renormalization procedure. 
  Using the shifted mass for the external chargino, but the tree-level mass 
  for internal charginos results in IR divergences.
  On the other hand, in general, inserting the shifted chargino mass 
  everywhere yields UV divergences.
  Consequently, we stick to our choice of imposing on-shell conditions 
  for the two charginos and one neutralino.
}

%%%%%%%%%%%%%%%%%%%%%%%%%%%%%%%%%%%%%%%%%%%%%%%%%%%%%%%%%%%%%%%%%%%%%%%%%%%%%%%

\pagebreak
\newpage

%%%%%%%%%%%%%%%%%%%%%%%%%%%%%%%%%%%%%%%%%%%%%%%%%%%%%%%%%%%%%%%%%%%%%%%%%%%%%%%
%%%%%%%%%%%%%%%%%%%%%%%%%%%%%%%%%%%%%%%%%%%%%%%%%%%%%%%%%%%%%%%%%%%%%%%%%%%%%%%

\section{Calculation of loop diagrams}
\label{sec:calc}

\begin{figure}[t!]
\begin{center}
\includegraphics[width=0.90\textwidth]{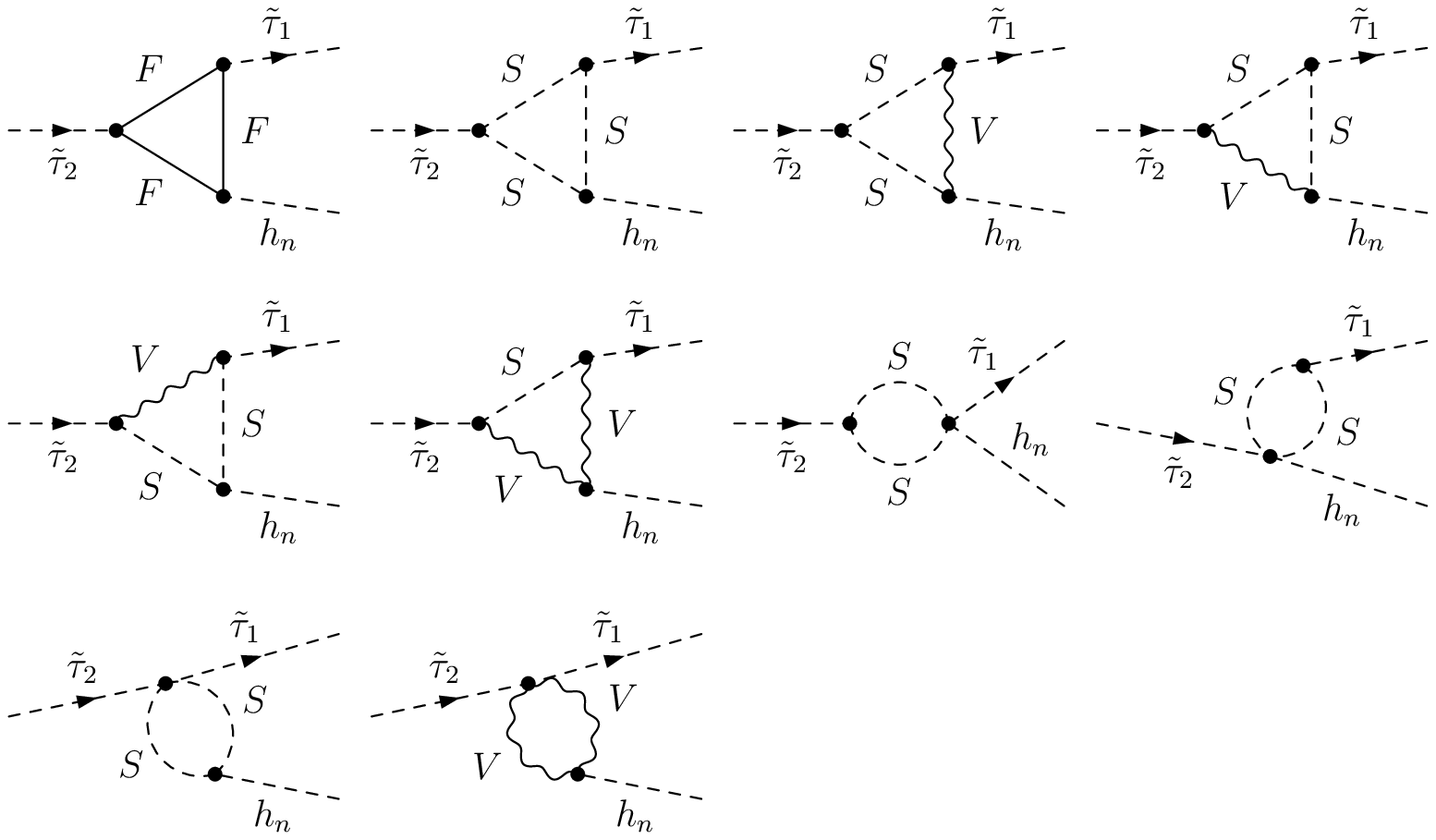}
\caption{
Generic Feynman diagrams for the decay $\decayhn$ ($n = 1, 2, 3$).
$F$ can be a tau, tau-neutrino, chargino or neutralino, 
$S$ can be a sfermion or a Higgs boson, $V$ can be a $\ga$, $Z$ or $W^\pm$. 
Not shown are the diagrams with a $Z$--$h_n$ or $G$--$h_n$ transition 
contribution on the external Higgs boson leg.
}
\label{fig:stau2stau1h}
\end{center}
\vspace{2em}
\end{figure}

\begin{figure}[htb!]
\begin{center}
\includegraphics[width=0.90\textwidth]{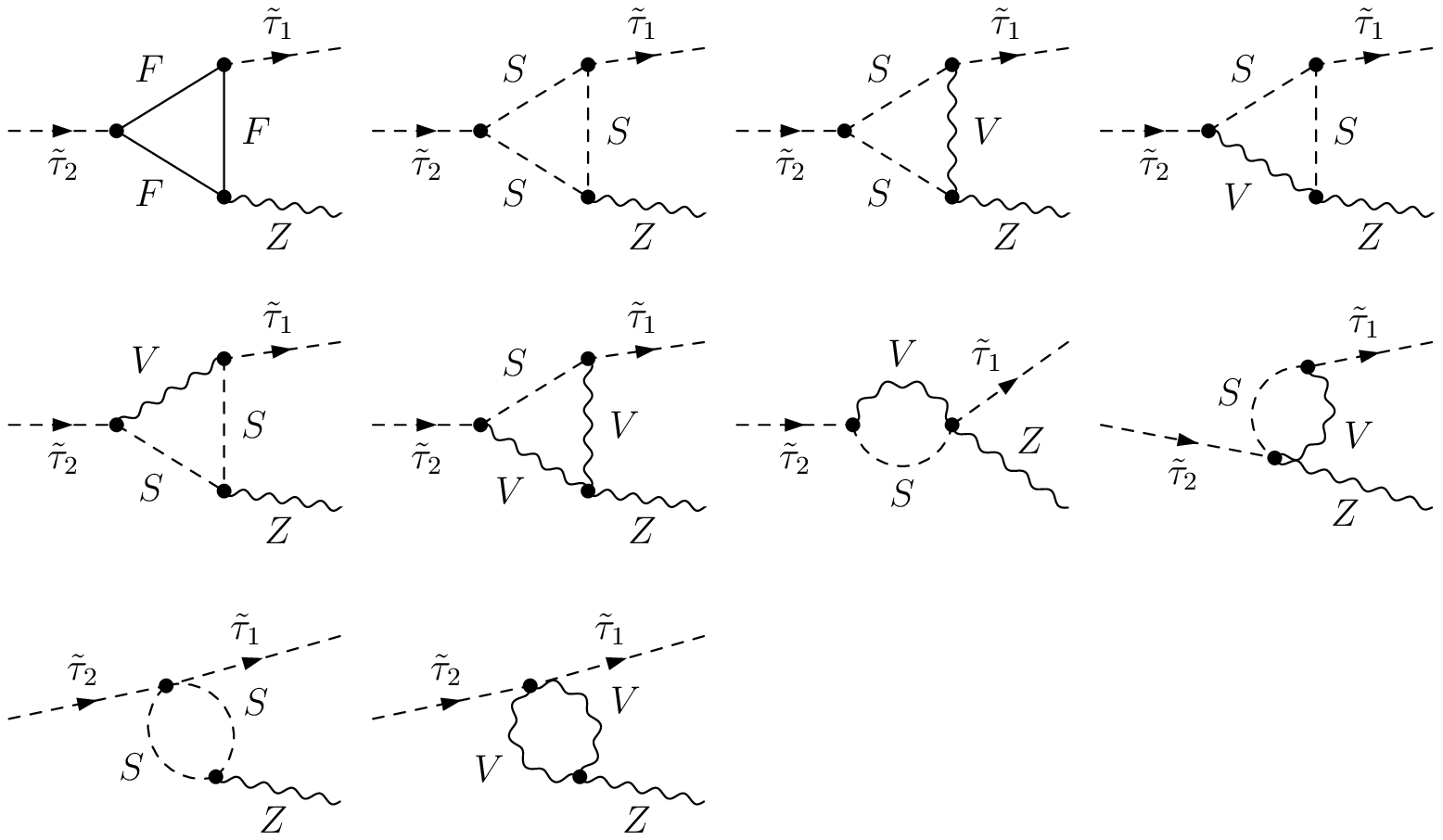}
\caption{
Generic Feynman diagrams for the decay $\decayZ$.
$F$ can be a tau, tau-neutrino, chargino, or neutralino, 
$S$ can be a sfermion or a Higgs boson, $V$ can be a $\ga$, $Z$ or $W^\pm$. 
}
\label{fig:stau2stau1Z}
\end{center}
\vspace{2em}
\end{figure}

\begin{figure}[htb!]
\begin{center}
\includegraphics[width=0.90\textwidth]{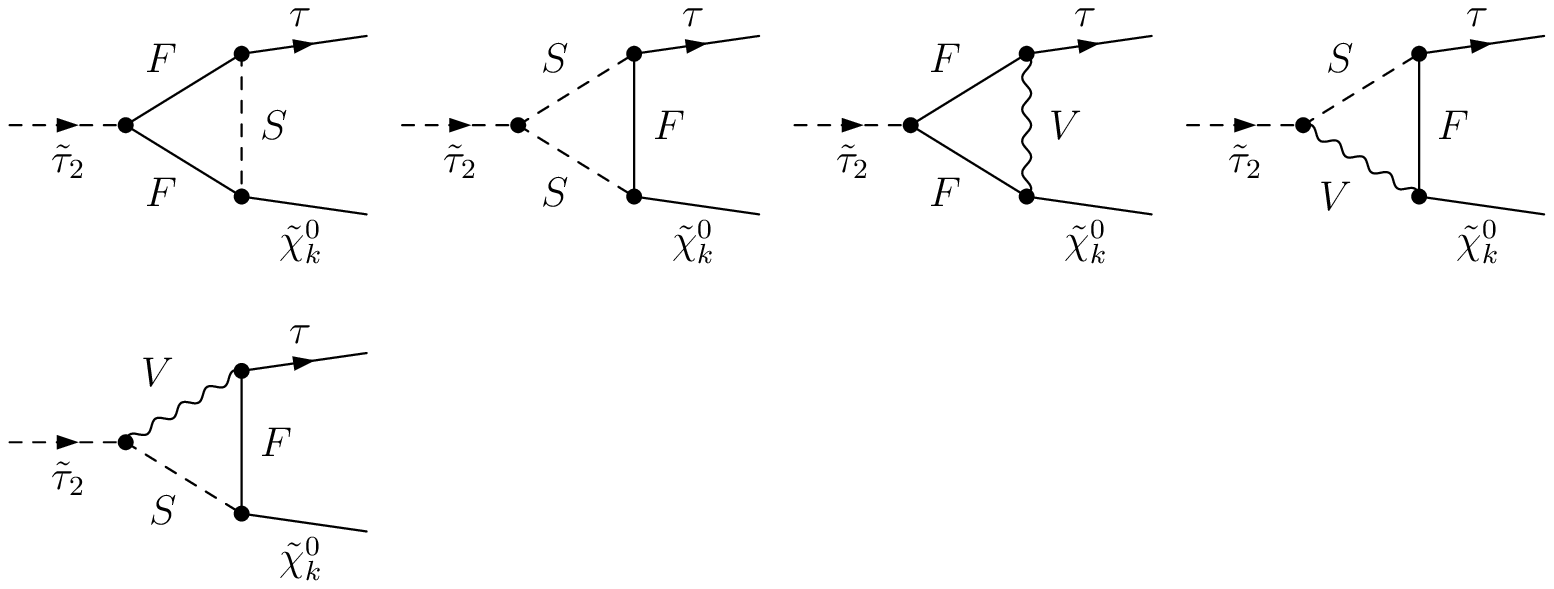}
\caption{
Generic Feynman diagrams for the decay $\decayNk$ ($k = 1,2,3,4$).
$F$ can be a tau, tau-neutrino, chargino or neutralino,
$S$ can be a stau, tau-sneutrino or a Higgs boson, 
$V$ can be a $\ga$, $Z$ or $W^\pm$. 
}
\label{fig:stau2tauneu1}
\end{center}
\vspace{4em}
\end{figure}

\begin{figure}[htb!]
\begin{center}
\includegraphics[width=0.90\textwidth]{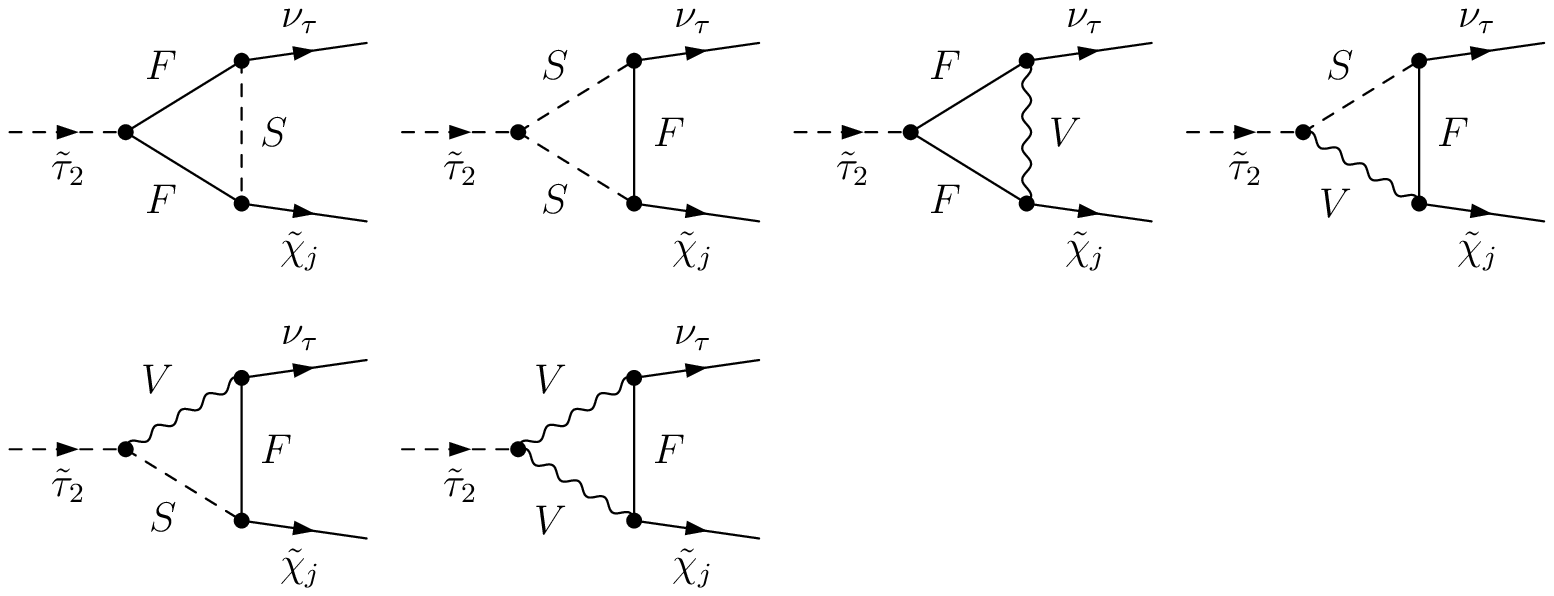}
\caption{
Generic Feynman diagrams for the decay $\decayCj$ ($j = 1,2$).
$F$ can be a tau, tau-neutrino, chargino or neutralino, 
$S$ can be a stau or a Higgs boson, $V$ can be a $\ga$, $Z$ or $W^\pm$. 
}
\label{fig:stau2ncha}
\end{center}
\vspace{4em}
\end{figure}

\begin{figure}[htb!]
\begin{center}
\includegraphics[width=0.90\textwidth]{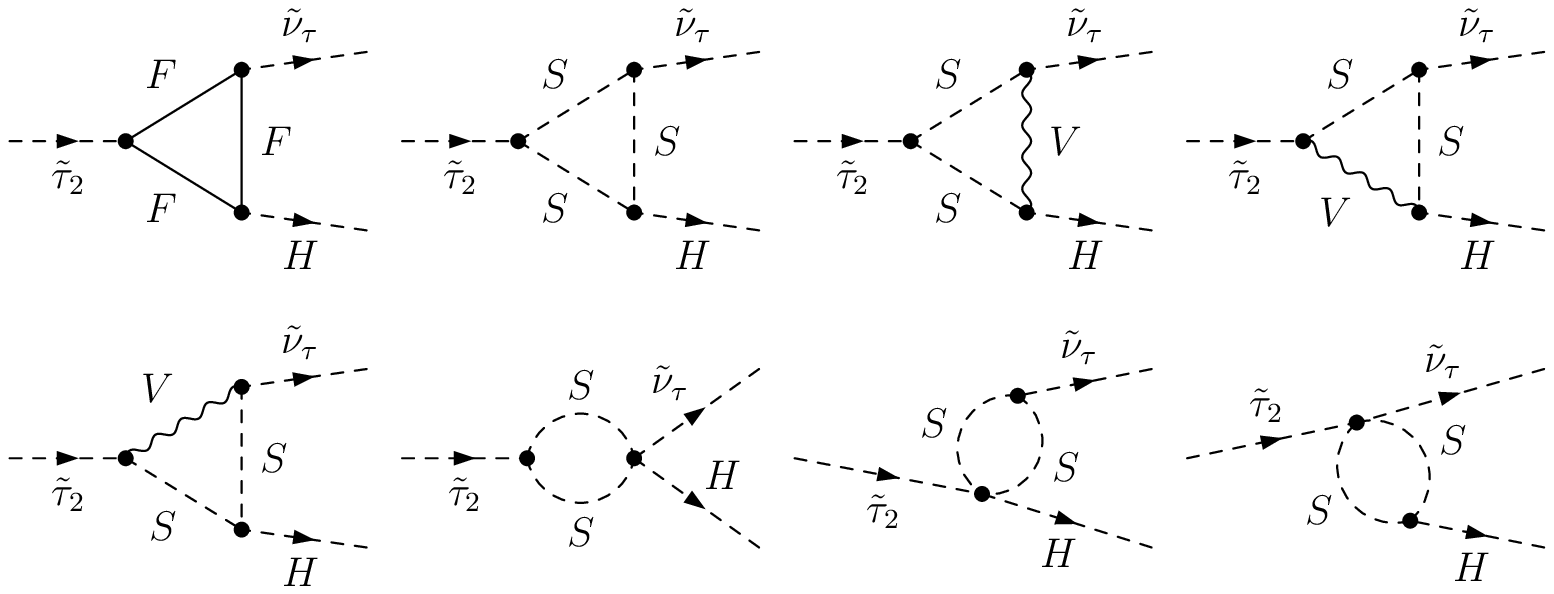}
\caption{
Generic Feynman diagrams for the decay $\decayHm$.
$F$ can be a tau, tau-neutrino, chargino or neutralino, $S$ can be a
sfermion or a Higgs boson, $V$ can be a $\ga$, $Z$ or $W^\pm$. 
Not shown are the diagrams with a $W^+$--$H^+$ or $G^+$--$H^+$ transition
contribution on the external Higgs boson leg. 
}
\label{fig:stau2snH}
\end{center}
\vspace{6em}
\end{figure}

\begin{figure}[htb!]
\begin{center}
\includegraphics[width=0.90\textwidth]{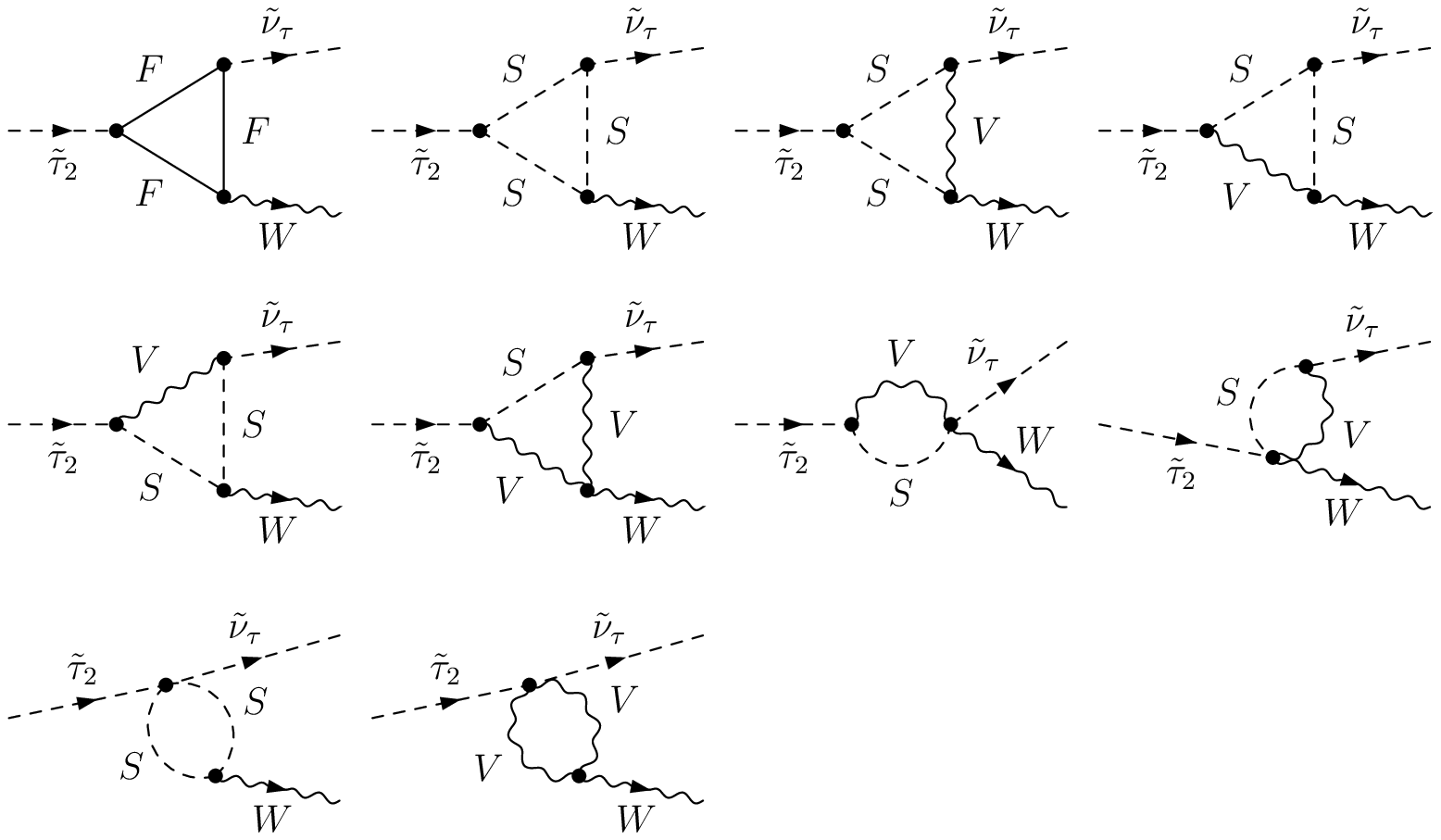}
\caption{
Generic Feynman diagrams for the decay $\decayW$.
$F$ can be a tau, tau-neutrino, chargino or neutralino, 
$S$ can be a sfermion or a Higgs boson, 
$V$ can be a $\ga$, $Z$ or $W^\pm$. 
}
\label{fig:stau2snW}
\end{center}
\vspace{2em}
\end{figure}

In this section we give some details about the calculation of the
higher-order corrections to the partial decay widths of scalar taus. 
Generic diagrams are shown in \reffis{fig:stau2stau1h} -- \ref{fig:stau2snW}.  
Not shown are the diagrams for real (hard or soft) photon radiation. 
They are obtained from the corresponding tree-level diagrams
by attaching a photon to the electrically charged particles. 
The internal generically depicted particles in
\reffis{fig:stau2stau1h} -- \ref{fig:stau2snW} are labeled as follows:
$F$ can be a tau~$\tau$, tau-neutrino $\nu_\tau$, 
chargino $\cha{j}$ or neutralino $\neu{k}$, 
$S$ can be a sfermion $\sfi$ or a Higgs boson $h_n$, 
$V$ can be a photon $\ga$ or a massive SM gauge boson, $Z$ or $W^\pm$. 
For internally appearing Higgs bosons no higher-order
corrections to their masses or couplings are taken into account; 
these corrections would correspond to effects beyond one-loop order.%
\footnote{
  We found that using loop corrected Higgs boson masses in the loops 
  leads to a UV divergent result.
}%
~For external Higgs bosons, as described in
\refse{sec:higgs}, the appropriate $\matr{Z}$~factors are applied and
on-shell masses (including higher order corrections) are used 
(as evaluated with 
{\tt FeynHiggs}~\cite{feynhiggs,mhiggslong,mhiggsAEC,mhcMSSMlong}).

Also not shown are the diagrams with a gauge--Higgs boson or
a Goldstone--Higgs boson self-energy contribution on the external 
Higgs boson leg. They appear in the decay $\Stauzm \to \Stauem h_n$, 
\reffi{fig:stau2stau1h}, with a $Z/G$--$h_n$ transition and in 
the decay $\Stauzm \to \Sneut H^-$, \reffi{fig:stau2snH}, with a
$W^-$/$G^-$--$H^-$ transition.%
\footnote{
  From a technical point of view, the $W^-$/$G^-$--$H^-$ transitions 
  have been absorbed into the respective counterterms, while the 
  $Z/G$--$h_n$ transitions have been calculated explicitly.
}%
~The corresponding self-energy diagram belonging to the process 
$\Stauzm \to \Stauem Z$ or $\Stauzm \to \Sneut W^-$, respectively, yields a
vanishing contribution for external on-shell gauge bosons due  
to $\eps \cdot p = 0$ for $p^2 = \MZ^2$ ($p^2 = \MW^2$), 
where $p$ denotes the external momentum and $\eps$ the polarization
vector of the gauge boson.

Furthermore, in general, in \reffis{fig:stau2stau1h} --
\ref{fig:stau2snW} we have  omitted diagrams with self-energy type
corrections of external (on-shell) particles. 
While the contributions from the real parts of the loop functions are
taken into account via the renormalization constants defined by on-shell
renormalization conditions, the contributions coming from the imaginary
part of the loop functions can result in an additional (real) correction
if multiplied by complex parameters (such as $\Atau$).
In the analytical and numerical evaluation, these diagrams have been taken 
into account via the prescription outlined in \refse{sec:cMSSM},
and their numerical contributions are included in the results
discussed in \refse{sec:numeval}.

Finally it should be noted that the purely loop induced decay channels 
$\Stauzm \to \Stauem \ga$ yield exactly zero due to the fact that the 
decay width is proportional to $\eps \cdot p$ and the photon is on-shell, 
i.e. $\eps \cdot p = 0$.

The diagrams and corresponding amplitudes have been obtained with 
\fa\ version~3.7~\cite{feynarts}. 
The model file, including the MSSM counterterms, 
is largely based on \citere{dissTF}, however adjusted to match exactly 
the renormalization prescription described in \refse{sec:cMSSM},
see also \citeres{SbotRen,Stop2decay,LHCxC,Gluinodecay}.
The further evaluation has been performed with \fc~version 7.3
(and \lt~version 2.7)~\cite{formcalc}.

%%%%%%%%%%%%%%%%%%%%%%%%% F I G U R E %%%%%%%%%%%%%%%%%%%%%%%%%%%%%%%%%%%%%%%%%
%%%%%%%%%%%%%%%%%%%%%%%%% F I G U R E %%%%%%%%%%%%%%%%%%%%%%%%%%%%%%%%%%%%%%%%%

%%%%%%%%%%%%%%%%%%%%%%%%% F I G U R E %%%%%%%%%%%%%%%%%%%%%%%%%%%%%%%%%%%%%%%%%
%%%%%%%%%%%%%%%%%%%%%%%%% F I G U R E %%%%%%%%%%%%%%%%%%%%%%%%%%%%%%%%%%%%%%%%%

%%%%%%%%%%%%%%%%%%%%%%%%% F I G U R E %%%%%%%%%%%%%%%%%%%%%%%%%%%%%%%%%%%%%%%%%
%%%%%%%%%%%%%%%%%%%%%%%%% F I G U R E %%%%%%%%%%%%%%%%%%%%%%%%%%%%%%%%%%%%%%%%%

%%%%%%%%%%%%%%%%%%%%%%%%% F I G U R E %%%%%%%%%%%%%%%%%%%%%%%%%%%%%%%%%%%%%%%%%
%%%%%%%%%%%%%%%%%%%%%%%%% F I G U R E %%%%%%%%%%%%%%%%%%%%%%%%%%%%%%%%%%%%%%%%%

%%%%%%%%%%%%%%%%%%%%%%%%% F I G U R E %%%%%%%%%%%%%%%%%%%%%%%%%%%%%%%%%%%%%%%%%
%%%%%%%%%%%%%%%%%%%%%%%%% F I G U R E %%%%%%%%%%%%%%%%%%%%%%%%%%%%%%%%%%%%%%%%%

%%%%%%%%%%%%%%%%%%%%%%%%% F I G U R E %%%%%%%%%%%%%%%%%%%%%%%%%%%%%%%%%%%%%%%%%
%%%%%%%%%%%%%%%%%%%%%%%%% F I G U R E %%%%%%%%%%%%%%%%%%%%%%%%%%%%%%%%%%%%%%%%%

\subsubsection*{Ultraviolet divergences}

As regularization scheme for the UV-divergences we have used 
constrained differential renormalization~\cite{cdr}, 
which has been shown to be equivalent to dimensional reduction~\cite{dred} 
at the \onel\ level~\cite{formcalc}. 
Thus the employed regularization scheme preserves SUSY~\cite{dredDS,dredDS2}
and guarantees that the SUSY relations are kept intact, e.g. that the 
gauge couplings of the SM vertices and the Yukawa couplings of the 
corresponding SUSY vertices also coincide to \onel\ order in the SUSY
limit. 
Therefore no additional shifts, which might occur when using a different 
regularization scheme, arise.
All UV-divergences cancel to all orders in the final result.

\subsubsection*{Infrared divergences}

The IR-divergences from diagrams with an internal photon have to cancel 
with the ones from the corresponding real soft radiation. 
They are included via analytical formulas following the description 
given in \citere{denner}.
The IR-divergences arising from the diagrams involving a $\ga$ 
are regularized by introducing a photon mass parameter, $\la$. 
All IR-divergences, i.e.\ all divergences in the limit $\la \to 0$, 
cancel to all orders
once virtual and real diagrams for one decay channel are added.

However, in order to achieve the all order cancellation,
special care has to be taken in the decay modes involving scalar 
neutrinos and a $W$~boson. 
Using tree-level stau masses yields a cancellation of 
IR divergences to all orders for all $\Stauzm$ decay modes. However,
inserting the one-loop corrected stau masses (see \refse{sec:slepton}),
as required for consistency, we found cancellation to all orders of the 
related IR divergences, except for the decay mode
$\decayW$. Within this decay the tree-level relation
required by the $SU(2)$ symmetry $M_{\Stau_L}(\Stau) = M_{\Stau_L}(\Sneut)$,
corresponding to
\begin{align}
\label{sleprel}
|U_{\Stau_{11}}|^2 \mstaue^2 + |U_{\Stau_{12}}|^2 \mstauz^2 =
        \msneut^2 + \mtau^2 - \MW^2 \CZb~,
\end{align}
has to be fulfilled to yield a cancellation of {\em all} IR
divergences.%
\footnote{
  \refeq{sleprel} has been deduced via
  \begin{align}
    M_{\Stau_L}^2(\Stau) &= |U_{\Stau_{11}}|^2 \mstaue^2 
                          + |U_{\Stau_{12}}|^2 \mstauz^2 
    - M_Z^2 c_{2\be} (I_\tau^3 - Q_\tau \sw^2) - \mtau^2 ~,\\
    M_{\Stau_L}^2(\Sneu) &= \msneut^2 - I_{\nu_\tau}^3 c_{2\be} \MZ^2 \non
  \end{align}
}
On the other hand, the requirement of on-shell stau masses as well as
an intact $SU(2)$ relation at the one-loop level leads to the necessity
of a shift in the scalar tau masses, see \refeq{MStaushift}.
Therefore \refeq{sleprel} is ``violated'' at the one-loop level, 
introducing a two-loop IR divergence in $\Ga(\decayW)$. 
In order to eliminate this two-loop IR divergence we introduced a 
counterterm in the $\Stauz\Sneut W$ vertex,
\begin{align}
\label{ir}
\dZ{\rm ir} &=
\KL 
  \KKL |U_{\Stau_{11}}|^2 \mstaue^2 + |U_{\Stau_{12}}|^2 \mstauz^2 \KKR
- \KKL |U_{\Stau_{11}}|^2 \mstaue^2 + |U_{\Stau_{12}}|^2 \mstauz^2 \KKR_{\rm shift}
\KR \times {\rm IR~div}
\end{align}
to restore the tree-level $SU(2)$~relation.
The left term in \refeq{ir} contains only `tree-level' values, while
the index `shift' refers to inserting the one-loop masses and mixing matrices.
The IR~divergence has been taken from Eq.~(B.5) of \citere{dittmaier}
(it can also be found in \citere{beenakker}), and reads (in our case):
\begin{align}
{\rm IR~div} &= - \frac{\al}{2 \pi}
                  \frac{x_\tau\, \ln(x_\tau)}{\mstauz \MW (1 - x_\tau^2)} 
                  \ln\KL\frac{\mstauz \MW}{\la^2}\KR
\end{align}
with
\begin{align}
x_\tau &= \frac{\sqrt{1 - 4\,\mstauz \MW/(\msneut^2+i0-(\MW - \mstauz)^2)} - 1}
              {\sqrt{1 - 4\,\mstauz \MW/(\msneut^2+i0-(\MW - \mstauz)^2)} + 1}~,
\end{align}
where $i0$ denotes an infinitesimally small imaginary part.
After including this tree-level relation restoring counterterm we find
an IR finite results to all orders as required.

\subsubsection*{Tree-level formulas}

For completeness we show here also the formulas for 
the tree-level decay widths:
\begin{align}
\Gamma^{\rm tree}(\decayhn) &= \frac{|C(\Stauzm, \Stauem, h_n)|^2\,
                              \la^{1/2}(\mstauz^2,\mstaue^2,m_{h_n}^2)}
                              {16\, \pi\, \mstauz^3}\qquad (n = 1,2,3)~, \\
\Gamma^{\rm tree}(\decayZ) &= \frac{|C(\Stauzm, \Stauem, Z)|^2\,
                             \la^{3/2}(\mstauz^2,\mstaue^2,\MZ^2)}
                             {16\, \pi\, \MZ^2\, \mstauz^3}~, \\
\Gamma^{\rm tree}(\decayNk) &= \Big[ \KL |C(\Stauzm, \tau^-, \neu{k})_{L}|^2
        + |C(\Stauzm, \tau^-, \neu{k})_R|^2 \KR 
          (\mstauz^2 - \mtau^2 - \mneu{k}^2) \non \\
&\qquad - 4\, \re \{C(\Stauzm, \tau^-, \neu{k})_L^*\, 
                    C(\Stauzm, \tau^-, \neu{k})_R \}\,
        \mtau\, \mneu{k} \Big] \times \non \\
&\qquad \frac{\la^{1/2}(\mstauz^2,\mtau^2,\mneu{k}^2)}
             {16\, \pi\, \mstauz^3}\qquad (k = 1,2,3,4)~, \\
\Gamma^{\rm tree}(\decayCmj) &= \KL |C(\Stauzm, \nu_\tau, \cham{j})_{L}|^2
        + |C(\Stauzm, \nu_\tau, \cham{j})_R|^2 \KR (\mstauz^2 - \mcha{j}^2)\, 
        \times \non \\
&\qquad \frac{\la^{1/2}(\mstauz^2, 0, \mcha{j}^2)}
             {16\, \pi\, \mstauz^3}\qquad (j = 1,2)~, \\
\Gamma^{\rm tree}(\decayHm) &= \frac{|C(\Stauzm, \Sneut, H^-)|^2\,
                                \la^{1/2}(\mstauz^2,\msneut^2,\MHp^2)}
                                {16\, \pi\, \mstauz^3}~, \\
\Gamma^{\rm tree}(\decayW) &= \frac{|C(\Stauzm, \Sneut, W^-)|^2\,
                                \la^{3/2}(\mstauz^2,\msneut^2,\MW^2)}
                                {16\, \pi\, \MW^2\, \mstauz^3}~,
\end{align}
where $\la(x,y,z) = (x - y - z)^2 - 4yz$, and the couplings 
$C(a, b, c)$ can be found in the \fa~model files~\cite{feynarts-mf}.
$C(a, b, c)_{L,R}$ denote the part of the coupling which
is proportional to $(\id \mp \ga_5)/2$.

%%%%%%%%%%%%%%%%%%%%%%%%%%%%%%%%%%%%%%%%%%%%%%%%%%%%%%%%%%%%%%%%%%%%%%%%%%%%%%%

\subsubsection*{Comparison with other calculations}

As discussed in the introduction, hardly any numerical
results for stau decays at the loop level are available in the
literature. We employed the program {\tt SFOLD}~\cite{sfold}
to obtain numerical results for scalar tau decays.
{\tt SFOLD} is based on a complete \DRbar\ renormalization
at one-loop order (restricted to the rMSSM), 
but with the possibility also having OS masses 
(instead of the \DRbar\ masses) internal and/or external.%
\footnote{
  It should be noted that we had to use \DRbar\ masses everywhere 
  for our comparison.
}
{\tt SFOLD} uses a running electromagnetic coupling~$\al(Q)$ 
with $Q$ denoting the \DRbar\ scale. 
This leads to a numerical value significantly higher than $\al(0)$, 
see \refeq{alphaNull} below. 
Consequently, our tree-level results differ substantially. 
However, at the loop-level the two results are in better agreement 
as expected. This agreement improves for lower values of $Q$,
but differences at the level of $5\%$ were found for $Q \sim 2 \tev$.

%%%%%%%%%%%%%%%%%%%%%%%%%%%%%%%%%%%%%%%%%%%%%%%%%%%%%%%%%%%%%%%%%%%%%%%%%%%%%%%
%%%%%%%%%%%%%%%%%%%%%%%%%%%%%%%%%%%%%%%%%%%%%%%%%%%%%%%%%%%%%%%%%%%%%%%%%%%%%%%

\section{Numerical analysis}
\label{sec:numeval}

In this section we present a numerical analysis of all 12 decay
channels. In the various figures below we show the partial decay 
widths and their relative correction at the tree-level (``tree'') 
and at the one-loop level (``full''), 
\begin{align}
\Ga^{\rm tree} \equiv \Ga^{\rm tree}(\decayxy)~, \quad
\Ga^{\rm full} \equiv \Ga^{\rm full}(\decayxy)~, \quad
\de\Ga/\Ga^{\rm tree} \equiv \frac{\Ga^{\rm full} - \Ga^{\rm tree}}
                                 {\Ga^{\rm tree}}~,
\label{Garel}
\end{align}
where xy denotes the specific final state.
The total decay width is defined as the sum of all 12 partial
decay widths,  
\begin{align}
\Ga_{\rm tot}^{\rm tree} \equiv \sum_{{\rm xy}} \Ga^{\rm tree}(\decayxy)~, \;\;
\Ga_{\rm tot}^{\rm full} \equiv \sum_{{\rm xy}} \Ga^{\rm full}(\decayxy)~, \;\;
\de\Ga_{\rm tot}/\Ga_{\rm tot}^{\rm tree} \equiv 
\frac{\Ga_{\rm tot}^{\rm full} - \Ga_{\rm tot}^{\rm tree}}
        {\Ga_{\rm tot}^{\rm tree}}~.
\end{align}
We also show the absolute and relative changes of the branching ratios,
\begin{align}
\br^{\rm tree} \equiv \frac{\Ga^{\rm tree}(\decayxy)}
                           {\Ga_{\rm tot}^{\rm tree}}~, \;\;
\br^{\rm full} \equiv \frac{\Ga^{\rm full}(\decayxy)}
                           {\Ga_{\rm tot}^{\rm full}}~, \;\;
\de\br/\br \equiv \frac{\br^{\rm full} - \br^{\rm tree}}{\br^{\rm full}}~. 
\label{brrel}
\end{align}
The last quantity is relevant for an analysis of the impact of the one-loop
corrections on the phenomenology at the LHC and the ILC.

%%%%%%%%%%%%%%%%%%%%%%%%%%%%%%%%%%%%%%%%%%%%%%%%%%%%%%%%%%%%%%%%%%%%%%%%%%%%%

\subsection{Parameter settings}
\label{sec:paraset}

The renormalization scale $\mu_R$ has been set to the mass of the 
decaying particle, i.e.\ $\mu_R = \mstauz$.
The SM parameters are chosen as follows, see also \cite{pdg}:
\begin{itemize}

\item Fermion masses\index{leptonmasses}:
\begin{align}
m_e   &= 0.51099891\mev~, & m_{\nu_e}     &= 0~, \non \\
m_\mu &= 105.658367\mev~, & m_{\nu_{\mu}}  &= 0~, \non \\
\mtau &= 1776.82\mev~,    & m_{\nu_{\tau}} &= 0~, \non \\
m_u &= 62.8\mev~,         & m_d &= 62.8\mev~, \non \\
m_c &= 1.27\gev~,         & m_s &= 101\mev~, \non \\
m_t &= 172.0\gev~,        & m_b &= 4.67\gev~.
\end{align}
$m_u$ and $m_d$ are effective parameters, calculated through the hadronic
contributions to:
\begin{align}
\Delta\alpha_{\text{had}}^{(5)}(\MZ) &= 
      \frac{\alpha}{\pi}\sum_{f = u,c,d,s,b}
      Q_f^2 \Bigl(\ln\frac{\MZ^2}{m_f^2} - \frac 53\Bigr) = 0.02793~.
\end{align}

\item The CKM matrix has been set to unity.

\item Gauge boson masses\index{gaugebosonmasses}:
\begin{align}
\MZ = 91.1876\gev~, \qquad \MW = 80.399\gev~.
\end{align}

\item Coupling constant\index{couplingconstants}:
\begin{align}
\al \equiv \al(0) = 1/137.035999679~.
\label{alphaNull}
\end{align}
\end{itemize}

The Higgs sector quantities (masses, mixings, etc.) have been
evaluated using {\tt FeynHiggs} (version 2.8.6)
\cite{feynhiggs,mhiggslong,mhiggsAEC,mhcMSSMlong}.%
\footnote{
  As default value within {\tt FeynHiggs}, $\mu_R = \mt$ is used.
}

We will show the results for some representative numerical examples. 
The parameters are chosen according to the scenario \SE,
shown in \refta{tab:para}, but with one of the parameters varied.
For the scalar quark sector we have chosen
$M_{\sq_L} = M_{\sq_R} = \tfrac{1}{2} A_{q} = 1000 \gev$ 
($q = u,c,t,d,s,b$) to yield $\MHe \simeq 120 \gev$.
The value of $\MOne$ is fixed via the GUT relation
$\MOne = \tfrac{5}{3} \tan^2\!\theta_{\rm w}\, \MTwo \approx \edz \MTwo$.
The scenarios are defined such that {\em all} decay modes are open
simultaneously to permit an analysis of all channels, i.e.\ not picking
specific parameters for each decay.
We will start with a variation of $\mstauz$, and show later the
results for varying $\phiatau$.
The scenarios are in agreement with the 
MSSM Higgs boson searches at LEP~\cite{LEPHiggsSM,LEPHiggsMSSM},
the Tevatron~\cite{TevHiggsSM} and the LHC~\cite{LHCHiggsSM}.
Furthermore the following exclusion limits \cite{pdg} hold in our scenario%
:
\begin{align}
\mneu{1} &> 46 \gev, \;
\mneu{2} > 62 \gev, \;
\mneu{3} > 100 \gev, \;
\mneu{4} > 116 \gev, \;
\mcha{1} > 94 \gev. %, \non \\
%m_{\Sele} &> 107 \gev, \;
%m_{\Smue} > 94 \gev,
%\mste > 95 \gev, \;
%\msbe > 89 \gev, \;
%\msq > 379 \gev.
\end{align}

%%%%%%%%%%%%%%%%%%%%% T A B L E %%%%%%%%%%%%%%%%%%%%%%%%%%%%%%%%%%%%%%%%%%%%%%
\begin{table}[tb!]
\renewcommand{\arraystretch}{1.5}
\BC
\begin{tabular}{|c||c|c|c|c|c|c|c|c|c|c|c|c|c|}
\hline
Scen.\ & $\tb$ & $\MHp$ & $\mstauz$ & $\mstaue$ & $M_{\sq_{L,R}}$
& $\mu$ & $A_l$ & $A_q$ & $M_1$ & $M_2$ & $M_3$
\\ \hline\hline
\SE & 5 & 200 & 550 & $\tfrac{1}{2} \mstauz$ & 1000 & 150
    & $\tfrac{9}{5} \mstauz$ & 2000 & $\sim\tfrac{1}{2} M_2$ & 250 & 1500
\\ \hline
\end{tabular}
\caption{MSSM input parameters for the initial numerical investigation; 
  all masses are in~GeV. 
  In our analysis $M_{\slep_L}$ and $M_{\slep_R}$ are chosen such that the 
  values of $\mstaue$ and $\mstauz$ are realized.
  For the $\Stau$~sector the shifts in $M_{\Stau_{L,R}}(\Stau)$ as 
  defined in \refeqs{MStaushift} and \eqref{backshift} are taken into account.
  $M_{\sq_{L,R}}$ denote the diagonal soft SUSY-breaking parameters in the
  scalar quark mass matrices, while $A_q$ is the trilinear squark Higgs
  coupling, and $M_3$ denotes the gluino mass parameter.
  The values for $A_f$ ($f = \tau, t, b, \ldots$) are chosen such 
  that charge- and color-breaking minima are avoided~\cite{ccb}.
}
\label{tab:para}
\EC
\renewcommand{\arraystretch}{1.0}
\end{table}
%%%%%%%%%%%%%%%%%%%%% T A B L E %%%%%%%%%%%%%%%%%%%%%%%%%%%%%%%%%%%%%%%%%%%%%%

%%%%%%%%%%%%%%%%%%%%% T A B L E %%%%%%%%%%%%%%%%%%%%%%%%%%%%%%%%%%%%%%%%%%%%%%
\begin{table}[tb!]
\renewcommand{\arraystretch}{1.5}
\BC
\begin{tabular}{|c|c|c||c|c|c|}
\hline
\multicolumn{3}{|c||}{Without shifts} &
\multicolumn{3}{|c|}{With shifts} \\
\hline
~$\mstaue$~ & ~$\mstauz$~ & ~$\msneut$~ &
~$\mstaue$~ & ~$\mstauz$~ & ~$\msneut$~
\\ \hline\hline
275.000 & 550.000 & 263.924 & 
274.478 & 550.000 & 263.924
\\ \hline
\end{tabular}
\caption{
  The stau and tau-sneutrino masses in \SE\ for the numerical investigation; 
  at the right-hand side of the table the shifts as defined in 
  \refeqs{MStaushift} and \eqref{backshift} have been taken into account.
  All masses are in GeV and rounded to one MeV.
}
\label{tab:stau}
\EC
\renewcommand{\arraystretch}{1.0}
\end{table}
%%%%%%%%%%%%%%%%%%%%% T A B L E %%%%%%%%%%%%%%%%%%%%%%%%%%%%%%%%%%%%%%%%%%%%%%

A few examples of the stau and sneutrino masses in \SE\ are 
shown in \refta{tab:stau}. We assume SUSY mass scales that allow
for the copious production of the colored particles at the LHC, 
with the subsequent cascade decay to uncolored particles we are
interested in.
Furthermore, in \SE\ the production of $\Stauzm$ at the ILC(1000), i.e.\ with 
$\sqrt{s} = 1000 \gev$, via $e^+e^- \to \aStaue\Stauzm$ will be possible, 
with all the subsequent decay modes (\ref{ststphi}) -- (\ref{stnucha}) 
being open. The clean environment of the ILC would permit a detailed study 
of the scalar tau decays.
We find, depending on the mixing in the stau sector, cross sections up to 
$\si(e^+e^- \to \aStaue\Stauzm) \sim 1~{\rm fb}$, where these larger cross 
sections are found for larger mixing. Even larger cross sections are found 
in the case of $\si(e^+e^- \to \aStaue\Stauem)$.
An integrated luminosity of $\sim 1\, \iab$ would yield up to~1000 scalar 
taus for $\si = 1~{\rm fb}$.
The ILC environment together with such high numbers of produced staus would 
result in an accuracy of the relative branching ratio~(\refeq{brrel}) close 
to the statistical uncertainty: a BR of 30\% could be determined down to 
$\sim 5\%$.
Depending on the combination of allowed decay channels a determination of 
the branching ratios at the few per-cent level might be achievable in the 
high-luminosity running of the ILC(1000).

The numerical results we will show in the next subsections are of course 
dependent on choice of the SUSY parameters. Nevertheless, they give an 
idea of the relevance of the full one-loop corrections.
As an example, the largest decay widths are $\Ga(\decayNi{1,2})$, 
dominating the total decay width, $\Ga_{\rm tot}$, and thus the various 
branching ratios. 
This is due to the strong bino component in $\neu{1,2}$ in combination 
with a relatively small mixing in the $\Stau$~sector.
For other choices of $\mu$, $M_1$, $M_2$, 
e.g.\ $\mu \ll M_{1,2}$ and/or larger mixing in the $\Stau$~sector,
the light neutralinos would be higgsino dominated and the decay widths would
turn out to be substantially smaller. Consequently, the corrections 
to the (other) decay widths would stay the same, but the branching
ratios would look very different. 
Channels (and their respective one-loop corrections) that may look 
unobservable due to the smallness of their BR in the plots shown below, 
could become important if other channels are kinematically forbidden.

%%%%%%%%%%%%%%%%%%%%%%%%%%%%%%%%%%%%%%%%%%%%%%%%%%%%%%%%%%%%%%%%%%%%%%%%%%%%%%

\subsection{Full one-loop results for varying \boldmath{$\mstauz$}}
\label{sec:full1L}

The results shown in this and the following subsections consist of 
``tree'', which denotes the tree-level value and of ``full'', which is
the partial decay width including {\em all} one-loop 
corrections as described in \refse{sec:calc}.
We start the numerical analysis with partial decay widths of $\Stauzm$
evaluated as a function of $\mstauz$, 
starting at $\mstauz = 220 \gev$ up to $\mstauz = 2 \tev$, which
roughly coincides with the reach of CLIC. 
The upper panels contain the results for the absolute
value of the various  partial decay widths, $\Ga(\decayxy)$ (left) and
the relative correction from the full one-loop contributions
(right). The lower panels show the same 
results for $\br(\decayxy)$.

Since in this section all parameters are chosen to be real no
contributions from 
absorptive parts of self-energy type corrections on external legs can
contribute. This will be different in \refse{sec:full1Lphiat}.

\begin{figure}[htb!]
\begin{center}
\begin{tabular}{c}
\includegraphics[width=0.49\textwidth,height=7.5cm]{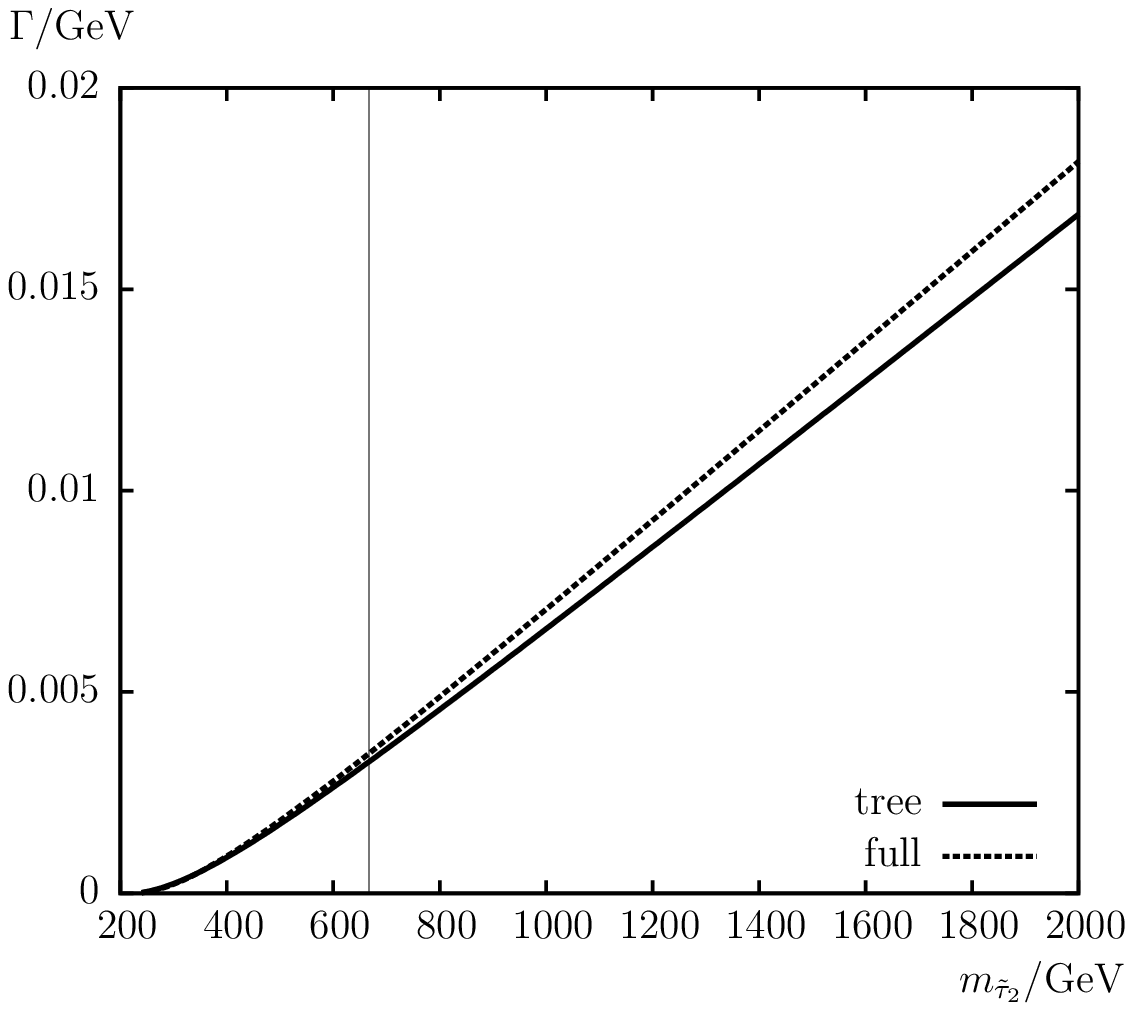}
\hspace{-4mm}
\includegraphics[width=0.49\textwidth,height=7.5cm]{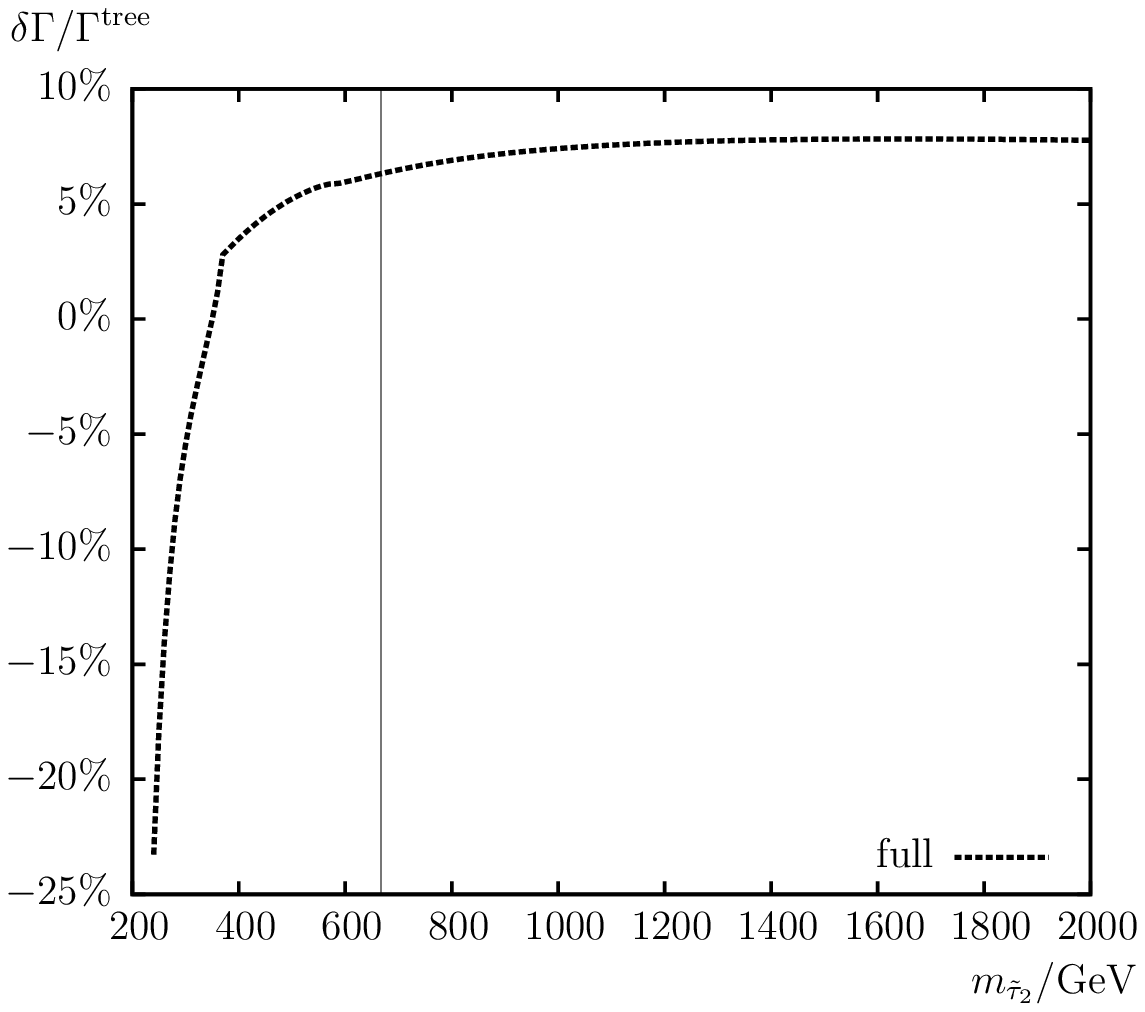} 
\\[4em]
\includegraphics[width=0.49\textwidth,height=7.5cm]{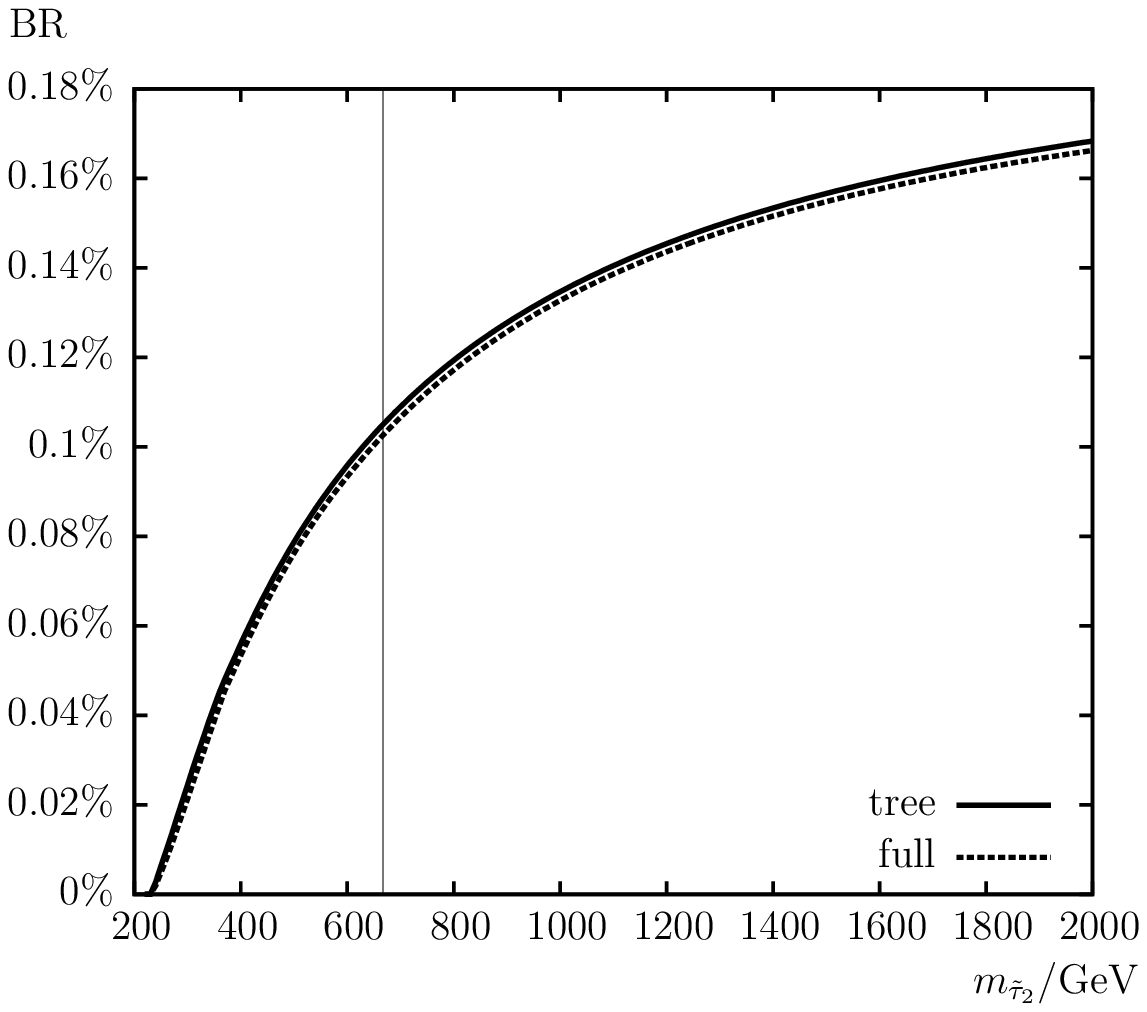}
\hspace{-4mm}
\includegraphics[width=0.49\textwidth,height=7.5cm]{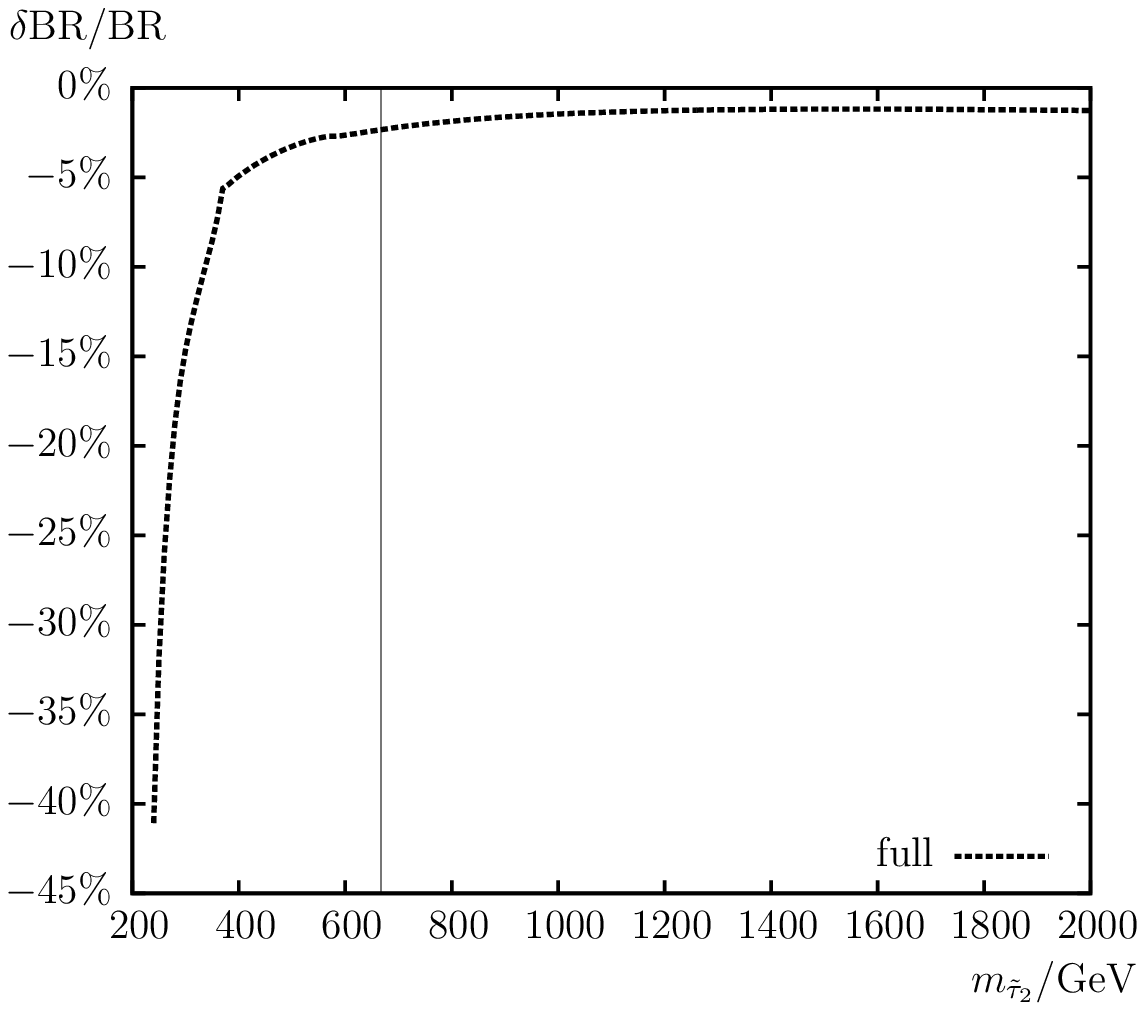}
\end{tabular}
\vspace{2em}
\caption{
  $\Ga(\decayh)$. Tree-level and full one-loop corrected partial decay widths 
  are shown with the parameters chosen according to \SE\
  (see \refta{tab:para}), with $\mstauz$ varied.
  The upper left plot shows the partial decay width, the upper right plot shows 
  the corresponding relative size of the corrections.
  The lower left plot shows the BR, the lower right plot shows 
  the relative correction of the BR.
  The vertical lines indicate where $\mstauz + \mstaue = 1000 \gev$, 
  i.e.\ the maximum reach of the ILC(1000).
}
\label{fig:mst2.stau2stau1h1}
\end{center}
\end{figure}

\begin{figure}[htb!]
\begin{center}
\begin{tabular}{c}
\includegraphics[width=0.49\textwidth,height=7.5cm]{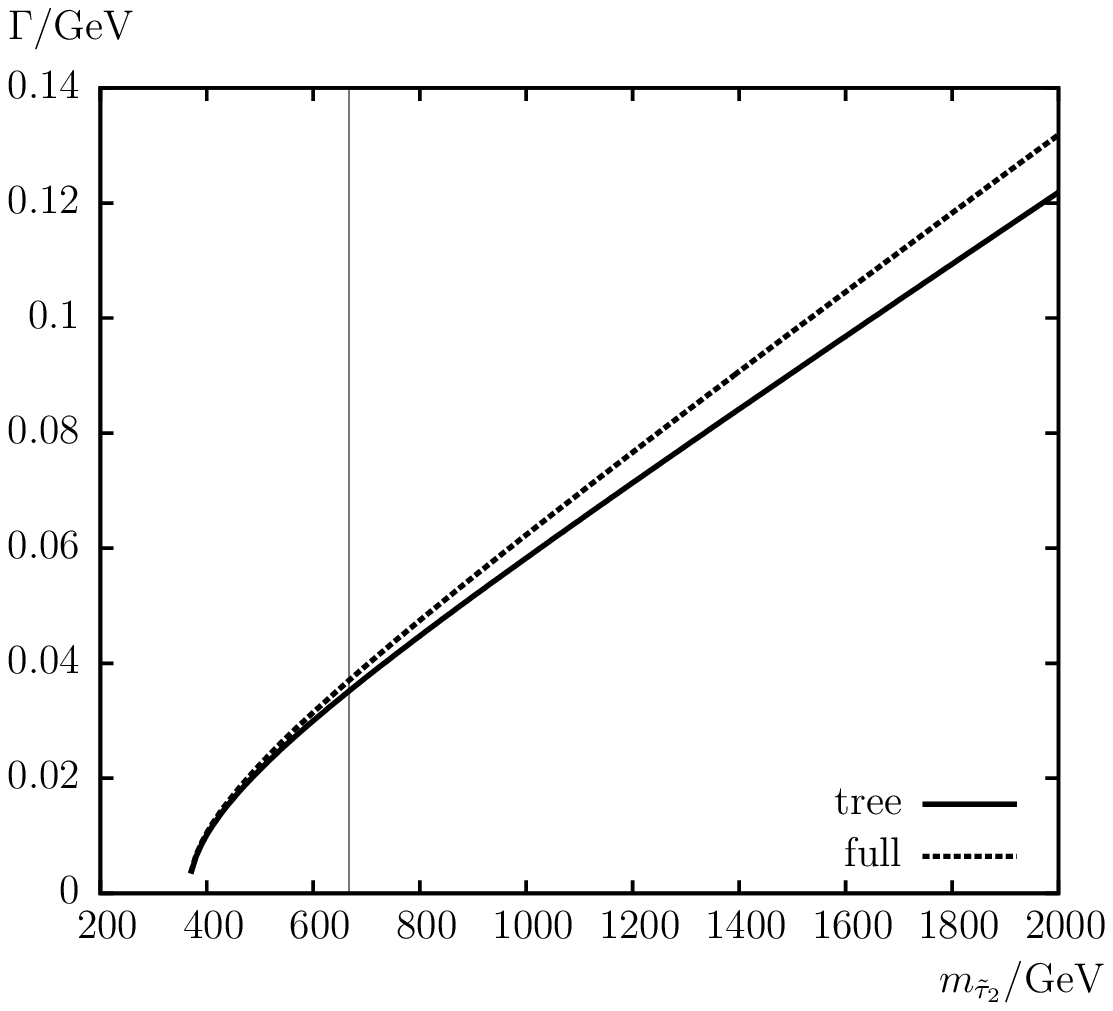}
\hspace{-4mm}
\includegraphics[width=0.49\textwidth,height=7.5cm]{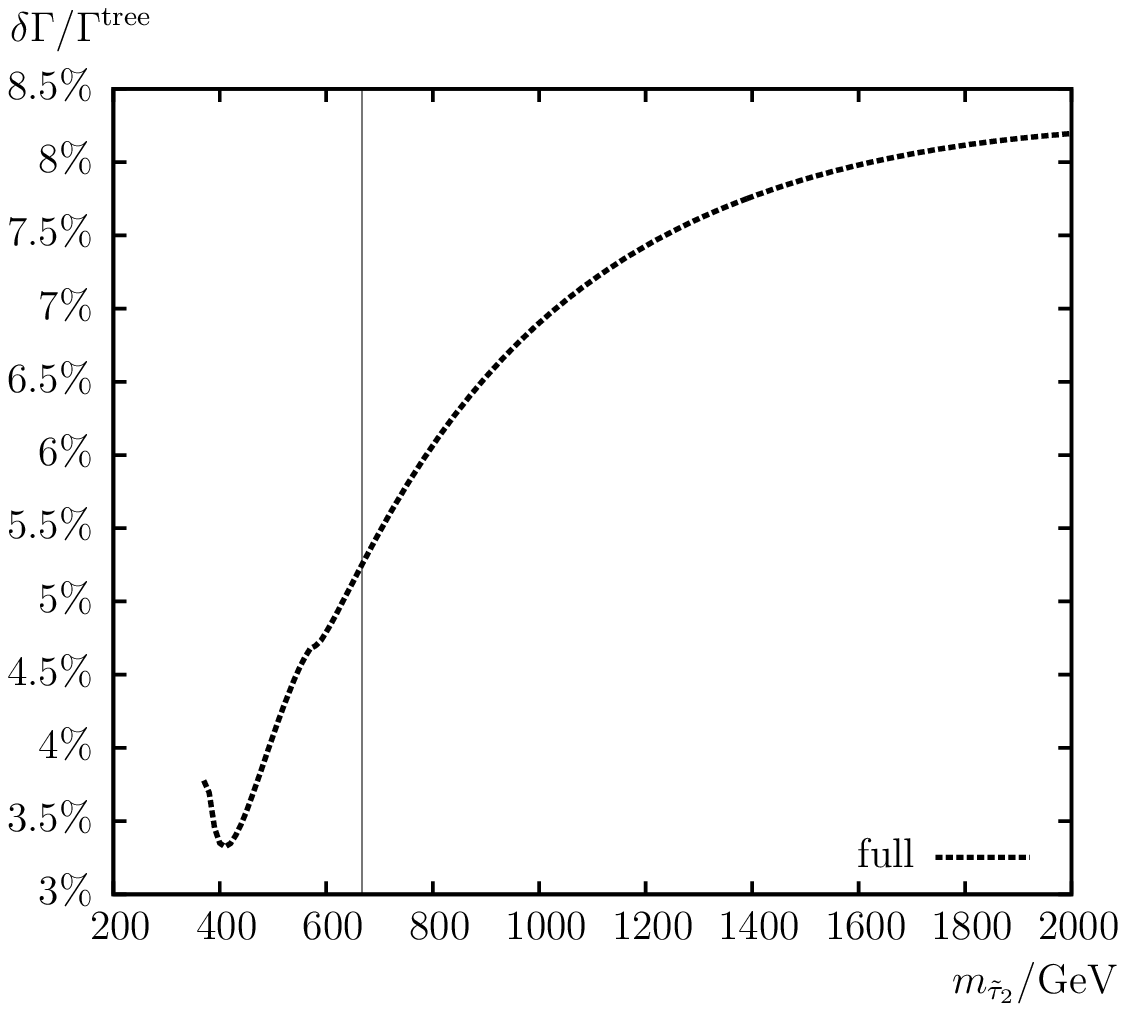} 
\\[4em]
\includegraphics[width=0.49\textwidth,height=7.5cm]{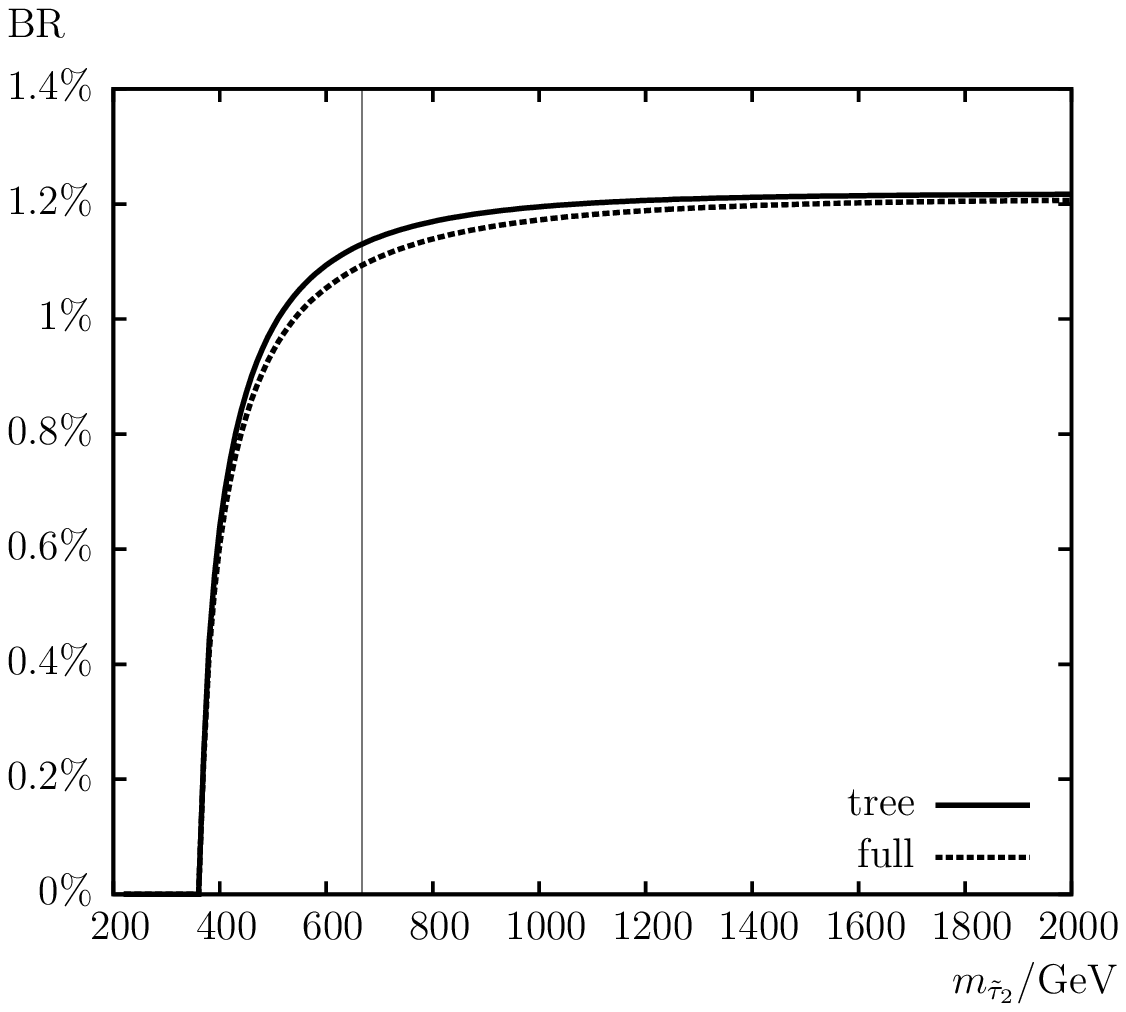}
\hspace{-4mm}
\includegraphics[width=0.49\textwidth,height=7.5cm]{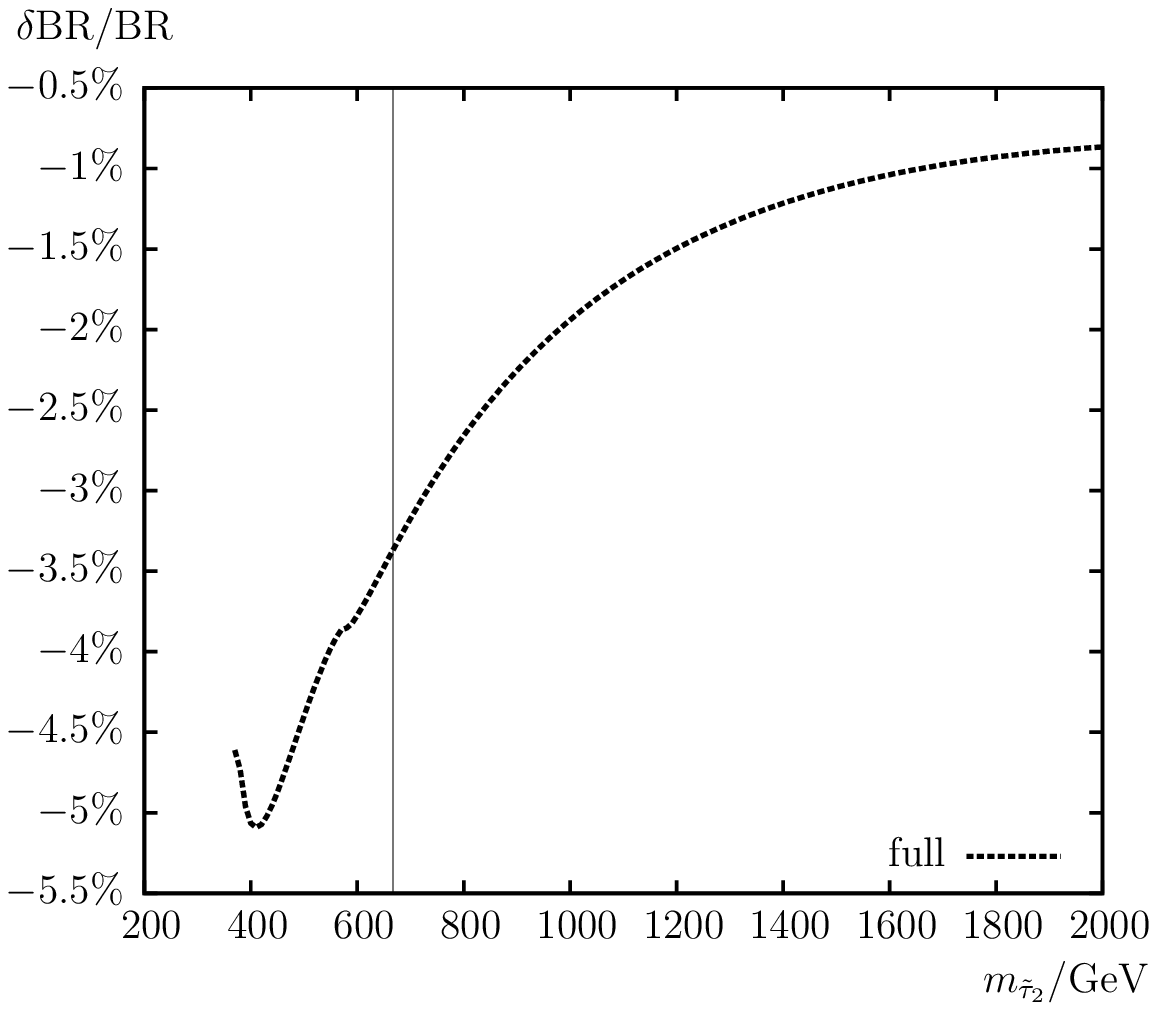}
\end{tabular}
\vspace{2em}
\caption{
  $\Ga(\decayH)$. Tree-level and full one-loop corrected partial decay widths 
  are shown with the parameters chosen according to \SE\ 
  (see \refta{tab:para}), with $\mstauz$ varied.
  The upper left plot shows the partial decay width, the upper right plot shows 
  the corresponding relative size of the corrections.
  The lower left plot shows the BR, the lower right plot shows 
  the relative correction of the BR.
  The vertical lines indicate where $\mstauz + \mstaue = 1000 \gev$, 
  i.e.\ the maximum reach of the ILC(1000).
}
\label{fig:mst2.stau2stau1h2}
\end{center}
\end{figure}

\begin{figure}[htb!]
\begin{center}
\begin{tabular}{c}
\includegraphics[width=0.49\textwidth,height=7.5cm]{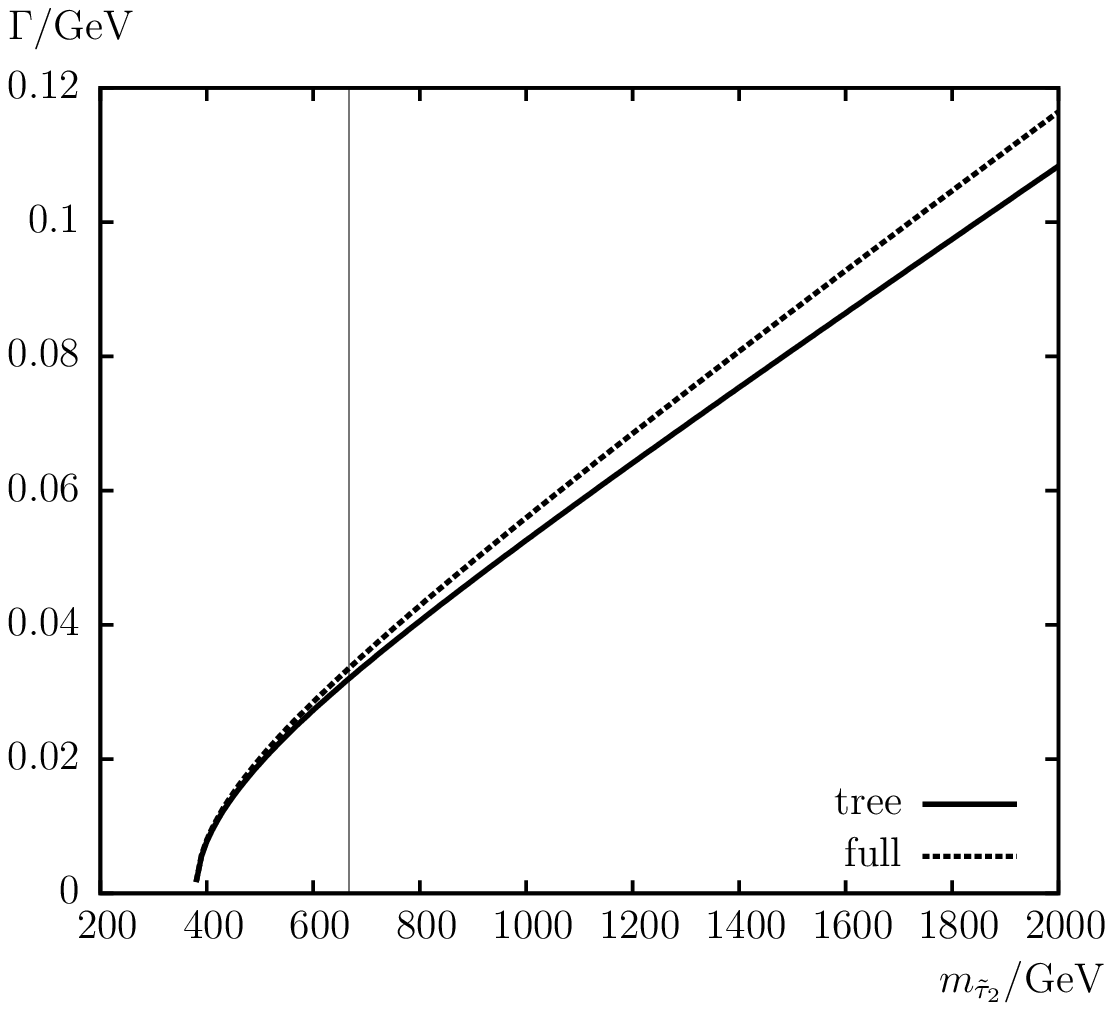}
\hspace{-4mm}
\includegraphics[width=0.49\textwidth,height=7.5cm]{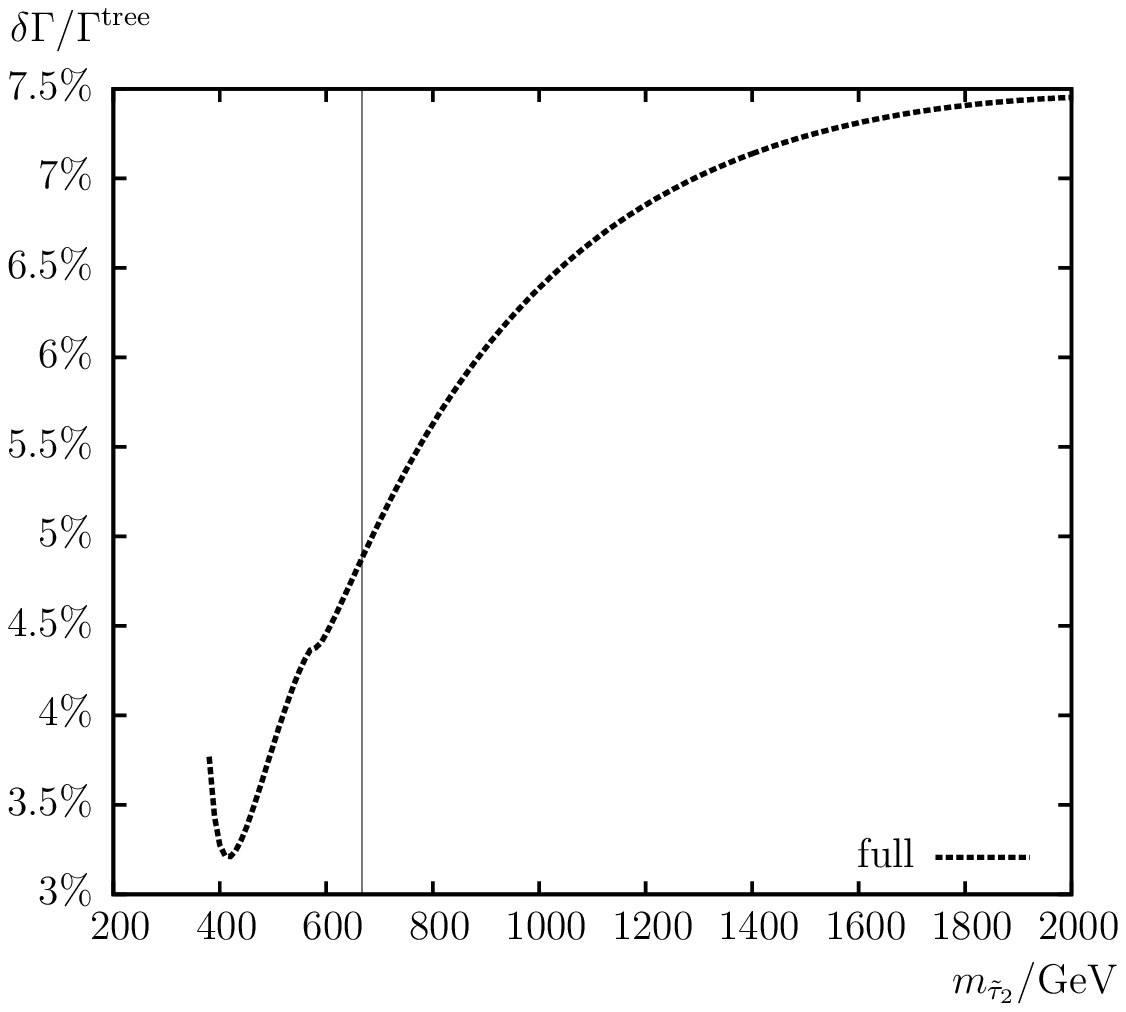} 
\\[4em]
\includegraphics[width=0.49\textwidth,height=7.5cm]{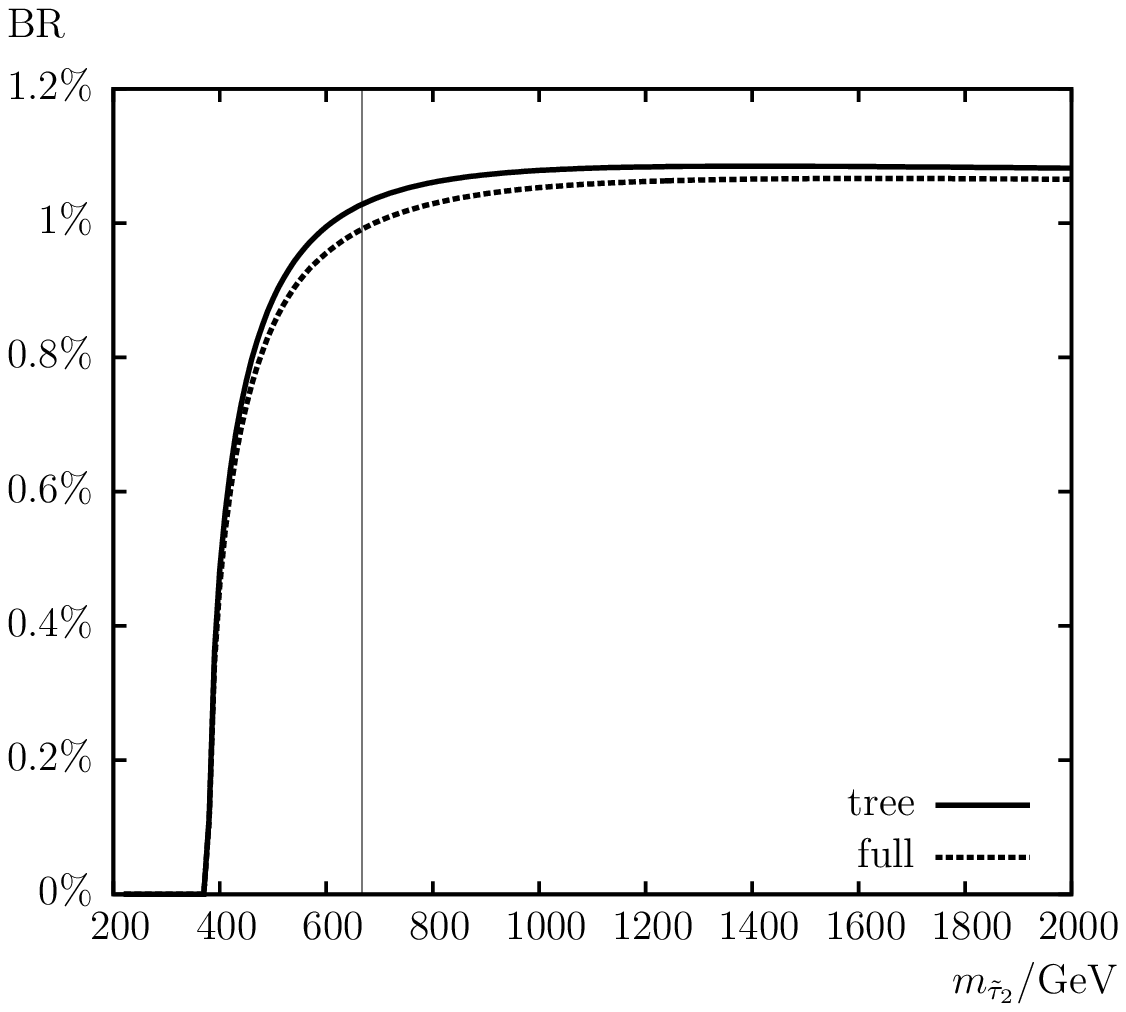}
\hspace{-4mm}
\includegraphics[width=0.49\textwidth,height=7.5cm]{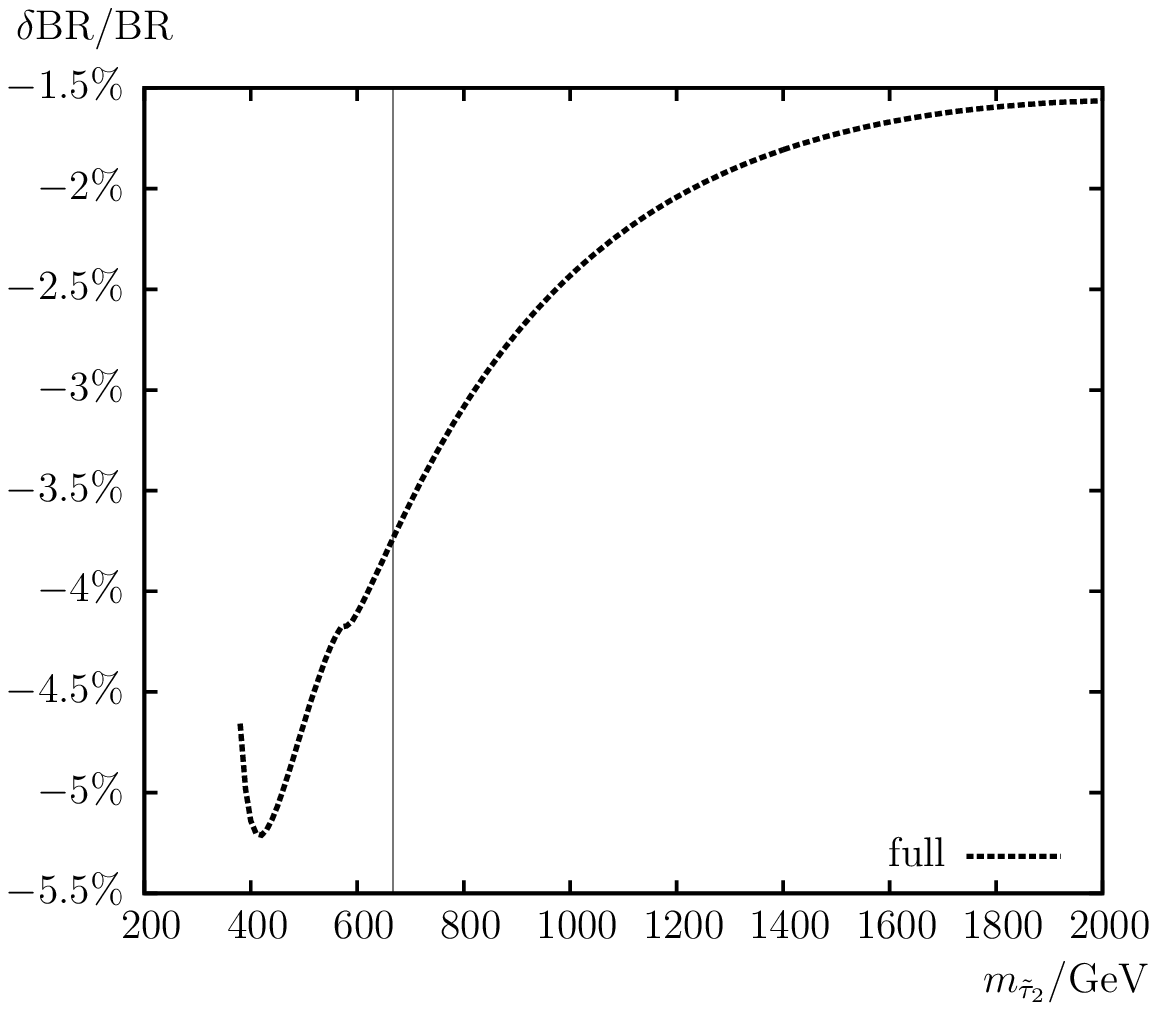}
\end{tabular}
\vspace{2em}
\caption{
  $\Ga(\decayA)$. Tree-level and full one-loop corrected partial decay widths 
  are shown with the parameters chosen according to \SE\ 
  (see \refta{tab:para}), with $\mstauz$ varied.
  The upper left plot shows the partial decay width, the upper right plot shows 
  the corresponding relative size of the corrections. 
  The lower left plot shows the BR, the lower right plot shows 
  the relative correction of the BR.
  The vertical lines indicate where $\mstauz + \mstaue = 1000 \gev$, 
  i.e.\ the maximum reach of the ILC(1000).
}
\label{fig:mst2.stau2stau1h3}
\end{center}
\end{figure}

In \reffi{fig:mst2.stau2stau1h1} -- \ref{fig:mst2.stau2stau1h3} we show the
results for the process $\decayhn$ ($n = 1,2,3$) as a function
of $\mstauz$. 
Here, as well as in the other channels some dips and peaks appear,
which are due to various thresholds in self-energy or vertex diagram
contributions. Three kinks that are present in principle in {\em all}
decays (with the partial exception of $\decayHm$, see below), 
but that are only partially visible, appear at%
\footnote{
  Here and below we round most of the values to one GeV.
}%
~$\mstauz \approx 364 \gev \approx \MA + \mstaue$, 
$366 \gev \approx \MHp + \msneut$, $373 \gev \approx \MH + \mstaue$. 
The thresholds appear in the stau self-energies and thus enter via
$\de\mstauz^2$, $\dZZm{\Stau}_{12,22}$ and $\de Y_\tau$. Visible
in the plots is only the kink at $\mstauz \approx 366 \gev$. 
In the decays $\decayhn$ at $\mstauz \approx 583 \gev$ we find a kink
due to $\mstaue \approx \mtau + \mneu{4}$. 
One can see that the size of the corrections of the partial decay widths
is especially large very close to the production threshold from which on
the considered decay mode is kinematically possible.%
\footnote{
  It should be noted that a calculation very close to threshold requires 
  the inclusion of additional (non-relativistic) contributions, which is 
  beyond the scope of this paper. Consequently, very close to threshold 
  our calculation (at the tree- or loop-level) does not provide a very 
  accurate description of the decay width.
}%
~Away from this threshold relative corrections of 
$\sim +5\%, +6\%, +6\%$ are found for $h_1, h_2, h_3$, respectively. 
In (all) the plots the value of $\mstauz$ for which 
$\mstaue + \mstauz = 1000 \gev$ is shown as a vertical line, 
i.e.\ the region where the heavier stau can be produced at the
ILC(1000). In these regions the size 
of the corrections is only slightly smaller than the numbers above.
The BRs are at the per-cent level for all three channels.
The relative change in the BRs for the 
masses accessible at the ILC(1000) are about $-5\%$, $-4\%$, $-4\%$
for $h_1$, $h_2$, $h_3$, respectively. For lager masses, only accessible at
CLIC, the one-loop corrections are even smaller.
Depending on the MSSM parameters (and the channels
kinematically allowed) the one-loop contributions presented here can be
relevant for analyses at the ILC and potentially as well as at the LHC.

\begin{figure}[htb!]
\begin{center}
\begin{tabular}{c}
\includegraphics[width=0.49\textwidth,height=7.5cm]{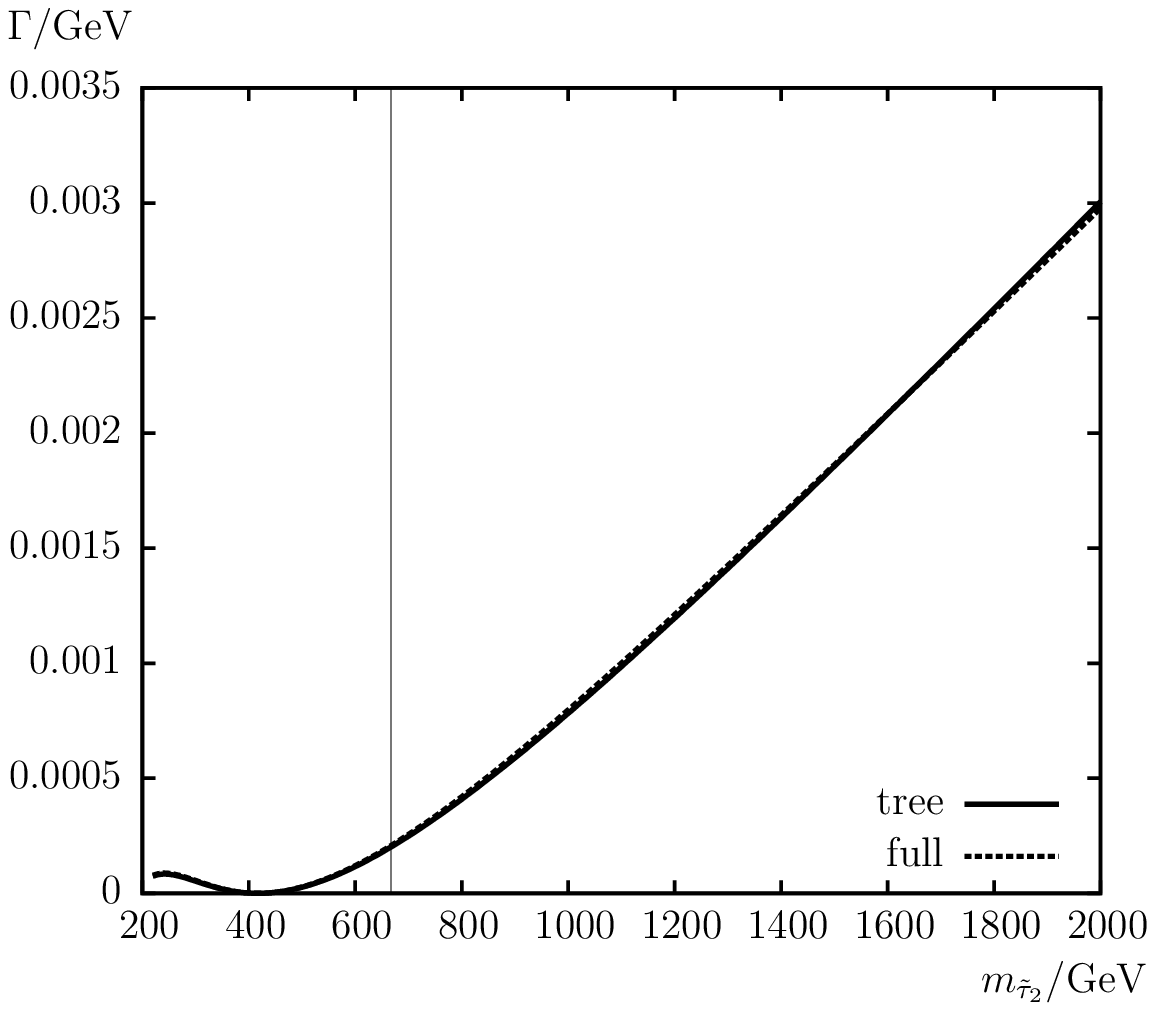}
\hspace{-4mm}
\includegraphics[width=0.49\textwidth,height=7.5cm]{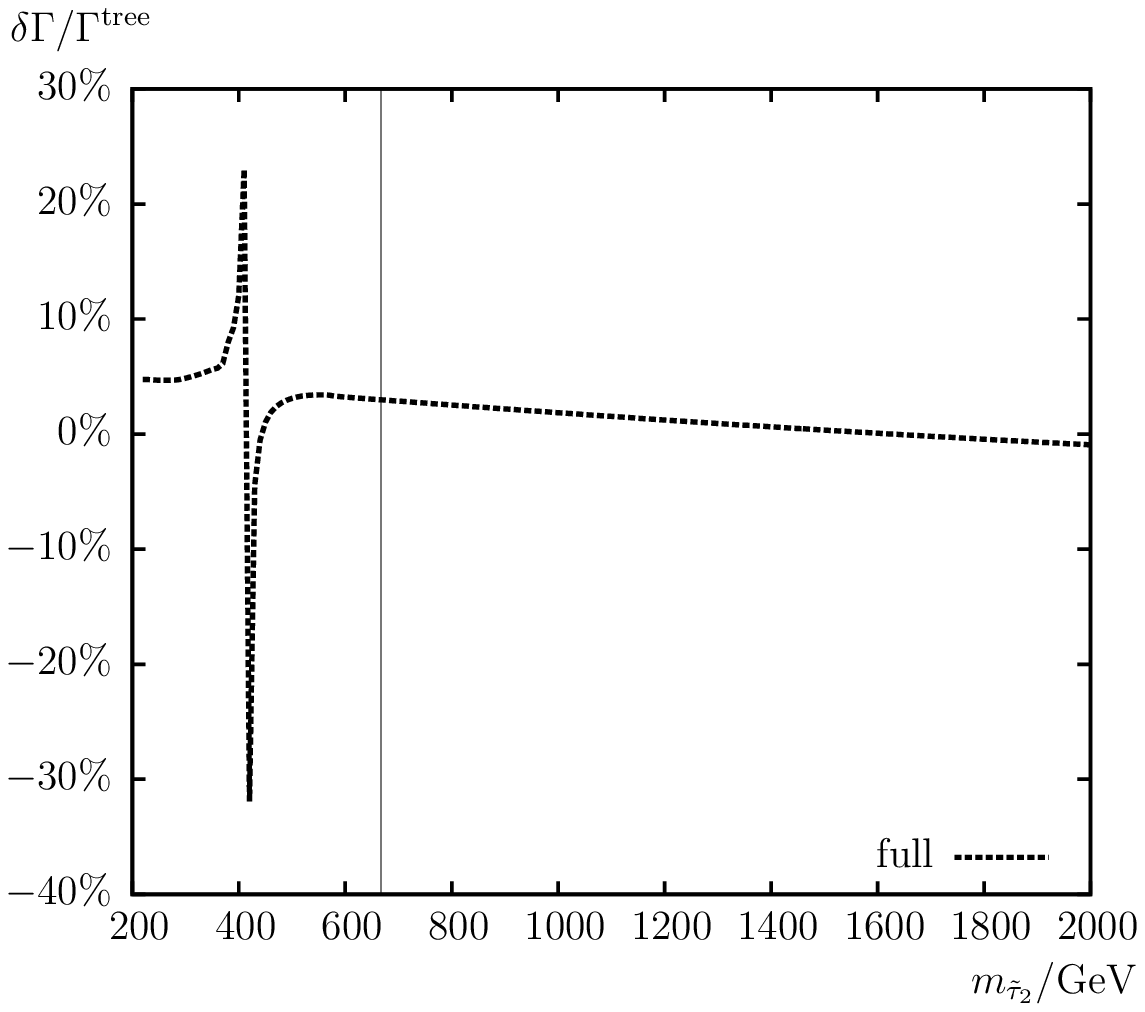} 
\\[4em]
\includegraphics[width=0.49\textwidth,height=7.5cm]{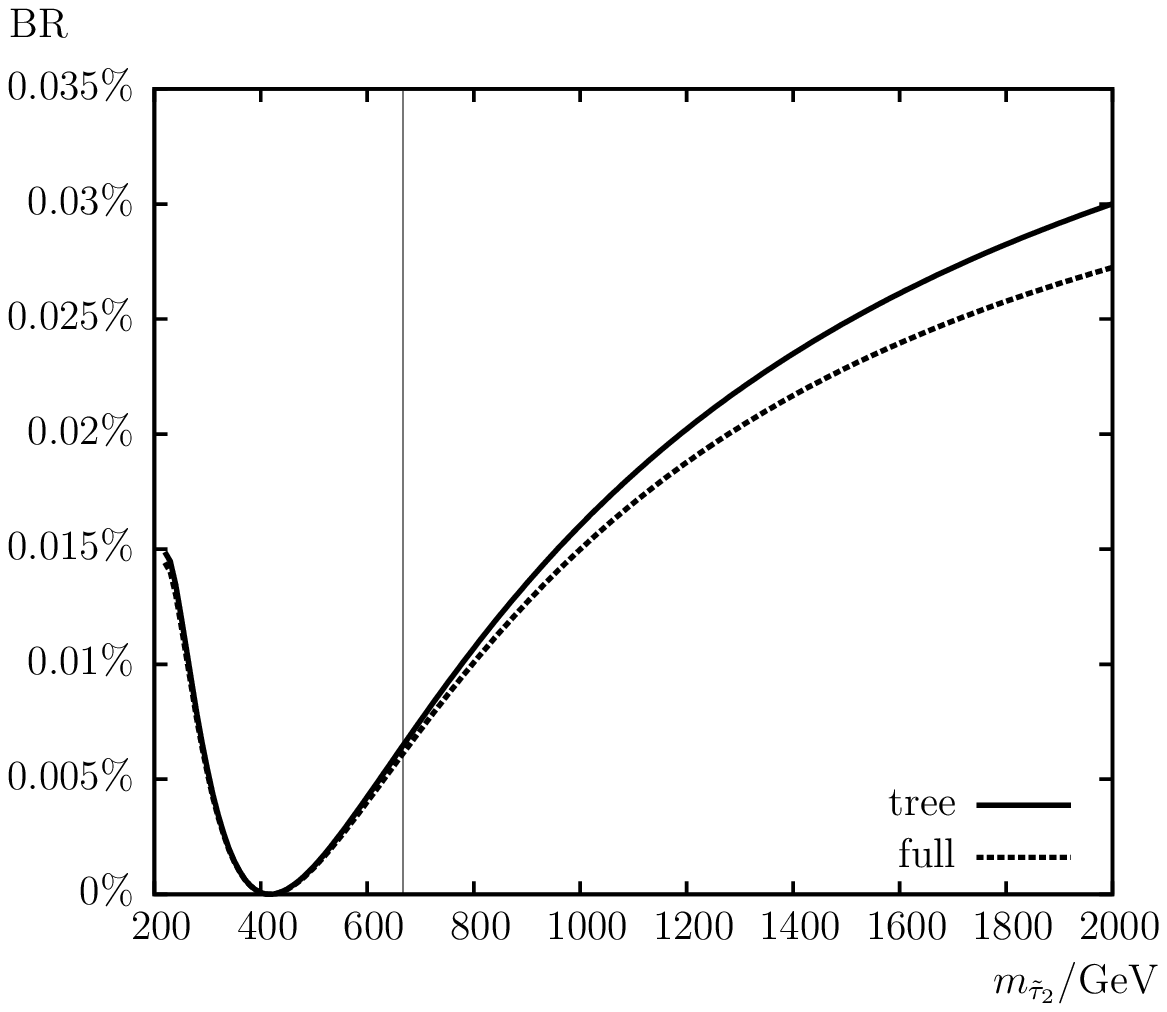}
\hspace{-4mm}
\includegraphics[width=0.49\textwidth,height=7.5cm]{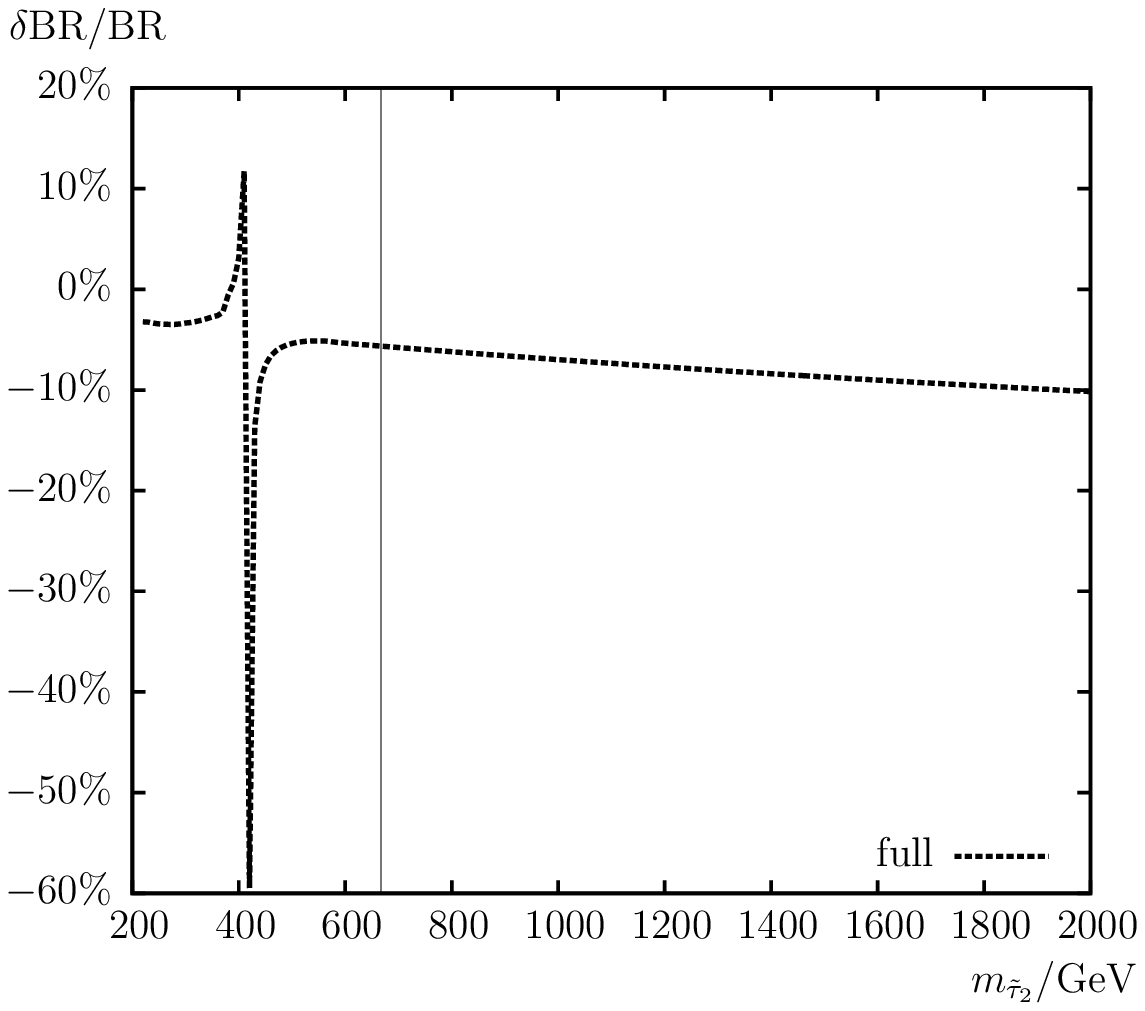}
\end{tabular}
\vspace{2em}
\caption{
  $\Ga(\decayZ)$. Tree-level and full one-loop corrected partial decay widths 
  are shown with the parameters chosen according to \SE\ 
  (see \refta{tab:para}), with $\mstauz$ varied.
  The upper left plot shows the partial decay width, the upper right plot shows 
  the corresponding relative size of the corrections. 
  The lower left plot shows the BR, the lower right plot shows 
  the relative correction of the BR.
  The vertical lines indicate where $\mstauz + \mstaue = 1000 \gev$, 
  i.e.\ the maximum reach of the ILC(1000).
}
\label{fig:mst2.stau2stau1Z}
\end{center}
\end{figure}

Next, in \reffi{fig:mst2.stau2stau1Z} we show results for the decay
$\Ga(\decayZ)$. The dips due to the thresholds in 
$\dZZm{\Stau}_{12,22}$, $\de Y_{\tau}$ and $\de\mstauz^2$ 
are the same as before. 
Furthermore at $\mstauz \approx 416.7 \gev$ a sign change in the stau
mixing matrix takes place, and the tree-level result comes out zero for 
$U_{\Stau} = \id$. Consequently, around this value the loop corrections
are substantially larger than the tree-level result, however, not
invalidating the perturbative series.
The relative corrections to the partial decay width in \SE\ range between 
+5\% at low $\mstauz$, i.e.\ in the ``ILC(1000) regime'',  
to about zero at large $\mstauz$, with the exception of the region around
$\mstauz \approx 416.7 \gev$.

\begin{figure}[htb!]
\begin{center}
\begin{tabular}{c}
\includegraphics[width=0.49\textwidth,height=7.5cm]{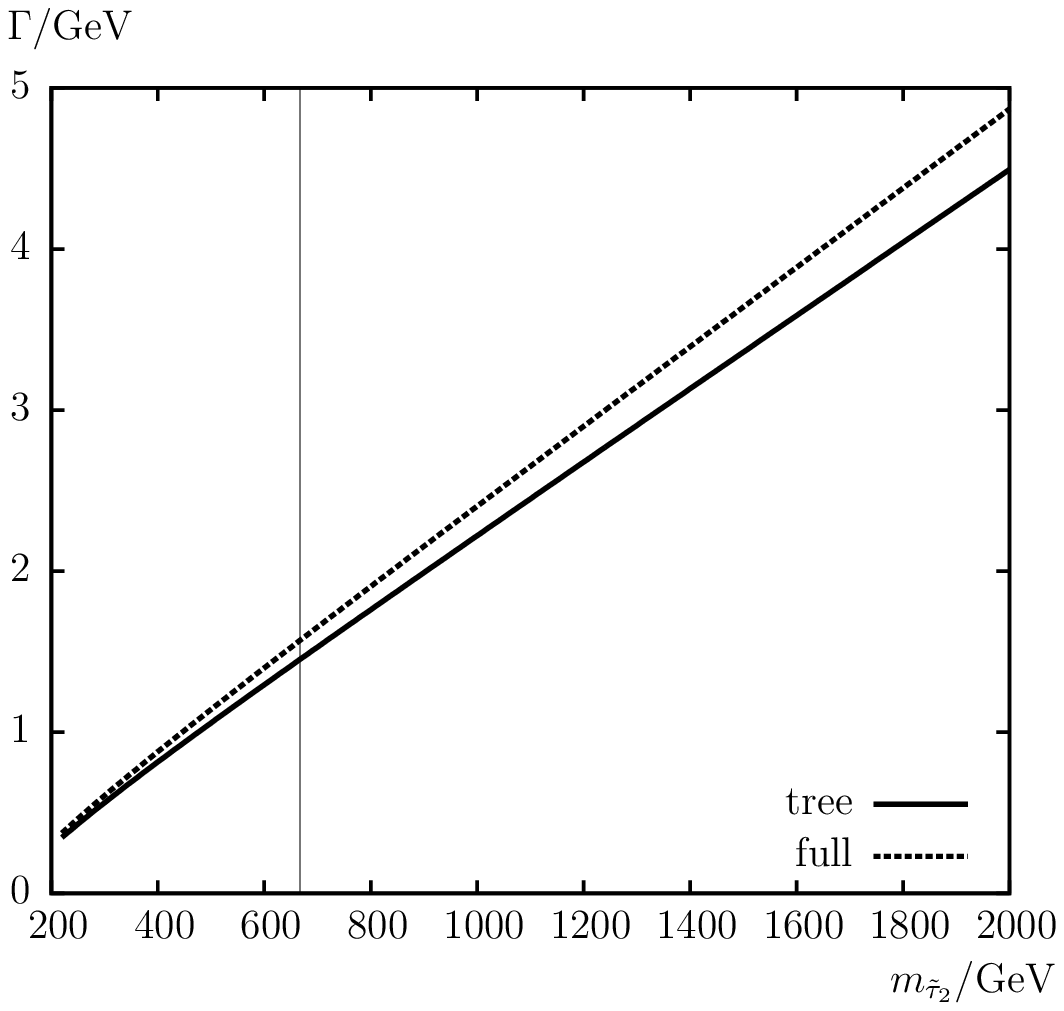}
\hspace{-4mm}
\includegraphics[width=0.49\textwidth,height=7.5cm]{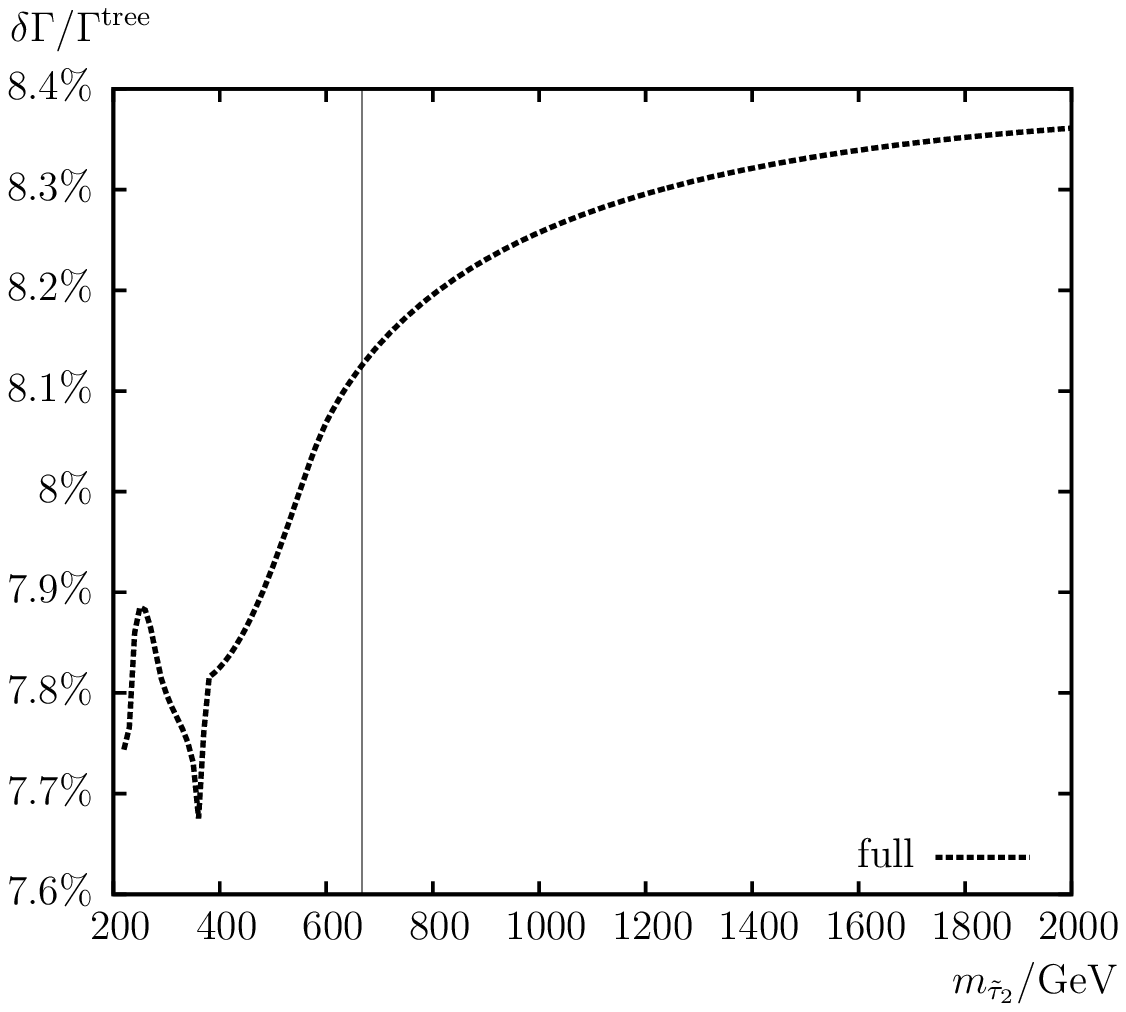}
\\[4em]
\includegraphics[width=0.49\textwidth,height=7.5cm]{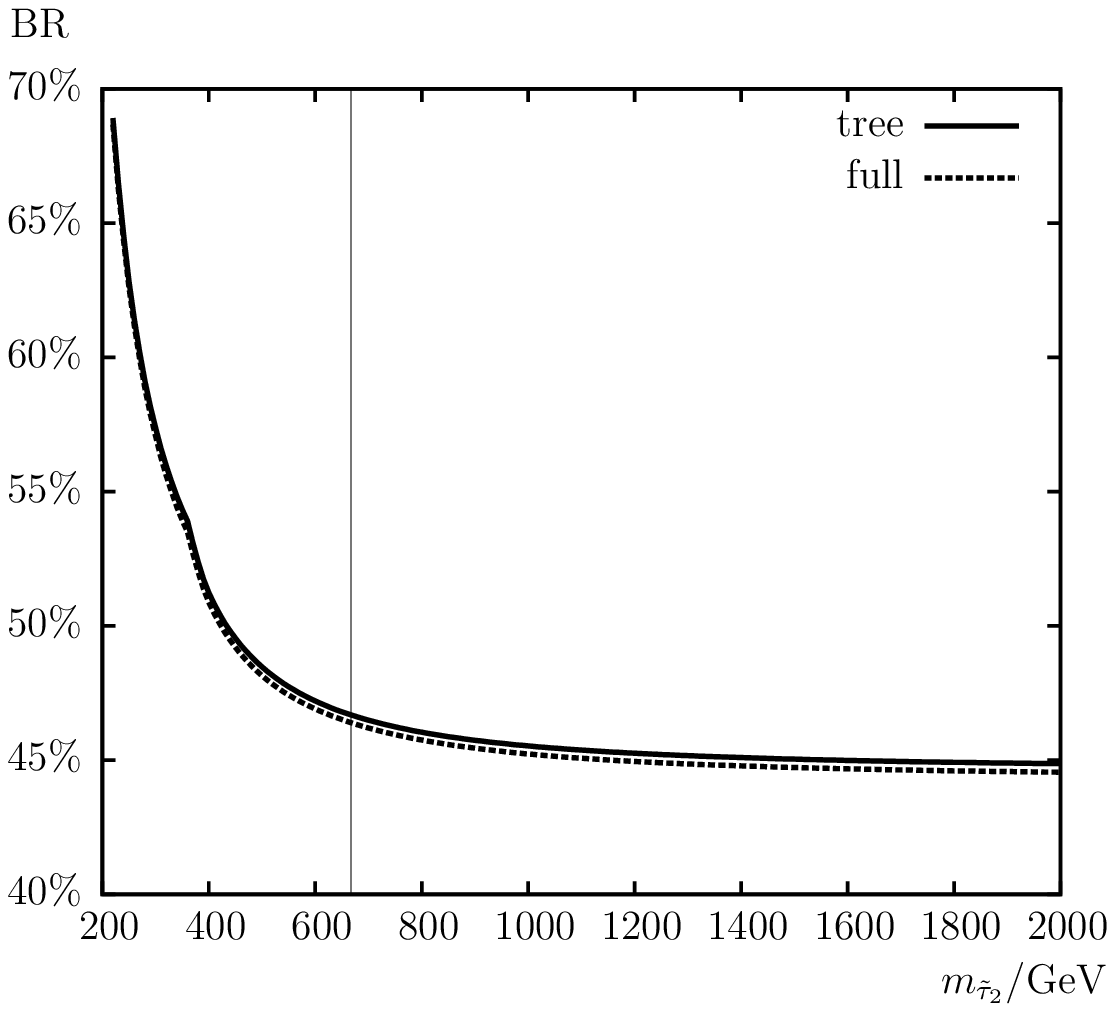}
\hspace{-4mm}
\includegraphics[width=0.49\textwidth,height=7.5cm]{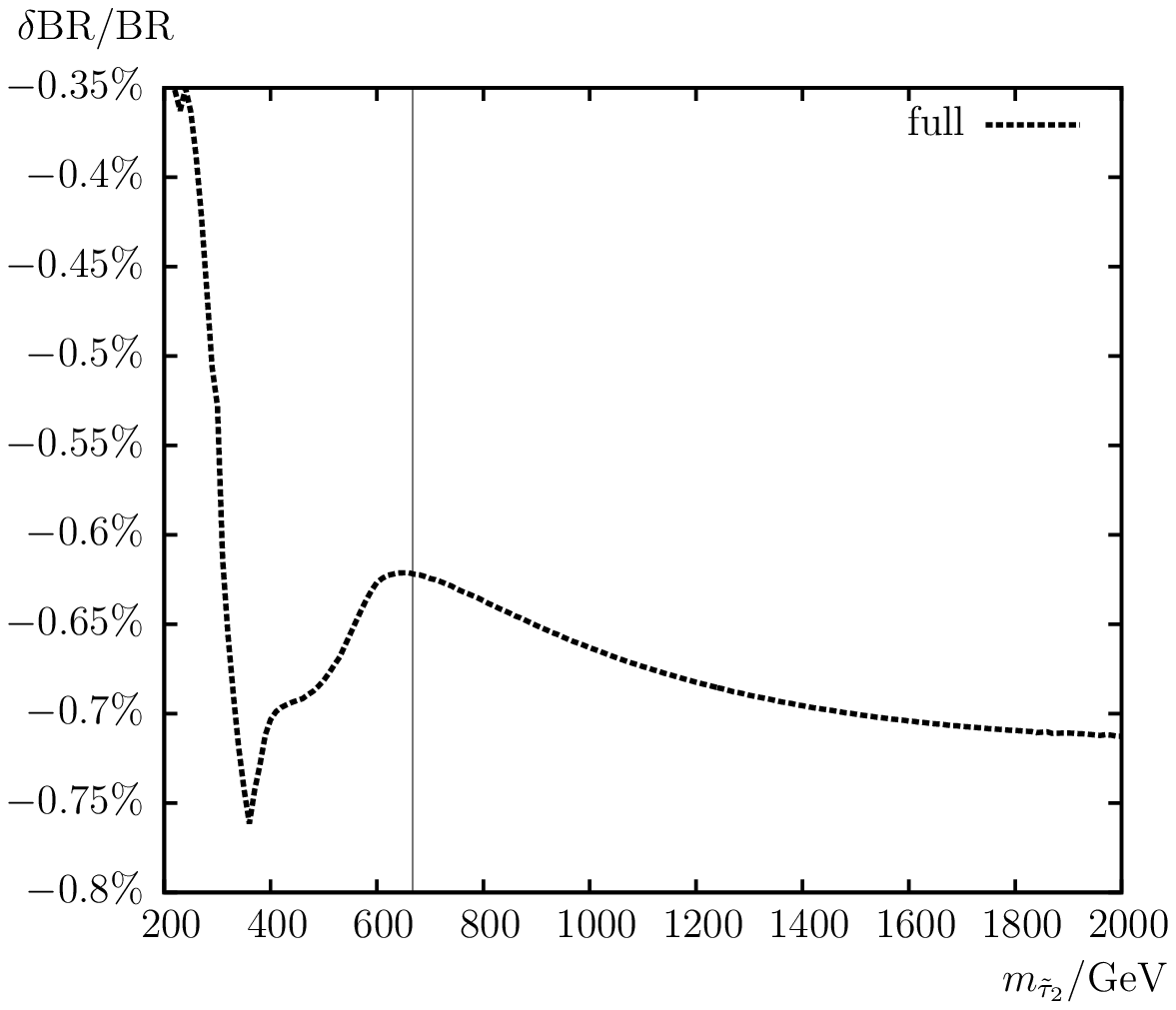}
\end{tabular}
\vspace{2em}
\caption{
  $\Ga(\decayNe)$. Tree-level and full one-loop corrected partial decay widths 
  are shown with the parameters chosen according to \SE\ 
  (see \refta{tab:para}), with $\mstauz$ varied.
  The upper left plot shows the partial decay width, the upper right plot shows 
  the corresponding relative size of the corrections.
  The lower left plot shows the BR, the lower right plot shows 
  the relative correction of the BR.
  The vertical lines indicate where $\mstauz + \mstaue = 1000 \gev$, 
  i.e.\ the maximum reach of the ILC(1000).
}
\label{fig:mst2.stau2tauneu1}
\end{center}
\end{figure}

\begin{figure}[htb!]
\begin{center}
\begin{tabular}{c}
\includegraphics[width=0.49\textwidth,height=7.5cm]{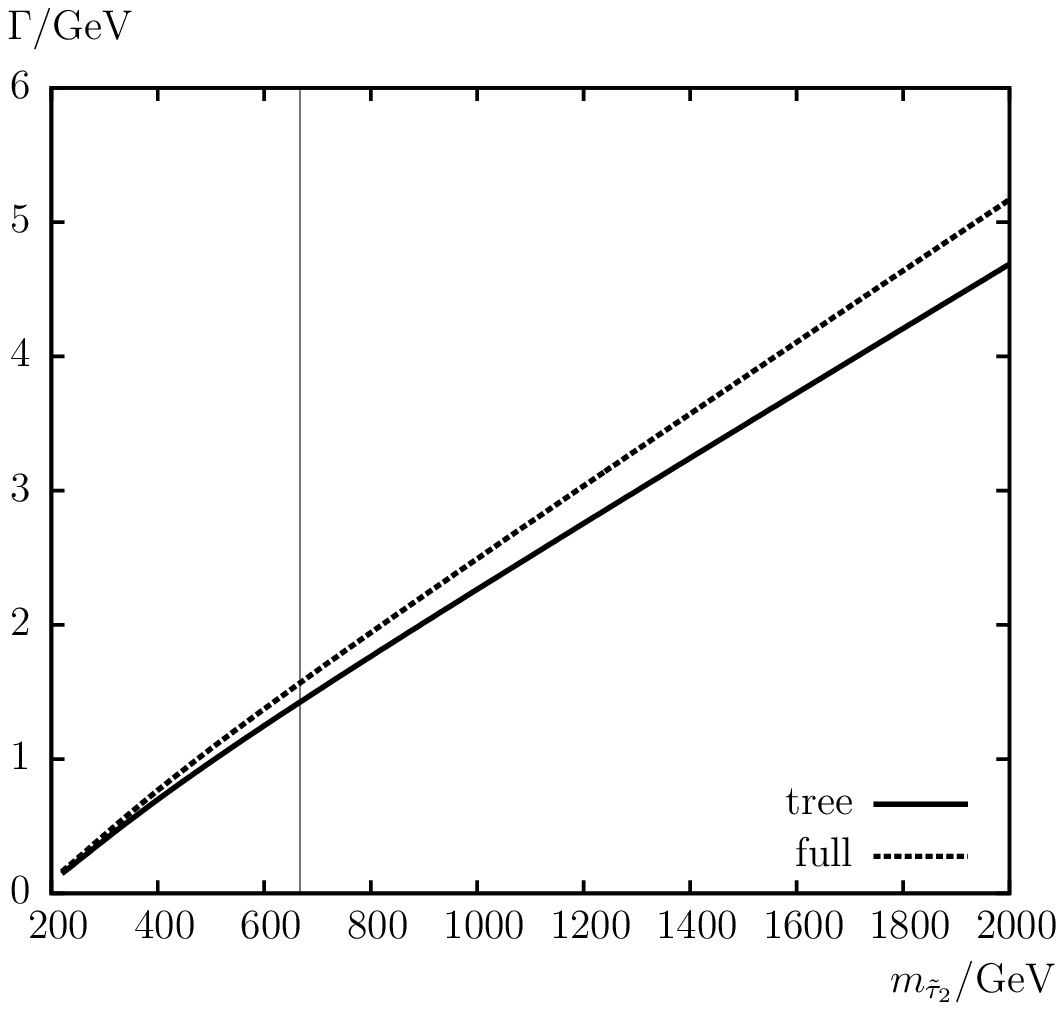}
\hspace{-4mm}
\includegraphics[width=0.49\textwidth,height=7.5cm]{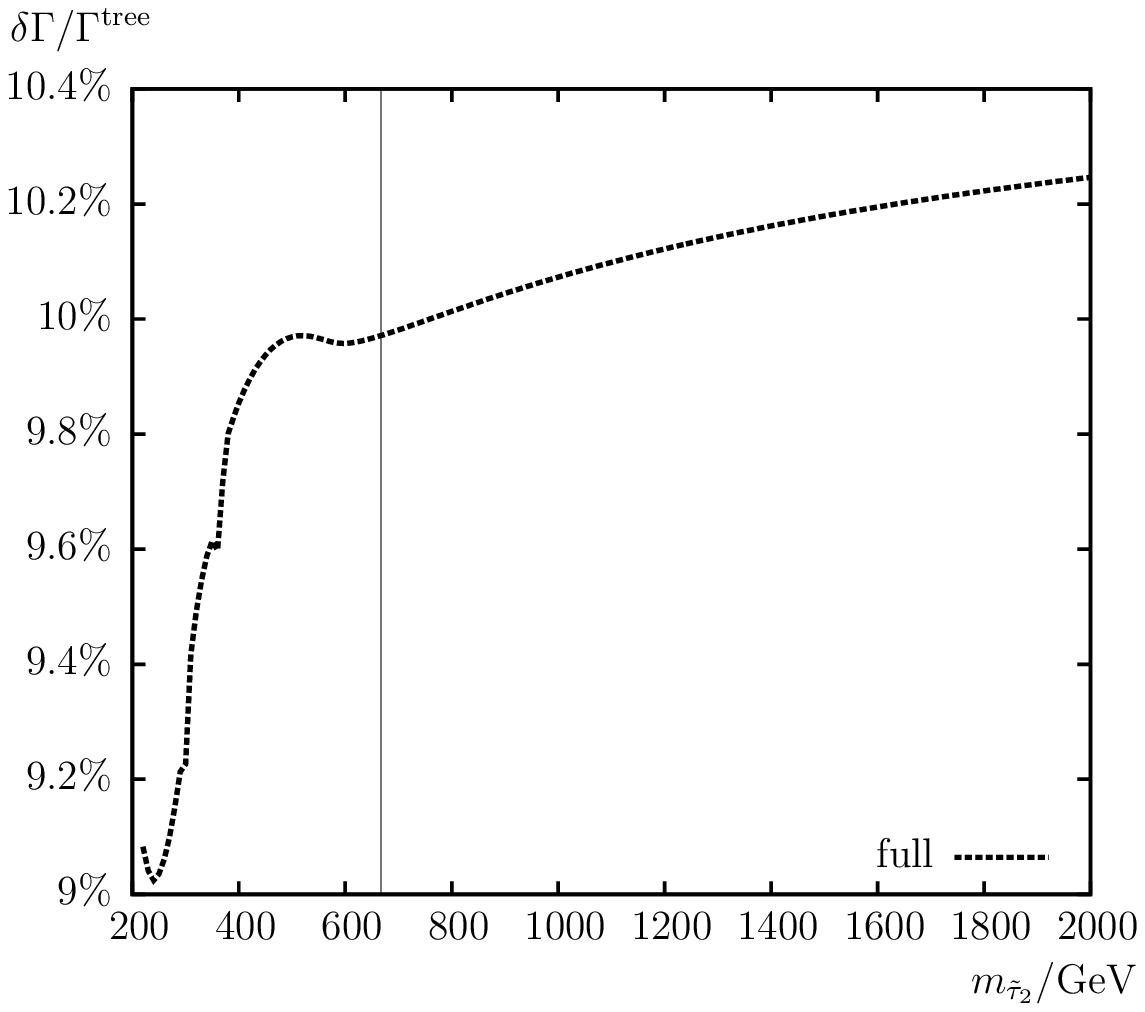}
\\[4em]
\includegraphics[width=0.49\textwidth,height=7.5cm]{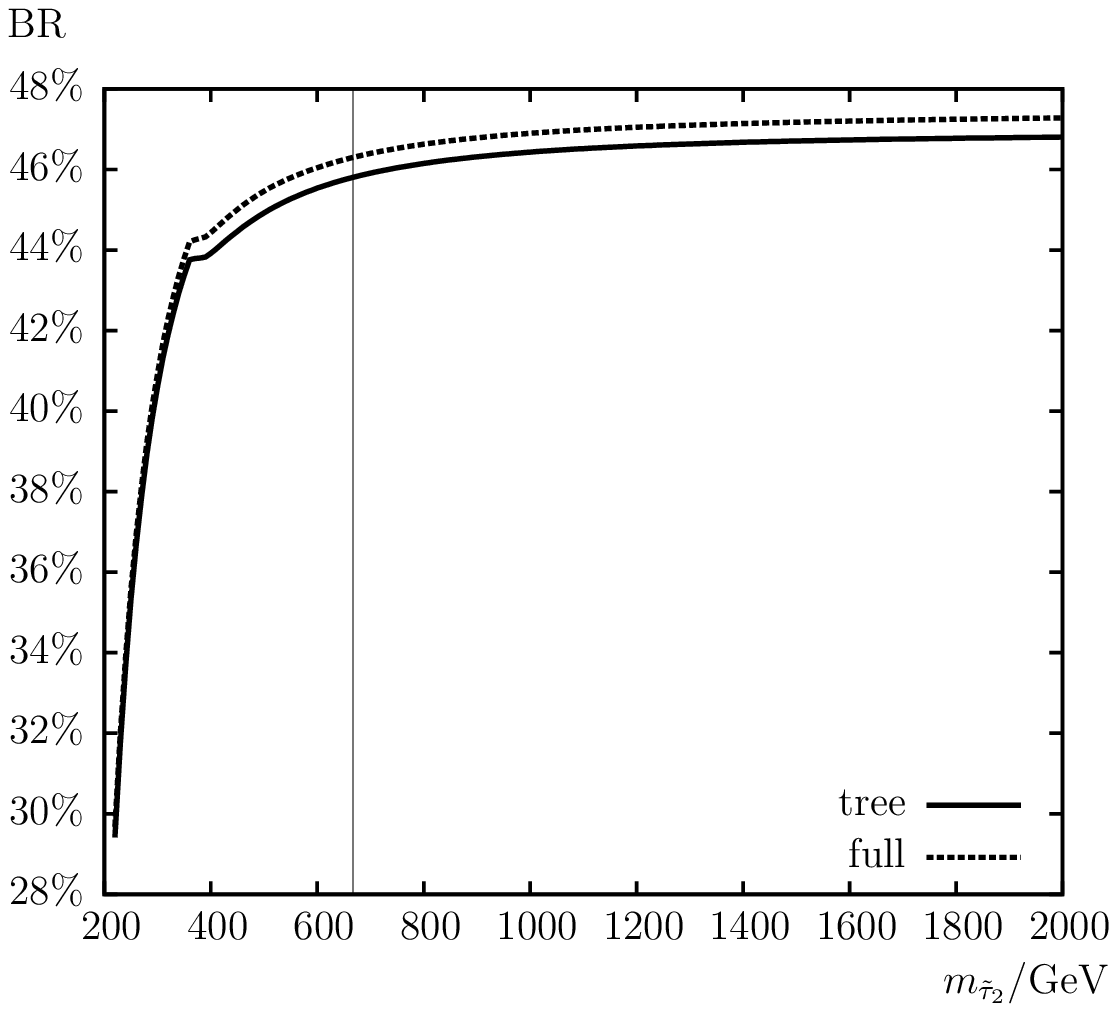}
\hspace{-4mm}
\includegraphics[width=0.49\textwidth,height=7.5cm]{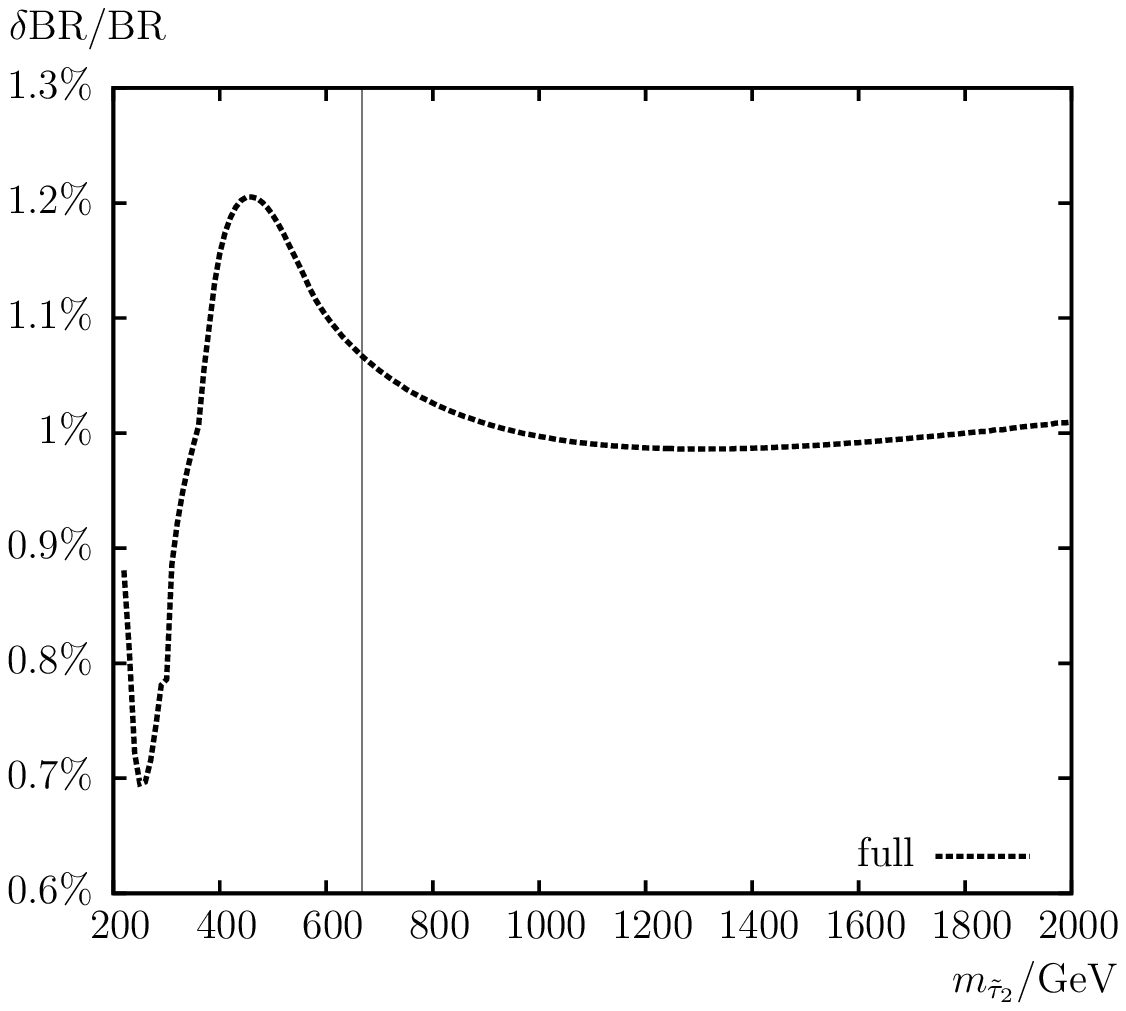}
\end{tabular}
\vspace{2em}
\caption{
  $\Ga(\decayNz)$. Tree-level and full one-loop corrected partial decay widths 
  are shown with the parameters chosen according to \SE\ 
  (see \refta{tab:para}), with $\mstauz$ varied.
  The upper left plot shows the partial decay width, the upper right plot shows 
  the corresponding relative size of the corrections.
  The lower left plot shows the BR, the lower right plot shows 
  the relative correction of the BR.
  The vertical lines indicate where $\mstauz + \mstaue = 1000 \gev$, 
  i.e.\ the maximum reach of the ILC(1000).
}
\label{fig:mst2.stau2tauneu2}
\end{center}
\end{figure}

\begin{figure}[htb!]
\begin{center}
\begin{tabular}{c}
\includegraphics[width=0.49\textwidth,height=7.5cm]{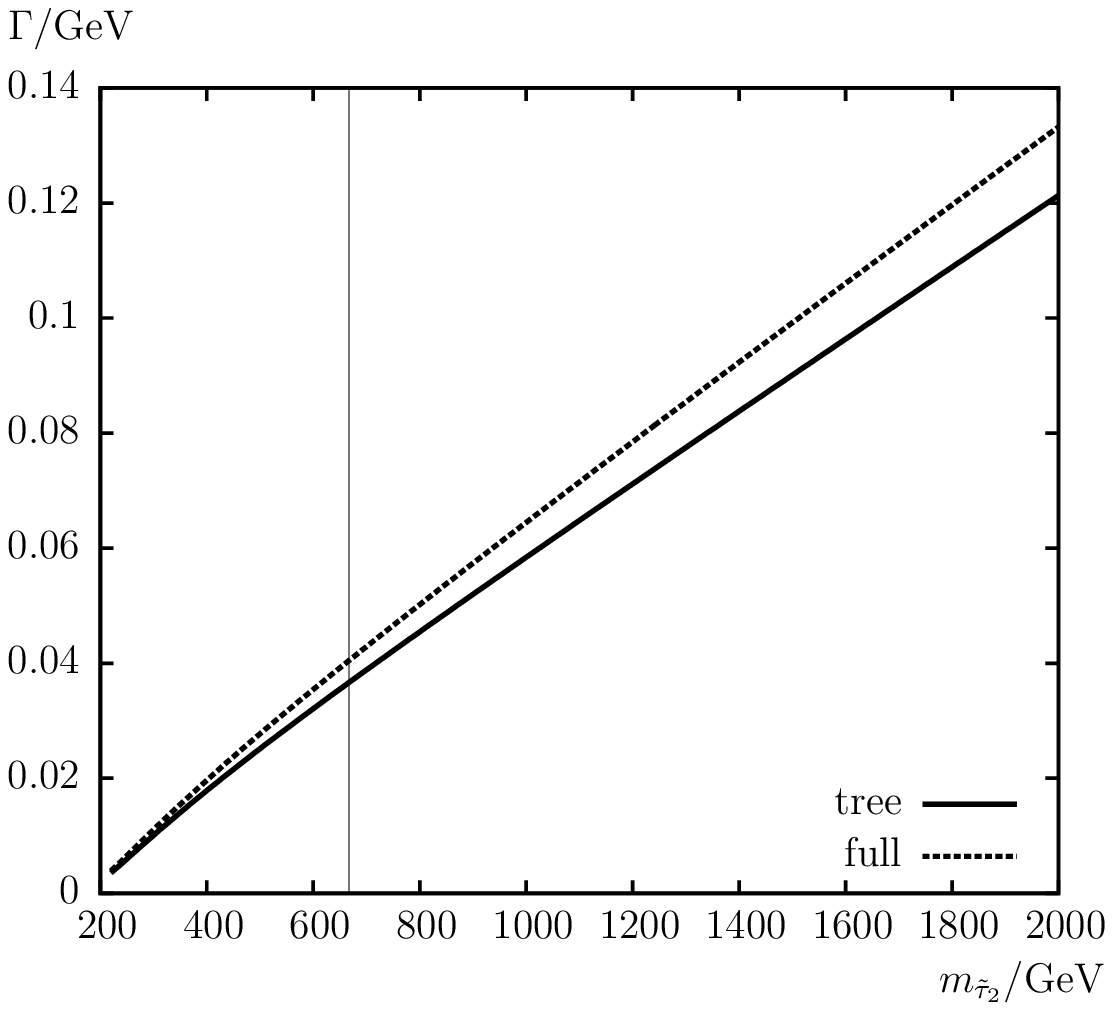}
\hspace{-4mm}
\includegraphics[width=0.49\textwidth,height=7.5cm]{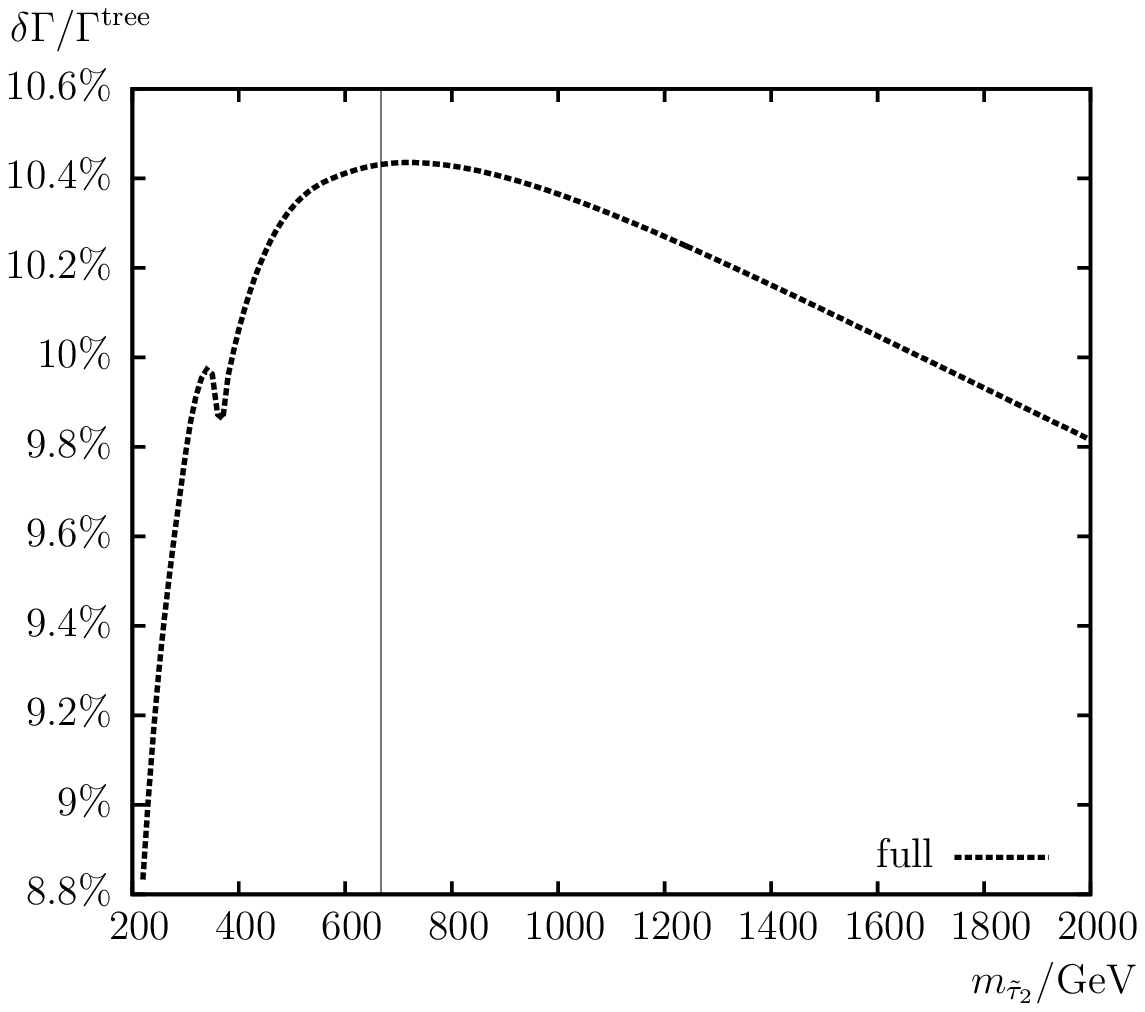}
\\[4em]
\includegraphics[width=0.49\textwidth,height=7.5cm]{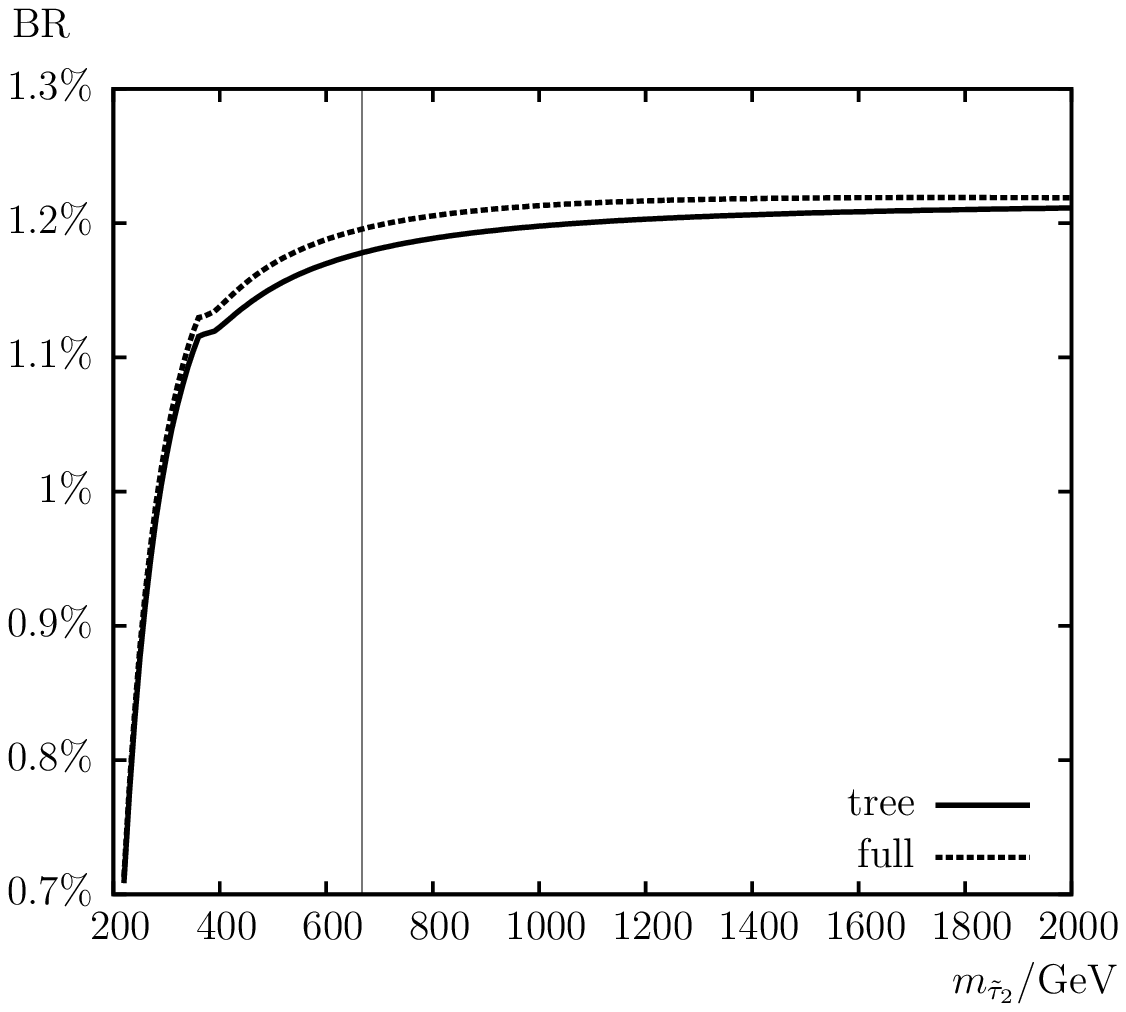}
\hspace{-4mm}
\includegraphics[width=0.49\textwidth,height=7.5cm]{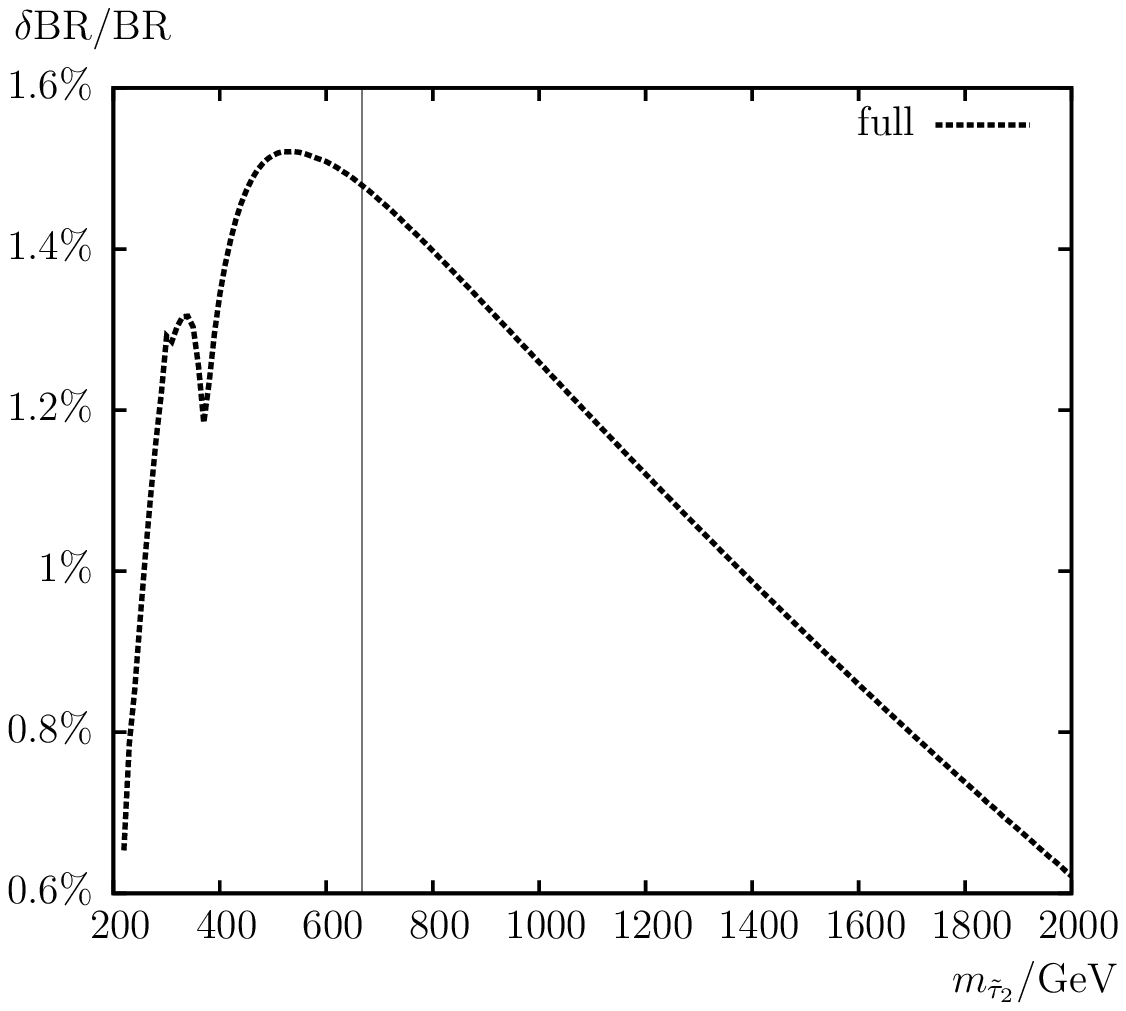}
\end{tabular}
\vspace{2em}
\caption{
  $\Ga(\decayNd)$. Tree-level and full one-loop corrected partial decay widths 
  are shown with the parameters chosen according to \SE\ 
  (see \refta{tab:para}), with $\mstauz$ varied.
  The upper left plot shows the partial decay width, the upper right plot shows 
  the corresponding relative size of the corrections.
  The lower left plot shows the BR, the lower right plot shows 
  the relative correction of the BR.
  The vertical lines indicate where $\mstauz + \mstaue = 1000 \gev$, 
  i.e.\ the maximum reach of the ILC(1000).
}
\label{fig:mst2.stau2tauneu3}
\end{center}
\end{figure}

\begin{figure}[htb!]
\begin{center}
\begin{tabular}{c}
\includegraphics[width=0.49\textwidth,height=7.5cm]{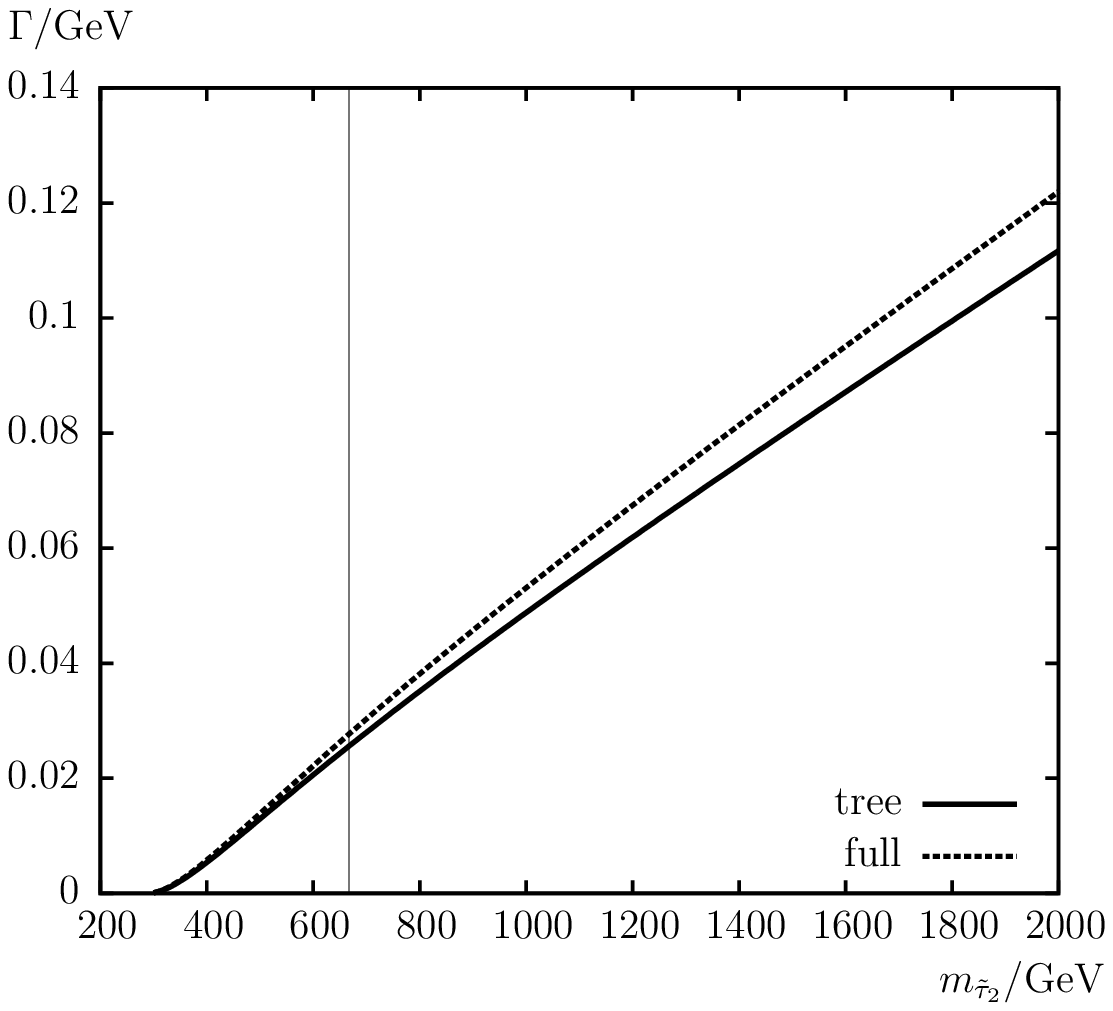}
\hspace{-4mm}
\includegraphics[width=0.49\textwidth,height=7.5cm]{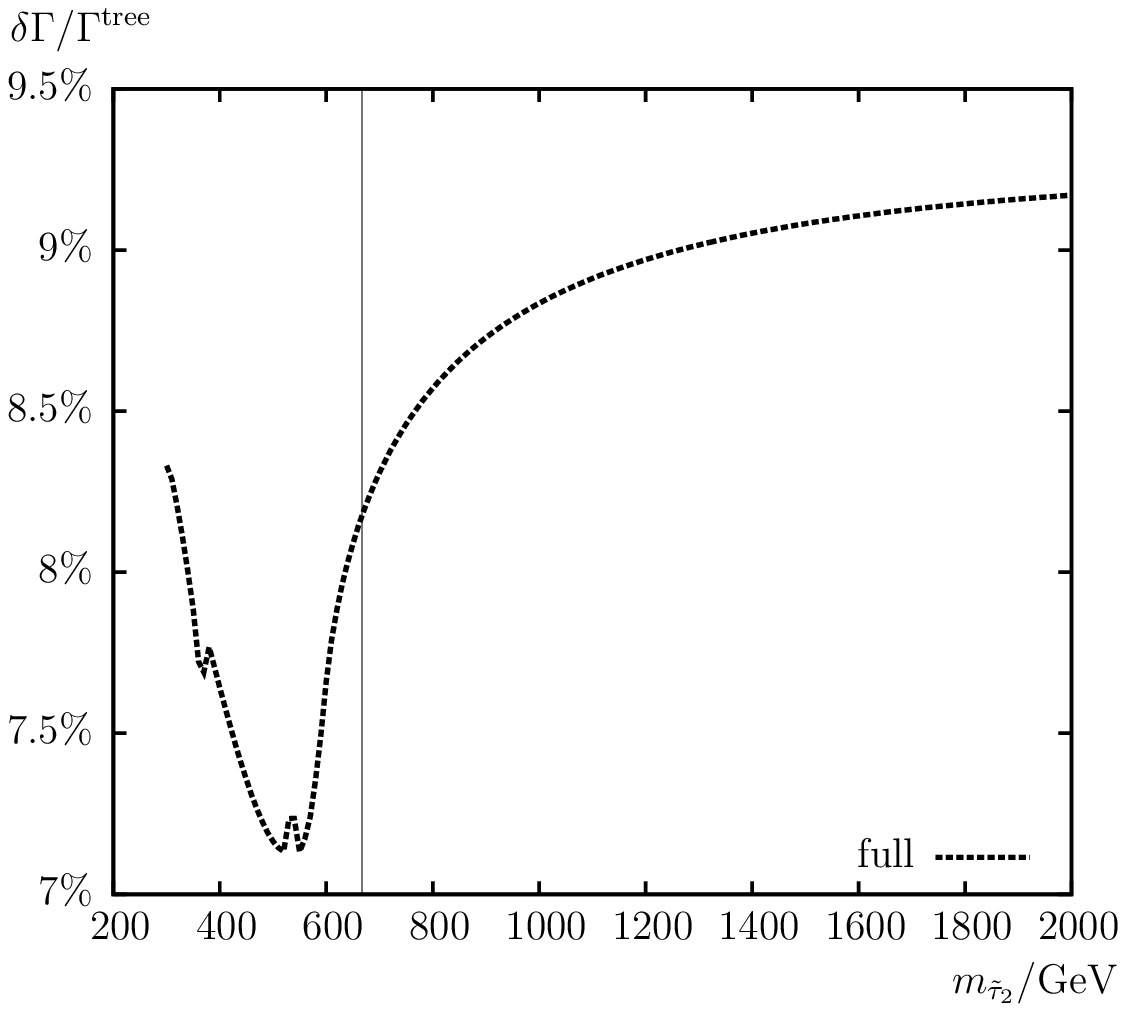}
\\[4em]
\includegraphics[width=0.49\textwidth,height=7.5cm]{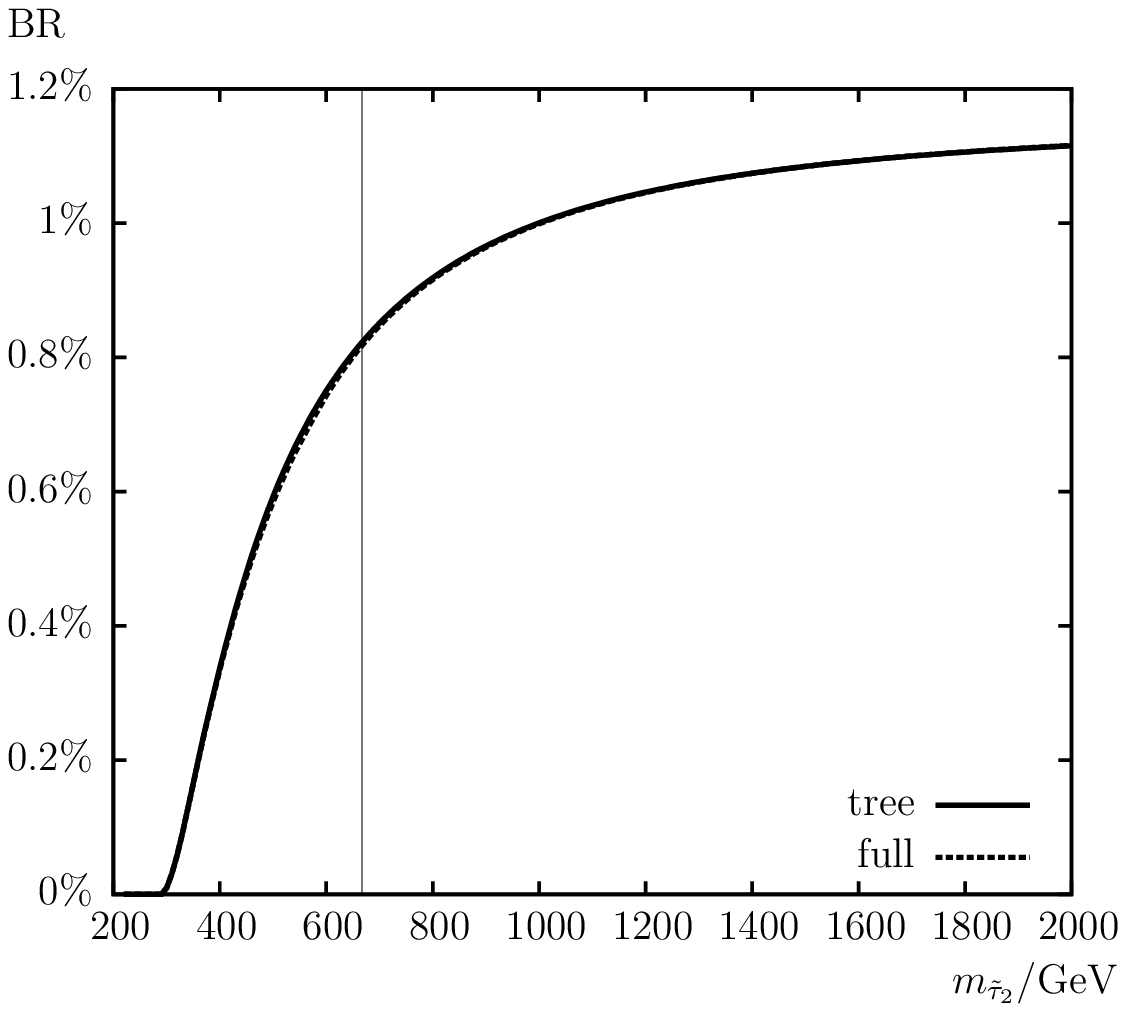}
\hspace{-4mm}
\includegraphics[width=0.49\textwidth,height=7.5cm]{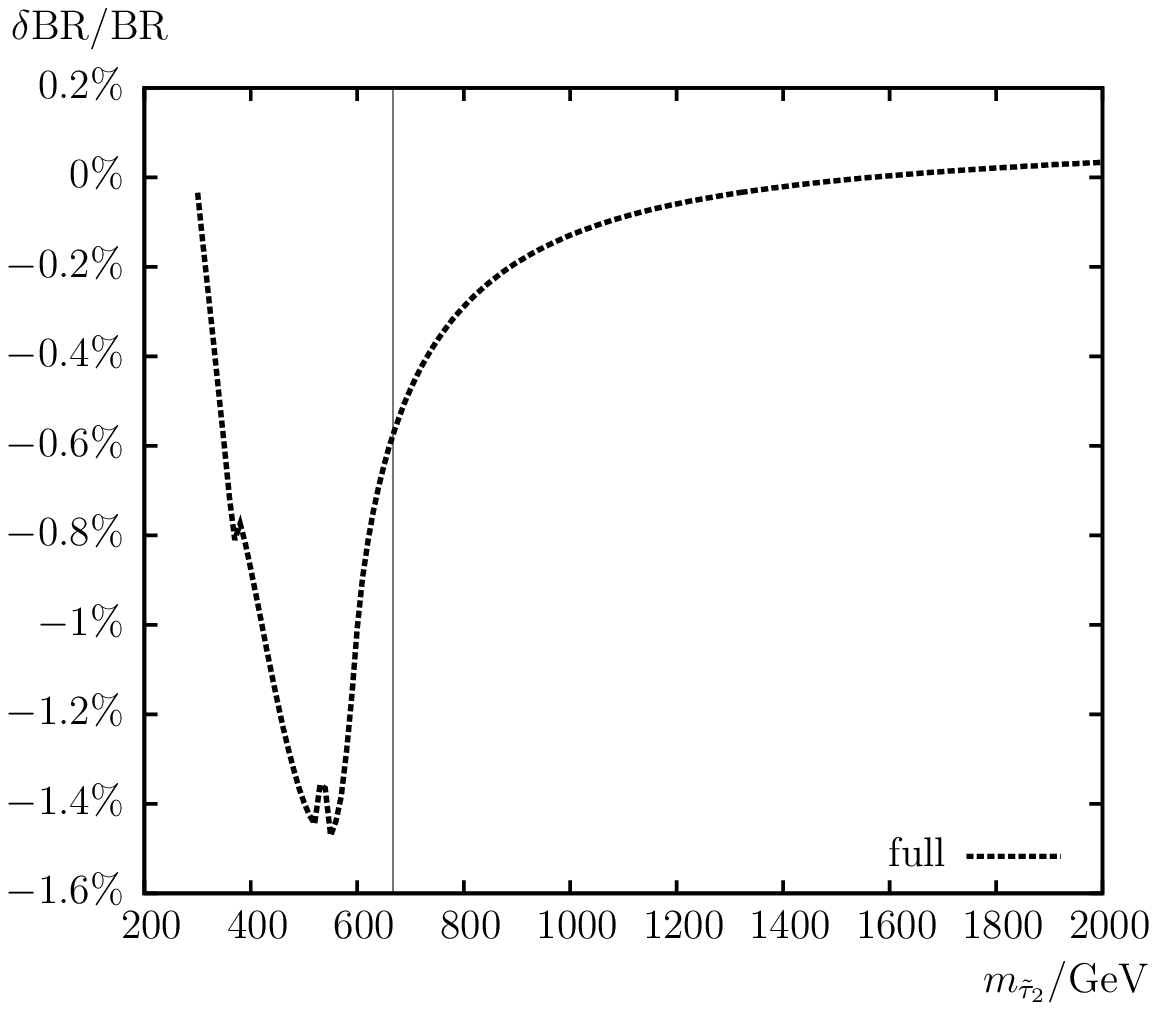}
\end{tabular}
\vspace{2em}
\caption{
  $\Ga(\decayNv)$. Tree-level and full one-loop corrected partial decay widths 
  are shown with the parameters chosen according to \SE\ 
  (see \refta{tab:para}), with $\mstauz$ varied.
  The upper left plot shows the partial decay width, the upper right plot shows 
  the corresponding relative size of the corrections.
  The lower left plot shows the BR, the lower right plot shows 
  the relative correction of the BR.
  The vertical lines indicate where $\mstauz + \mstaue = 1000 \gev$, 
  i.e.\ the maximum reach of the ILC(1000).
}
\label{fig:mst2.stau2tauneu4}
\end{center}
\end{figure}

Now we turn to the decays $\decayNk$ ($k = 1,2,3,4$), with the results shown
in \reffis{fig:mst2.stau2tauneu1} -- \ref{fig:mst2.stau2tauneu4}. 
Since $\mu$, $M_1$ and $M_2$ are roughly of the same order, the four
states are a mixture of gauginos and higgsinos, however, the two lighter
states carry a substantial bino component, which is the only one in the 
case of small $\Stau$~mixing, which is not suppressed by small lepton masses.
Consequently, these two partial decay widths, $\Ga(\decayNi{1,2})$
are found to be roughly the same and dominating above all other decay
widths. Also $\Ga(\decayNi{3,4})$ are roughly the same but significantly 
smaller, due to their small bino component, where the dominating higgsino 
component is proportional to the (suppressed) Yukawa coupling.
Apart from the
above mentioned general dips and thresholds, we find another threshold
for $\Ga(\decayNi{2,4})$ for $\mstauz = 300, 576 \gev$, where
$\mneu{2,4} = \mstaue + \mtau$, respectively. This threshold appears in
the stau self-energies and enters via the field renormalization of the
neutralinos~\cite{Stop2decay}. 
Small steps can be observed at $\mstauz = 524, 528, 544 \gev$ in the
decay $\decayNi{4}$. At this values of the heavy stau mass kinks
in the vertex loop function occurs.
The larger partial decay widths in \SE\ for the decay modes $\decayNk$ with 
$k = 1,2$ go up to $\sim 5 \gev$ for 
$\mstauz = 2 \tev$. The radiative corrections are at the $8\%\, (10\%)$
level for $\decayNi{1 (2)}$, respectively. Since these decays
are dominating the total width the effect in the corresponding branching
ratios is small. For $\Ga(\decayNi{3 (4)})$ the corrections are around 
$10\%\, (8\%)$, respectively. All four decay modes show a substantial
one-loop correction in the region accessible at the ILC(1000). 

\begin{figure}[htb!]
\begin{center}
\begin{tabular}{c}
\includegraphics[width=0.49\textwidth,height=7.5cm]{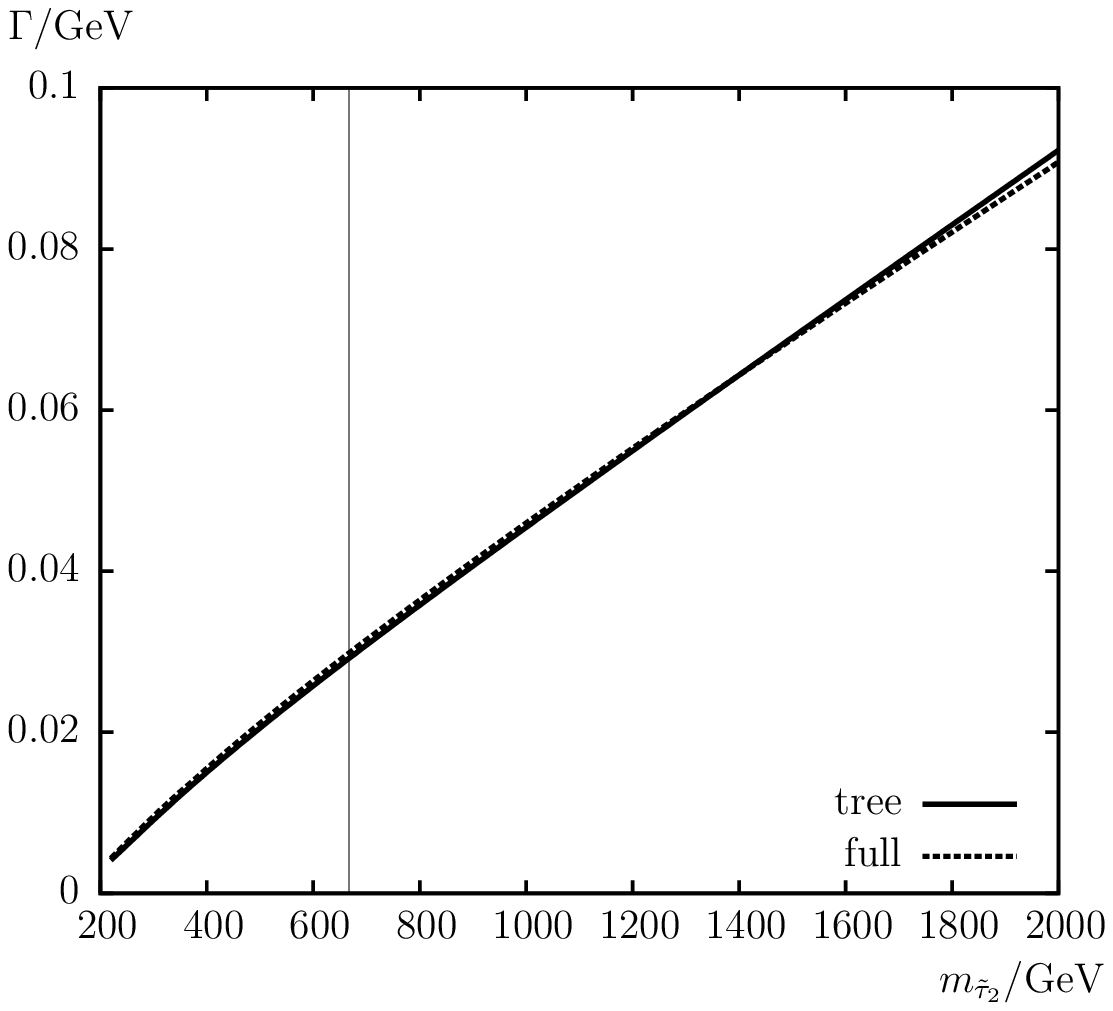}
\hspace{-4mm}
\includegraphics[width=0.49\textwidth,height=7.5cm]{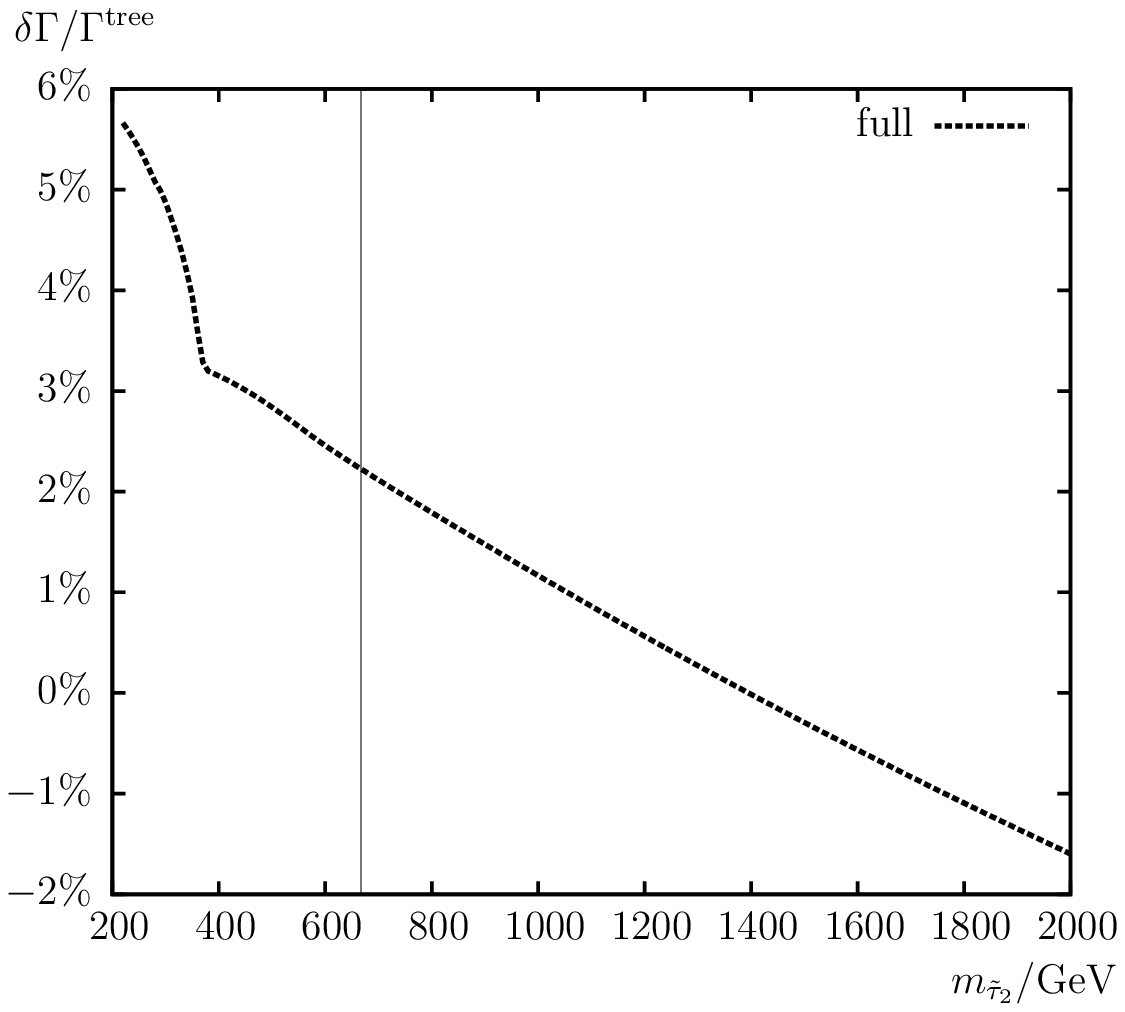}
\\[4em]
\includegraphics[width=0.49\textwidth,height=7.5cm]{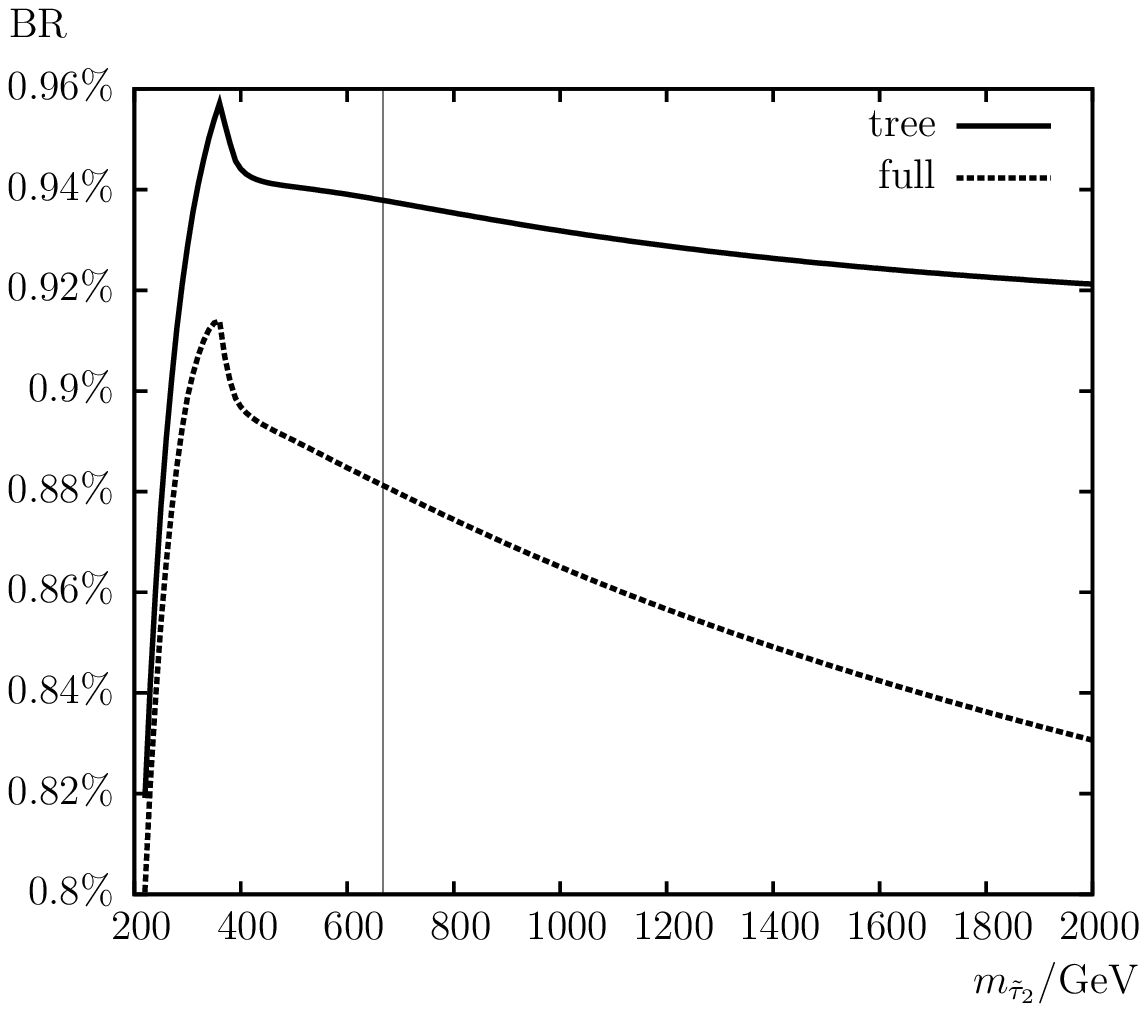}
\hspace{-4mm}
\includegraphics[width=0.49\textwidth,height=7.5cm]{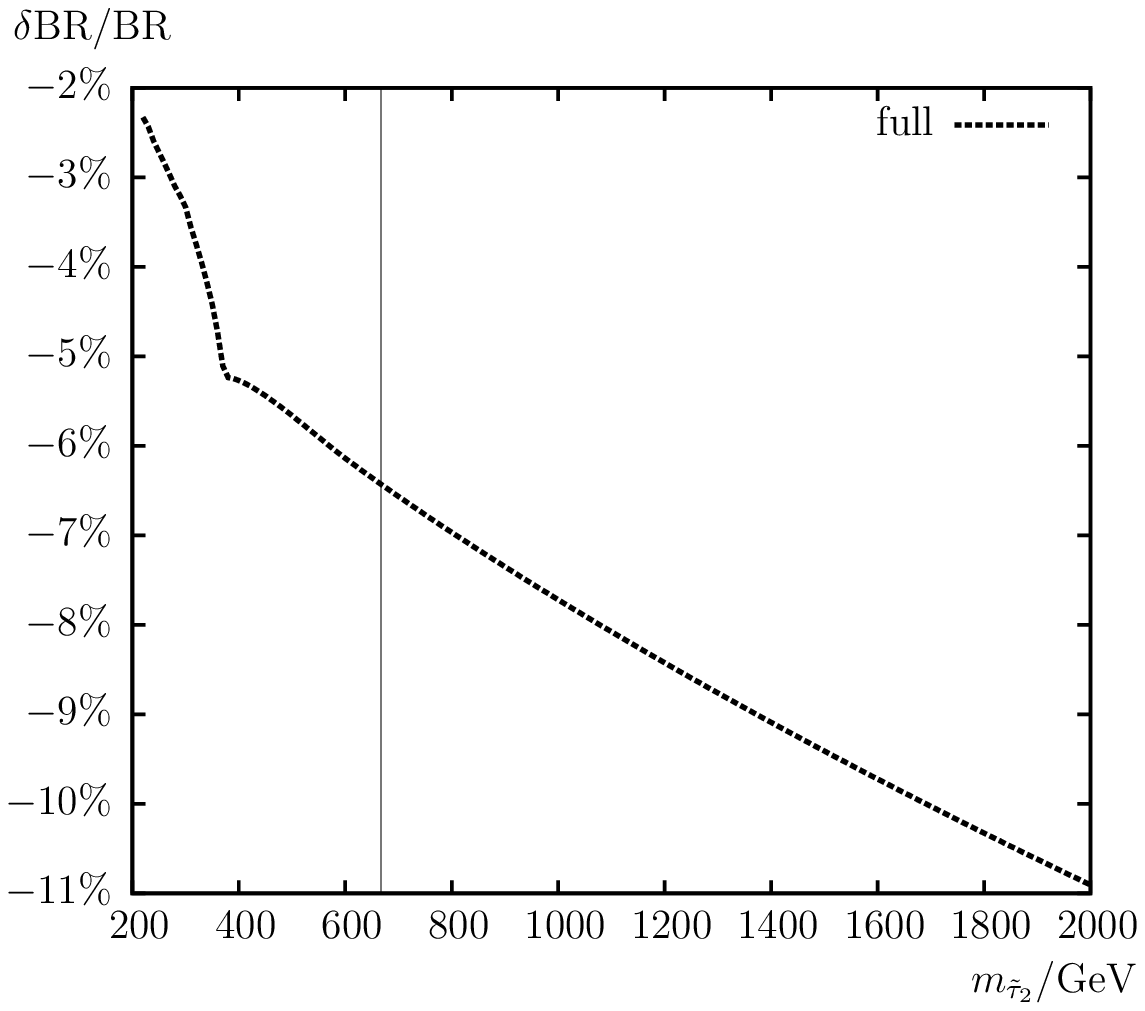}
\end{tabular}
\vspace{2em}
\caption{
  $\Ga(\decayCme)$. Tree-level and full one-loop corrected partial decay widths 
  are shown with the parameters chosen according to \SE\ 
  (see \refta{tab:para}), with $\mstauz$ varied.
  The upper left plot shows the partial decay width, the upper right plot shows 
  the corresponding relative size of the corrections.
  The lower left plot shows the BR, the lower right plot shows 
  the relative correction of the BR.
  The vertical lines indicate where $\mstauz + \mstaue = 1000 \gev$, 
  i.e.\ the maximum reach of the ILC(1000).
}
\label{fig:mst2.stau2ncha1}
\end{center}
\end{figure}

\begin{figure}[htb!]
\begin{center}
\begin{tabular}{c}
\includegraphics[width=0.49\textwidth,height=7.5cm]{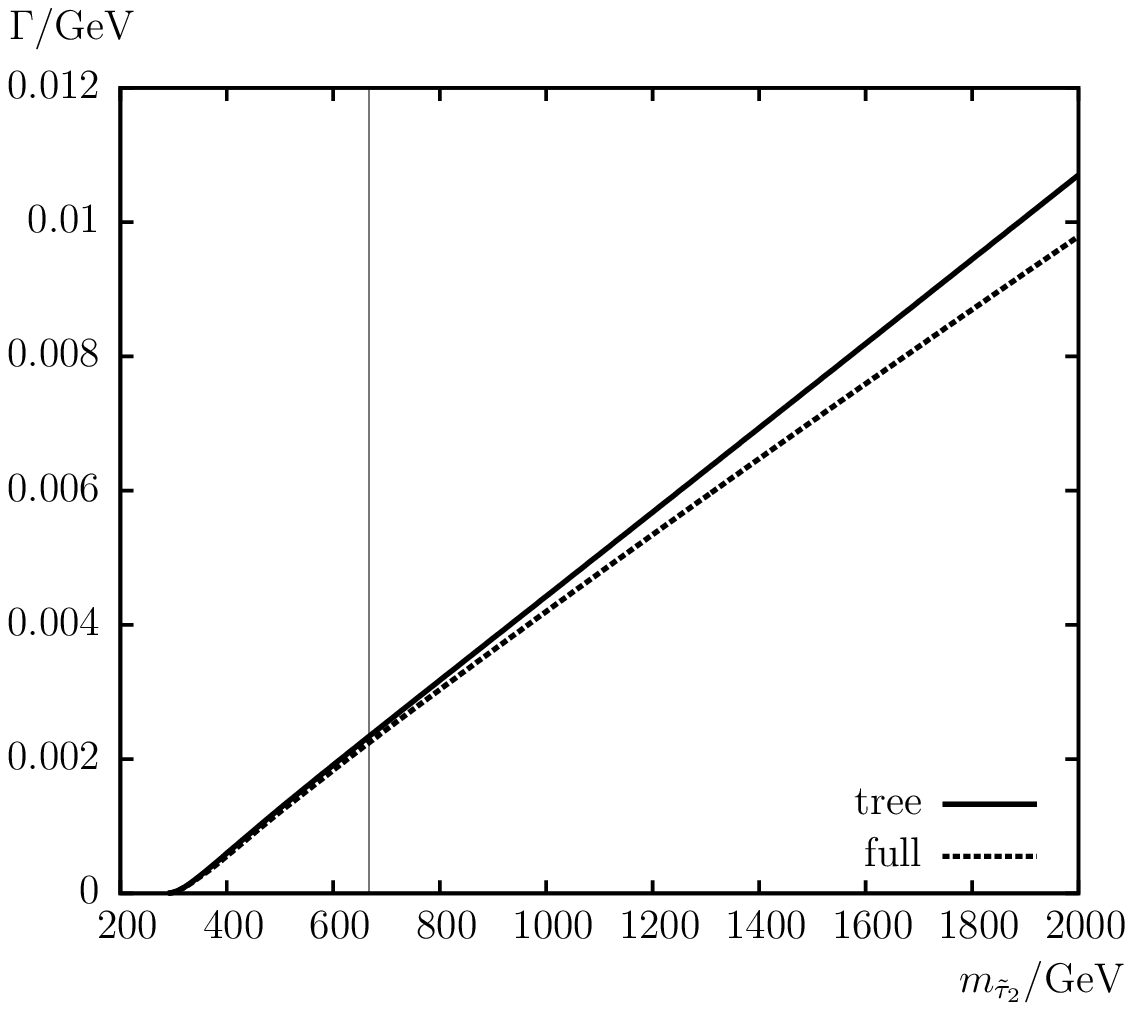}
\hspace{-4mm}
\includegraphics[width=0.49\textwidth,height=7.5cm]{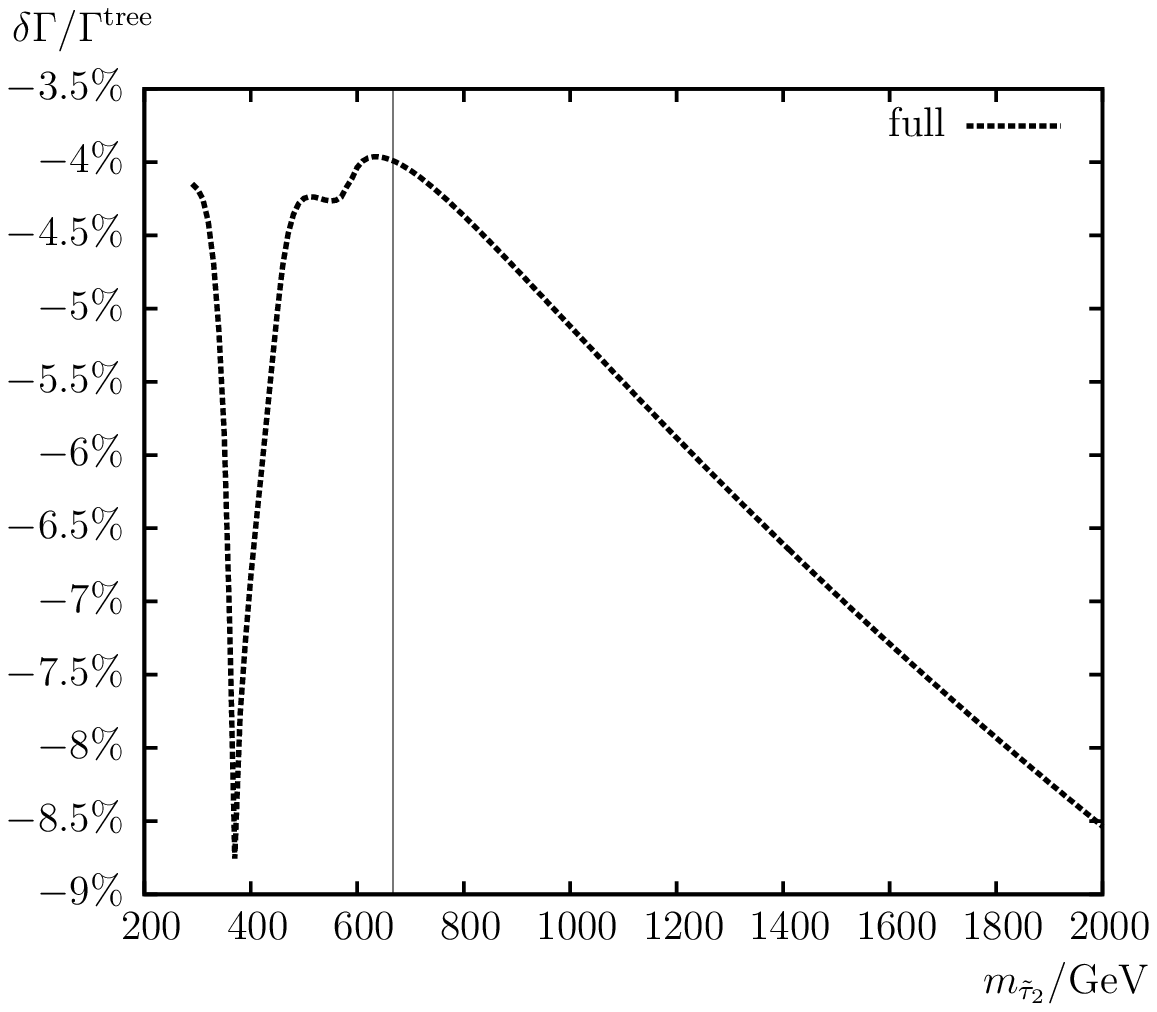}
\\[4em]
\includegraphics[width=0.49\textwidth,height=7.5cm]{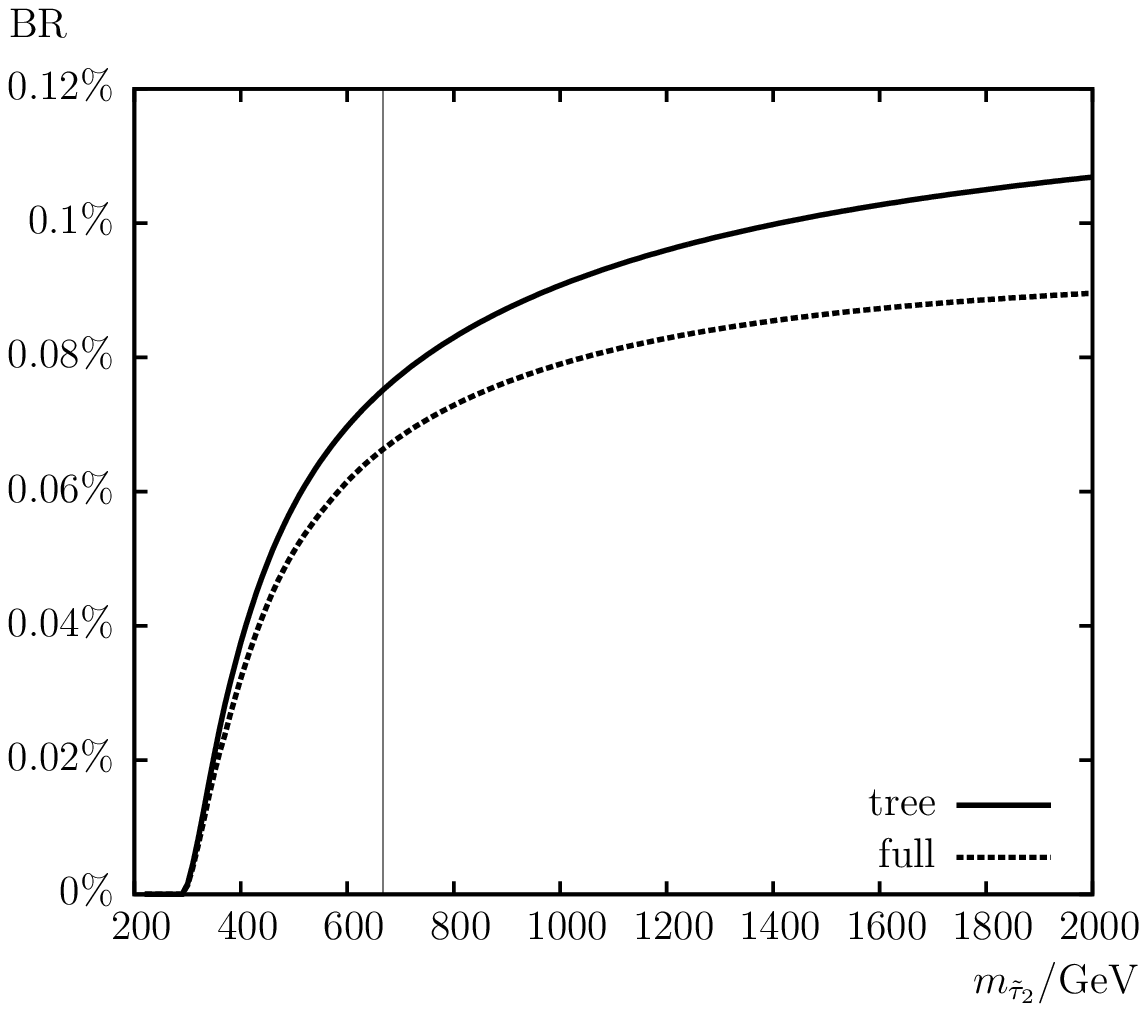}
\hspace{-4mm}
\includegraphics[width=0.49\textwidth,height=7.5cm]{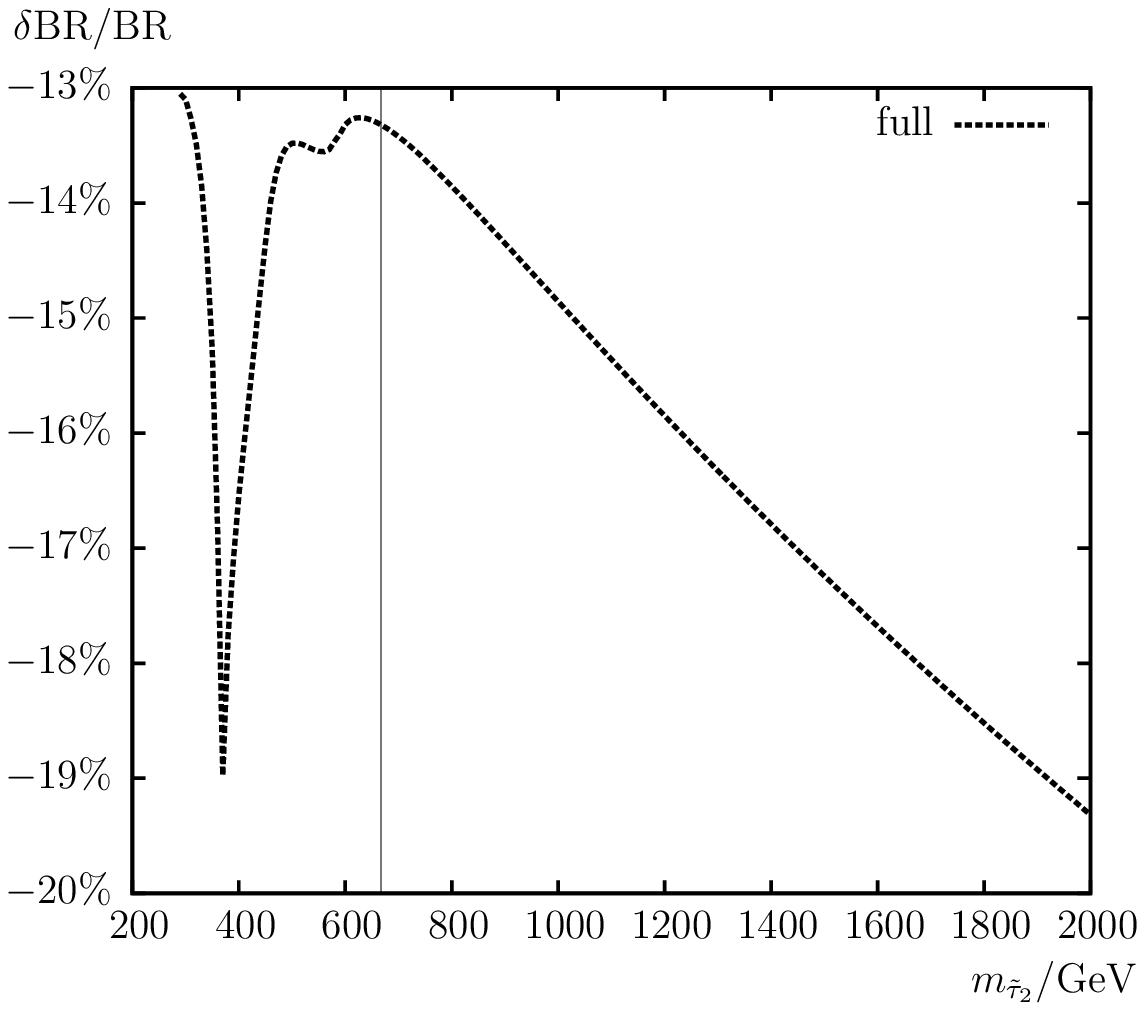}
\end{tabular}
\vspace{2em}
\caption{
  $\Ga(\decayCmz)$. Tree-level and full one-loop corrected partial decay widths 
  are shown with the parameters chosen according to \SE\
  (see \refta{tab:para}), with $\mstauz$ varied.
  The upper left plot shows the partial decay width, the upper right plot shows 
  the corresponding relative size of the corrections. 
  The lower left plot shows the BR, the lower right plot shows 
  the relative correction of the BR.
  The vertical lines indicate where $\mstauz + \mstaue = 1000 \gev$, 
  i.e.\ the maximum reach of the ILC(1000).
}
\label{fig:mst2.stau2ncha2}
\end{center}
\end{figure}

Next in \reffis{fig:mst2.stau2ncha1}, \ref{fig:mst2.stau2ncha2} we present
the results for $\decayCmj$ ($j = 1,2$). 
The size of the partial decay widths and branching ratios for 
$\decayCme$ ($\decayCmz$) are relatively small, below $0.1\, (0.01) \gev$,
respectively. Apart from the general thresholds we find two additional
ones for each decay, located at 
$\mstauz \approx 284.8, 287.6\, (593.8, 597.0) \gev$, where
$\mcha{1} = \mtau + \msneut$, $\mcha{1} = \mmu + \msneum$
($\mcha{2} = \mtau + \msneut$, $\mcha{2} = \mmu + \msneum$).
The thresholds occur in the respective chargino self-energies and thus enter
via $\de \MTwo$, $\de\mu$, $\de\mcha{1,2}$ and the field renormalization
constants. 
The corrections to the decay widths are $\sim +5\%, -5\%$ for 
$\decayCme$ and $\decayCmz$ in the ILC(1000) relevant region and thus
potentially relevant. For larger $\mstauz$ they become smaller in the
case of the lighter chargino and larger for the heavier chargino.

\begin{figure}[htb!]
\begin{center}
\begin{tabular}{c}
\includegraphics[width=0.49\textwidth,height=7.5cm]{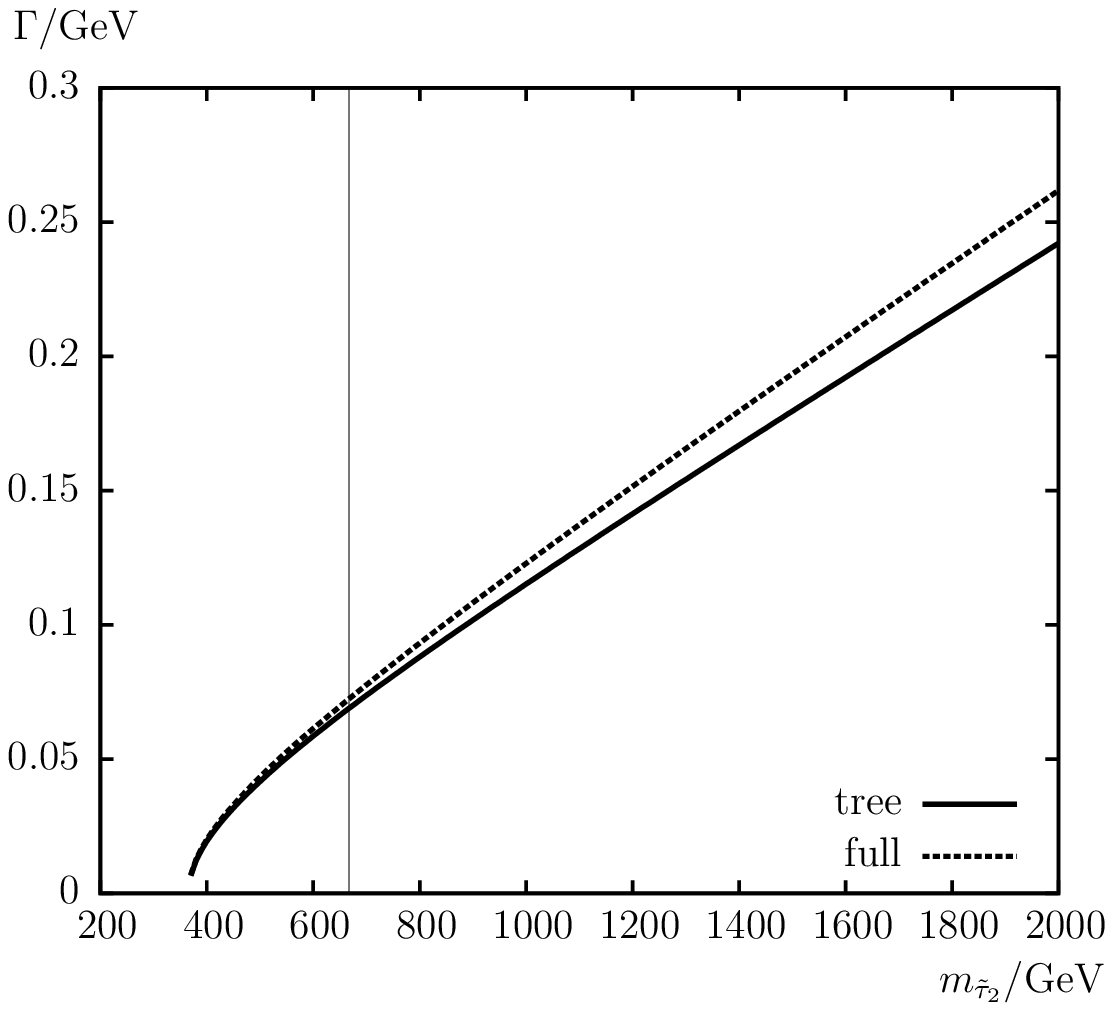}
\hspace{-4mm}
\includegraphics[width=0.49\textwidth,height=7.5cm]{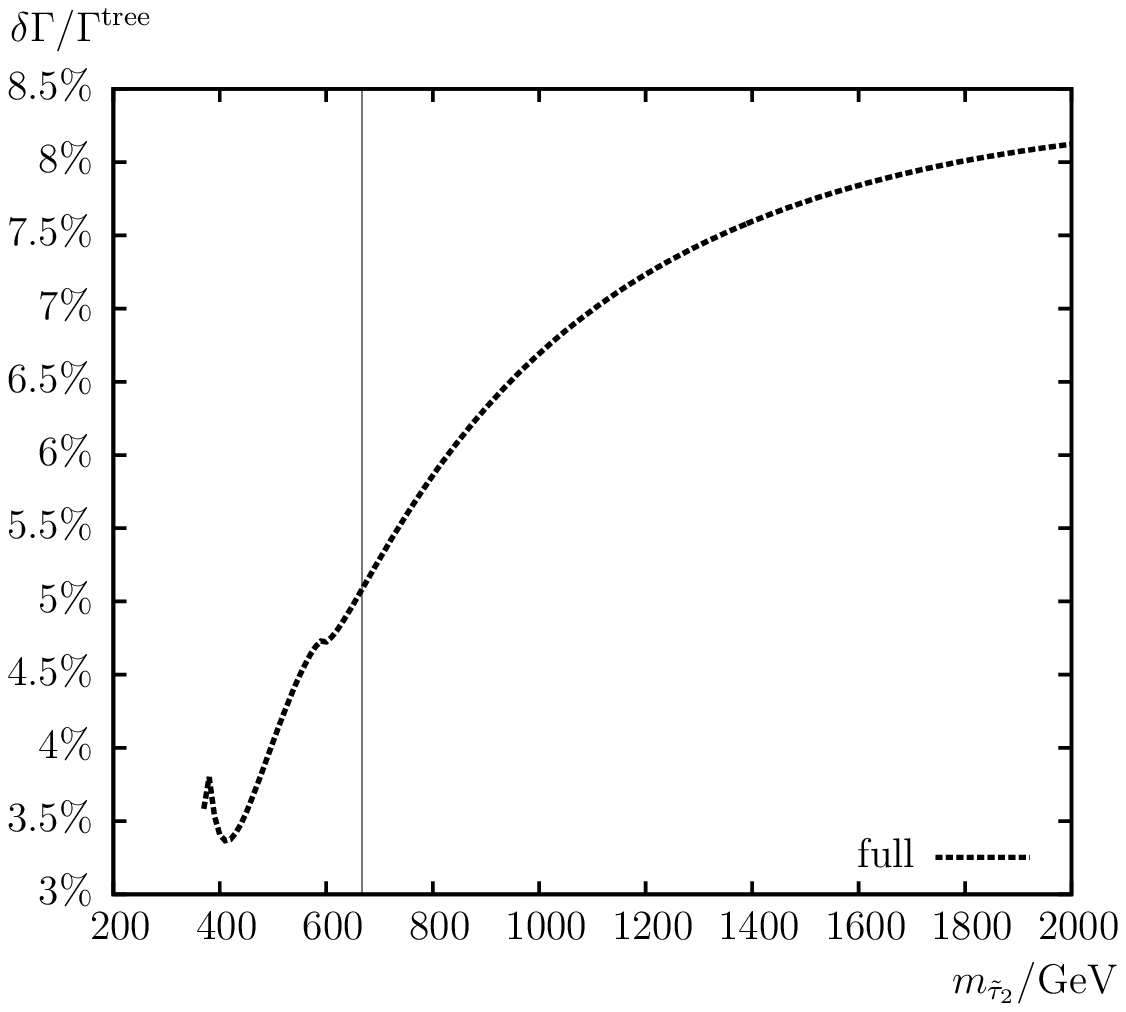}
\\[4em]
\includegraphics[width=0.49\textwidth,height=7.5cm]{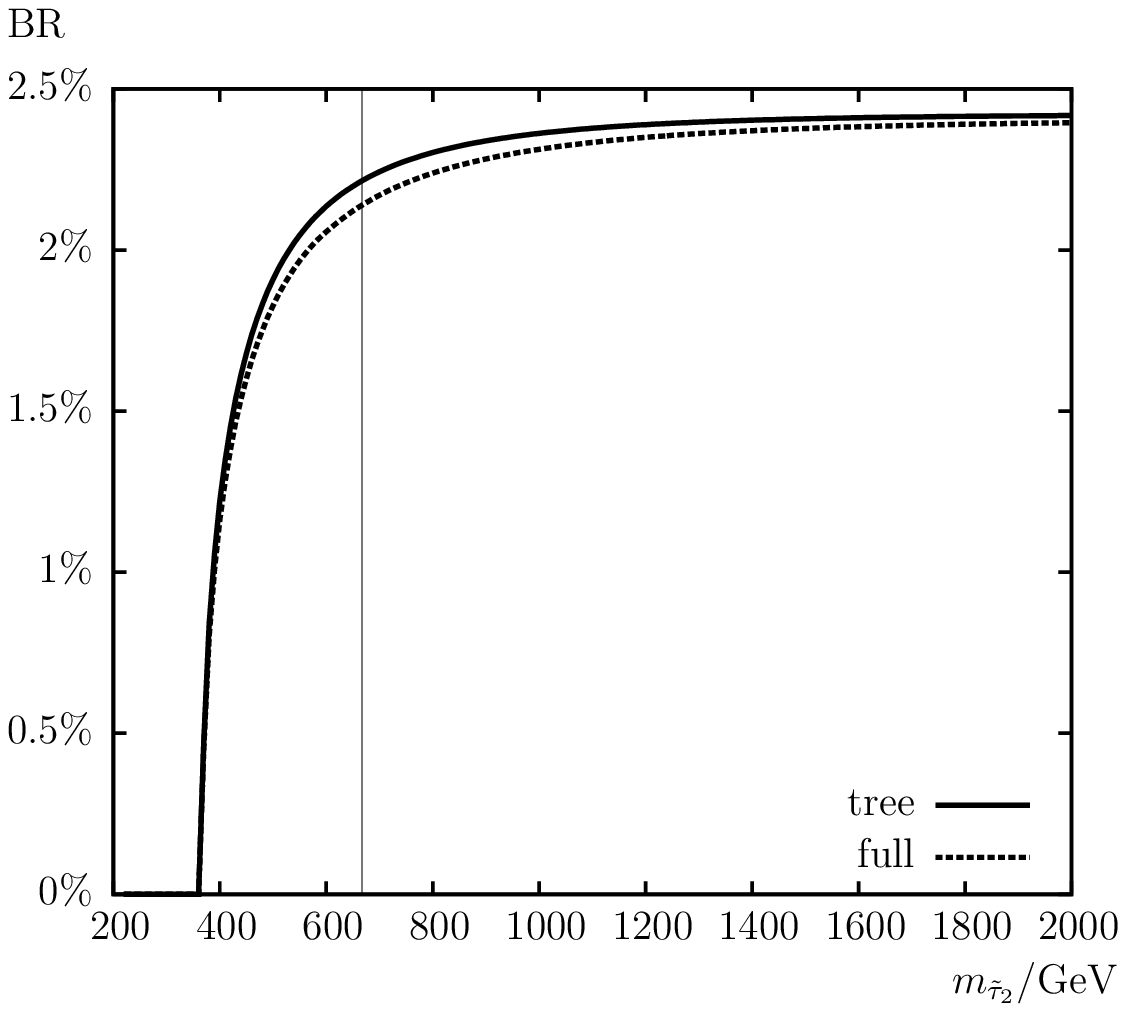}
\hspace{-4mm}
\includegraphics[width=0.49\textwidth,height=7.5cm]{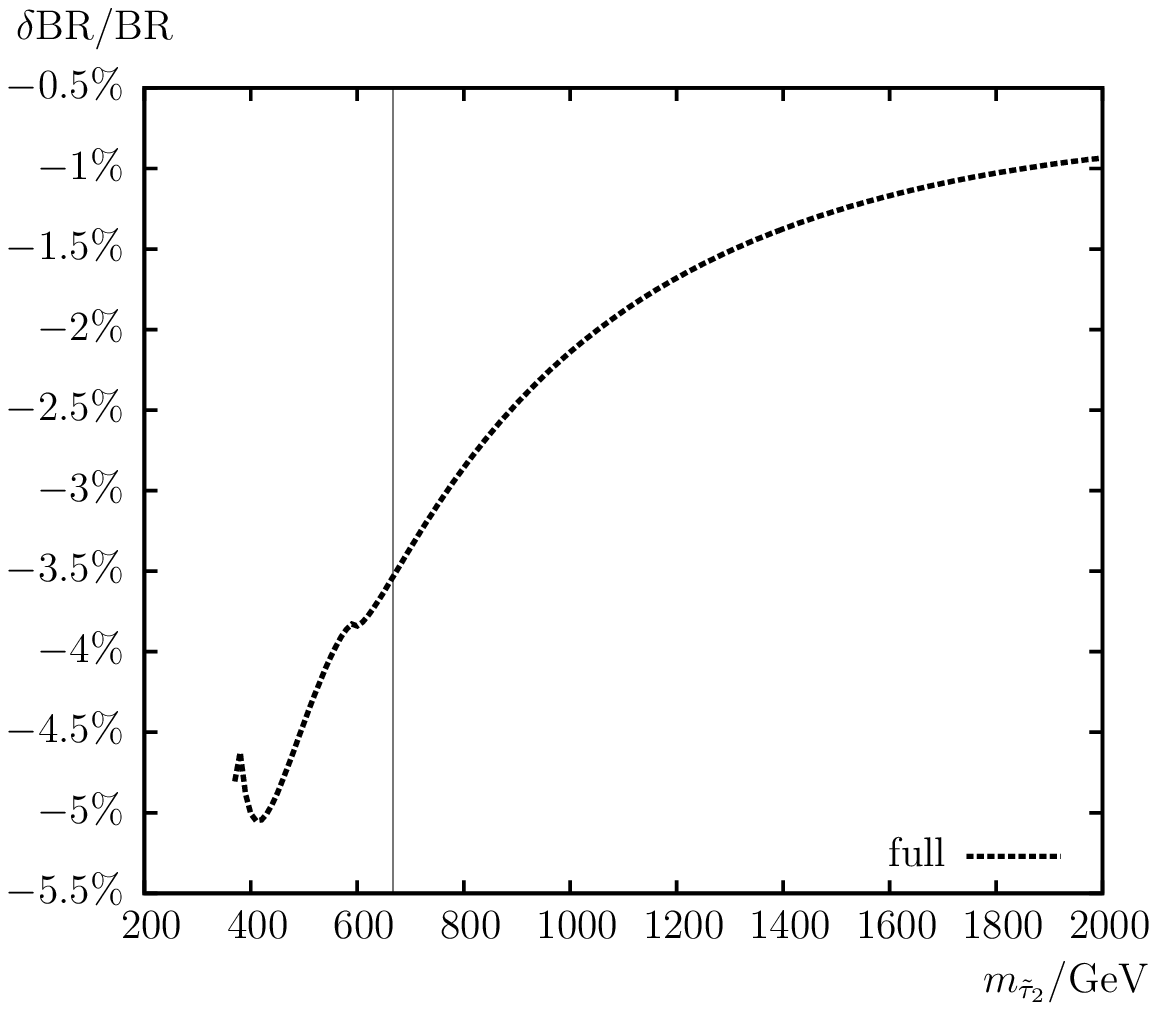}
\end{tabular}
\vspace{2em}
\caption{
  $\Ga(\decayHm)$. Tree-level and full one-loop corrected partial decay widths 
  are shown with the parameters chosen according to \SE\
  (see \refta{tab:para}), with $\mstauz$ varied.
  The upper left plot shows the partial decay width, the upper right plot shows 
  the corresponding relative size of the corrections.
  The lower left plot shows the BR, the lower right plot shows 
  the relative correction of the BR.
  The vertical lines indicate where $\mstauz + \mstaue = 1000 \gev$, 
  i.e.\ the maximum reach of the ILC(1000).
}
\label{fig:mst2.stau2snH}
\end{center}
\end{figure}

We now turn to the decay mode $\decayHm$, which is
shown in \reffi{fig:mst2.stau2snH}.
In addition to the general dips and thresholds another one can be found
at $\mstauz \approx 601 \gev$, where $\msneut = \mtau + \mcha{2}$ is
realized, i.e.\ the threshold enters via the sneutrino field
renormalization. 
The corrections to $\Ga(\decayHm)$ is $\sim 4.5\%$ in the ILC(1000)
regions, and thus potentially relevant, and rises to $\sim 8\%$ for
large $\mstauz$. The relative correction to $\br(\decayHm)$ reaches 
$-5\%$ for low $\mstauz$ values. 

\begin{figure}[htb!]
\begin{center}
\begin{tabular}{c}
\includegraphics[width=0.49\textwidth,height=7.5cm]{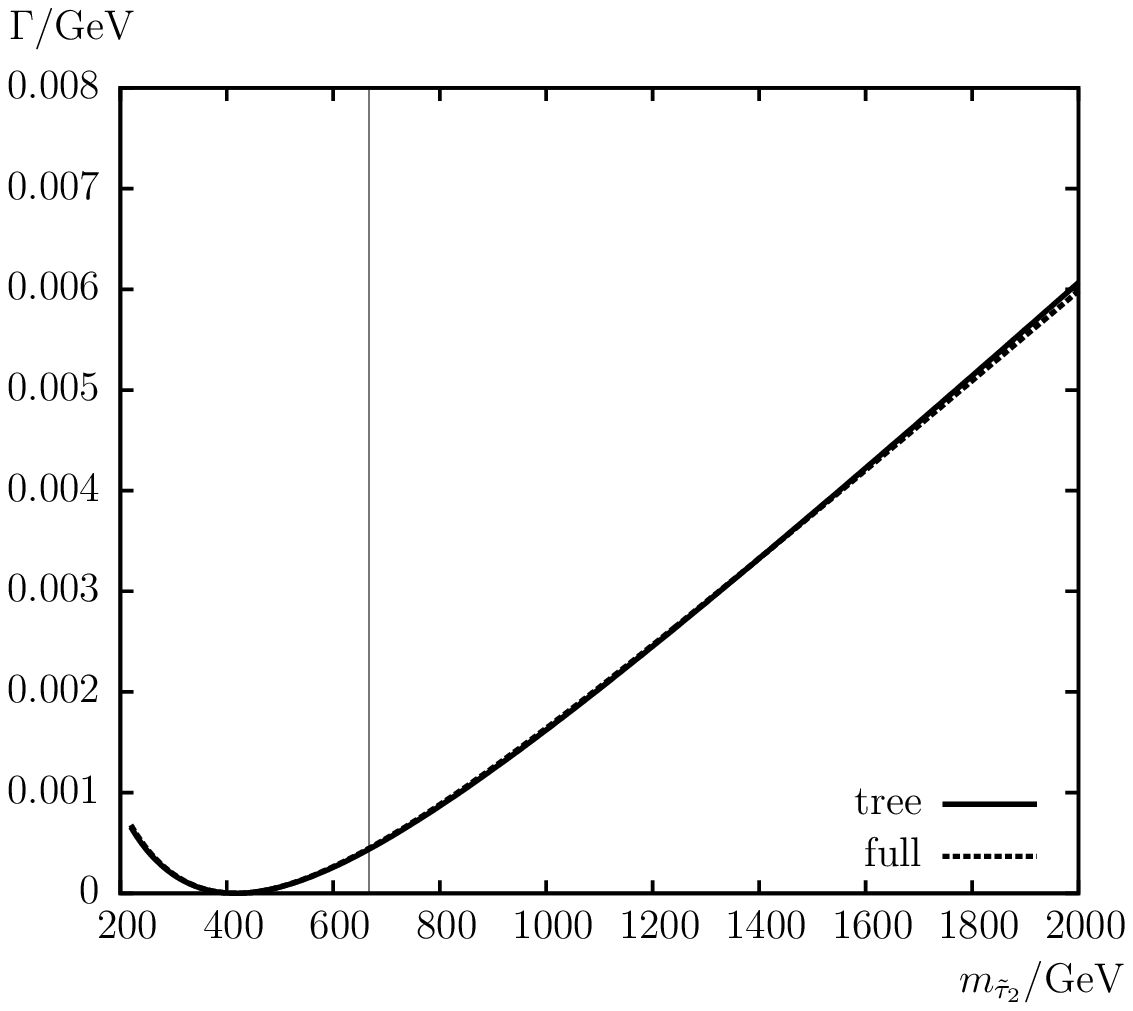}
\hspace{-4mm}
\includegraphics[width=0.49\textwidth,height=7.5cm]{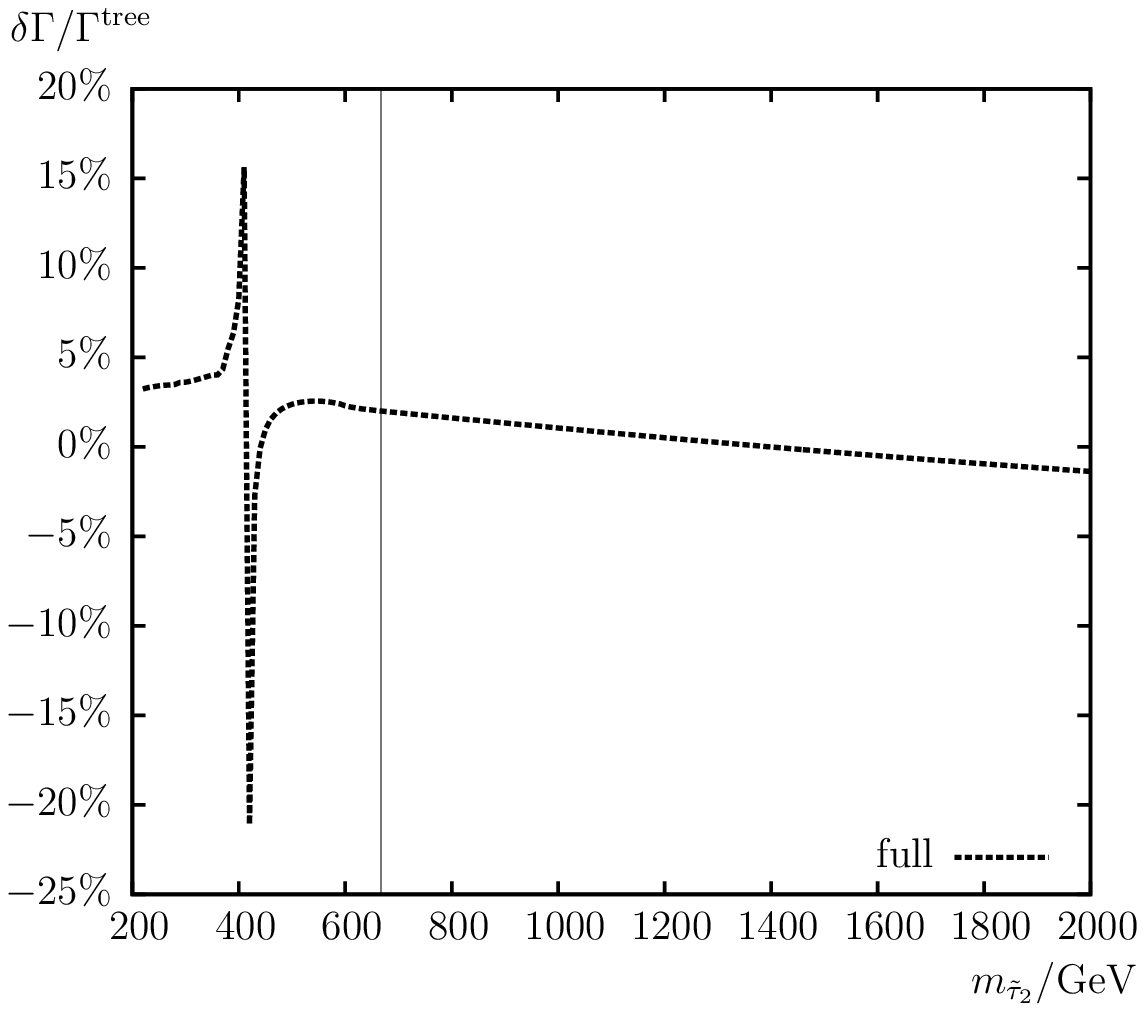}
\\[4em]
\includegraphics[width=0.49\textwidth,height=7.5cm]{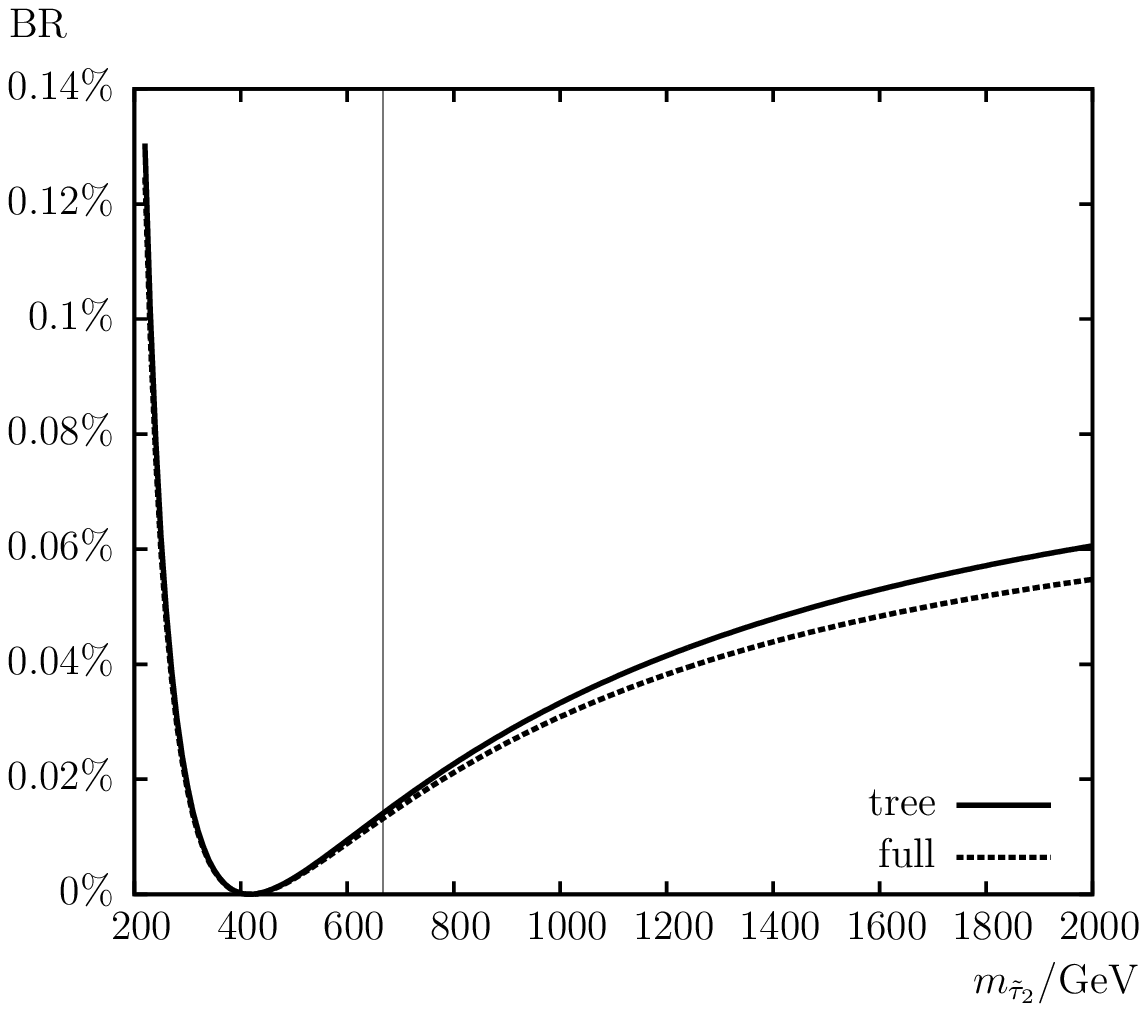}
\hspace{-4mm}
\includegraphics[width=0.49\textwidth,height=7.5cm]{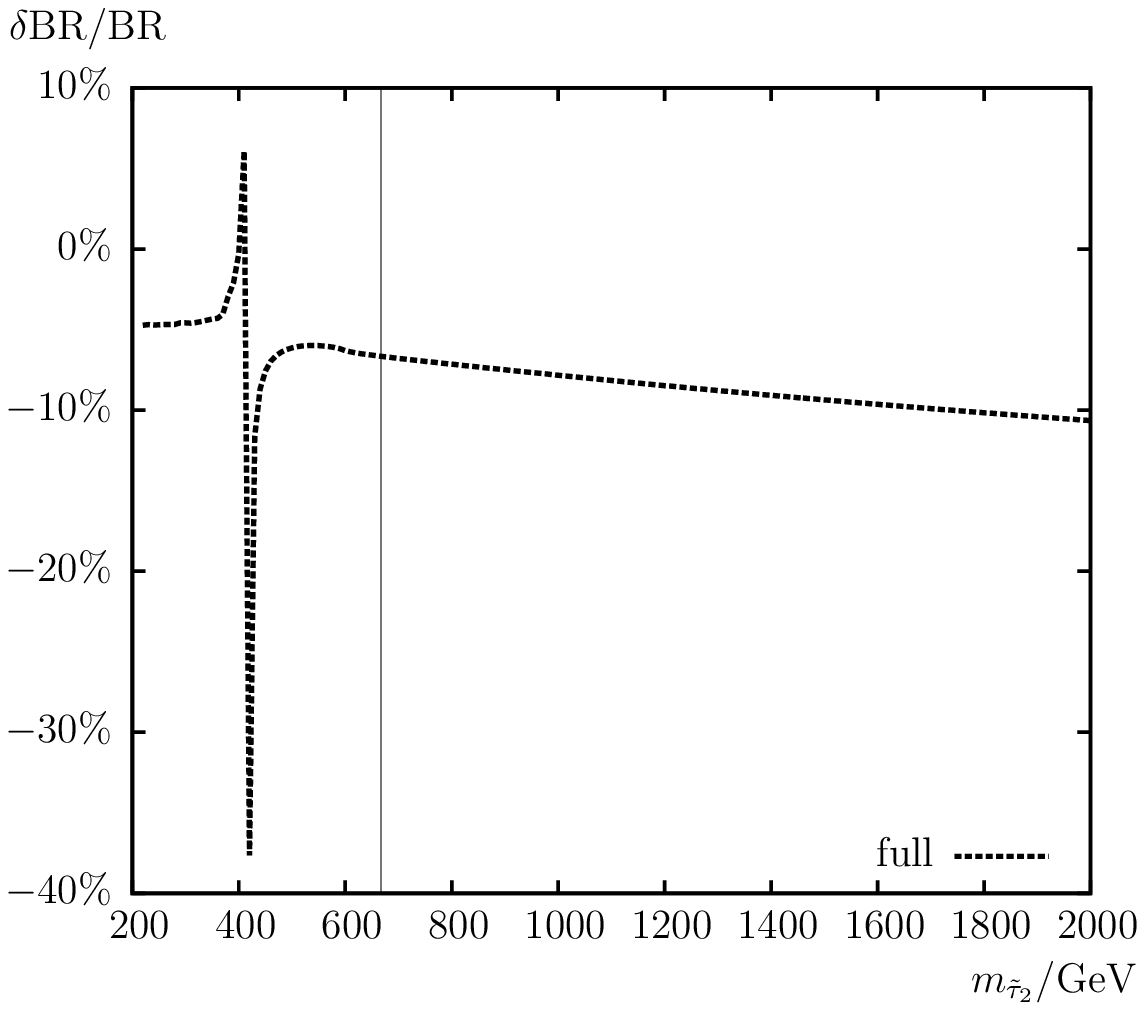}
\end{tabular}
\vspace{2em}
\caption{
  $\Ga(\decayW)$. Tree-level and full one-loop corrected partial decay widths 
  are shown with the parameters chosen according to \SE\
  (see \refta{tab:para}), with $\mstauz$ varied.
  The upper left plot shows the partial decay width, the upper right plot shows 
  the corresponding relative size of the corrections.
  The lower left plot shows the BR, the lower right plot shows 
  the relative correction of the BR.
  The vertical lines indicate where $\mstauz + \mstaue = 1000 \gev$, 
  i.e.\ the maximum reach of the ILC(1000).
}
\label{fig:mst2.stau2snW}
\end{center}
\end{figure}

Finally, results for the other decay mode involving scalar neutrinos, 
$\decayW$, are shown in \reffi{fig:mst2.stau2snW}.
We find two additional dips due to thresholds at $\mstauz = 291, 601 \gev$,
where $\msneut = \mtau + \mcha{1,2}$. Furthermore, as in the decay 
$\decayZ$, at $\mstauz \approx 416.7 \gev$ a sign change in the stau
mixing matrix takes place, and the tree-level result comes out zero for 
$U_{\Stau} = \id$. Consequently, around this value the loop corrections
are substantially larger than the tree-level result, however, not
invalidating the perturbative series (see above).
The one-loop corrections to the decay width are found to be below the
$3\%$~level, except around the threshold at $\mstauz \approx 416.7 \gev$.

\clearpage
\newpage

%%%%%%%%%%%%%%%%%%%%%%%%%% F I G U R E %%%%%%%%%%%%%%%%%%%%%%%%%%%%%%%%%%%%%%%%%
%%%%%%%%%%%%%%%%%%%%%%%%%% F I G U R E %%%%%%%%%%%%%%%%%%%%%%%%%%%%%%%%%%%%%%%%%

\newpage

%%%%%%%%%%%%%%%%%%%%%%%%%% F I G U R E %%%%%%%%%%%%%%%%%%%%%%%%%%%%%%%%%%%%%%%%%
%%%%%%%%%%%%%%%%%%%%%%%%%% F I G U R E %%%%%%%%%%%%%%%%%%%%%%%%%%%%%%%%%%%%%%%%%

\newpage

%%%%%%%%%%%%%%%%%%%%%%%%%% F I G U R E %%%%%%%%%%%%%%%%%%%%%%%%%%%%%%%%%%%%%%%%%
%%%%%%%%%%%%%%%%%%%%%%%%%% F I G U R E %%%%%%%%%%%%%%%%%%%%%%%%%%%%%%%%%%%%%%%%%

\newpage

%%%%%%%%%%%%%%%%%%%%%%%%%% F I G U R E %%%%%%%%%%%%%%%%%%%%%%%%%%%%%%%%%%%%%%%%%
%%%%%%%%%%%%%%%%%%%%%%%%%% F I G U R E %%%%%%%%%%%%%%%%%%%%%%%%%%%%%%%%%%%%%%%%%

\newpage

%%%%%%%%%%%%%%%%%%%%%%%%%% F I G U R E %%%%%%%%%%%%%%%%%%%%%%%%%%%%%%%%%%%%%%%%%
%%%%%%%%%%%%%%%%%%%%%%%%%% F I G U R E %%%%%%%%%%%%%%%%%%%%%%%%%%%%%%%%%%%%%%%%%

\newpage

%%%%%%%%%%%%%%%%%%%%%%%%%% F I G U R E %%%%%%%%%%%%%%%%%%%%%%%%%%%%%%%%%%%%%%%%%
%%%%%%%%%%%%%%%%%%%%%%%%%% F I G U R E %%%%%%%%%%%%%%%%%%%%%%%%%%%%%%%%%%%%%%%%%

\newpage

%%%%%%%%%%%%%%%%%%%%%%%%%% F I G U R E %%%%%%%%%%%%%%%%%%%%%%%%%%%%%%%%%%%%%%%%%
%%%%%%%%%%%%%%%%%%%%%%%%%% F I G U R E %%%%%%%%%%%%%%%%%%%%%%%%%%%%%%%%%%%%%%%%%

\newpage

%%%%%%%%%%%%%%%%%%%%%%%%%% F I G U R E %%%%%%%%%%%%%%%%%%%%%%%%%%%%%%%%%%%%%%%%%
%%%%%%%%%%%%%%%%%%%%%%%%%% F I G U R E %%%%%%%%%%%%%%%%%%%%%%%%%%%%%%%%%%%%%%%%%

\newpage

%%%%%%%%%%%%%%%%%%%%%%%%%% F I G U R E %%%%%%%%%%%%%%%%%%%%%%%%%%%%%%%%%%%%%%%%%
%%%%%%%%%%%%%%%%%%%%%%%%%% F I G U R E %%%%%%%%%%%%%%%%%%%%%%%%%%%%%%%%%%%%%%%%%

\newpage

%%%%%%%%%%%%%%%%%%%%%%%%%% F I G U R E %%%%%%%%%%%%%%%%%%%%%%%%%%%%%%%%%%%%%%%%%
%%%%%%%%%%%%%%%%%%%%%%%%%% F I G U R E %%%%%%%%%%%%%%%%%%%%%%%%%%%%%%%%%%%%%%%%%

\newpage

%%%%%%%%%%%%%%%%%%%%%%%%%% F I G U R E %%%%%%%%%%%%%%%%%%%%%%%%%%%%%%%%%%%%%%%%%
%%%%%%%%%%%%%%%%%%%%%%%%%% F I G U R E %%%%%%%%%%%%%%%%%%%%%%%%%%%%%%%%%%%%%%%%%

\newpage

%%%%%%%%%%%%%%%%%%%%%%%%%% F I G U R E %%%%%%%%%%%%%%%%%%%%%%%%%%%%%%%%%%%%%%%%%
%%%%%%%%%%%%%%%%%%%%%%%%%% F I G U R E %%%%%%%%%%%%%%%%%%%%%%%%%%%%%%%%%%%%%%%%%

\clearpage
\newpage

%%%%%%%%%%%%%%%%%%%%%%%%%%%%%%%%%%%%%%%%%%%%%%%%%%%%%%%%%%%%%%%%%%%%%%%%%%%%%%

\subsection{Full one-loop results for varying \boldmath{$\phiatau$}}
\label{sec:full1Lphiat}

In this subsection we analyze the various partial decay widths%
\footnote{
  Again we note, that we do not investigate the decays of $\aStauz$ here, 
  which would correspond to an analysis of $\cp$-asymmetries, which is 
  beyond the scope of this paper.
} 
and branching ratios as a function of $\phiatau$. 
The other parameters are chosen according to \refta{tab:para}. 
Thus, within \SE\ we have $\mstaue + \mstauz = 825 \gev$, i.e.\ 
the production channel $e^+e^- \to \aStaue\Stauzm$ 
is open at the ILC(1000). 
Consequently, the accuracy of the prediction of the various partial decay 
widths and branching ratios should be at the same level (or better) as 
the anticipated ILC precision.
It should be noted that already the tree-level prediction depends on 
$\phiatau$ via the stau mixing matrix.

When performing an analysis involving complex parameters it should be 
noted that the results for physical observables are affected only by 
certain combinations of the complex phases of the parameters $\mu$, the 
trilinear couplings $A_f$ ($f = \tau, t, b, \ldots$) and the gaugino 
mass parameters $M_1$, $M_2$, $M_3$~\cite{MSSMcomplphasen,SUSYphases}.
It is possible, for instance, to rotate the phase $\phiMz$ away.
Experimental constraints on the (combinations of) complex phases 
arise in particular from their contributions to electric dipole moments
of the electron and the neutron (see \citeres{EDMrev2,EDMPilaftsis} and
references therein), of the deuteron~\cite{EDMRitz} and of
heavy quarks~\cite{EDMDoink}.
While SM contributions enter only at the three-loop level, due to its
complex phases the MSSM can contribute already at one-loop order.
Large phases in the first two generations of sfermions
can only be accommodated if these generations are assumed to be very
heavy~\cite{EDMheavy} or large cancellations occur~\cite{EDMmiracle},
see however the discussion in \citere{EDMrev1}. 
A recent review can be found in \citere{EDMrev3}.
Accordingly (using the convention that $\phiMz = 0$, as done in this paper), 
in particular the phase $\phimu$ is tightly constrained~\cite{plehnix}, 
while the bounds on the phases of the third generation trilinear couplings 
are much weaker.
The phases of $\mu$ and $A_{\tau,t,b}$ enter only in the combinations 
$(\varphi_{A_{\tau,t,b}} + \phimu)$ (or in different combinations
together with phases of $M_1$ or $M_3$). 
Setting $\phimu = 0$ (see above) as well as $\phiMe = 0$ 
(we do not consider this phase in this paper)
leaves us with the trilinear couplings as the only complex valued
parameters. The dependence on $\phiab$ and $\phiat$ on the
partial decay widths involving scalar bottom and top
quarks has been analyzed in detail in \citeres{SbotRen,Stop2decay},
and these phases only enter via loop corrections into the prediction for
the stau decays, whereas $\Atau$ enters at the tree-level.
Consequently, we focus on a complex~$\Atau$ and keep $\At$ and
$\Ab$ real. 

Since now a complex $\Atau$ can appear in the couplings, contributions 
from absorptive parts of self-energy type corrections on external legs can
arise, and they are included in the numerical results shown as
``full''.
The corresponding formulas for an inclusion of these absorptive 
contributions via finite wave function correction factors can be found 
in \refse{sec:cMSSM}.

\begin{figure}[htb!]
\begin{center}
\begin{tabular}{c}
\includegraphics[width=0.49\textwidth,height=7.5cm]{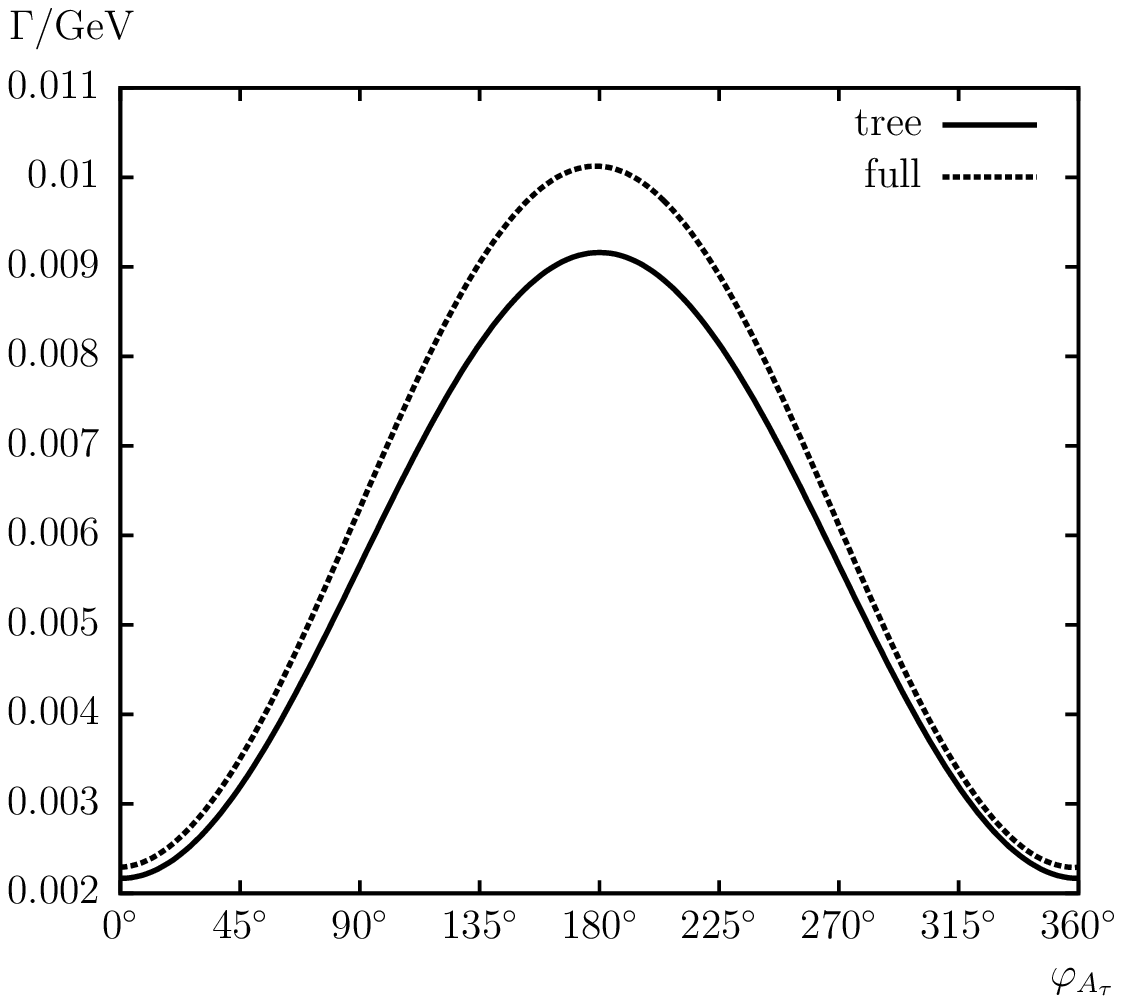}
\hspace{-4mm}
\includegraphics[width=0.49\textwidth,height=7.5cm]{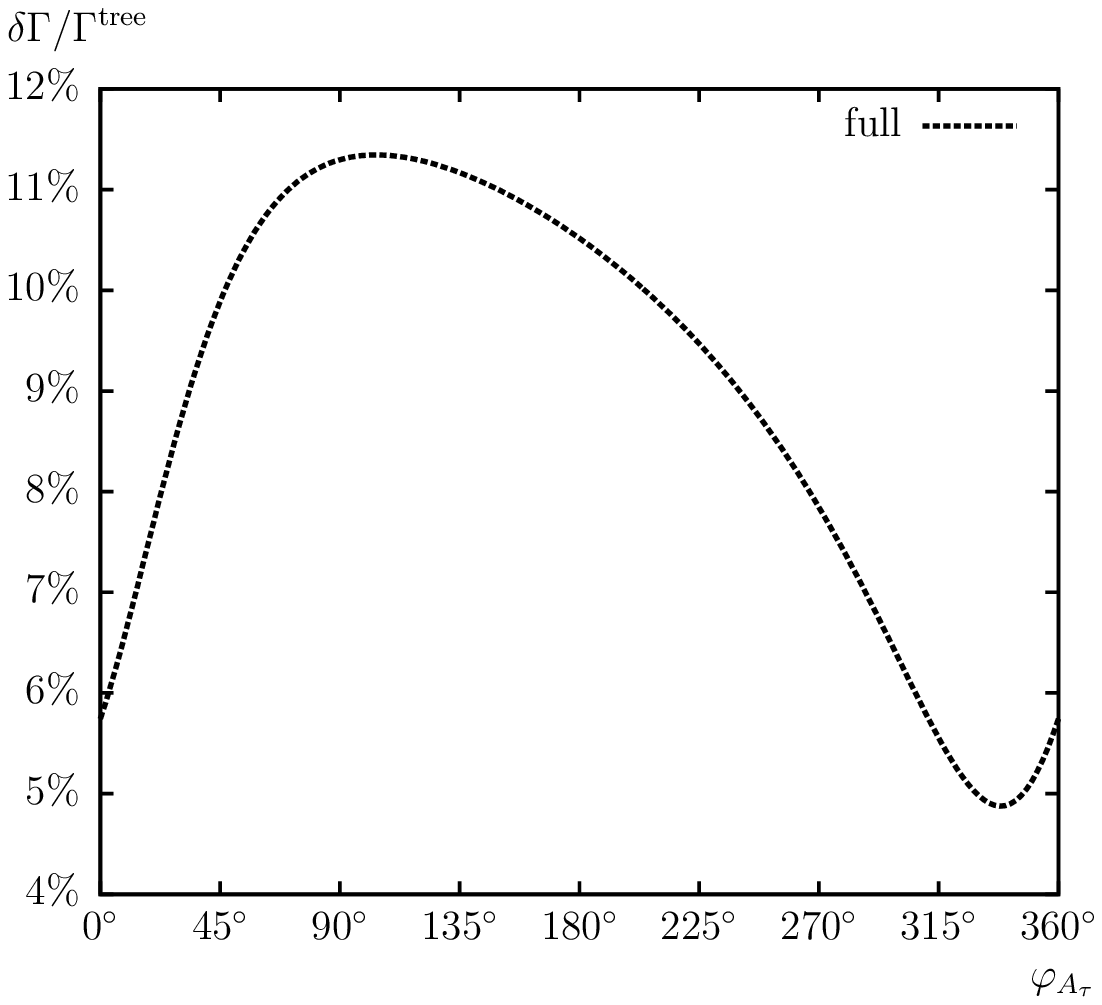}
\\[4em]
\includegraphics[width=0.49\textwidth,height=7.5cm]{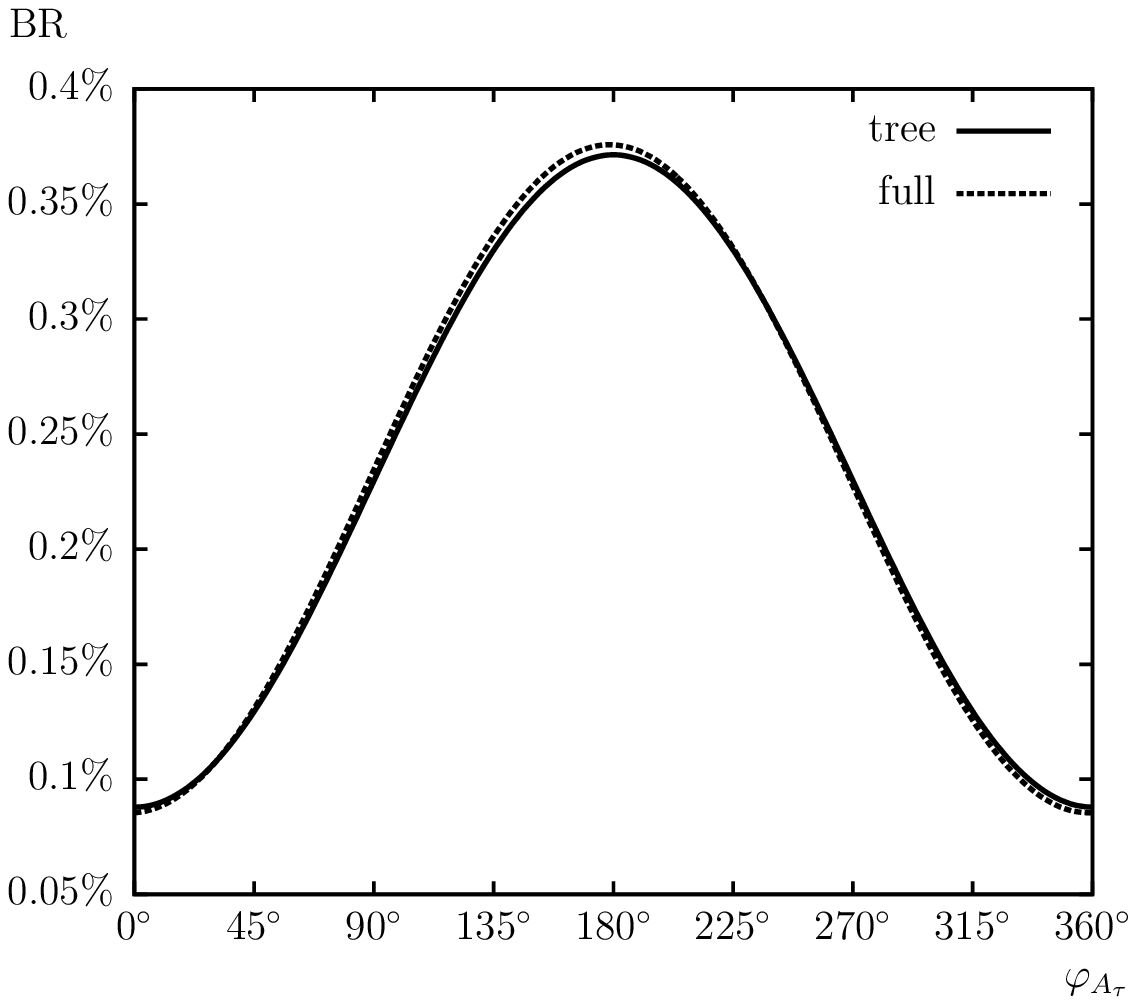}
\hspace{-4mm}
\includegraphics[width=0.49\textwidth,height=7.5cm]{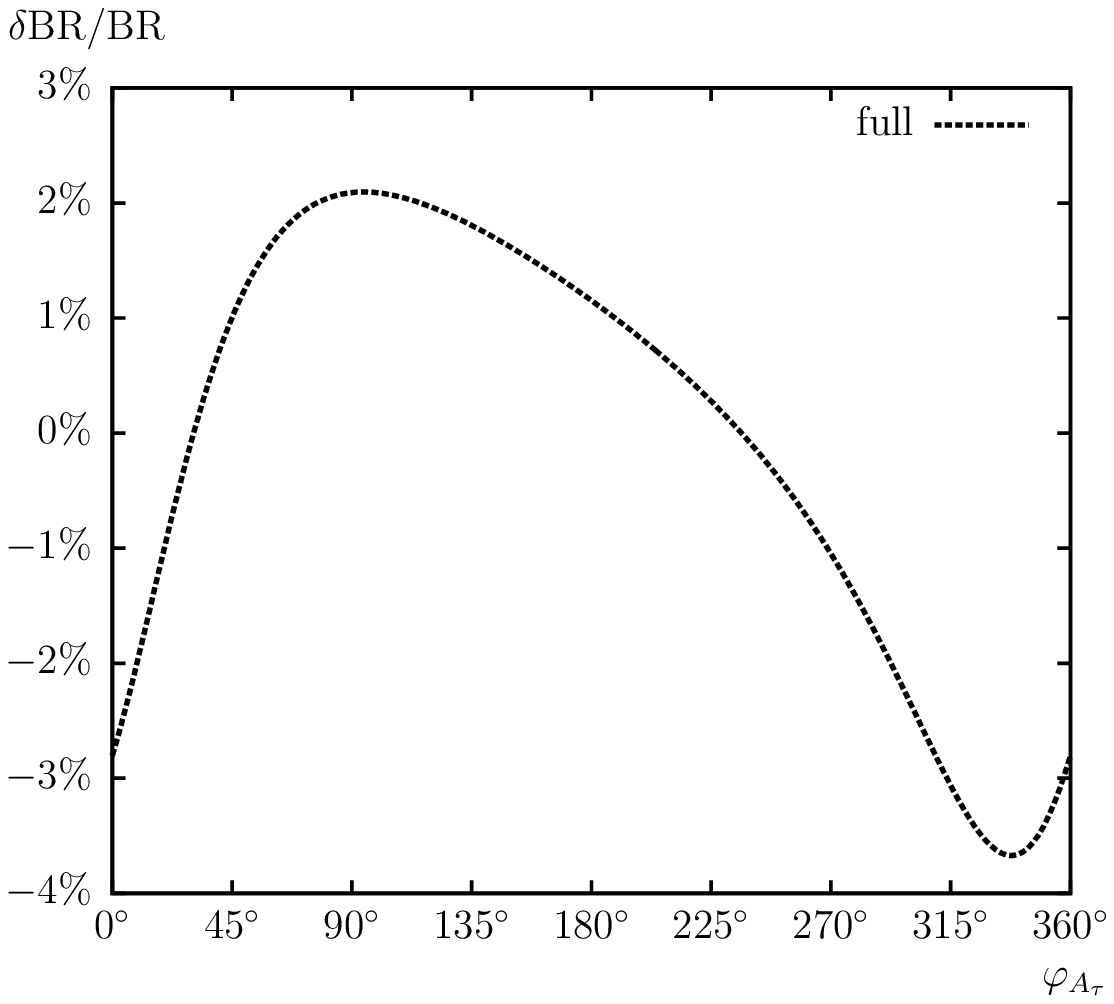}
\end{tabular}
\vspace{2em}
\caption{$\Ga(\decayh)$. 
  Tree-level (``tree'') and full one-loop (``full'') corrected partial decay 
  widths (including absorptive self-energy contributions) are shown.
  The parameters are chosen according to \SE\ (see \refta{tab:para}), 
  with $\phiatau$ varied.
  The upper left plot shows the partial decay width, the upper right plot 
  the corresponding relative size of the corrections. 
  The lower left plot shows the BR, the lower right plot 
  the relative correction of the BR.
}
\label{fig:PhiAt.stau2stau1h1}
\end{center}
\end{figure}

\begin{figure}[htb!]
\begin{center}
\begin{tabular}{c}
\includegraphics[width=0.49\textwidth,height=7.5cm]{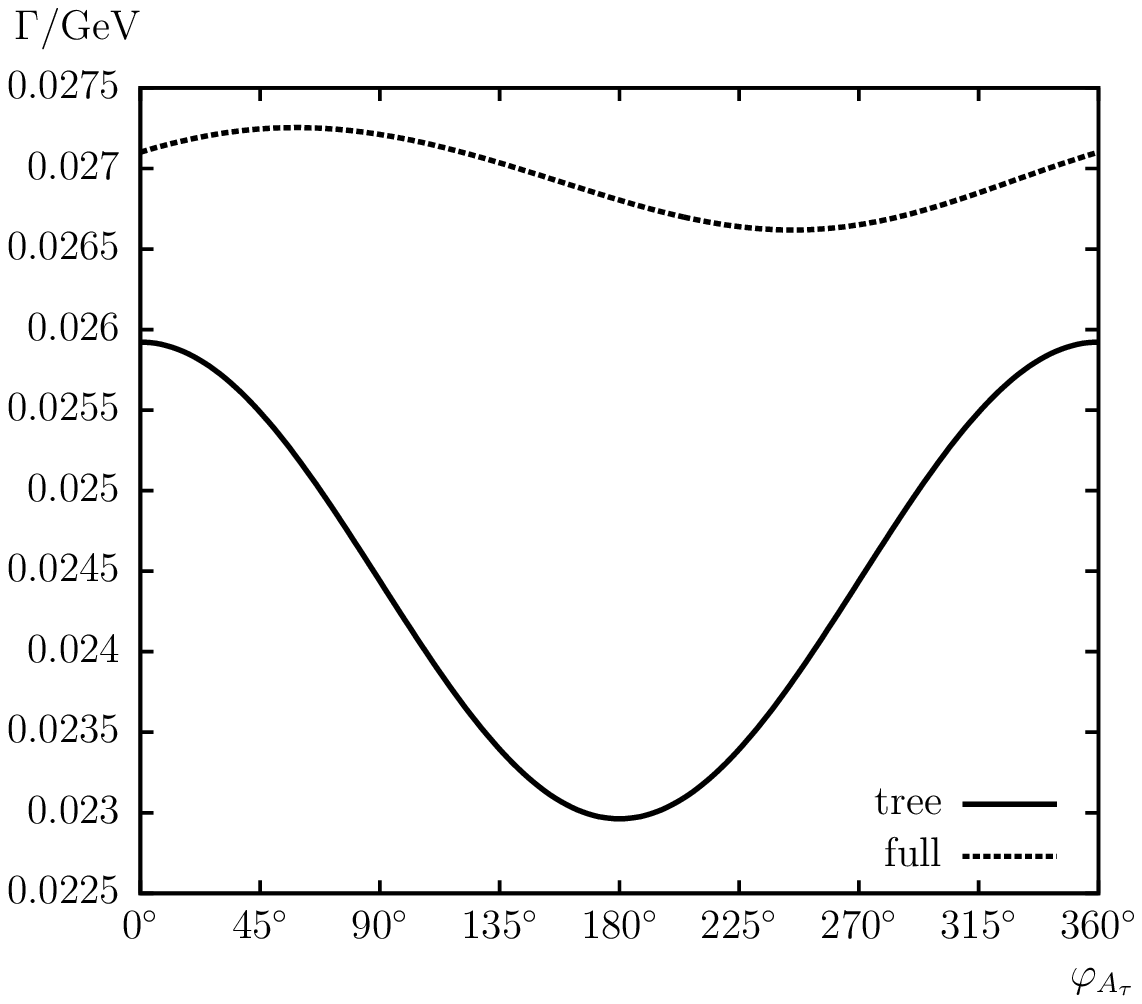}
\hspace{-4mm}
\includegraphics[width=0.49\textwidth,height=7.5cm]{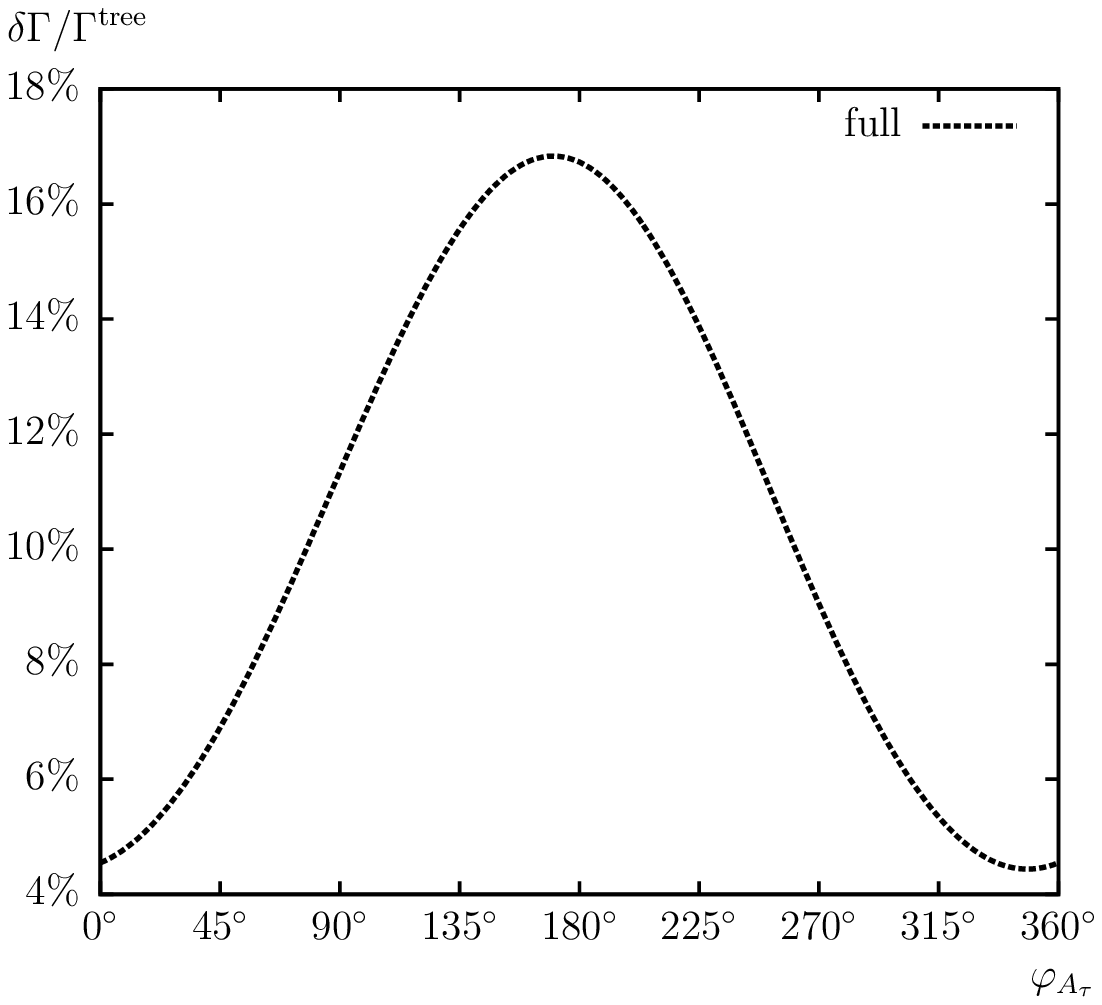}
\\[4em]
\includegraphics[width=0.49\textwidth,height=7.5cm]{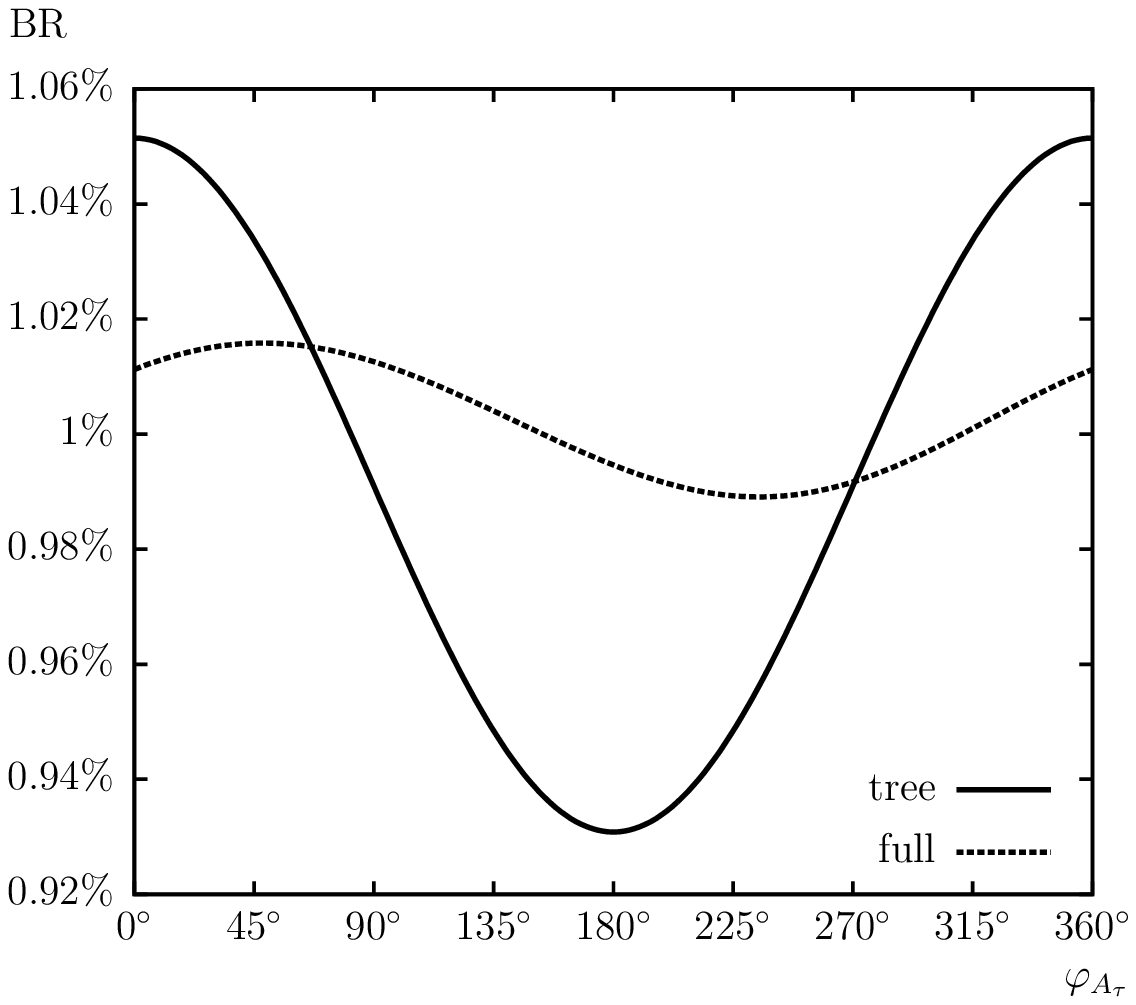}
\hspace{-4mm}
\includegraphics[width=0.49\textwidth,height=7.5cm]{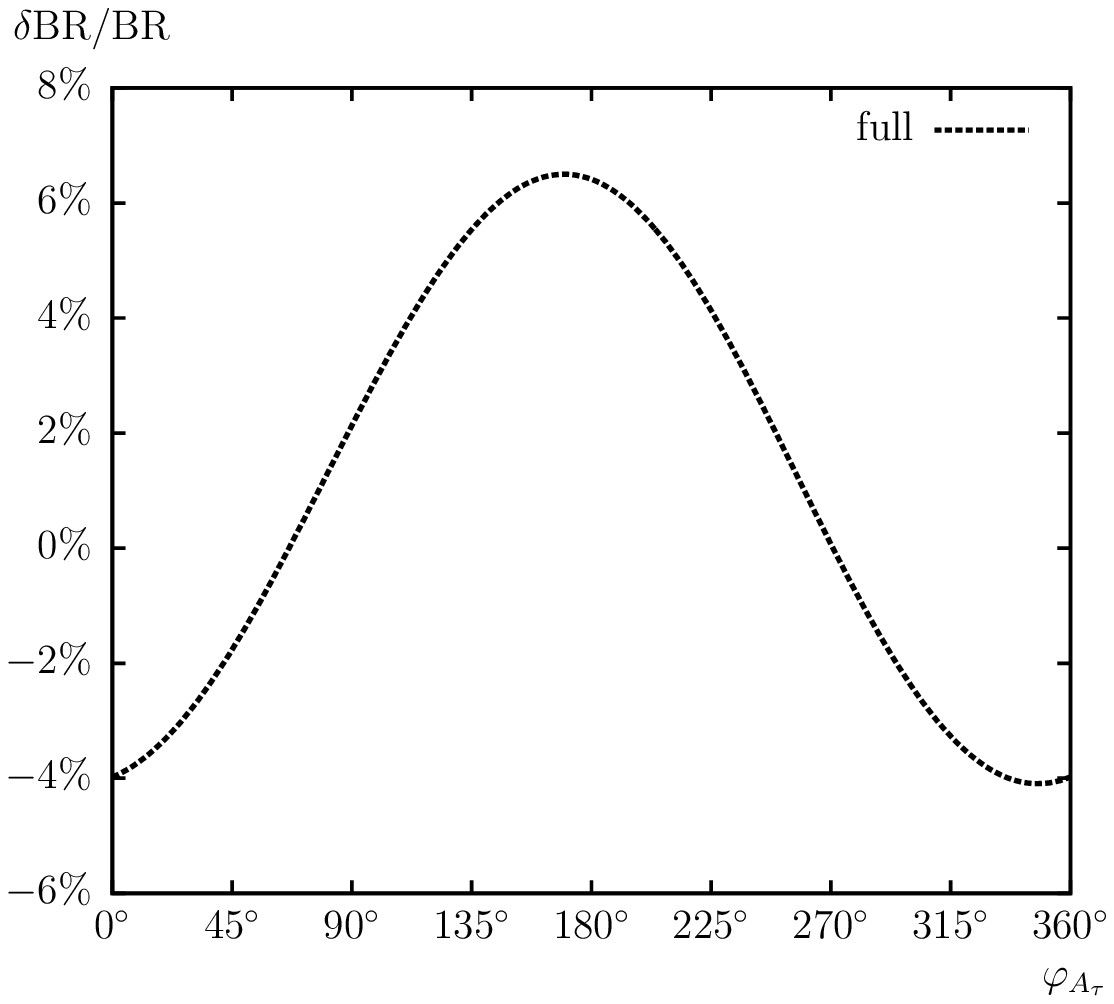}
\end{tabular}
\vspace{2em}
\caption{$\Ga(\decayH)$. 
  Tree-level (``tree'') and full one-loop (``full'') corrected partial decay 
  widths (including absorptive self-energy contributions) are shown. 
  The parameters are chosen according to \SE\ 
  (see \refta{tab:para}), with $\phiatau$ varied.
  The upper left plot shows the partial decay width, the upper right plot 
  the corresponding relative size of the corrections. 
  The lower left plot shows the BR, the lower right plot  
  the relative correction of the BR.
}
\label{fig:PhiAt.stau2stau1h2}
\end{center}
\end{figure}

\begin{figure}[htb!]
\begin{center}
\begin{tabular}{c}
\includegraphics[width=0.49\textwidth,height=7.5cm]{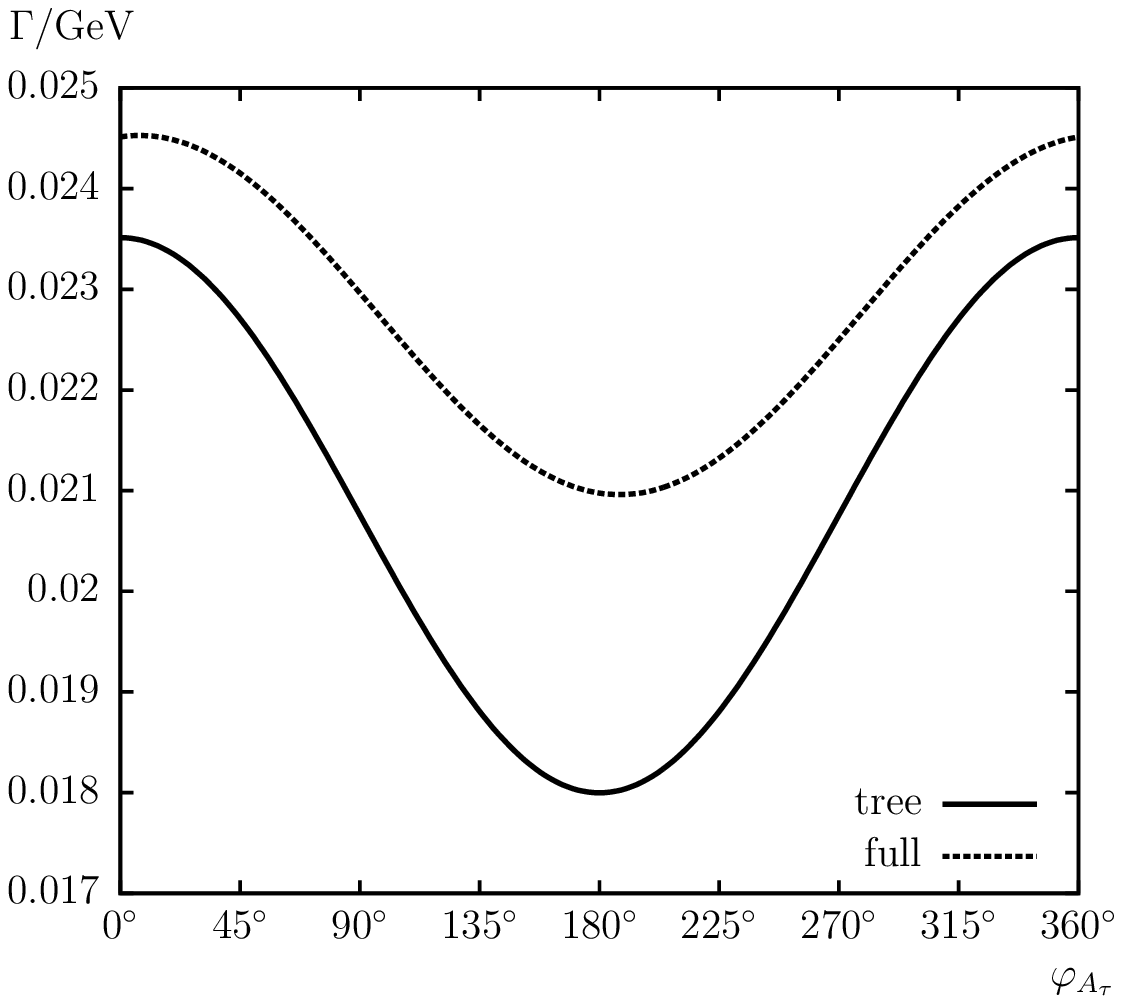}
\hspace{-4mm}
\includegraphics[width=0.49\textwidth,height=7.5cm]{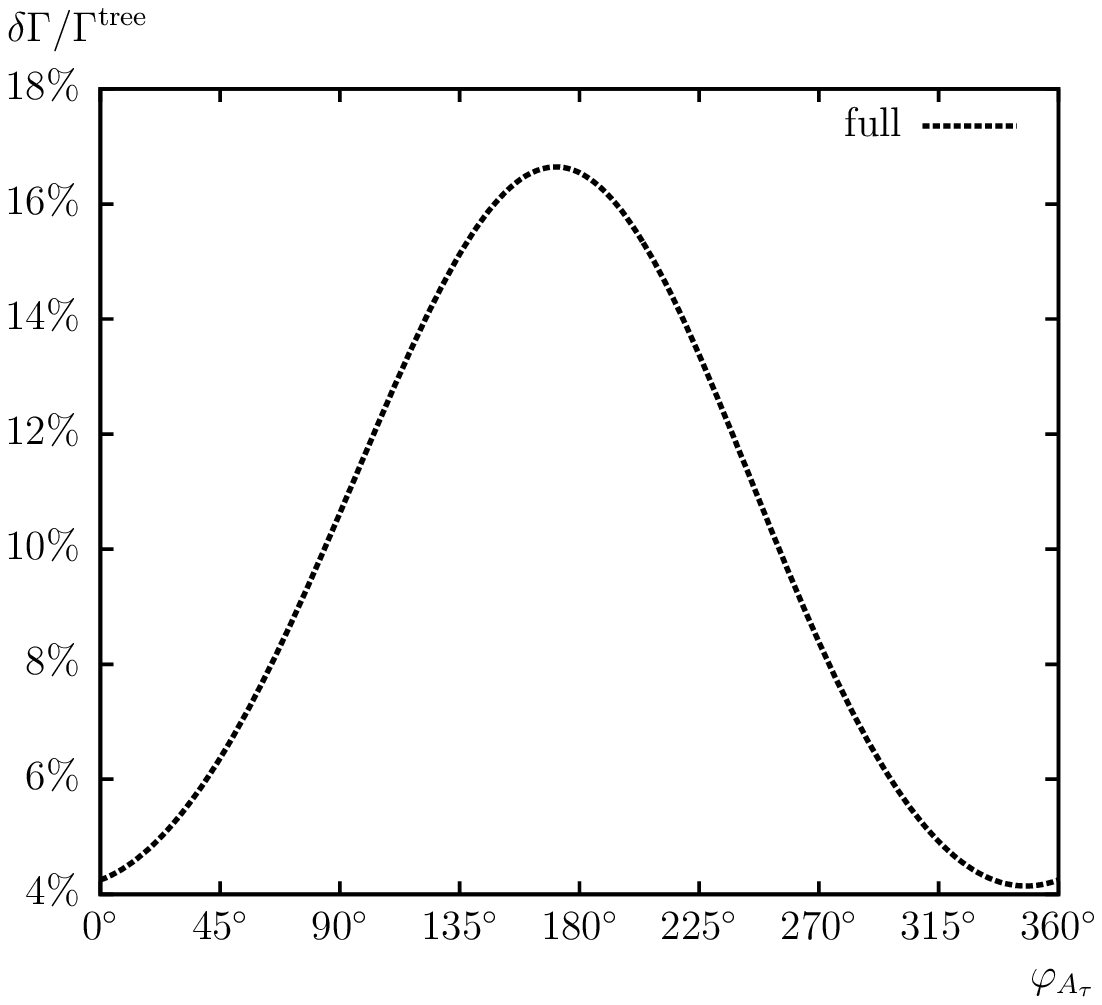}
\\[4em]
\includegraphics[width=0.49\textwidth,height=7.5cm]{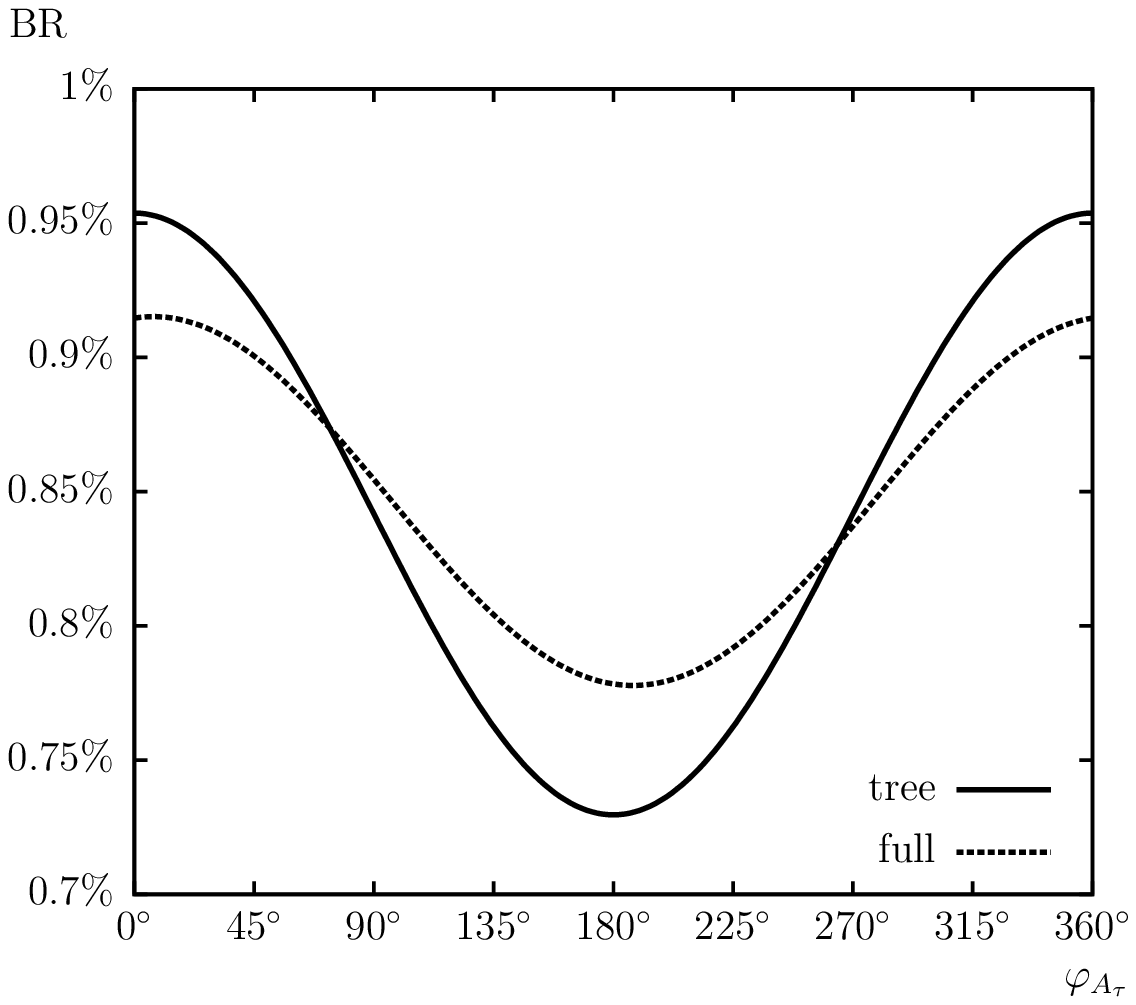}
\hspace{-4mm}
\includegraphics[width=0.49\textwidth,height=7.5cm]{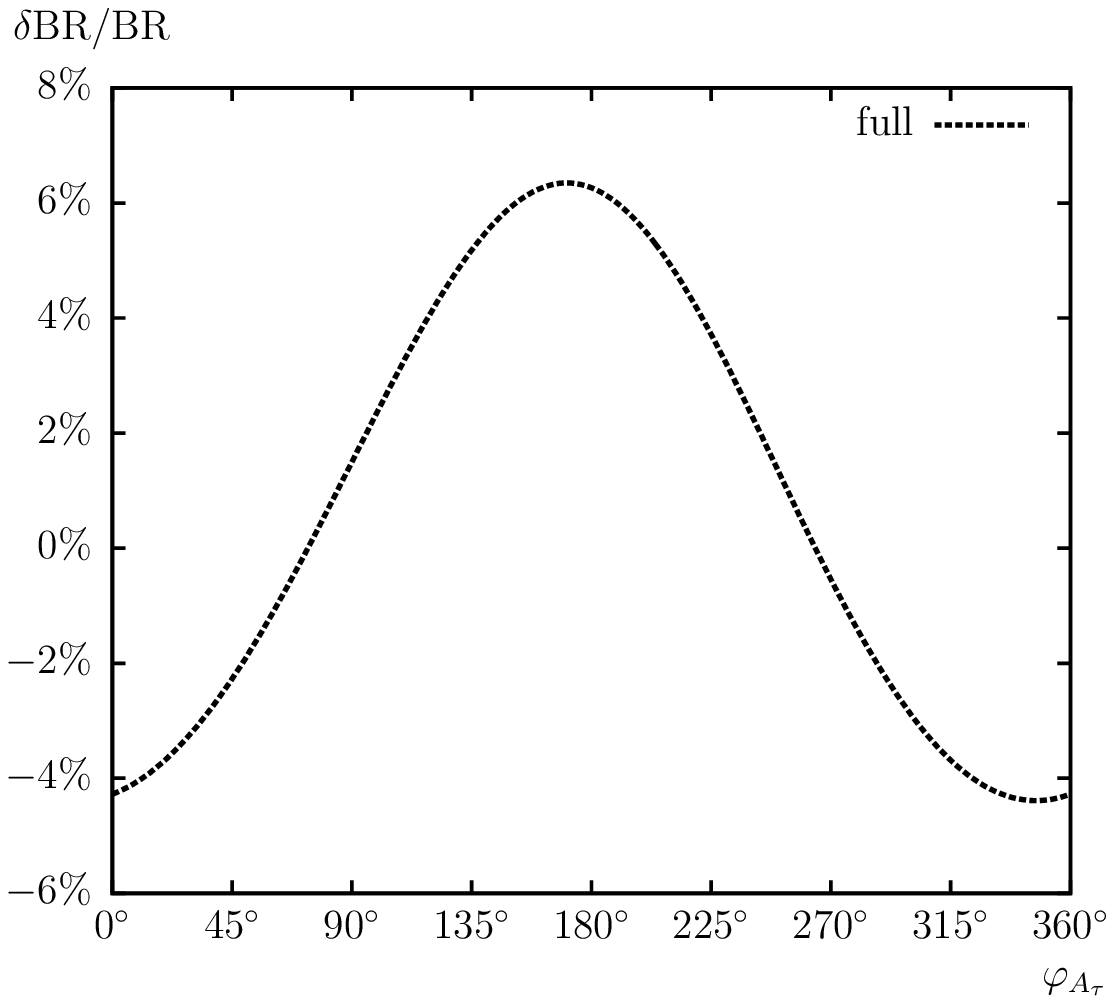}
\end{tabular}
\vspace{2em}
\caption{$\Ga(\decayA)$. 
  Tree-level (``tree'') and full one-loop (``full'') corrected partial decay 
  widths (including absorptive self-energy contributions) are shown. 
  The parameters are chosen according to \SE\ (see \refta{tab:para}), 
  with $\phiatau$ varied.
  The upper left plot shows the partial decay width, the upper right plot  
  the corresponding  relative size of the corrections. 
  The lower left plot shows the BR, the lower right plot  
  the relative correction of the BR.
}
\label{fig:PhiAt.stau2stau1h3}
\end{center}
\end{figure}

As before we start with the decays to Higgs bosons, $\decayhn$ ($n = 1,2,3$)
shown in \reffi{fig:PhiAt.stau2stau1h1} -- \ref{fig:PhiAt.stau2stau1h3}.
The arrangement of the panels is the same as in the previous subsection.
In \reffi{fig:PhiAt.stau2stau1h1}, where the  partial decay width
$\Ga(\decayh)$ is given as a 
function of $\phiatau$, one can see that the size of the correction to
the partial decay width varies substantially with $\phiatau$. 
The one-loop effects range from $+5\%$ to $+11\%$ in \SE.
It should be kept in mind that the parameters are chosen such that 
$e^+e^- \to \aStaue\Stauzm$ is kinematically possible 
at the ILC(1000) in 
\SE, where the knowledge of such a large variation can be very important. 
For $\decayH$, shown in \reffi{fig:PhiAt.stau2stau1h2}, the variation
with $\phiatau$ is even larger, ranging from $+5\%$ to $+16\%$ with
similar conclusions for the ILC(1000) as above.
The results for $\decayA$ can be found in \reffi{fig:PhiAt.stau2stau1h3}.
Also here the size of the corrections shows a large variation with
$\phiatau$, similar to $\Ga(\decayH)$.
For the two heavier Higgs bosons, also the variation of the respective
branching ratios with $\phiatau$ is substantial in \SE, ranging from
$-4\%$ to $+6\%$, potentially exceeding the ILC precision.

\begin{figure}[htb!]
\begin{center}
\begin{tabular}{c}
\includegraphics[width=0.49\textwidth,height=7.5cm]{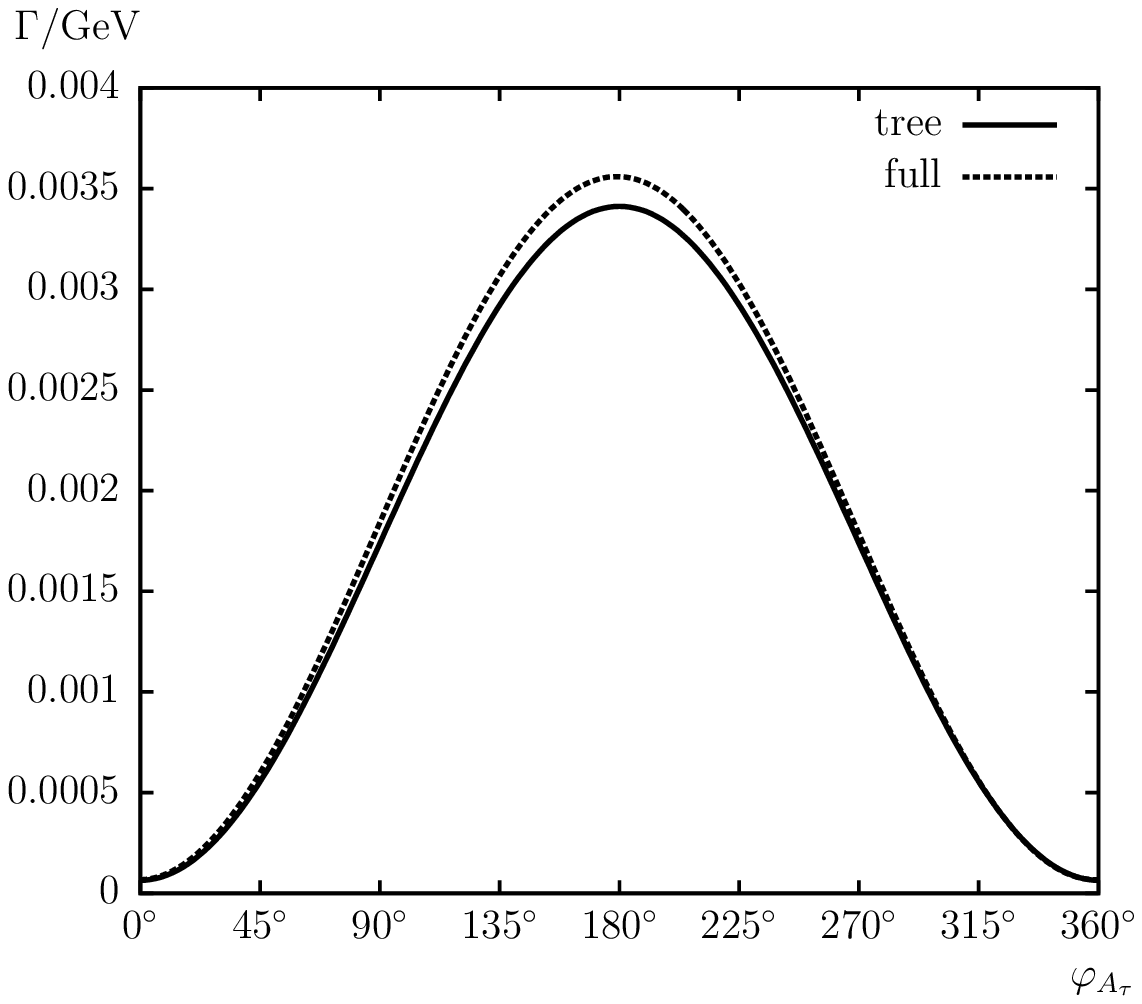}
\hspace{-4mm}
\includegraphics[width=0.49\textwidth,height=7.5cm]{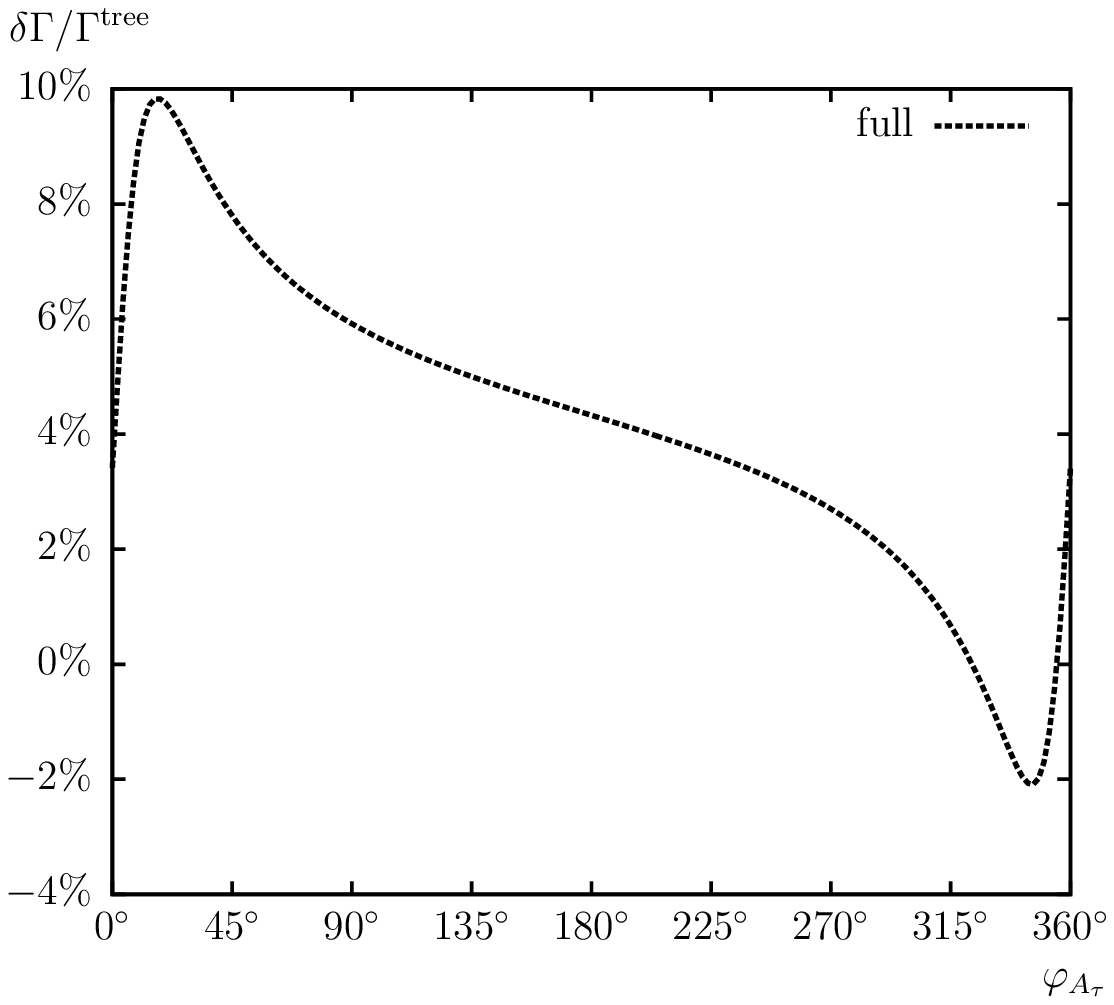}
\\[4em]
\includegraphics[width=0.49\textwidth,height=7.5cm]{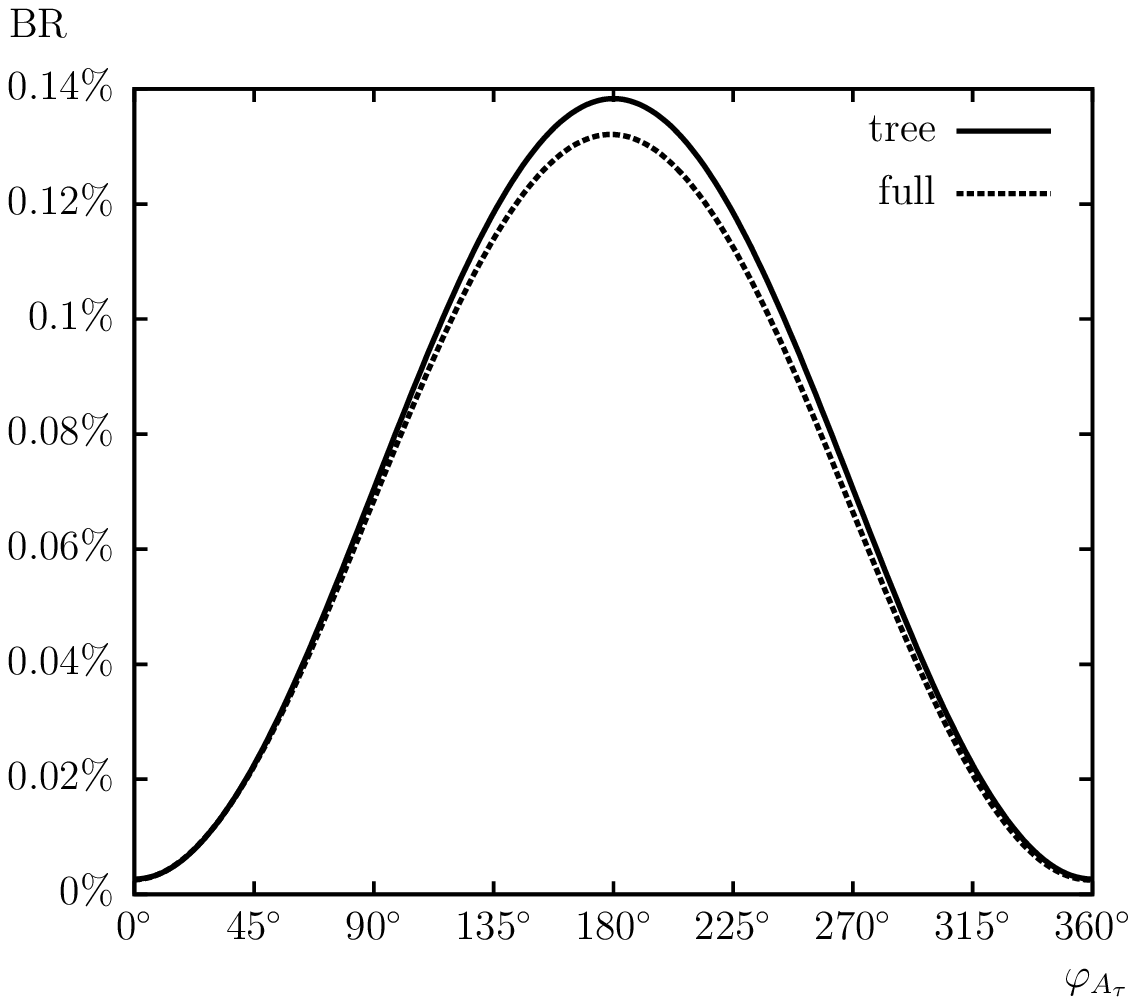}
\hspace{-4mm}
\includegraphics[width=0.49\textwidth,height=7.5cm]{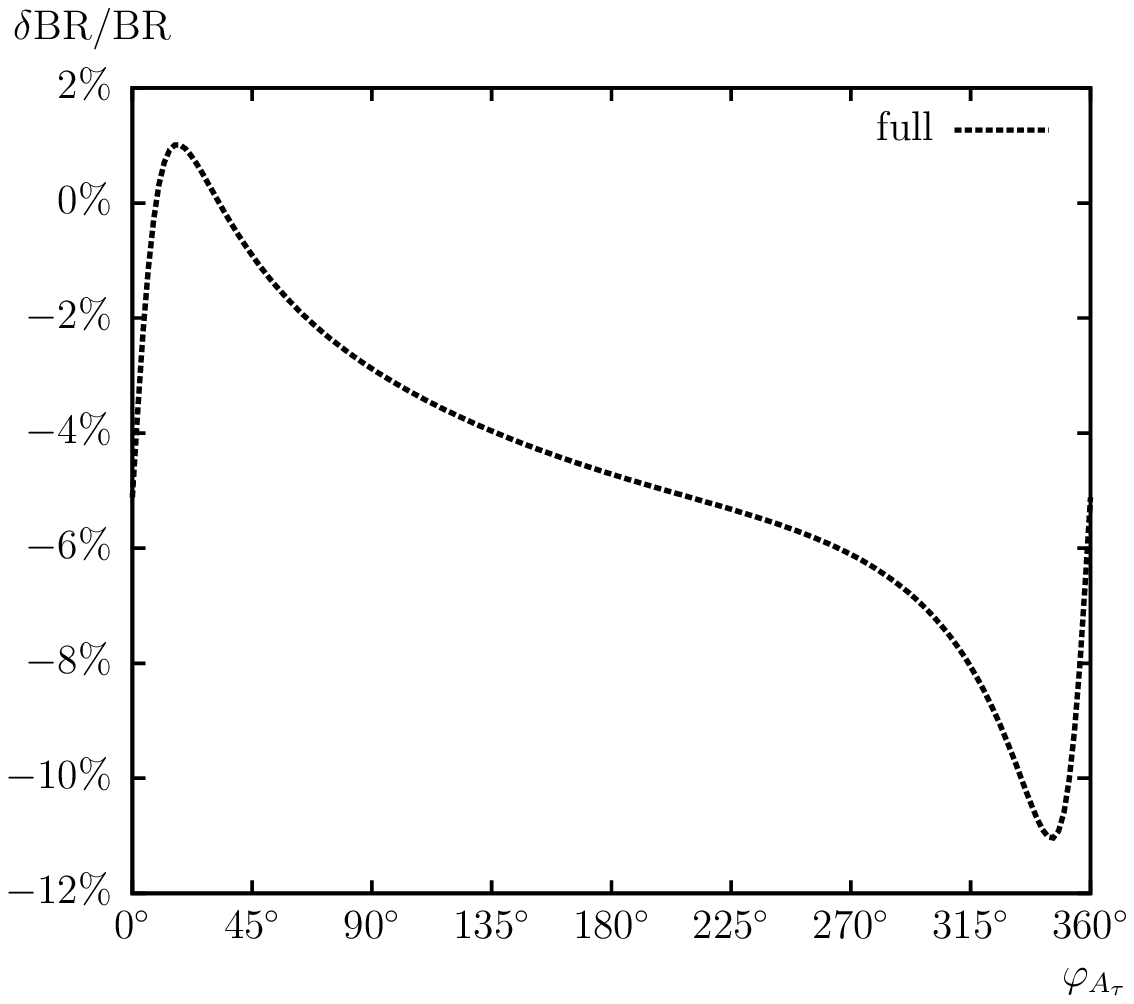}
\end{tabular}
\vspace{2em}
\caption{$\Ga(\decayZ)$. 
  Tree-level (``tree'') and full one-loop (``full'') corrected partial decay 
  widths (including absorptive self-energy contributions) are shown. 
  The parameters are chosen according to \SE\ (see \refta{tab:para}), 
  with $\phiatau$ varied.
  The upper left plot shows the partial decay width, the upper right plot  
  the corresponding  relative size of the corrections. 
  The lower left plot shows the BR, the lower right plot  
  the relative correction of the BR.
}
\label{fig:PhiAt.stau2stau1Z}
\end{center}
\end{figure}

In \reffi{fig:PhiAt.stau2stau1Z} we present the phase dependence for the
decay mode $\decayZ$. In our scenario \SE\ the effect of the one-loop
corrections to $\Ga(\decayZ)$ varies from $\sim -2\%$ to $\sim +10\%$,
again relevant for the ILC precision. An effect of similar size can be
observed for $\br(\decayZ)$.

\begin{figure}[htb!]
\begin{center}
\begin{tabular}{c}
\includegraphics[width=0.49\textwidth,height=7.5cm]{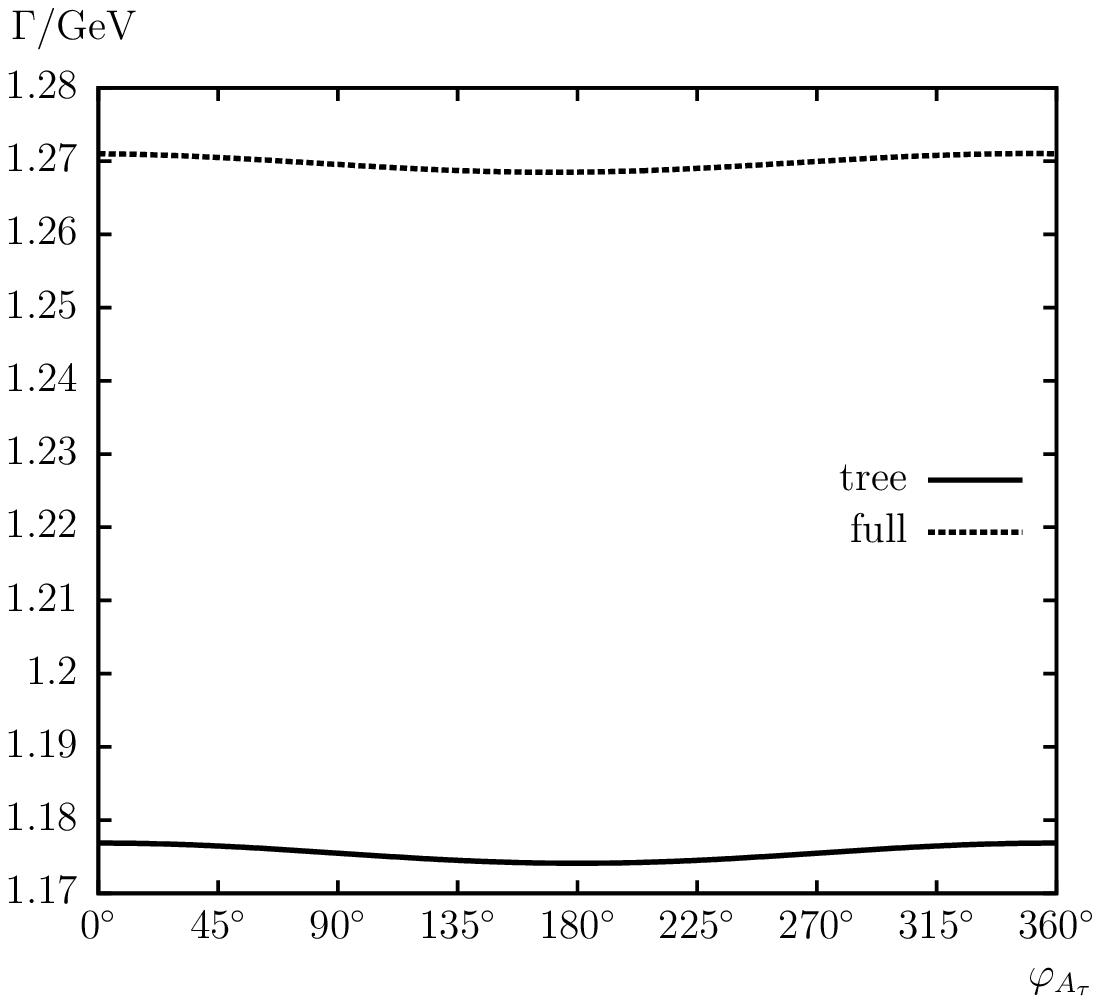}
\hspace{-4mm}
\includegraphics[width=0.49\textwidth,height=7.5cm]{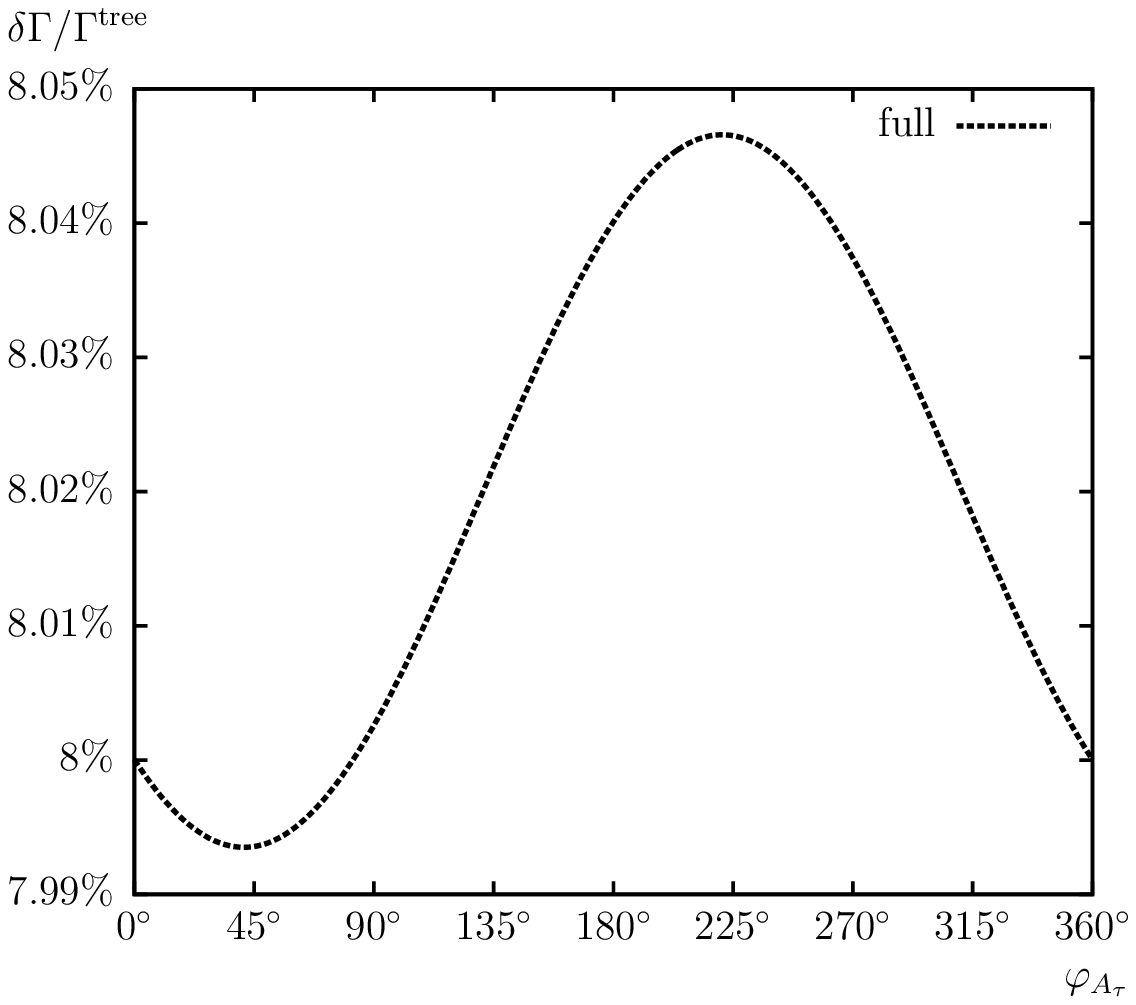}
\\[4em]
\includegraphics[width=0.49\textwidth,height=7.5cm]{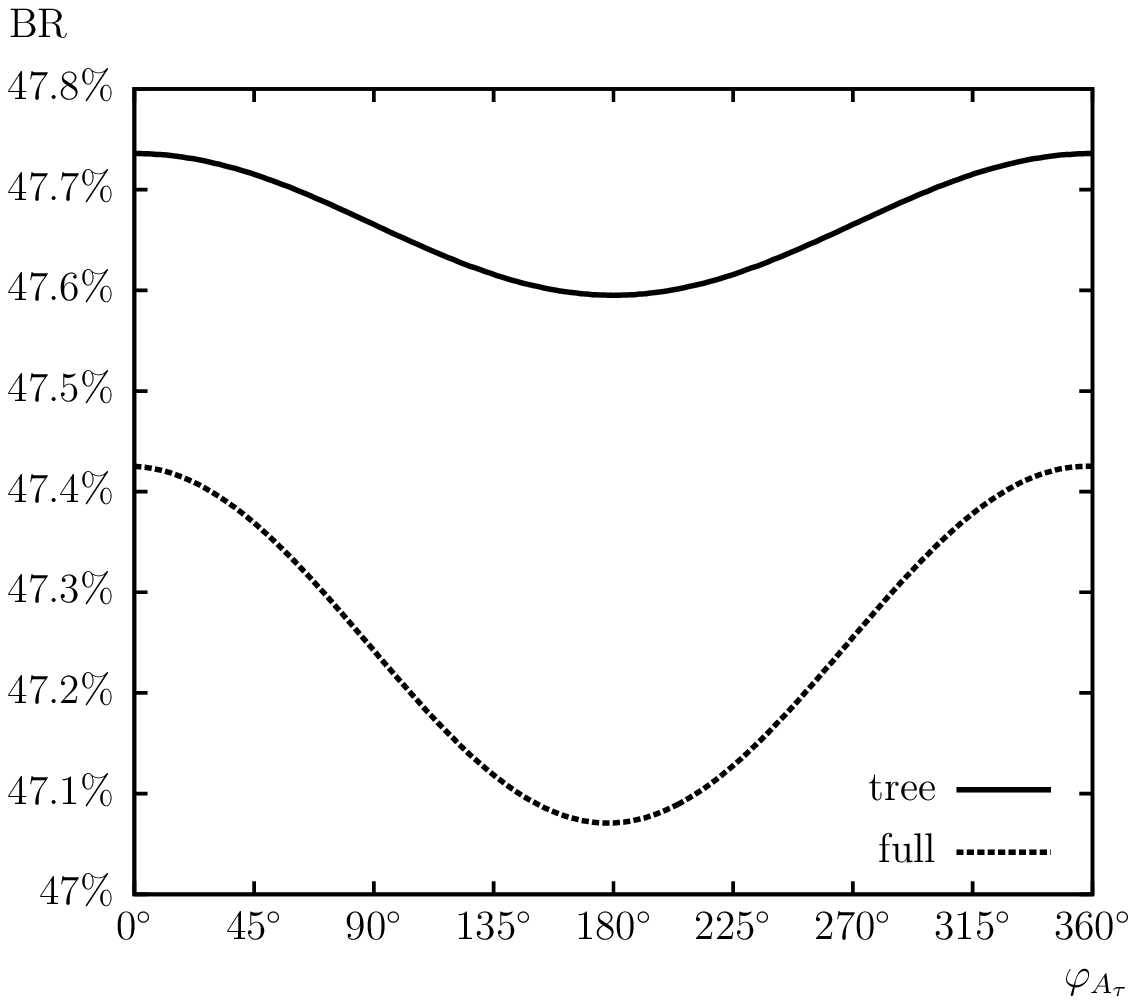}
\hspace{-4mm}
\includegraphics[width=0.49\textwidth,height=7.5cm]{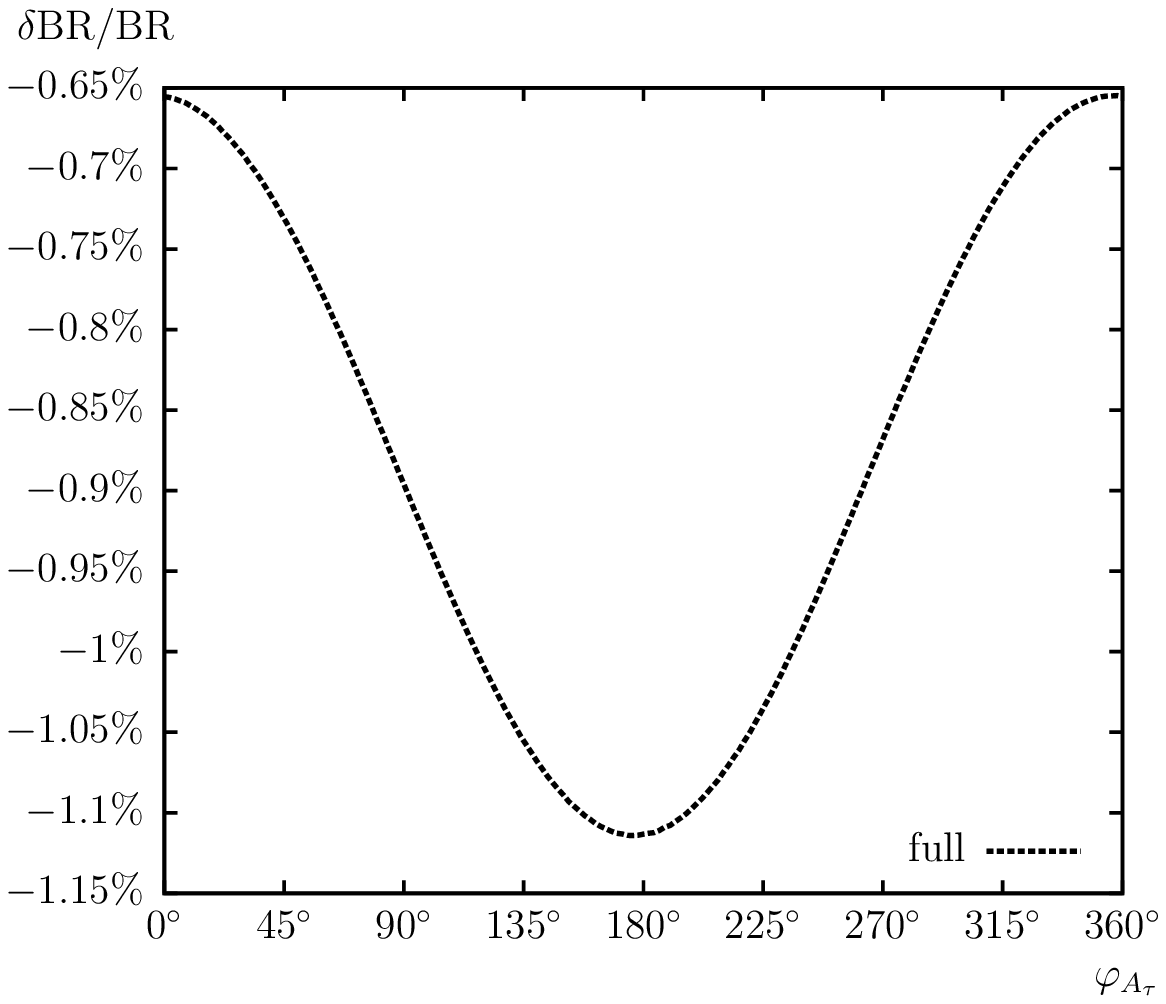}
\end{tabular}
\vspace{2em}
\caption{$\Ga(\decayNe)$. 
  Tree-level (``tree'') and full one-loop (``full'') corrected partial decay 
  widths (including absorptive self-energy contributions) are shown.  
  The parameters are chosen according to \SE\ (see \refta{tab:para}), 
  with $\phiatau$ varied.
  The upper left plot shows the partial decay width, the upper right plot  
  the corresponding  relative size of the corrections. 
  The lower left plot shows the BR, the lower right plot  
  the relative correction of the BR.
}
\label{fig:PhiAt.stau2tauneu1}
\end{center}
\end{figure}

\begin{figure}[htb!]
\begin{center}
\begin{tabular}{c}
\includegraphics[width=0.49\textwidth,height=7.5cm]{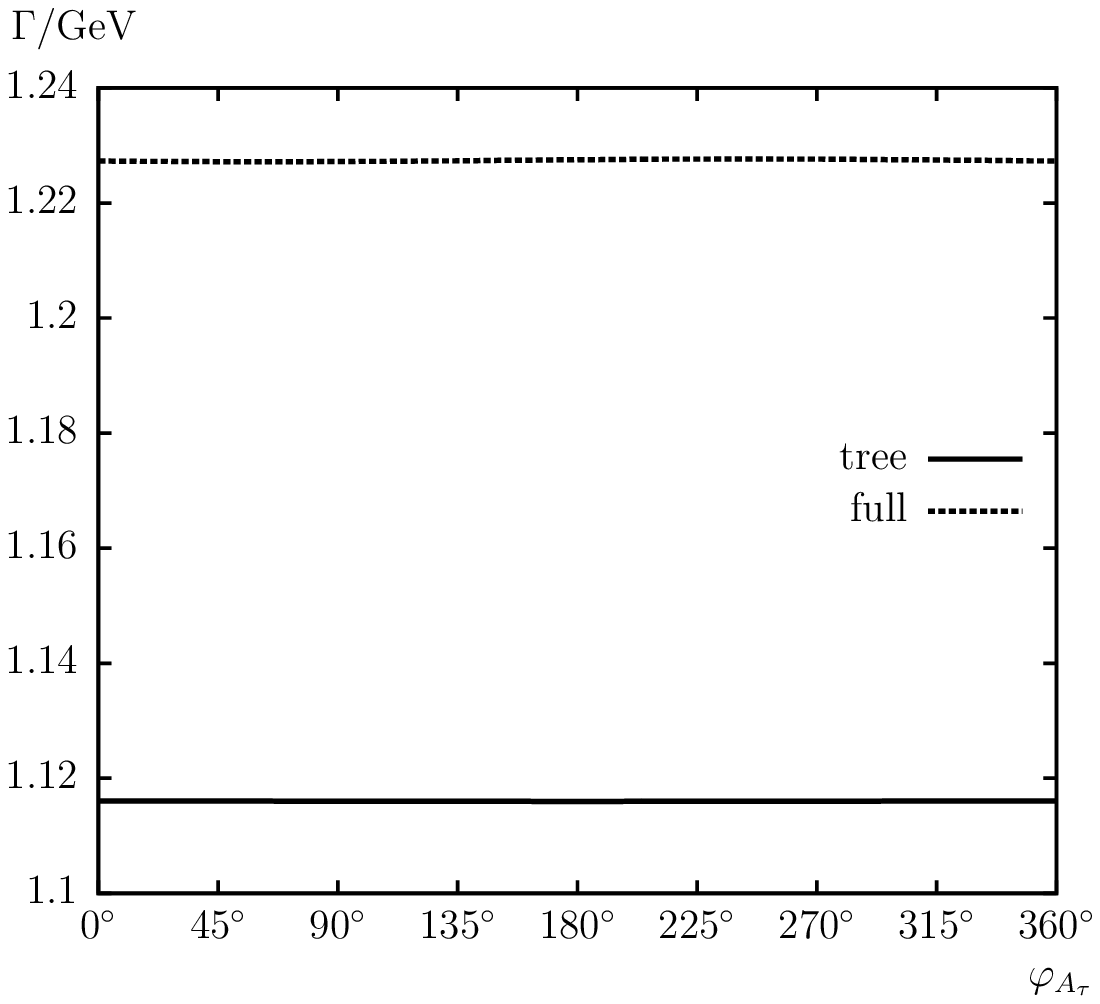}
\hspace{-4mm}
\includegraphics[width=0.49\textwidth,height=7.5cm]{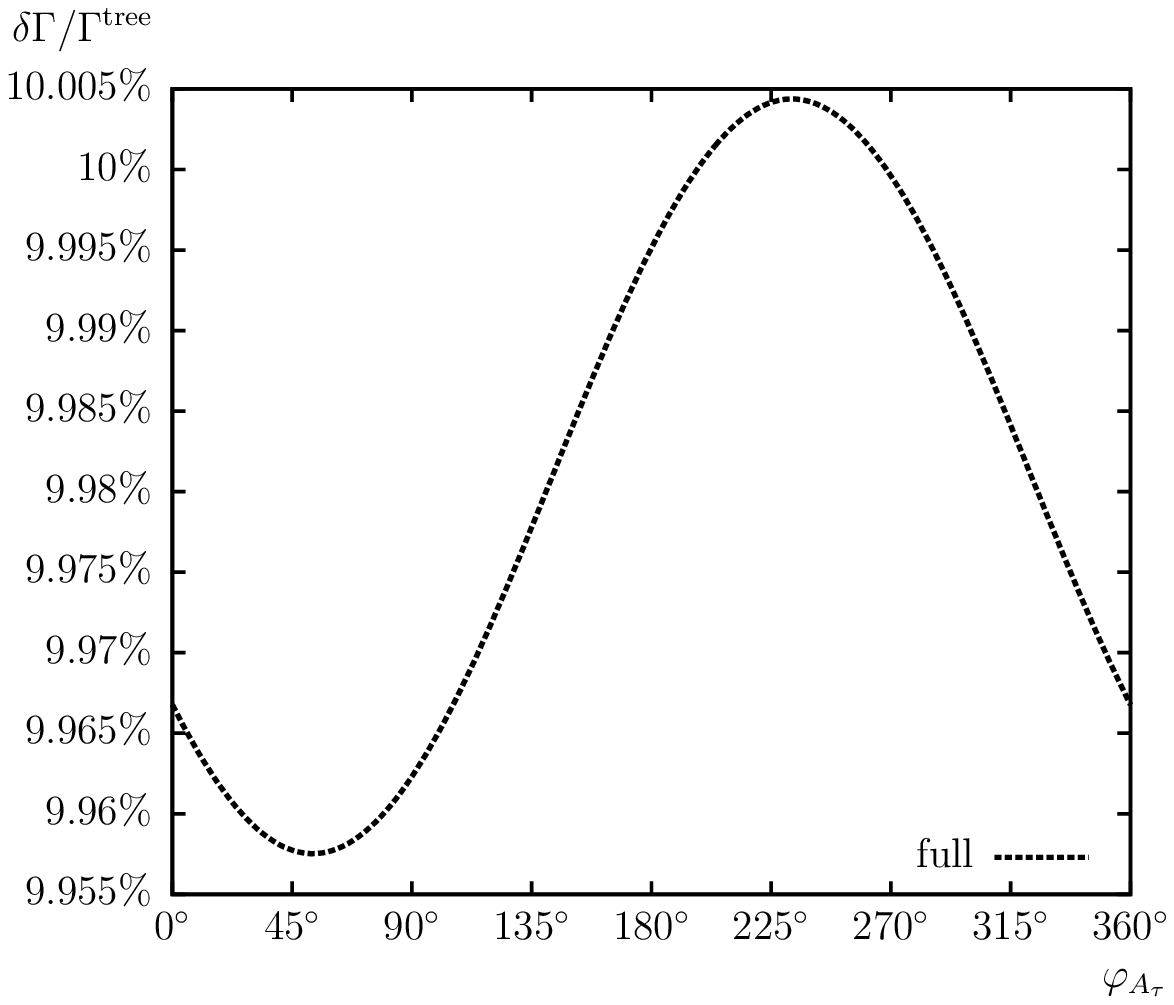}
\\[4em]
\includegraphics[width=0.49\textwidth,height=7.5cm]{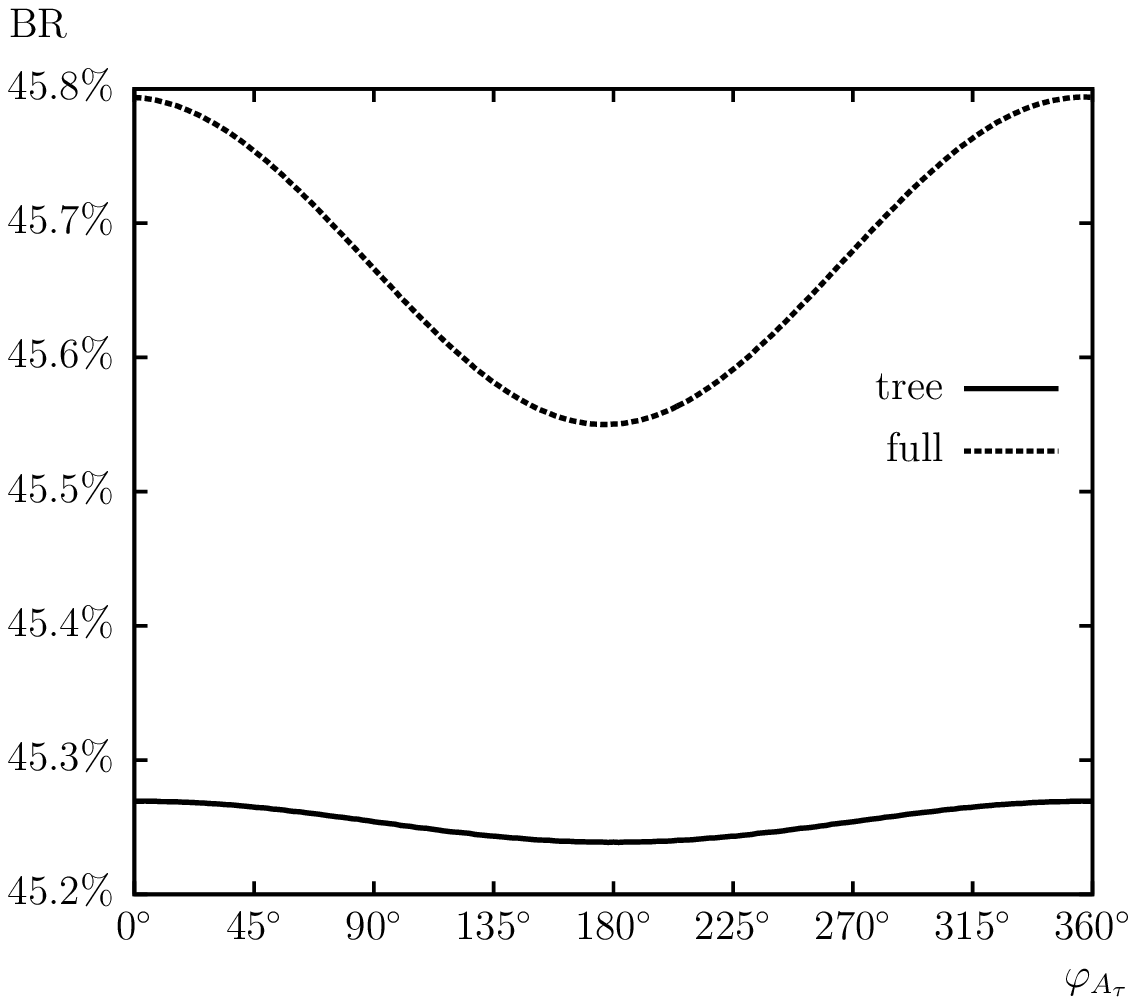}
\hspace{-4mm}
\includegraphics[width=0.49\textwidth,height=7.5cm]{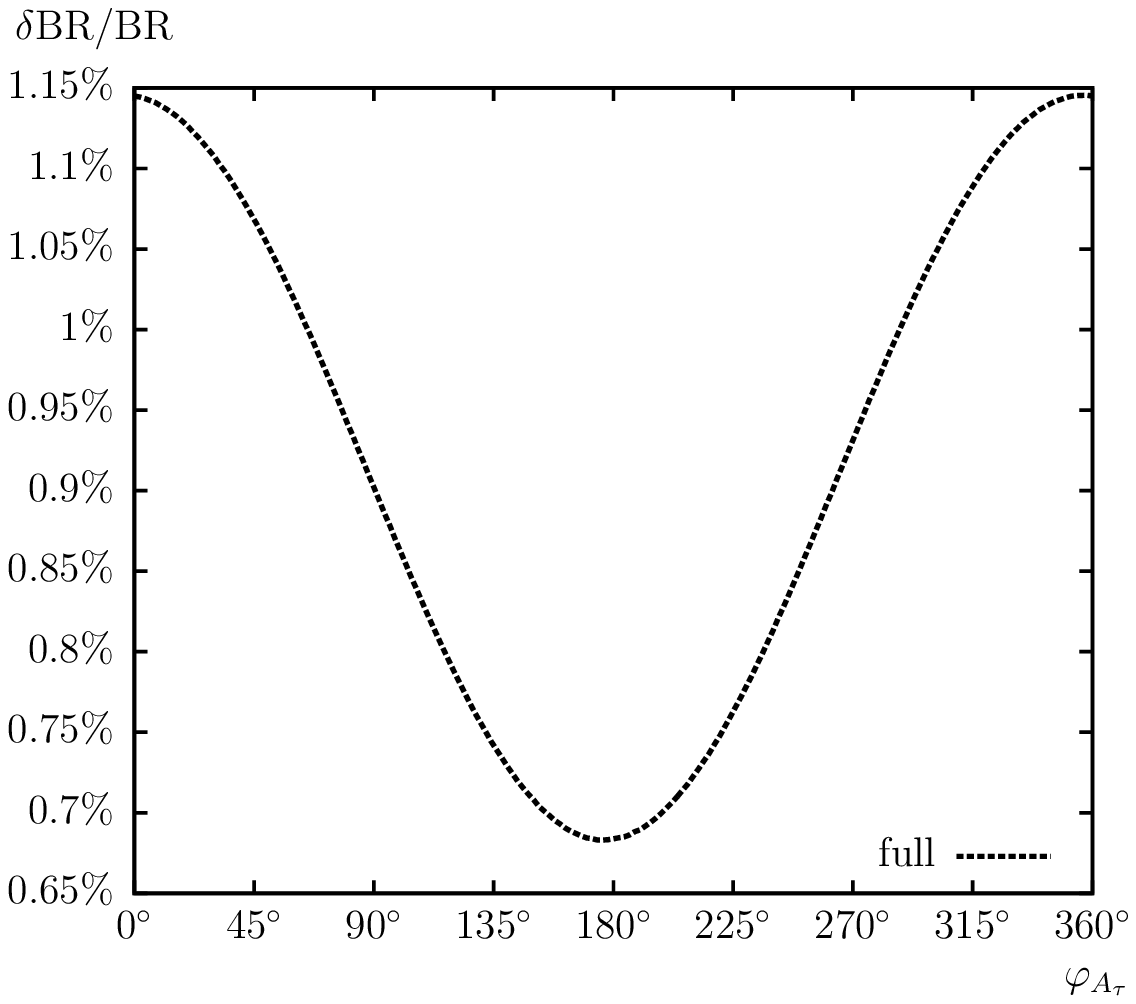}
\end{tabular}
\vspace{2em}
\caption{$\Ga(\decayNz)$.
  Tree-level (``tree'') and full one-loop (``full'') corrected partial decay 
  widths (including absorptive self-energy contributions) are shown.
  The parameters are chosen according to \SE\ (see \refta{tab:para}), 
  with $\phiatau$ varied.
  The upper left plot shows the partial decay width, the upper right plot  
  the corresponding  relative size of the corrections. 
  The lower left plot shows the BR, the lower right plot  
  the relative correction of the BR.
}
\label{fig:PhiAt.stau2tauneu2}
\end{center}
\end{figure}

\begin{figure}[htb!]
\begin{center}
\begin{tabular}{c}
\includegraphics[width=0.49\textwidth,height=7.5cm]{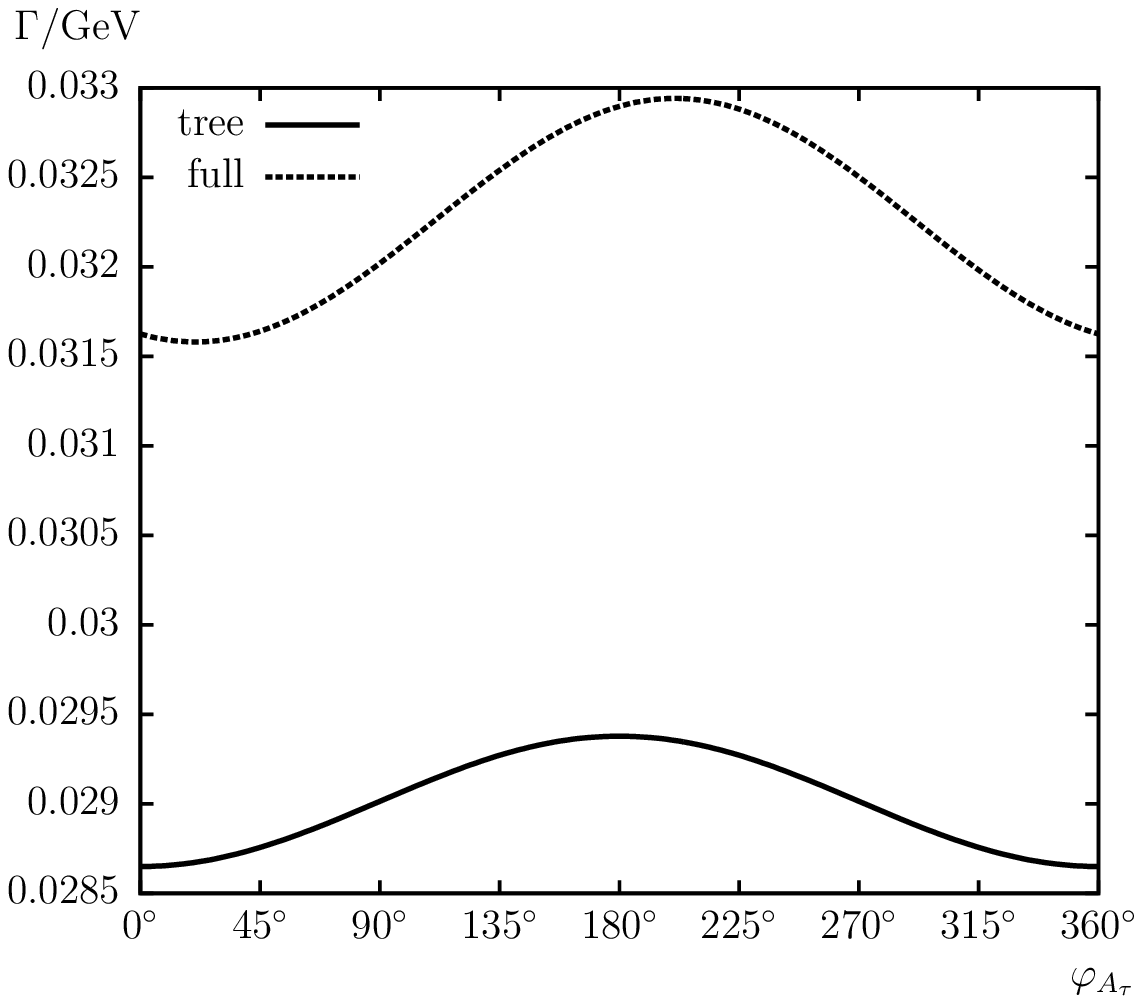}
\hspace{-4mm}
\includegraphics[width=0.49\textwidth,height=7.5cm]{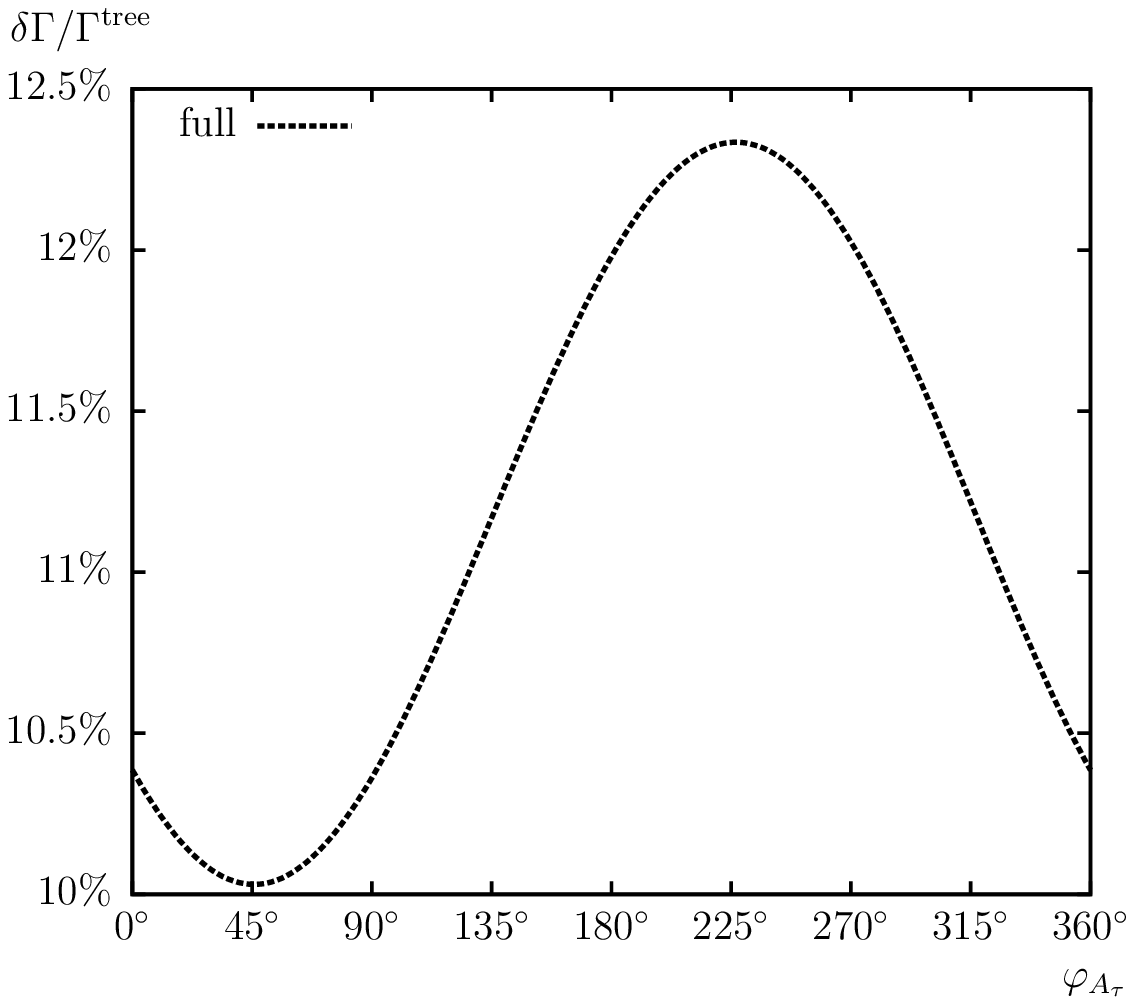}
\\[4em]
\includegraphics[width=0.49\textwidth,height=7.5cm]{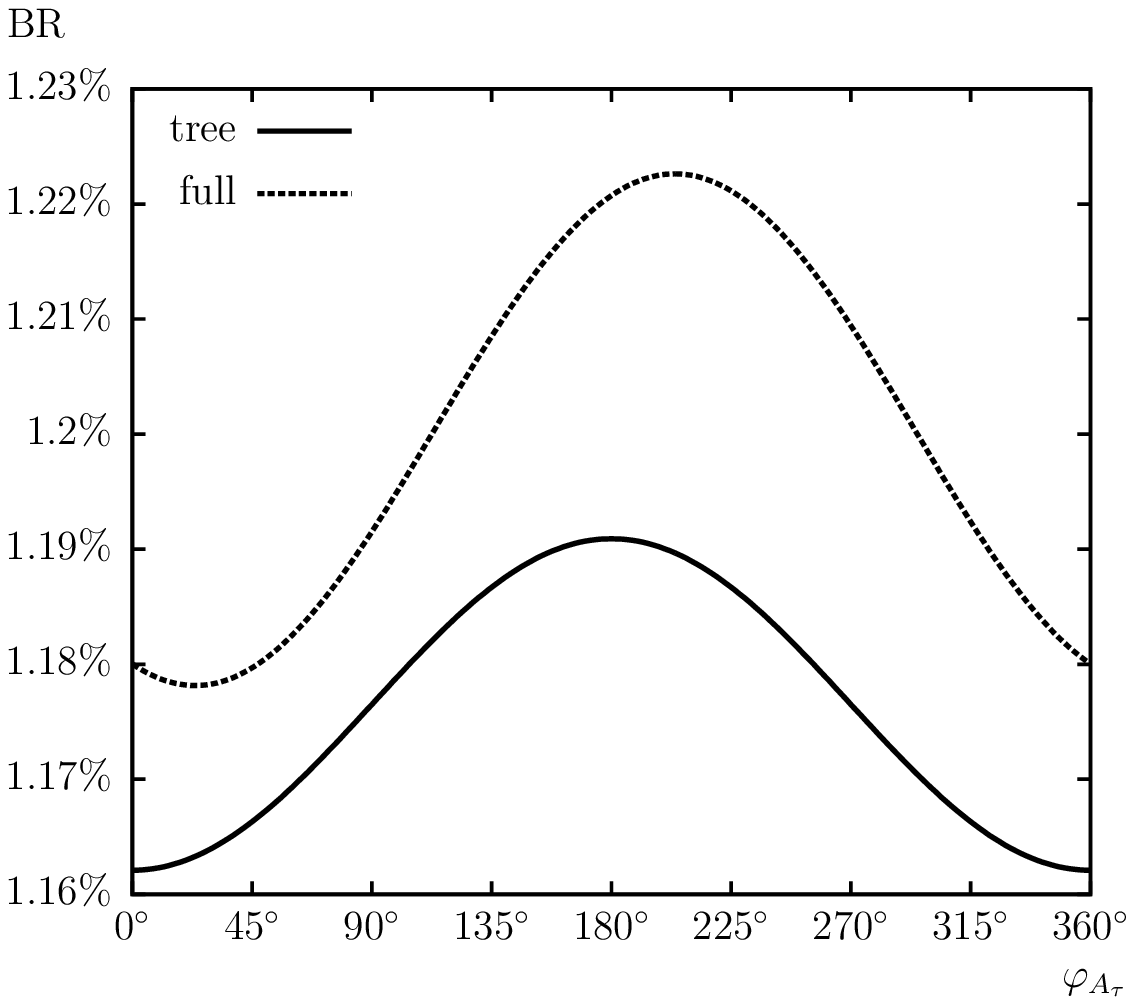}
\hspace{-4mm}
\includegraphics[width=0.49\textwidth,height=7.5cm]{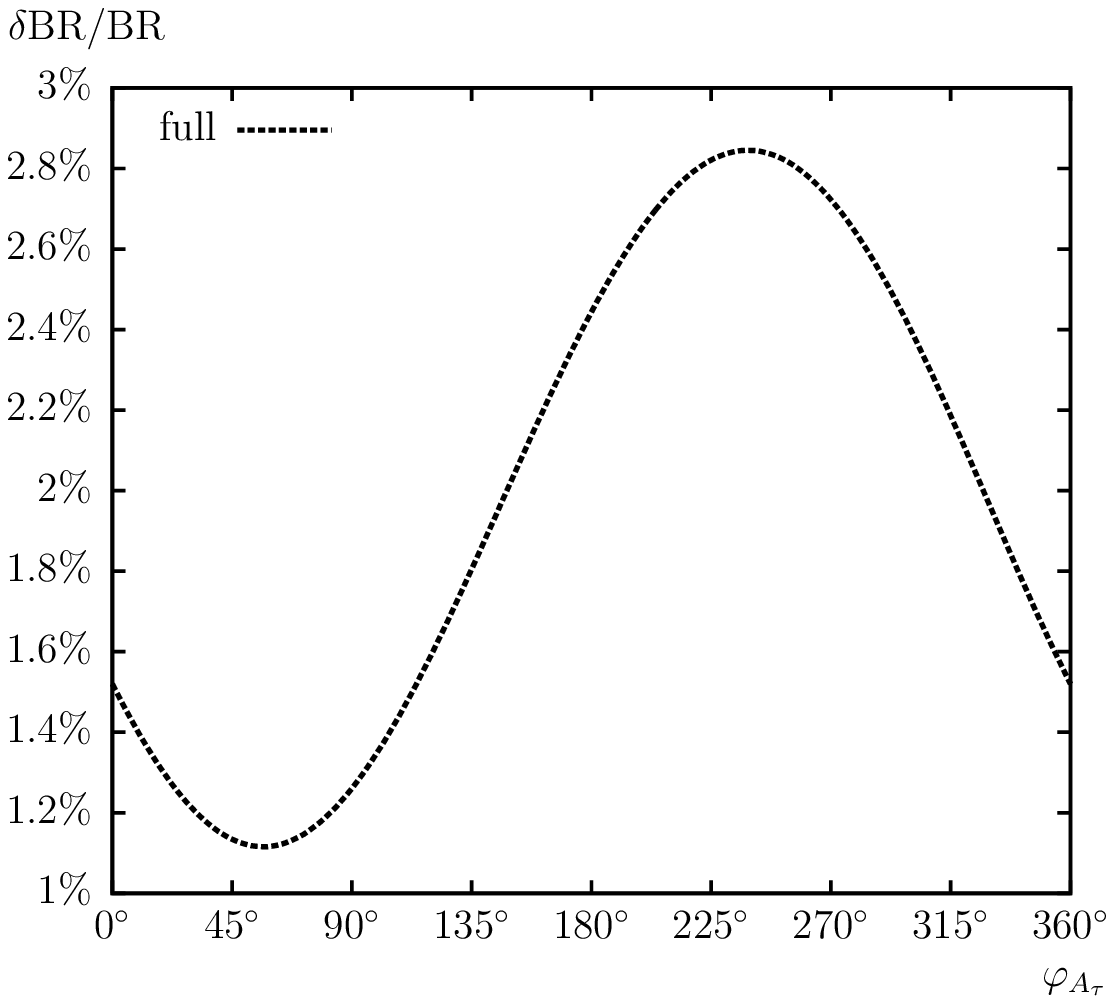}
\end{tabular}
\vspace{2em}
\caption{$\Ga(\decayNd)$.
  Tree-level (``tree'') and full one-loop (``full'') corrected partial decay 
  widths (including absorptive self-energy contributions) are shown.
  The parameters are chosen according to \SE\ (see \refta{tab:para}), 
  with $\phiatau$ varied.
  The upper left plot shows the partial decay width, the upper right plot  
  the corresponding  relative size of the corrections. 
  The lower left plot shows the BR, the lower right plot  
  the relative correction of the BR.
}
\label{fig:PhiAt.stau2tauneu3}
\end{center}
\end{figure}

\begin{figure}[htb!]
\begin{center}
\begin{tabular}{c}
\includegraphics[width=0.49\textwidth,height=7.5cm]{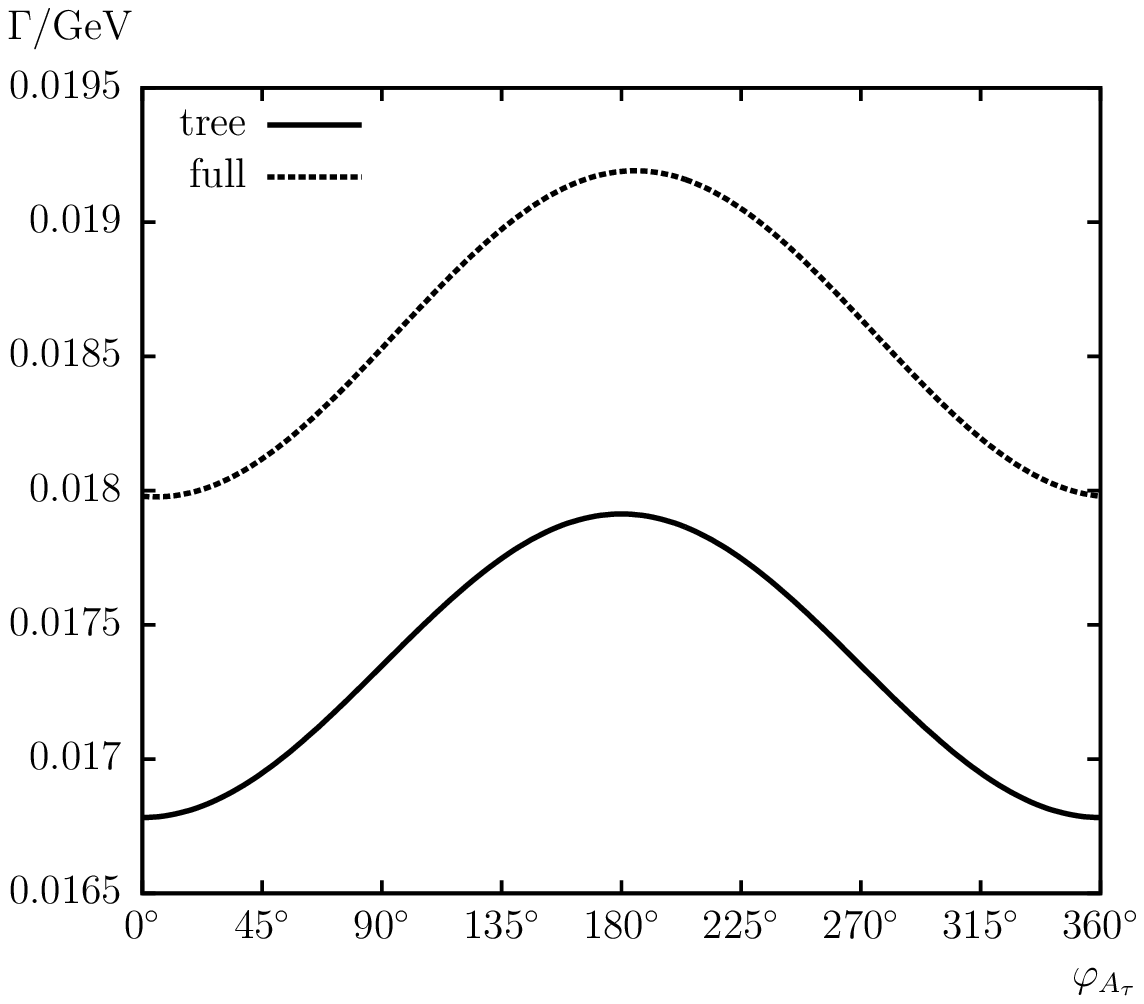}
\hspace{-4mm}
\includegraphics[width=0.49\textwidth,height=7.5cm]{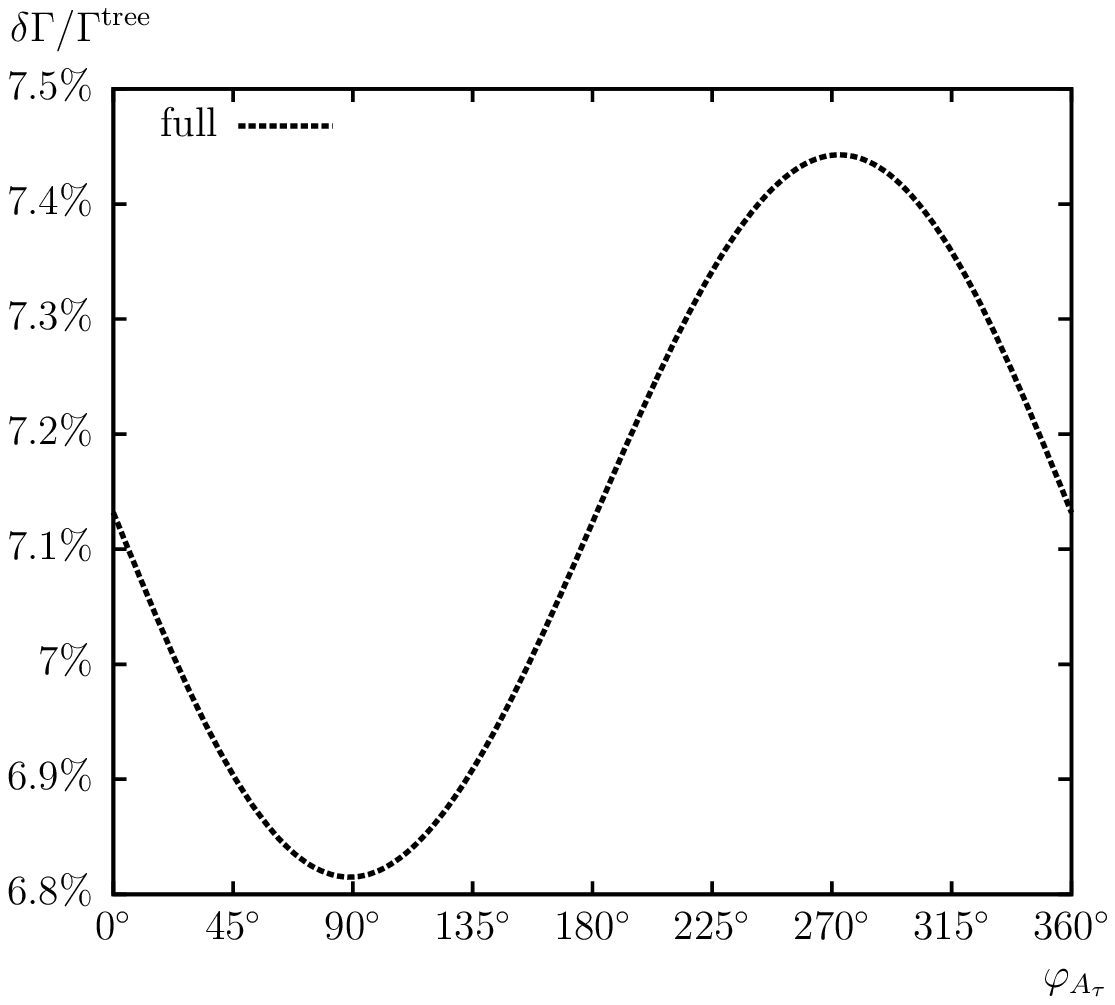}
\\[4em]
\includegraphics[width=0.49\textwidth,height=7.5cm]{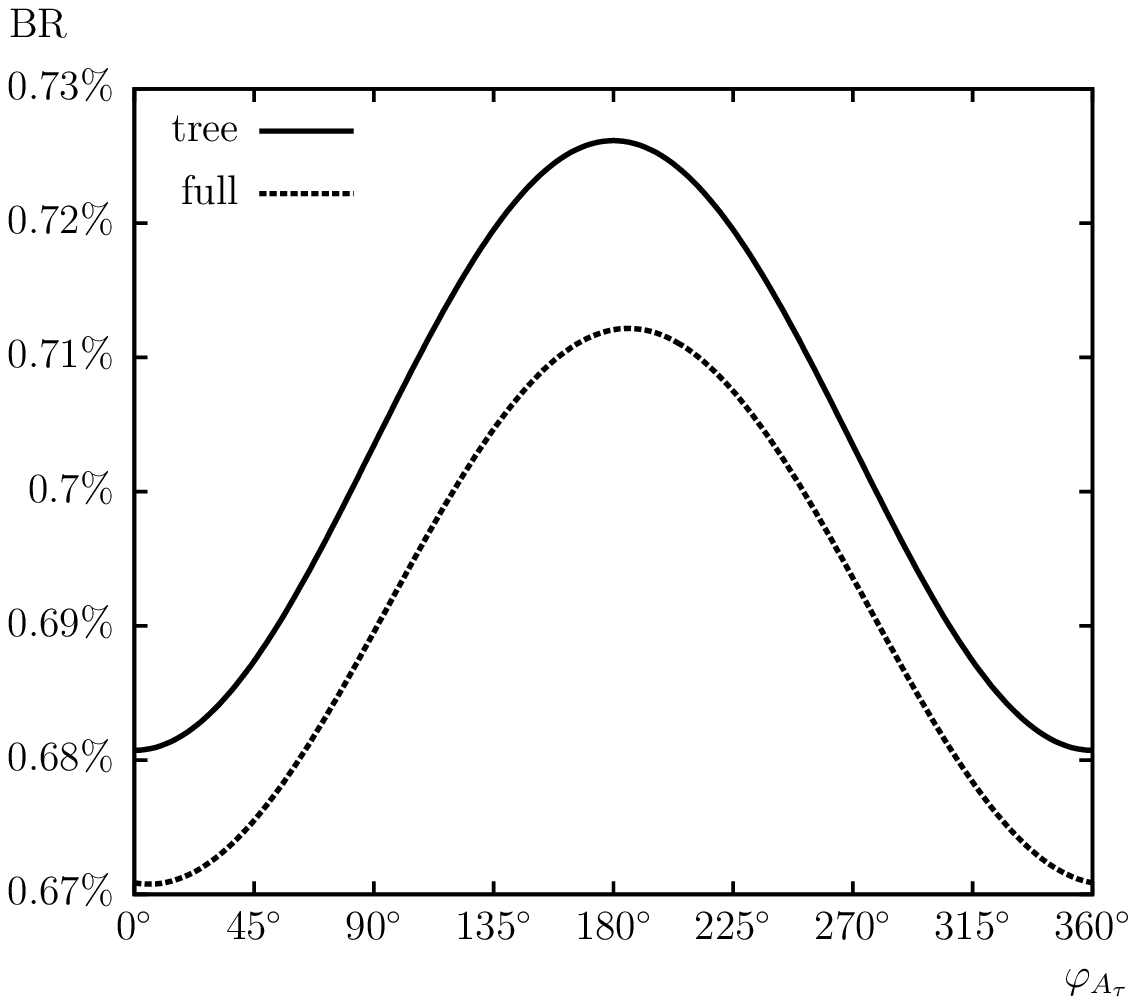}
\hspace{-4mm}
\includegraphics[width=0.49\textwidth,height=7.5cm]{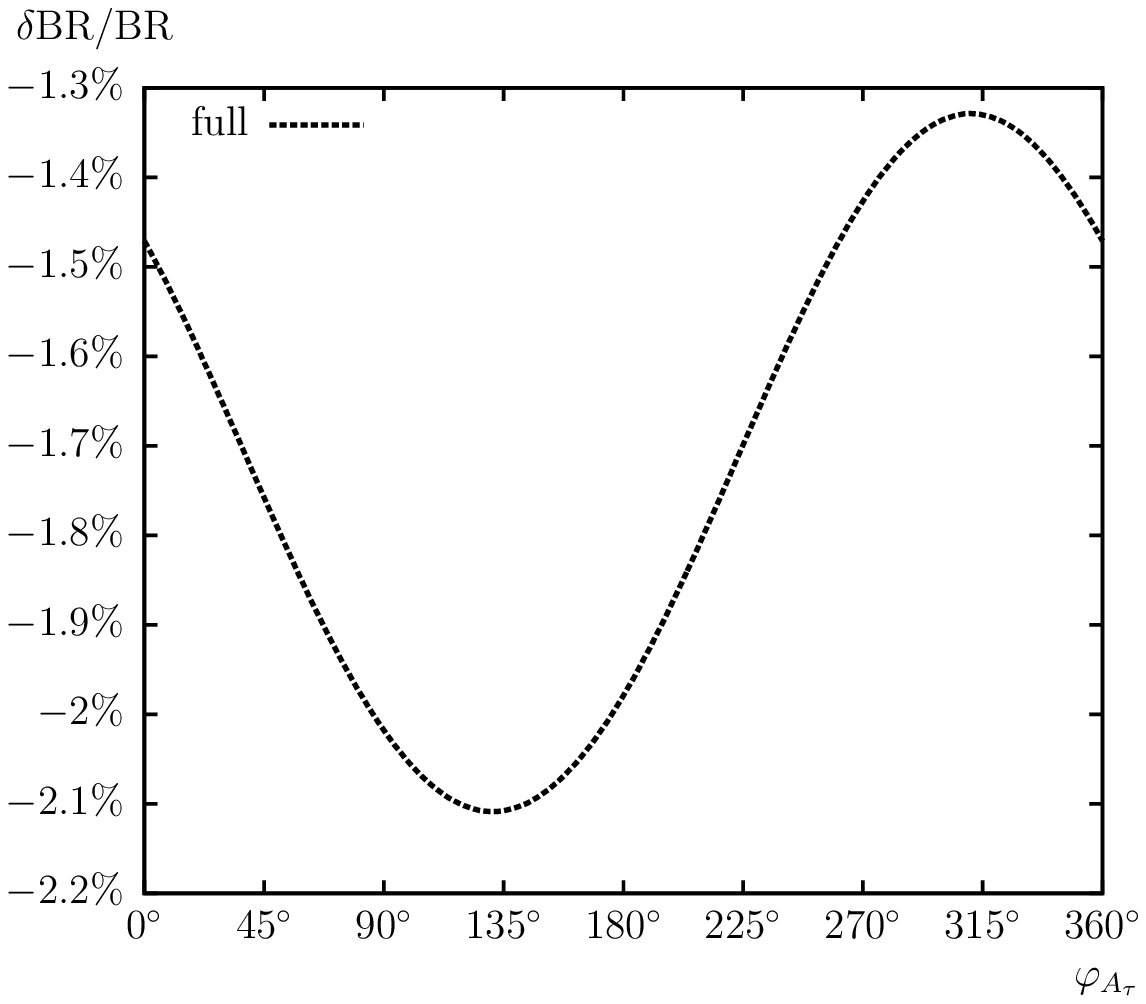}
\end{tabular}
\vspace{2em}
\caption{$\Ga(\decayNv)$.
  Tree-level (``tree'') and full one-loop (``full'') corrected partial decay 
  widths (including absorptive self-energy contributions) are shown.
  The parameters are chosen according to \SE\ (see \refta{tab:para}), 
  with $\phiatau$ varied.
  The upper left plot shows the partial decay width, the upper right plot  
  the corresponding  relative size of the corrections. 
  The lower left plot shows the BR, the lower right plot  
  the relative correction of the BR.
}
\label{fig:PhiAt.stau2tauneu4}
\end{center}
\end{figure}

In \reffis{fig:PhiAt.stau2tauneu1} -- \ref{fig:PhiAt.stau2tauneu4} we present
the variation of $\Ga(\decayNk)$ ($k = 1,2,3,4$) as a function of
$\phiatau$. 
As for the variation with $\mstauz$ also here for $k = 1,2$
larger values of the partial decay width are found  in
\SE\ with a similar size as before, again dominating the total width
(see the discussions above). 
The one-loop effects on $\Ga(\decayNi{1,2})$ are about $8\%, 10\%$
with a small variation with $\phiatau$. 
For $\Ga(\decayNi{3,4})$, which are substantially smaller, the one-loop
effects are of similar size, $\sim 11\%, 7\%$, respectively, 
again with a small phase variation. 
Within \SE, i.e.\ with the ILC(1000) accessible parameter space, the
one-loop corrections to the various branching ratios is smaller than the
effects on the partial widths. However, as discussed above, with a
different combination of $\mu$, $\MOne$ and $\MTwo$ the one-loop effects
can still exceed the potential ILC precision, where, however, only a
moderate phase dependence is observed.

\begin{figure}[htb!]
\begin{center}
\begin{tabular}{c}
\includegraphics[width=0.49\textwidth,height=7.5cm]{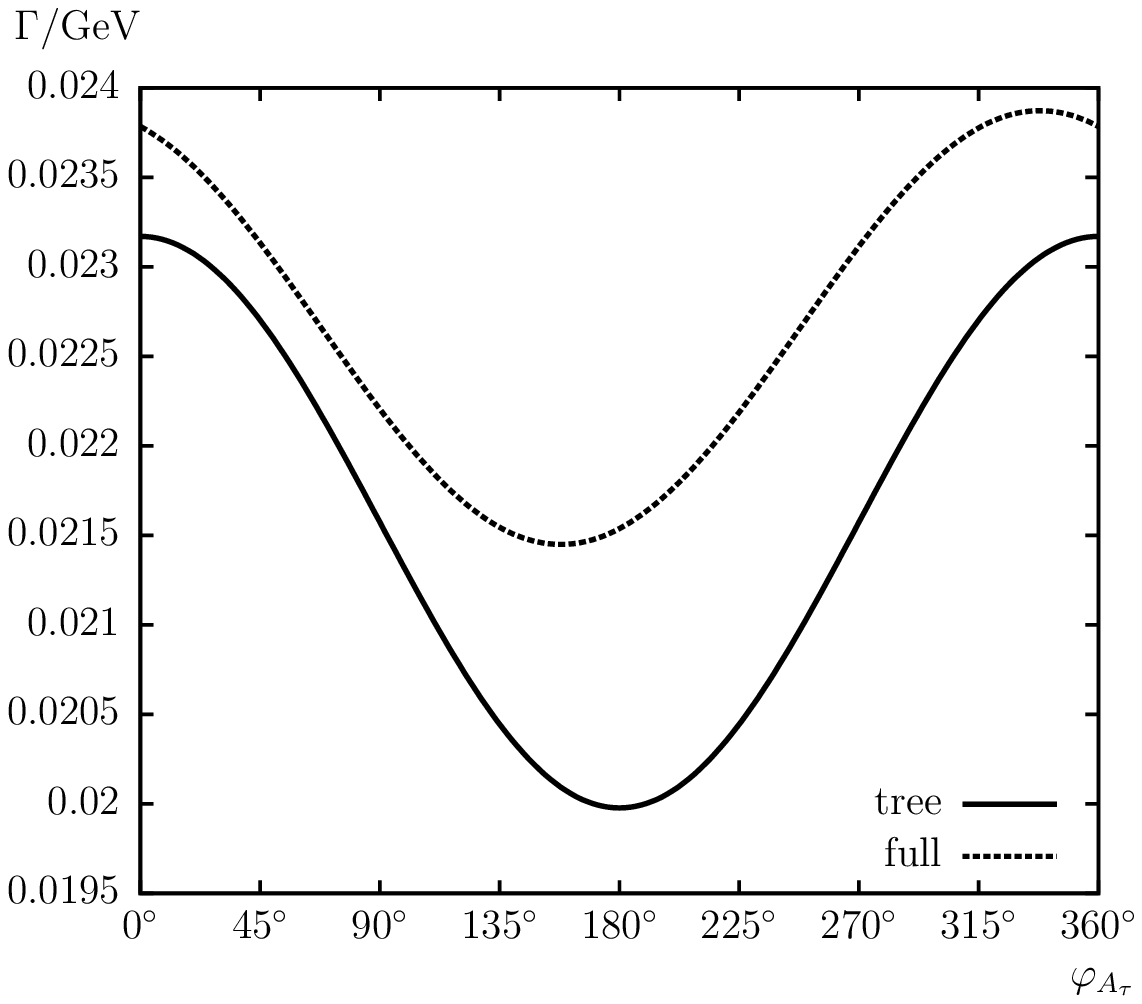}
\hspace{-4mm}
\includegraphics[width=0.49\textwidth,height=7.5cm]{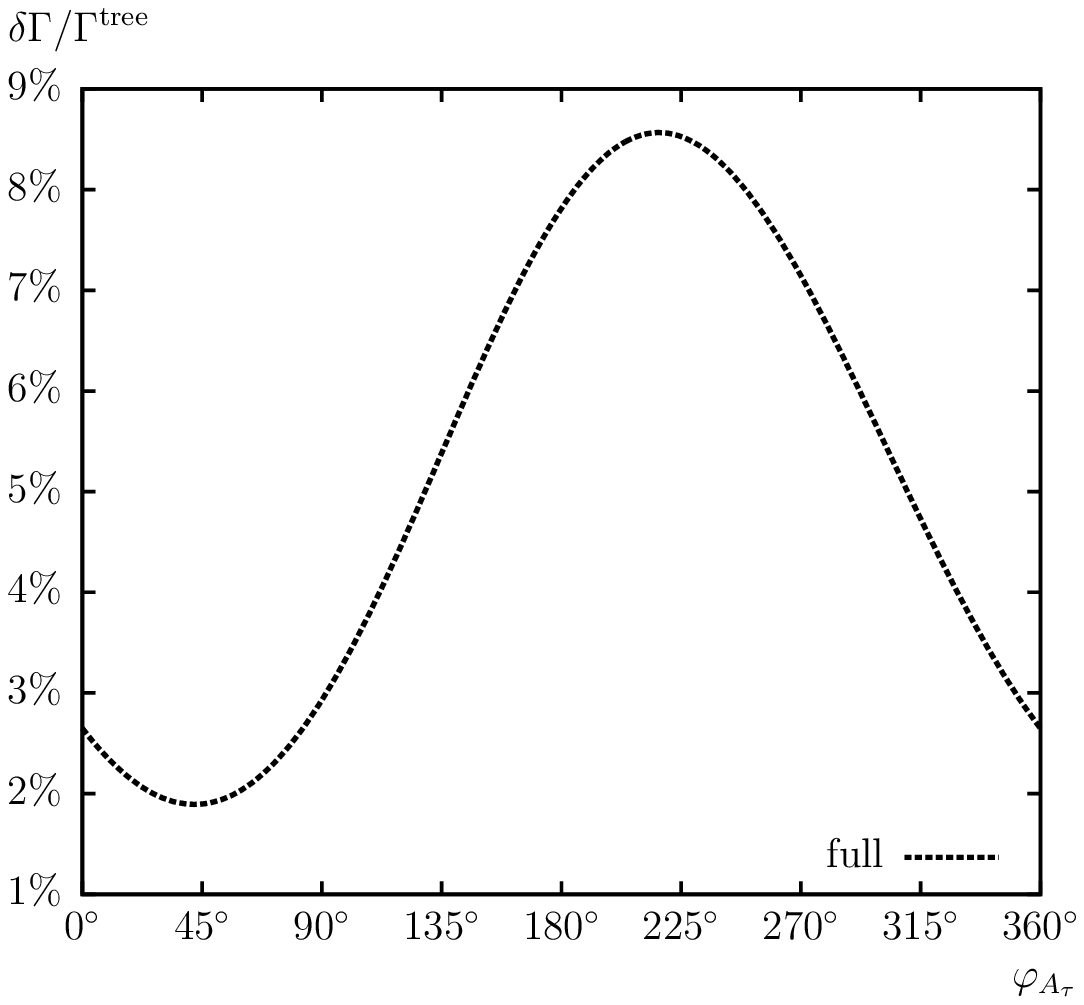}
\\[4em]
\includegraphics[width=0.49\textwidth,height=7.5cm]{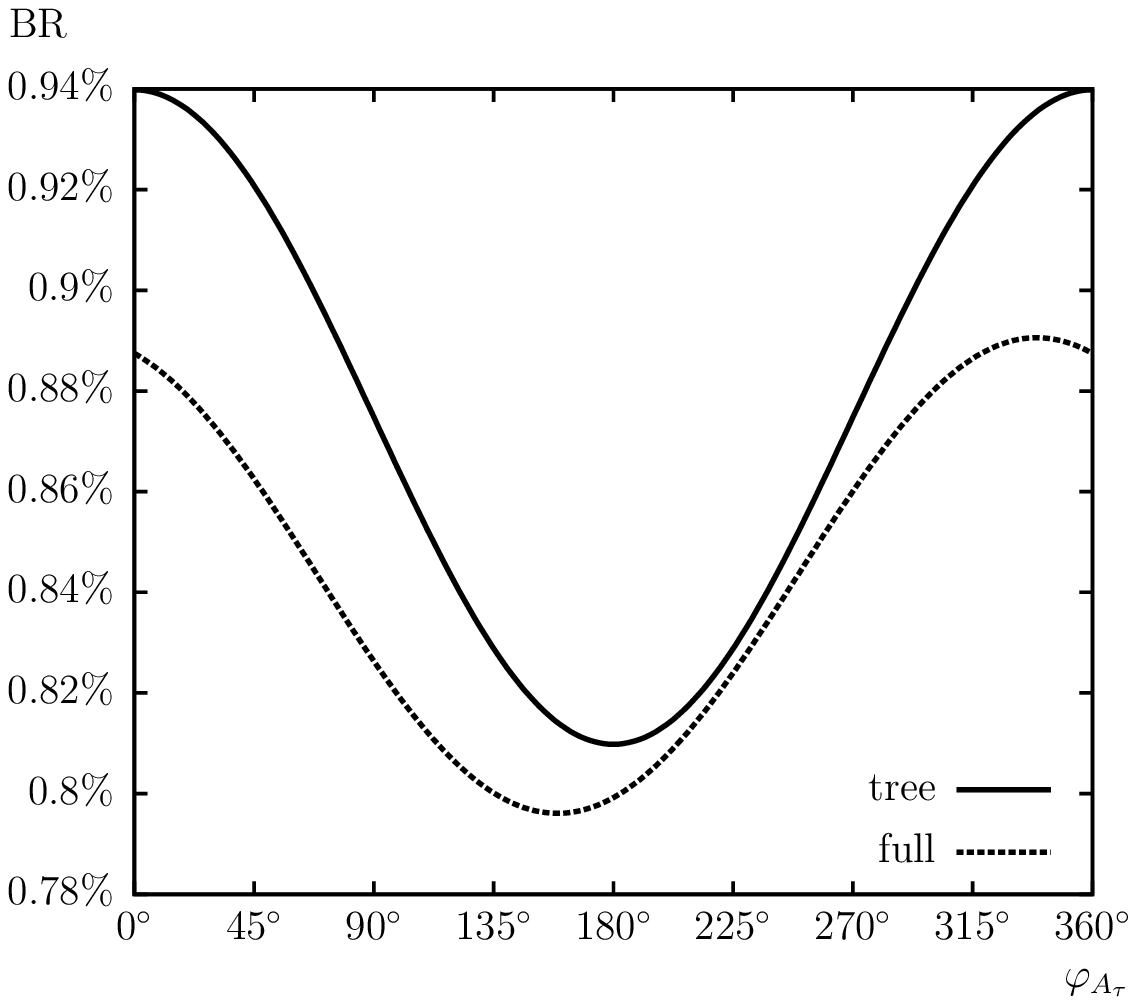}
\hspace{-4mm}
\includegraphics[width=0.49\textwidth,height=7.5cm]{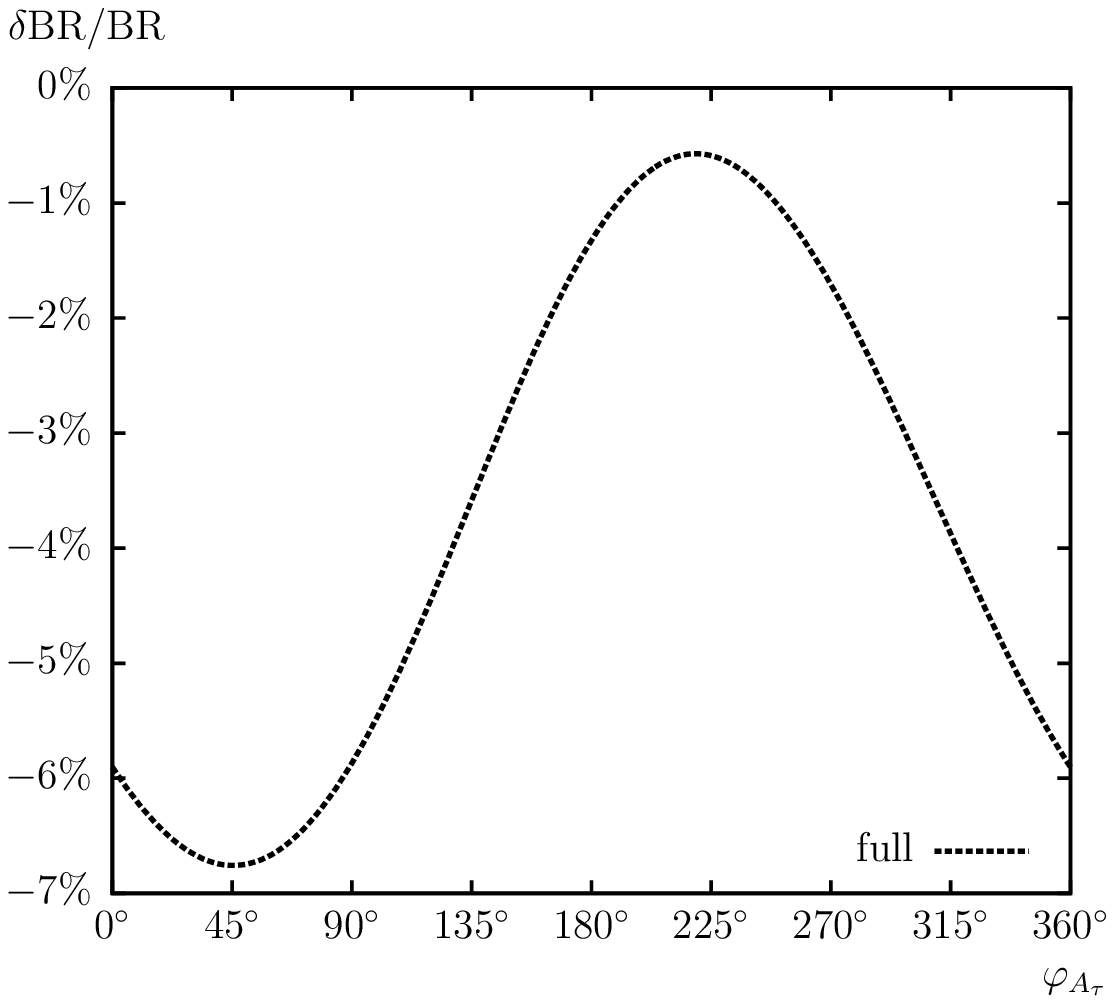}
\end{tabular}
\vspace{2em}
\caption{$\Ga(\decayCme)$. 
  Tree-level (``tree'') and full one-loop (``full'') corrected partial decay 
  widths (including absorptive self-energy contributions) are shown. 
  The parameters are chosen according to \SE\ (see \refta{tab:para}), 
  with $\phiatau$ varied.
  The upper left plot shows the partial decay width, the upper right plot  
  the corresponding  relative size of the corrections. 
  The lower left plot shows the BR, the lower right plot  
  the relative correction of the BR.
}
\label{fig:PhiAt.stau2ncha1}
\end{center}
\end{figure}

\begin{figure}[htb!]
\begin{center}
\begin{tabular}{c}
\includegraphics[width=0.49\textwidth,height=7.5cm]{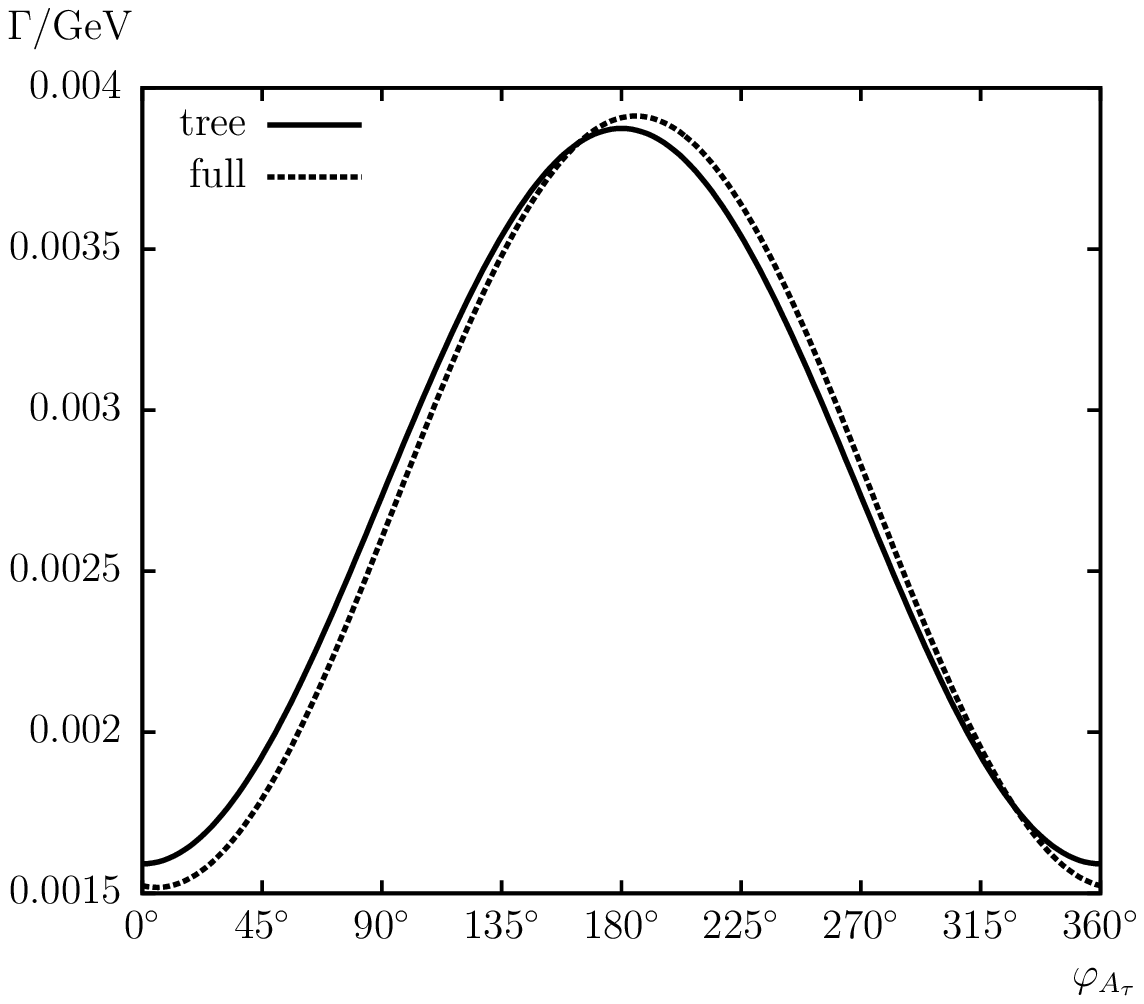}
\hspace{-4mm}
\includegraphics[width=0.49\textwidth,height=7.5cm]{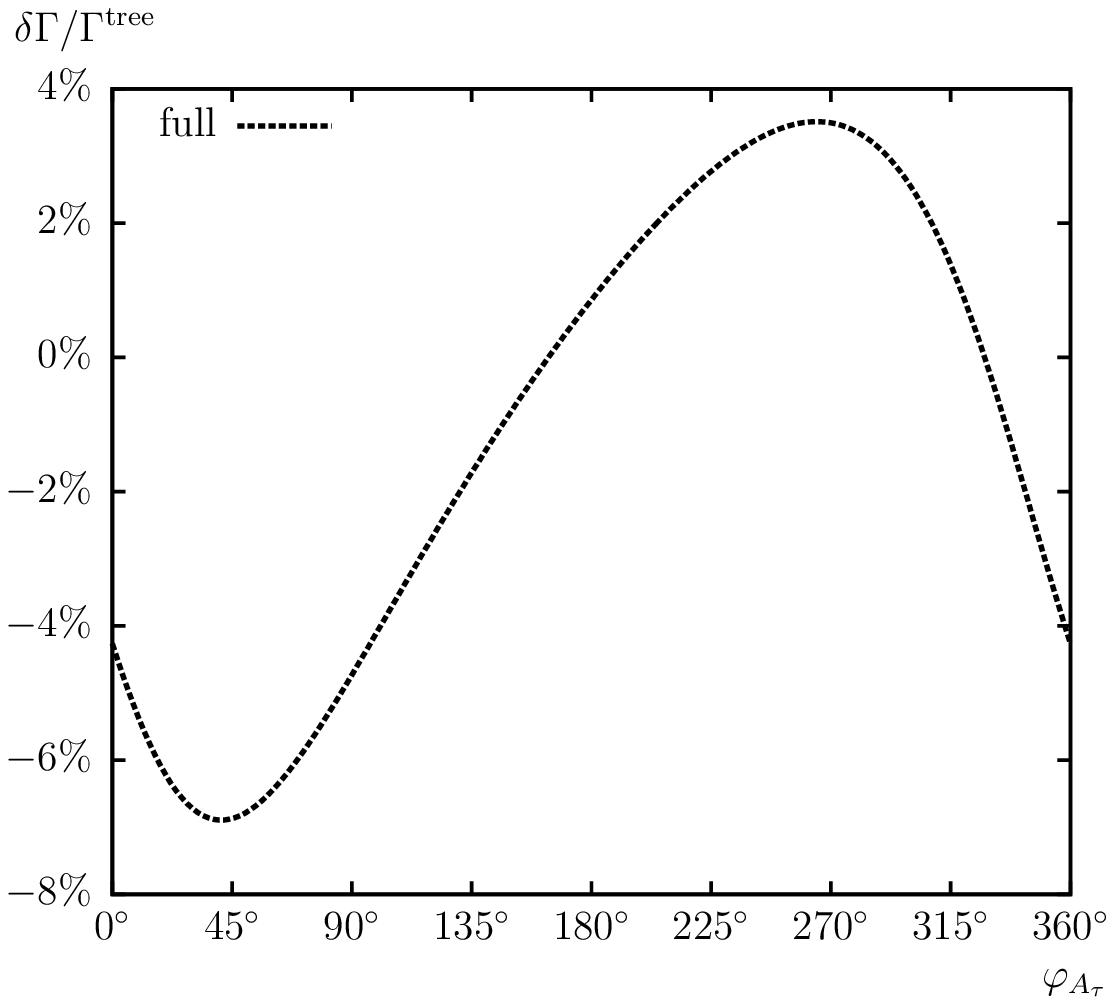}
\\[4em]
\includegraphics[width=0.49\textwidth,height=7.5cm]{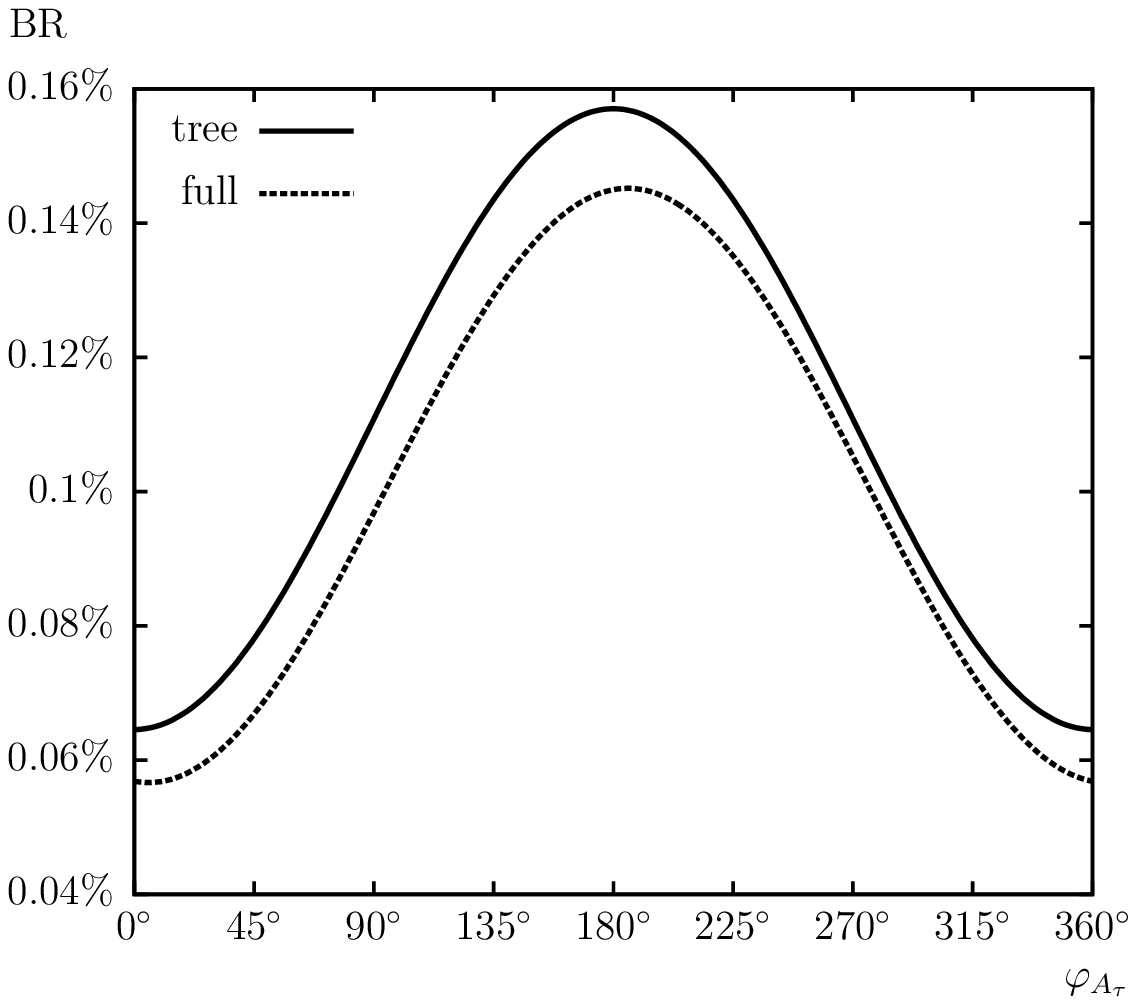}
\hspace{-4mm}
\includegraphics[width=0.49\textwidth,height=7.5cm]{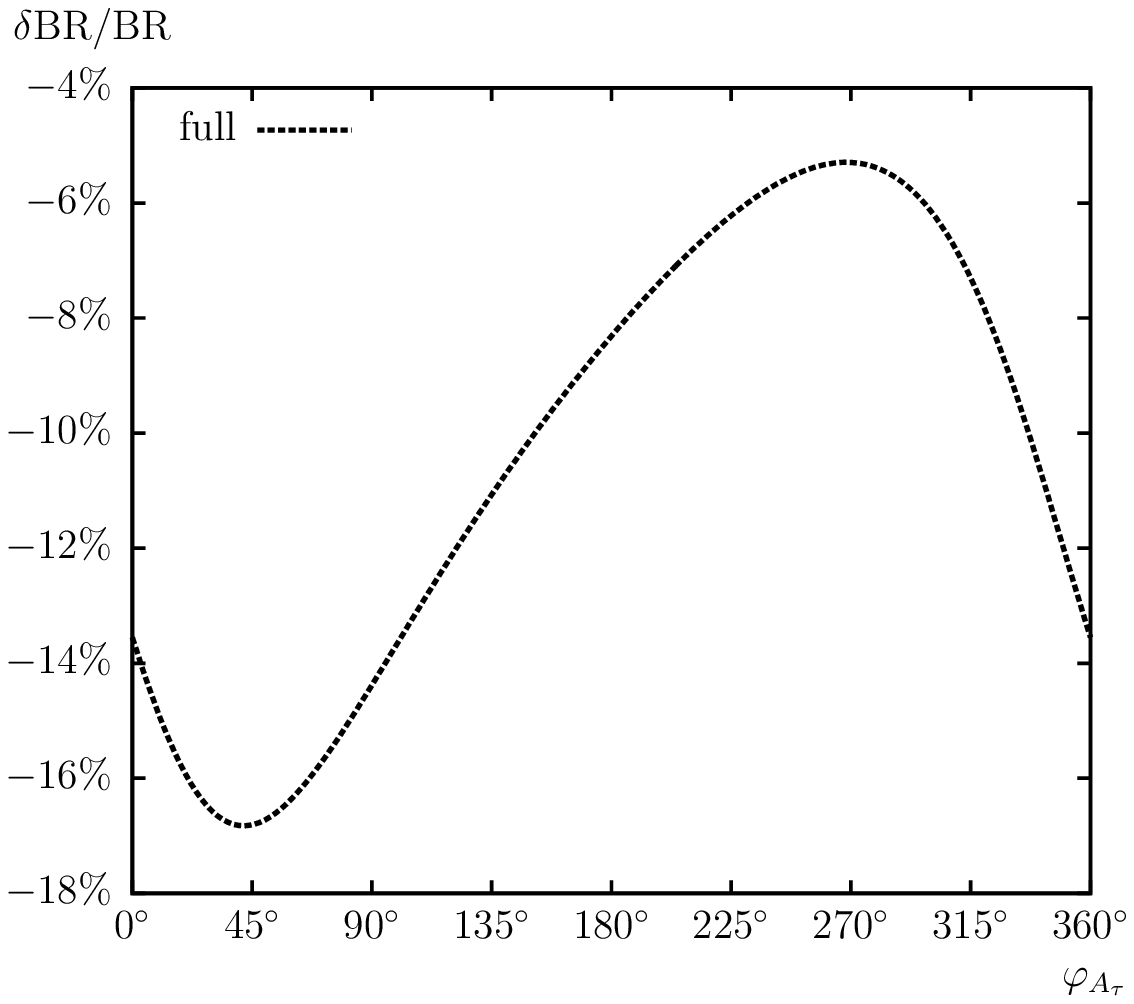}
\end{tabular}
\vspace{2em}
\caption{$\Ga(\decayCmz)$.
  Tree-level (``tree'') and full one-loop (``full'') corrected partial decay 
  widths (including absorptive self-energy contributions) are shown. 
  The parameters are chosen according to \SE\ (see \refta{tab:para}), 
  with $\phiatau$ varied.
  The upper left plot shows the partial decay width, the upper right plot  
  the corresponding  relative size of the corrections. 
  The lower left plot shows the BR, the lower right plot  
  the relative correction of the BR.
}
\label{fig:PhiAt.stau2ncha2}
\end{center}
\end{figure}

The results for $\Ga(\decayCmj)$ ($j = 1,2$) are shown in
\reffis{fig:PhiAt.stau2ncha1}, \ref{fig:PhiAt.stau2ncha2}. Both 
decay widths change substantially with $\phiatau$.
The relative corrections are  between $+2\%$ and $+9\%$ for
$\decayCme$ and between $-7\%$ and $+3\%$ for $\decayCmz$. 
Within \SE\ the variation of the branching ratios is slightly smaller
for $\decayCme$, and substantially larger, up to $-16\%$ for
$\decayCmz$, again with a strong variation with $\phiatau$, 
which can then be relevant for the ILC.

\begin{figure}[htb!]
\begin{center}
\begin{tabular}{c}
\includegraphics[width=0.49\textwidth,height=7.5cm]{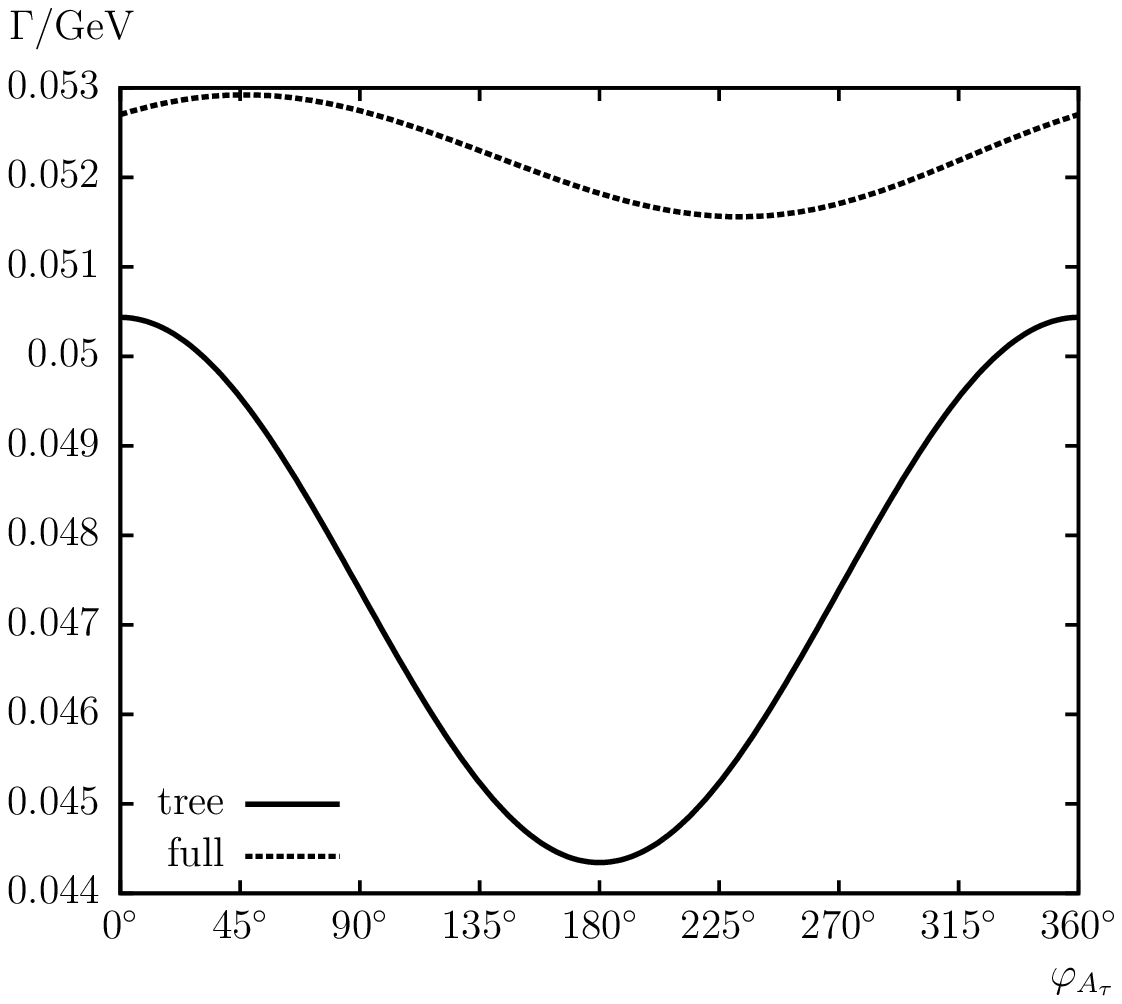}
\hspace{-4mm}
\includegraphics[width=0.49\textwidth,height=7.5cm]{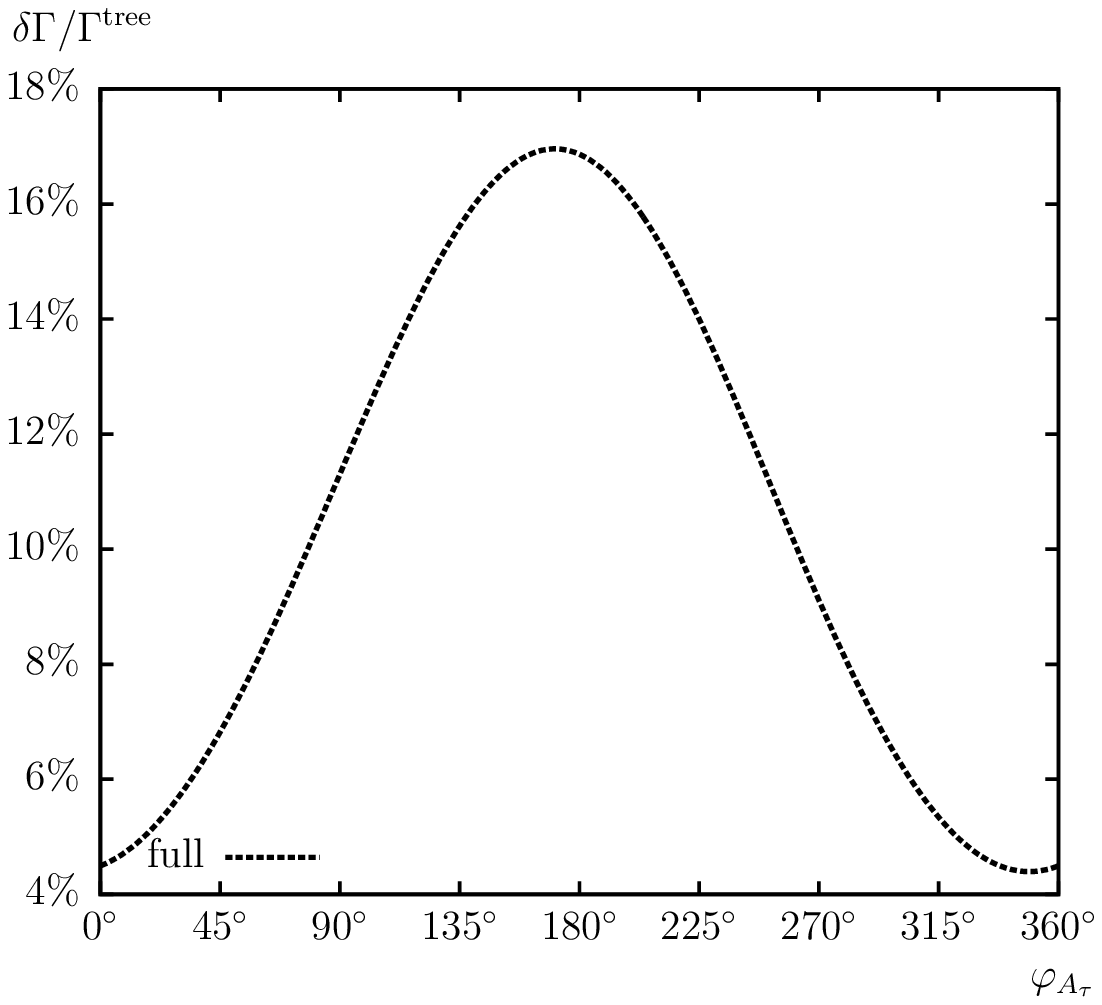}
\\[4em]
\includegraphics[width=0.49\textwidth,height=7.5cm]{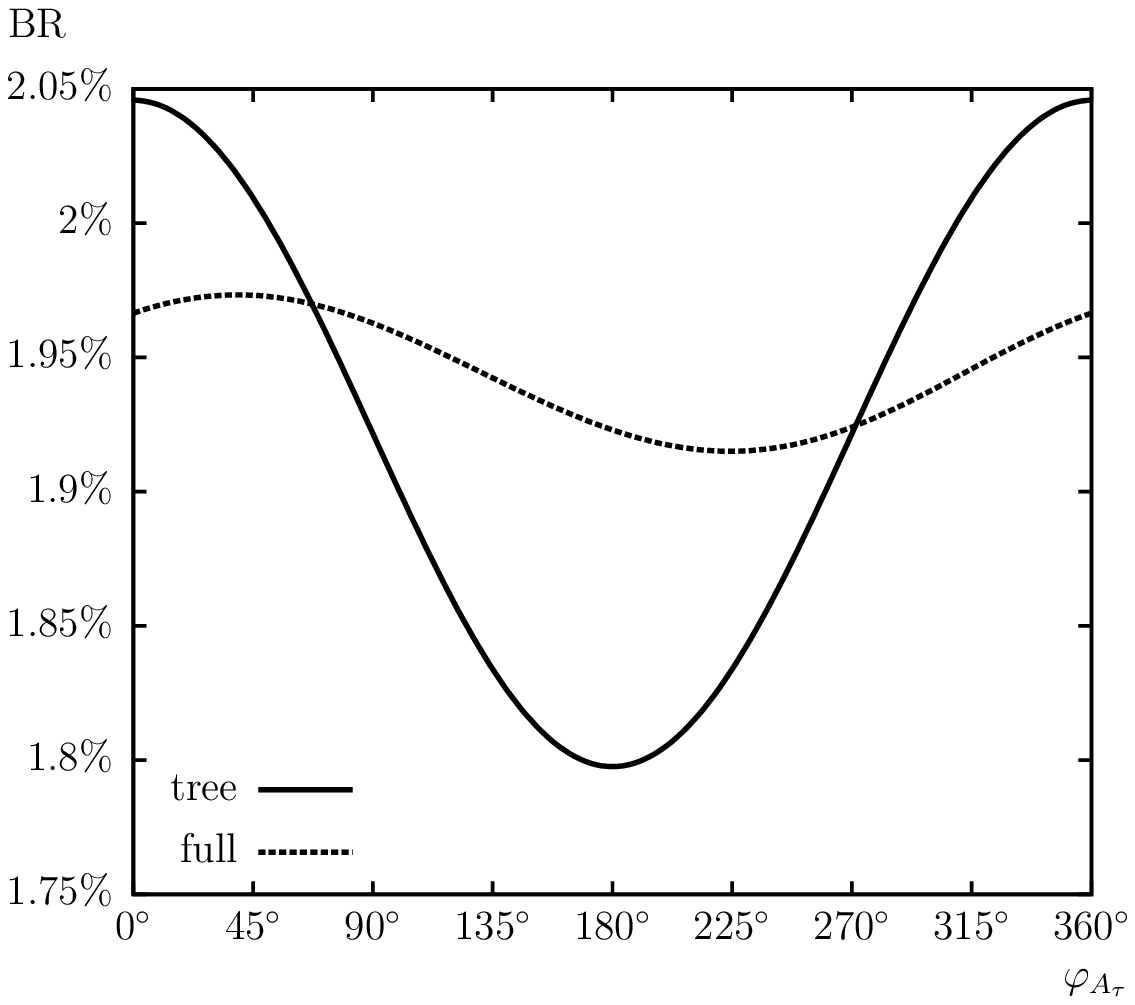}
\hspace{-4mm}
\includegraphics[width=0.49\textwidth,height=7.5cm]{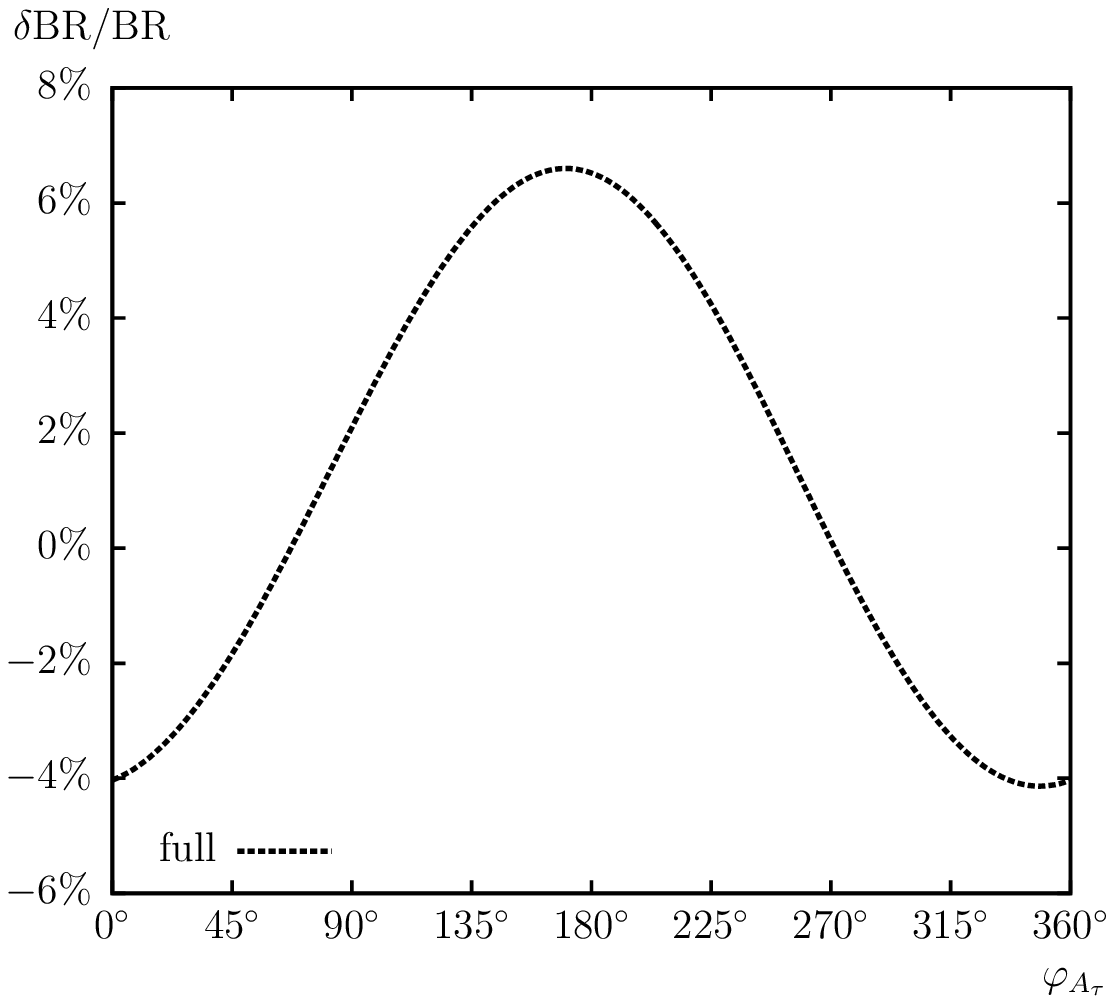}
\end{tabular}
\vspace{2em}
\caption{$\Ga(\decayHm)$. 
  Tree-level (``tree'') and full one-loop (``full'') corrected partial decay 
  widths (including absorptive self-energy contributions) are shown.
  The parameters are chosen according to \SE\ (see \refta{tab:para}), 
  with $\phiatau$ varied.
  The upper left plot shows the partial decay width, the upper right plot  
  the corresponding  relative size of the corrections. 
  The lower left plot shows the BR, the lower right plot  
  the relative correction of the BR.
}
\label{fig:PhiAt.stau2snH}
\end{center}
\end{figure}

Finally we turn to the decay modes involving scalar neutrinos.
In \reffi{fig:PhiAt.stau2snH} the results for $\Ga(\decayHm)$ are presented. 
The relative correction to the decay width varies strongly between
$+4\%$ and $+17\%$ with $\phiatau$, whereas $\br(\decayHm)$ varies
within \SE\ between $-4\%$ and $+7\%$, potentially exceeding the ILC
precision. 

\begin{figure}[htb!]
\begin{center}
\begin{tabular}{c}
\includegraphics[width=0.49\textwidth,height=7.5cm]{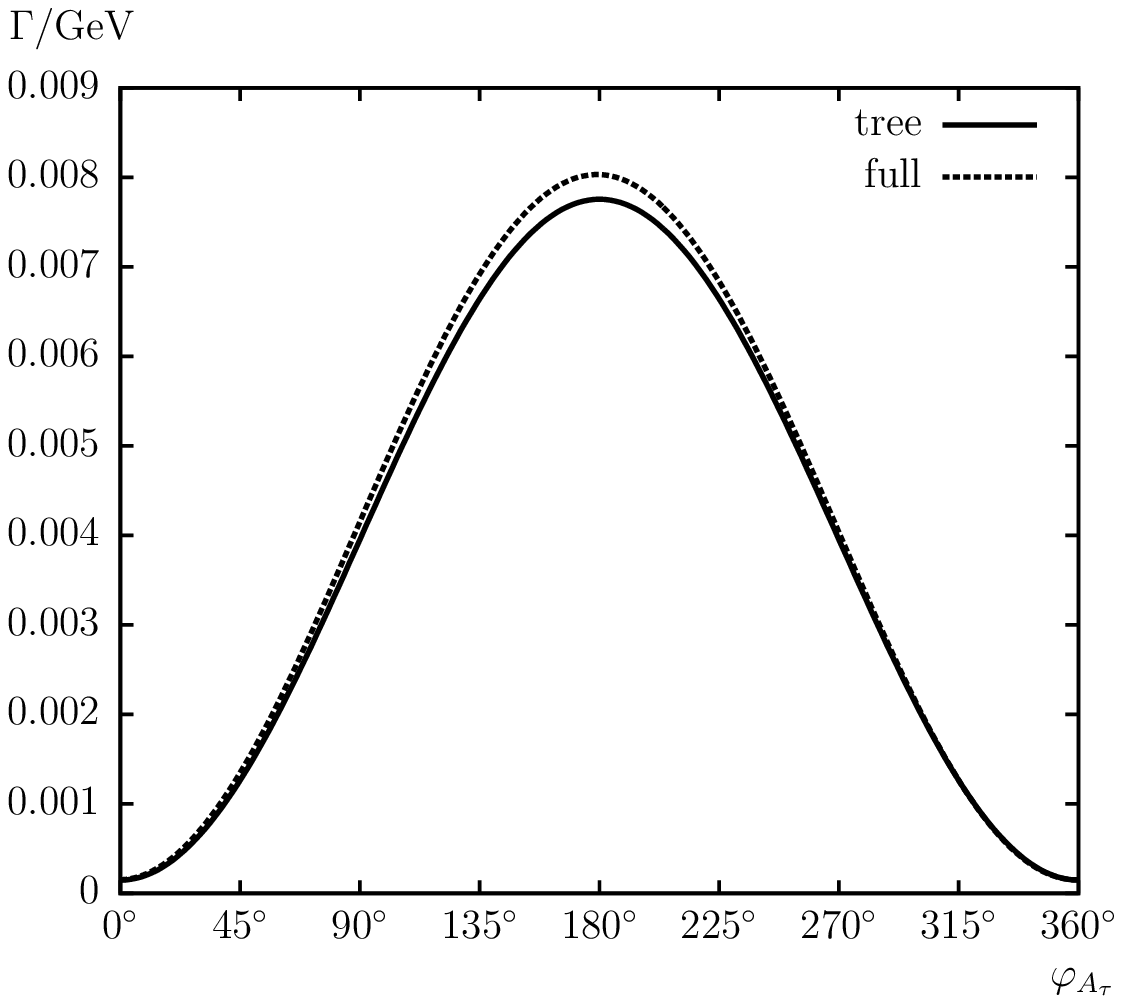}
\hspace{-4mm}
\includegraphics[width=0.49\textwidth,height=7.5cm]{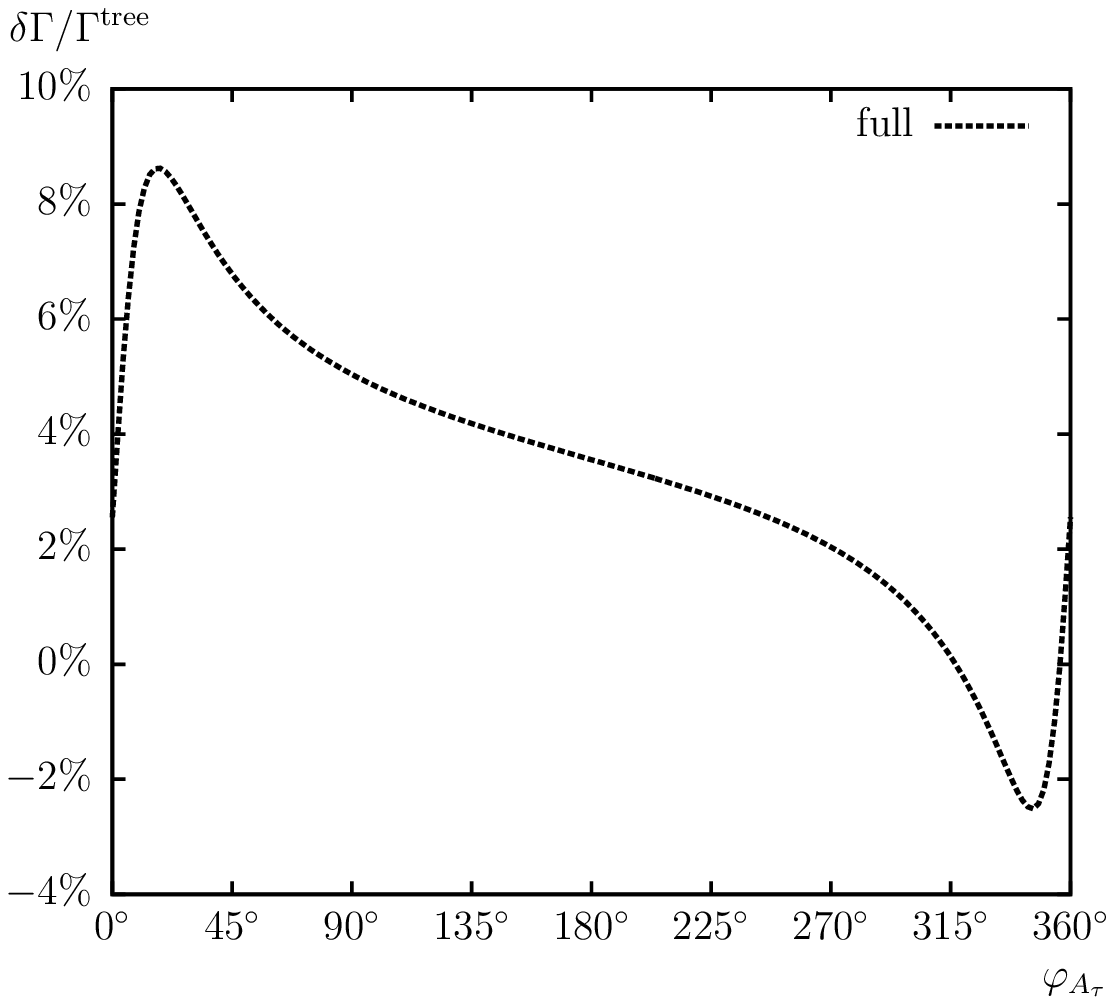}
\\[4em]
\includegraphics[width=0.49\textwidth,height=7.5cm]{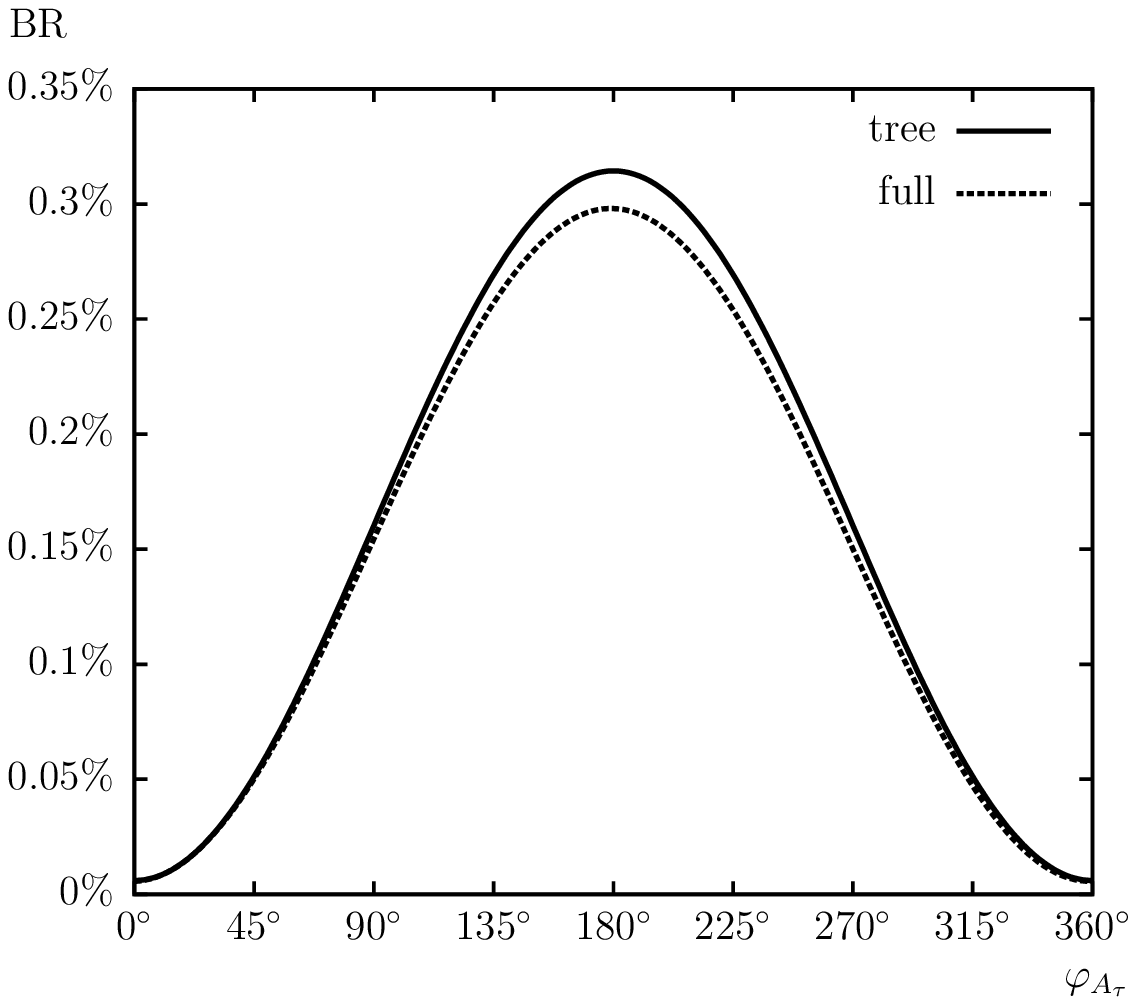}
\hspace{-4mm}
\includegraphics[width=0.49\textwidth,height=7.5cm]{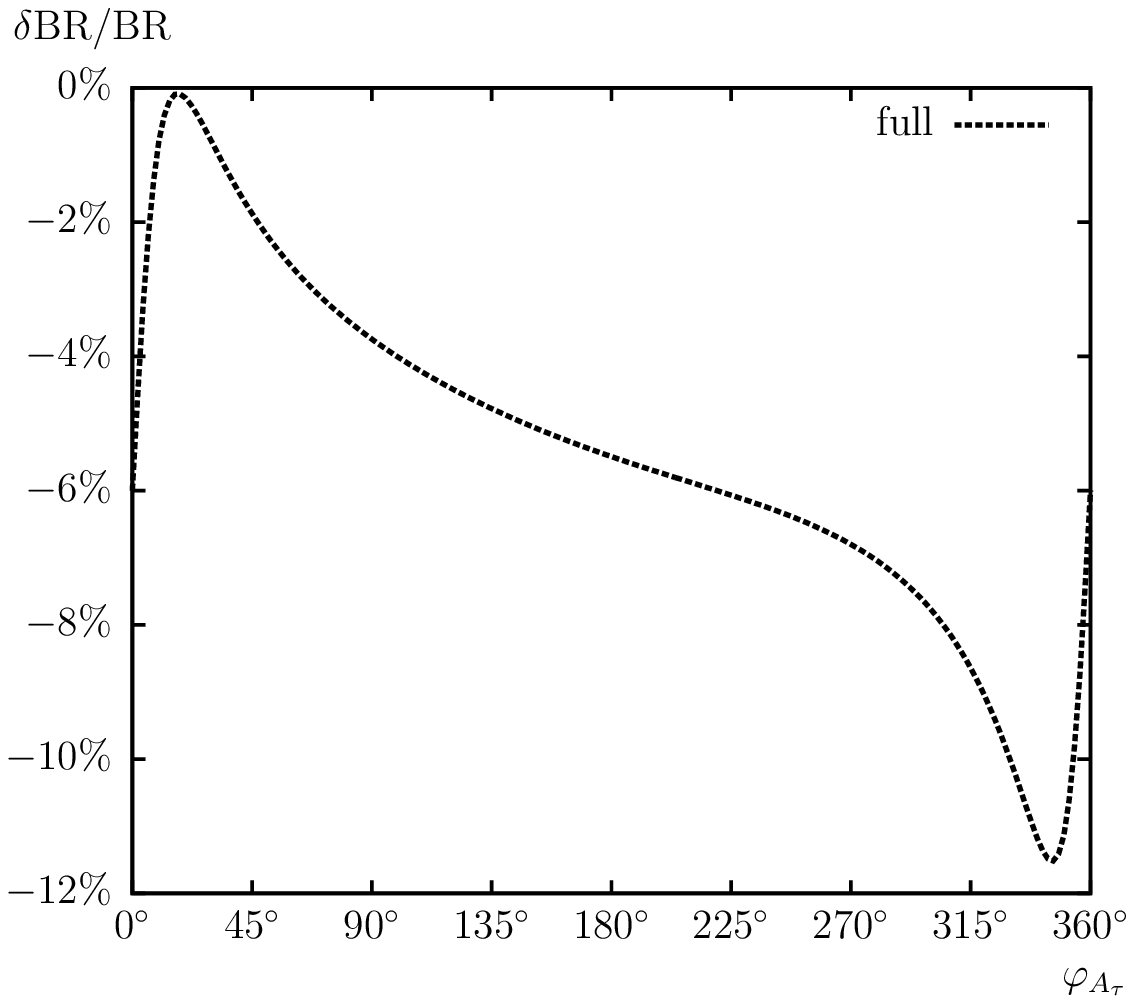}
\end{tabular}
\vspace{2em}
\caption{$\Ga(\decayW)$. 
  Tree-level (``tree'') and full one-loop (``full'') corrected partial decay 
  widths (including absorptive self-energy contributions) are shown.
  The parameters are chosen according to \SE\ (see \refta{tab:para}), 
  with $\phiatau$ varied.
  The upper left plot shows the partial decay width, the upper right plot  
  the corresponding  relative size of the corrections. 
  The lower left plot shows the BR, the lower right plot  
  the relative correction of the BR. 
}
\label{fig:PhiAt.stau2snW}
\end{center}
\end{figure}

The other decay mode involving scalar neutrinos, $\decayW$ is
analyzed in \reffi{fig:PhiAt.stau2snW}. 
The size of the relative correction to the decay width is similar to the
$\decayHm$ channel, varying between $+8\%$ and $-2\%$ with $\phiatau$,
with a corresponding variation in the branching ratio between~0~and
$-11\%$, which is potentially important for physics at the ILC.

\clearpage
\newpage

%%%%%%%%%%%%%%%%%%%%%%%%%% F I G U R E %%%%%%%%%%%%%%%%%%%%%%%%%%%%%%%%%%%%%%%%%
%%%%%%%%%%%%%%%%%%%%%%%%%% F I G U R E %%%%%%%%%%%%%%%%%%%%%%%%%%%%%%%%%%%%%%%%%

\newpage

%%%%%%%%%%%%%%%%%%%%%%%%%% F I G U R E %%%%%%%%%%%%%%%%%%%%%%%%%%%%%%%%%%%%%%%%%
%%%%%%%%%%%%%%%%%%%%%%%%%% F I G U R E %%%%%%%%%%%%%%%%%%%%%%%%%%%%%%%%%%%%%%%%%

\newpage

%%%%%%%%%%%%%%%%%%%%%%%%%% F I G U R E %%%%%%%%%%%%%%%%%%%%%%%%%%%%%%%%%%%%%%%%%
%%%%%%%%%%%%%%%%%%%%%%%%%% F I G U R E %%%%%%%%%%%%%%%%%%%%%%%%%%%%%%%%%%%%%%%%%

\newpage

%%%%%%%%%%%%%%%%%%%%%%%%%% F I G U R E %%%%%%%%%%%%%%%%%%%%%%%%%%%%%%%%%%%%%%%%%
%%%%%%%%%%%%%%%%%%%%%%%%%% F I G U R E %%%%%%%%%%%%%%%%%%%%%%%%%%%%%%%%%%%%%%%%%

\newpage

%%%%%%%%%%%%%%%%%%%%%%%%%% F I G U R E %%%%%%%%%%%%%%%%%%%%%%%%%%%%%%%%%%%%%%%%%
%%%%%%%%%%%%%%%%%%%%%%%%%% F I G U R E %%%%%%%%%%%%%%%%%%%%%%%%%%%%%%%%%%%%%%%%%

\newpage

%%%%%%%%%%%%%%%%%%%%%%%%%% F I G U R E %%%%%%%%%%%%%%%%%%%%%%%%%%%%%%%%%%%%%%%%%
%%%%%%%%%%%%%%%%%%%%%%%%%% F I G U R E %%%%%%%%%%%%%%%%%%%%%%%%%%%%%%%%%%%%%%%%%

\newpage

%%%%%%%%%%%%%%%%%%%%%%%%%% F I G U R E %%%%%%%%%%%%%%%%%%%%%%%%%%%%%%%%%%%%%%%%%
%%%%%%%%%%%%%%%%%%%%%%%%%% F I G U R E %%%%%%%%%%%%%%%%%%%%%%%%%%%%%%%%%%%%%%%%%

\newpage

%%%%%%%%%%%%%%%%%%%%%%%%%% F I G U R E %%%%%%%%%%%%%%%%%%%%%%%%%%%%%%%%%%%%%%%%%
%%%%%%%%%%%%%%%%%%%%%%%%%% F I G U R E %%%%%%%%%%%%%%%%%%%%%%%%%%%%%%%%%%%%%%%%%

\newpage

%%%%%%%%%%%%%%%%%%%%%%%%%% F I G U R E %%%%%%%%%%%%%%%%%%%%%%%%%%%%%%%%%%%%%%%%%
%%%%%%%%%%%%%%%%%%%%%%%%%% F I G U R E %%%%%%%%%%%%%%%%%%%%%%%%%%%%%%%%%%%%%%%%%

\newpage

%%%%%%%%%%%%%%%%%%%%%%%%%% F I G U R E %%%%%%%%%%%%%%%%%%%%%%%%%%%%%%%%%%%%%%%%%
%%%%%%%%%%%%%%%%%%%%%%%%%% F I G U R E %%%%%%%%%%%%%%%%%%%%%%%%%%%%%%%%%%%%%%%%%

\newpage

%%%%%%%%%%%%%%%%%%%%%%%%%% F I G U R E %%%%%%%%%%%%%%%%%%%%%%%%%%%%%%%%%%%%%%%%%
%%%%%%%%%%%%%%%%%%%%%%%%%% F I G U R E %%%%%%%%%%%%%%%%%%%%%%%%%%%%%%%%%%%%%%%%%

\newpage

%%%%%%%%%%%%%%%%%%%%%%%%%% F I G U R E %%%%%%%%%%%%%%%%%%%%%%%%%%%%%%%%%%%%%%%%%
%%%%%%%%%%%%%%%%%%%%%%%%%% F I G U R E %%%%%%%%%%%%%%%%%%%%%%%%%%%%%%%%%%%%%%%%%

\clearpage
\newpage

%%%%%%%%%%%%%%%%%%%%%%%%%%%%%%%%%%%%%%%%%%%%%%%%%%%%%%%%%%%%%%%%%%%%%%%%%%%%%%%

\subsection{The total decay width}
\label{sec:totdecay}

\begin{figure}[htb!]
\begin{center}
\begin{tabular}{c}
\includegraphics[width=0.49\textwidth,height=7.5cm]{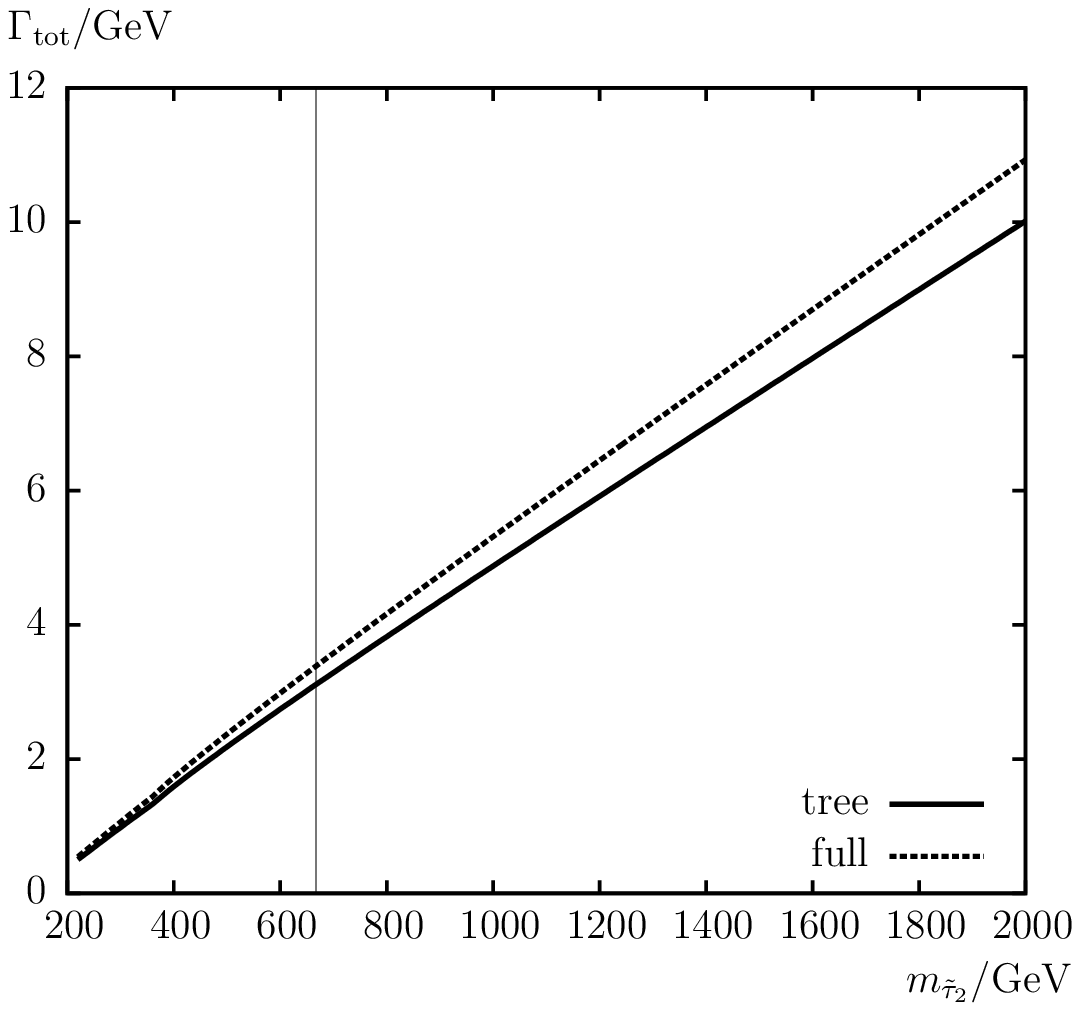}
\hspace{-4mm}
\includegraphics[width=0.49\textwidth,height=7.5cm]{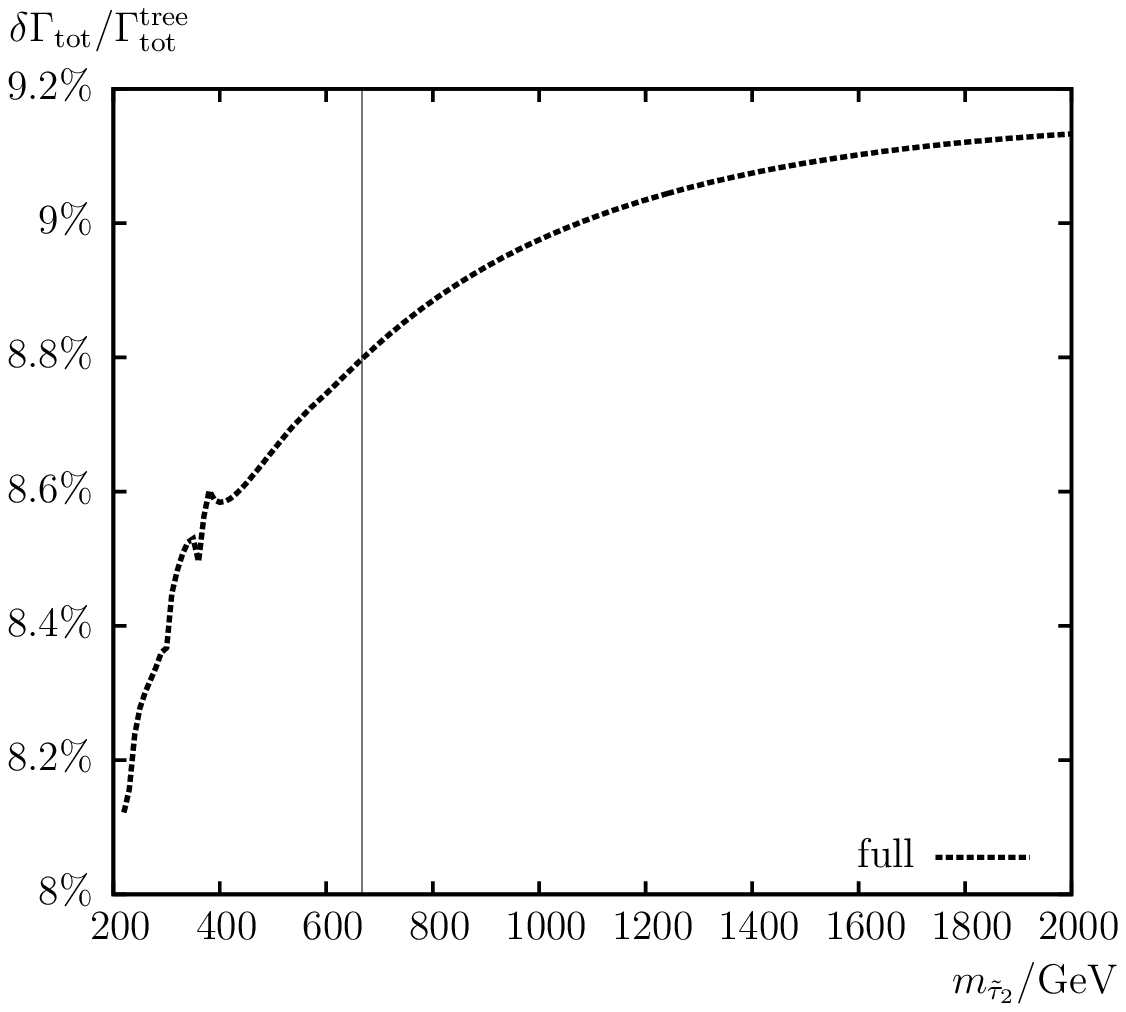}
\\[4em]
\includegraphics[width=0.49\textwidth,height=7.5cm]{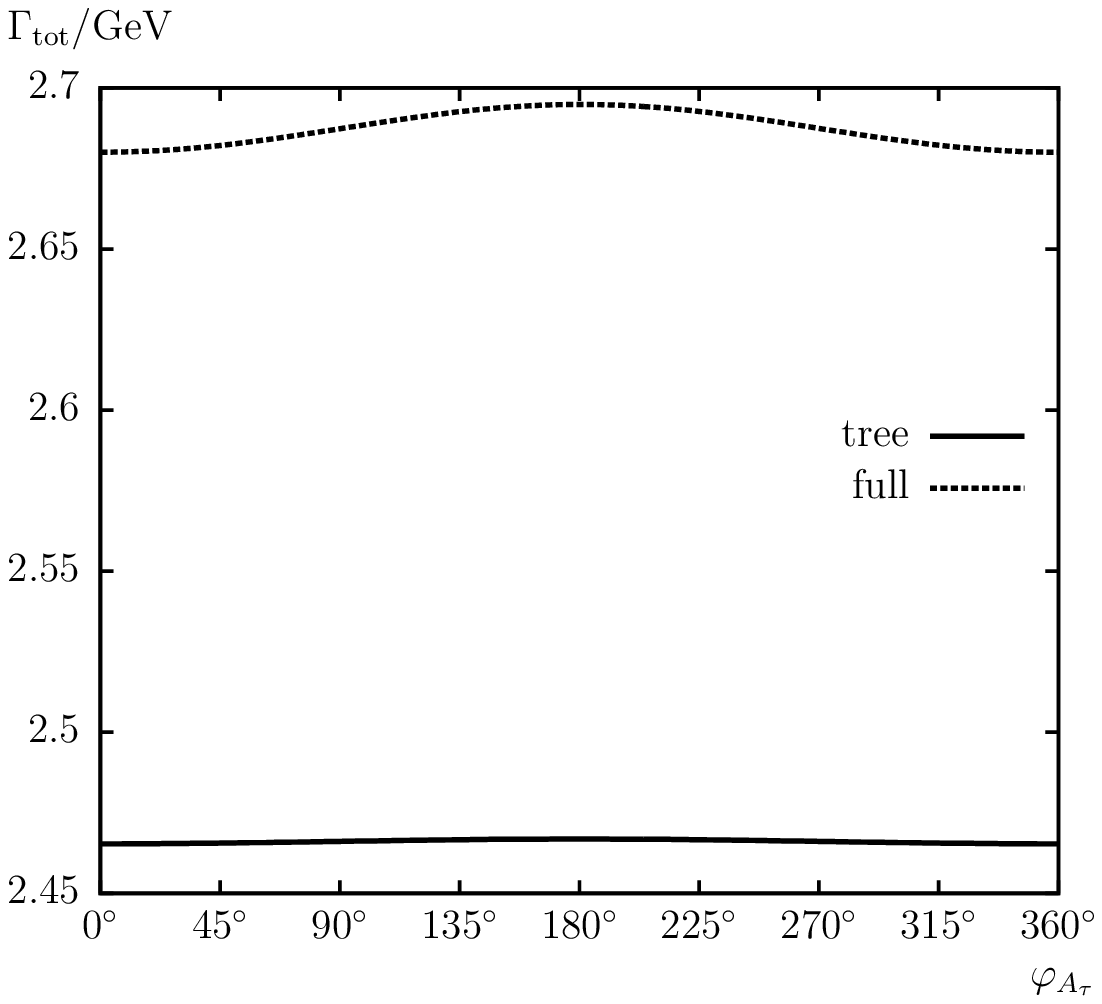}
\hspace{-4mm}
\includegraphics[width=0.49\textwidth,height=7.5cm]{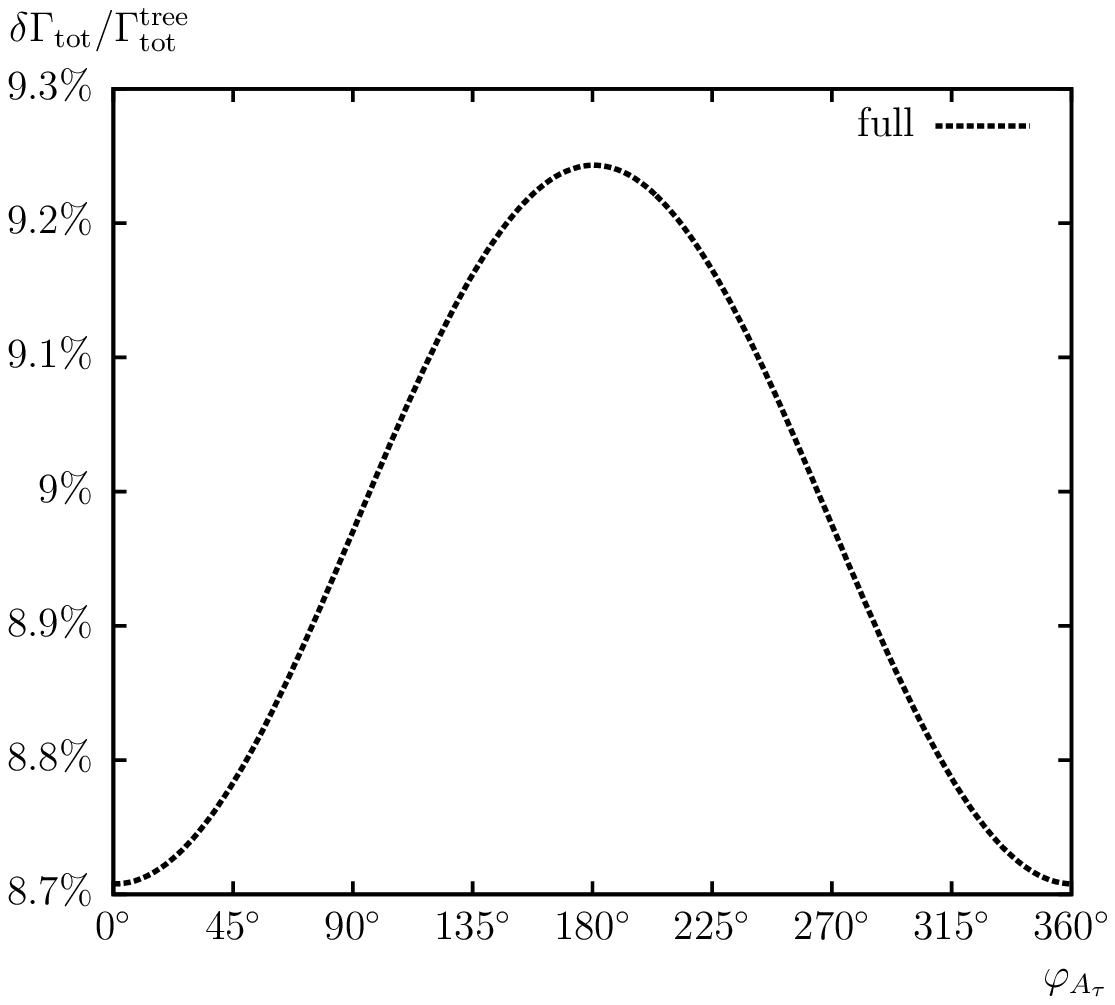}
\end{tabular}
\vspace{2em}
\caption{$\Ga_{\rm tot}$.
  The tree level (``tree'') and full one-loop (``full'') corrected 
  total decay widths shown with the parameters chosen according to 
  \SE\ (see \refta{tab:para}).
  The upper left plot shows the total decay width, the upper right plot 
  the corresponding relative size of the total corrections, with
  $\mstauz$ varied.  
  The vertical lines indicate where $\mstauz + \mstaue = 1000 \gev$, 
  i.e.\ the maximum reach of the ILC(1000).
  The lower plots show the same but with $\phiatau$ varied 
  (including absorptive self-energy contributions).
}
\label{fig:GammaTot}
\end{center}
%\vspace{2em}
\end{figure}

Finally we show the results for the total decay width of $\Stauzm$. In
\reffi{fig:GammaTot} the upper panels show the absolute and relative
variation with $\mstauz$. The lower panels depict the result for varying
$\phiatau$. The dips and peaks visible (best) in the upper right panel
have been described in \refse{sec:full1L}. 
In \SE\ the size of the relative corrections of $\Ga_{\rm tot}$ ranges
between about $+8\%$ close to threshold and goes up to above $+9\%$ for
large values of $\mstauz$. Such an effect should be detectable at the
ILC(1000) or CLIC. 
The variation with $\phiatau$ is found to be small in our numerical 
scenario, due to the dominance of $\Ga(\decayNi{1,2})$, which shows a small 
variation with $\phiatau$, see \refse{sec:full1Lphiat}.
The overall size of the effect of $\phiatau$, shown in the lower row,
are around $+9\%$, again a value that should be detectable at a
future~LC.

%%%%%%%%%%%%%%%%%%%%%%%%%% F I G U R E %%%%%%%%%%%%%%%%%%%%%%%%%%%%%%%%%%%%%%%%
%%%%%%%%%%%%%%%%%%%%%%%%%% F I G U R E %%%%%%%%%%%%%%%%%%%%%%%%%%%%%%%%%%%%%%%%

%%%%%%%%%%%%%%%%%%%%%%%%%%%%%%%%%%%%%%%%%%%%%%%%%%%%%%%%%%%%%%%%%%%%%%%%%%%%%%%
%%%%%%%%%%%%%%%%%%%%%%%%%%%%%%%%%%%%%%%%%%%%%%%%%%%%%%%%%%%%%%%%%%%%%%%%%%%%%%%

\section{Conclusions}
\label{sec:conclusions}

We evaluate all partial decay widths corresponding to a two-body decay
of the heavy scalar tau in 
the Minimal Supersymmetric Standard Model with complex parameters (cMSSM). 
The decay modes are given in \refeqs{ststphi} -- (\ref{stnucha}).
The evaluation is based on a complete one-loop calculation of all 
decay channels, also including soft and hard QED radiation. 
Such a calculation is necessary to derive a reliable prediction of any 
two-body decay branching ratio.
Three-body decay modes can become sizable only if all the two-body decay
channels are kinematically (nearly) closed and have thus been neglected
throughout the paper. 

We first reviewed the one-loop renormalization procedure of the 
$\tau/\Stau$ and $\nu_\tau/\Sneut$ sector (according to the analyses in 
\citeres{SbotRen,LHCxC}) in the cMSSM, which is relevant for our calculation. 
The details for the Higgs boson and chargino/neutralino sector
renormalization can be found in \citere{Stop2decay}.
We have discussed the calculation of the one-loop diagrams, the
treatment of UV- and IR-divergences that are canceled by the inclusion
of soft QED radiation.

Our calculation set-up can easily be extended to other two-body decay 
modes in the cMSSM. 

For the numerical analysis we have chosen a parameter set that allows
simultaneously {\em all} two-body decay modes, i.e.\ {\em not} to
maximize any loop effects.
The masses of the scalar taus in these scenarios are $275$
and $550 \gev$ for the lighter and the heavier stau, respectively. 
The production of colored particles at the LHC lead to the subsequent
cascade decay production of scalar taus at the LHC.
A decay of the heavy stau to a lighter
stau (or sneutrino) and a neutral (or charged) Higgs boson can 
serve as a source of Higgs bosons
at the LHC, thus a precise knowledge of stau branching ratios is 
desirable. The scenario also allows $\aStaue\Stauzm$
production at the ILC(1000) or at CLIC, 
where statistically dominated experimental
measurements of the heavy stau branching ratios will be possible
(depending on the details of the MSSM parameters).
Depending on the integrated luminosity a precision at the few per-cent
level could be achievable.

In our numerical analysis we have shown results for varying $\mstauz$ and
$\phiatau$, the phase of the trilinear coupling~$\Atau$. In the results with
varied $\mstauz$ only the lighter values allow $\aStaue\Stauzm$ 
production at the ILC(1000), whereas the results with varied $\phiatau$ have
sufficiently light scalar taus to permit 
$e^+e^- \to \aStaue\Stauzm$. In the 
numerical scenario we compared the tree-level partial widths with the
one-loop corrected  partial decay widths. In the analysis with $\phiatau$
varied we explicitly included the effect of the absorptive parts of
self-energy type corrections on external legs.
We also analyzed the relative change of the partial decay widths
to demonstrate the size of the loop corrections on each individual
channel. In order to see the effect on the experimentally accessible
quantities we also show the various branching ratios at tree-level (all
channels are evaluated at tree-level) and at the one-loop level (with
all channels evaluated including the full one-loop
contributions). Furthermore we presented the relative change of the BRs
that can directly be compared with the anticipated experimental
accuracy.

We found sizable, roughly \order{5-10\%}, corrections in most of the
channels. For some parts of the parameter space (not only close to
thresholds) also larger corrections up to $15\%$~or even up to~$20\%$
have been observed. 
The size of the full one-loop corrections to the partial decay widths 
and the branching ratios also depends strongly on $\phiatau$. 
The one-loop contributions, again being roughly of \order{5-10\%}, 
often vary by a factor of $2$ as a function of $\phiatau$. 
All results are given in detail in \refses{sec:full1L}
and~\ref{sec:full1Lphiat}.  

The numerical results we have shown are, of course, dependent on the 
choice of the MSSM parameters. Nevertheless, they give an idea of 
the relevance of the full one-loop corrections. 
The largest partial decay widths are $\Ga(\decayNi{1,2})$ in our scenario,  
dominating the total decay width, $\Ga_{\rm tot}$, and thus the various 
branching ratios. 
This is due to the strong bino component in $\neu{1,2}$.
For other choices of $\mu$, $M_1$, $M_2$, e.g.\ $\mu \ll M_{1,2}$, 
the light neutralinos would be higgsino dominated and the decay widths would
turn out to be substantially smaller. Consequently,
corrections to the partial decay widths would stay the same, but the
branching ratios would look very different. 
Decay channels (and their respective one-loop corrections) that may look 
unobservable due to the smallness of their BR in our numerical examples
could become important if other channels are kinematically forbidden.

Following our analysis it is evident that the full one-loop corrections
are mandatory for a precise prediction of the various branching ratios.
The results for the scalar tau decays will be implemented into the
Fortran code {\tt FeynHiggs}.

%%%%%%%%%%%%%%%%%%%%%%%%%%%%%%%%%%%%%%%%%%%%%%%%%%%%%%%%%%%%%%%%%%%%%%%%%%%%%%

\subsection*{Acknowledgements}

We thank
F. Campanario,
T.~Hahn, 
W.~Hollik,
O.~Kittel,
K.~Kovarik,
F.~von~der~Pahlen,
H.~Rzehak
and
G.~Weiglein
for helpful discussions.
We furthermore thank H.~Eberl for assistance with the code 
{\tt SFOLD} and corresponding discussions.
The work of S.H.\ was supported in part by CICYT 
(grant FPA 2010--22163-C02-01) and by the Spanish MICINN's 
Consolider-Ingenio 2010 Program under grant MultiDark CSD2009-00064.

%%%%%%%%%%%%%%%%%%%%%%%%%%%%%%%%%%%%%%%%%%%%%%%%%%%%%%%%%%%%%%%%%%%%%%%%%%%%%%%
%%%%%%%%%%%%%%%%%%%%%%%%%%%%%%%%%%%%%%%%%%%%%%%%%%%%%%%%%%%%%%%%%%%%%%%%%%%%%%%

\newpage
\pagebreak


\begin{thebibliography}{99}
\bibitem{feynarts-mf} The couplings can be found in the files
                      {\tt MSSM.ps.gz} and {\tt HMix.ps.gz} 
                      as part of the \fa~package~\cite{feynarts}.
\end{thebibliography}
\end{document}